\documentclass[fleqn,usenatbib]{mnras}

\usepackage[T1]{fontenc}

\DeclareRobustCommand{\VAN}[3]{#2}
\let\VANthebibliography\thebibliography
\def\thebibliography{\DeclareRobustCommand{\VAN}[3]{##3}\VANthebibliography}

\usepackage{graphicx}	
\usepackage{amsmath}	
\usepackage{amssymb}
\usepackage{orcidlink} 

\newcommand{\beq}{\begin{equation}}
\newcommand{\eeq}{\end{equation}}
\newcommand{\fracp}[2]{\left(\dfrac{#1}{#2}\right)}
\newcommand{\vv}[1]{\bmath{#1} }
\newcommand{\ort}[1]{ \bmath{e}_{#1} }

\newcommand{\vpr}[2]{\bmath{#1} \!\times\! \bmath{#2}}
\newcommand{\vdiv}[1]{\bmath\nabla\!\cdot\!\bmath{#1}}

\newcommand{\mBr}[1]{$\langle \hat{B}_r \rangle$ }

\title[Kink instability of jet-wind outflow]{Magnetic kink instability of jet-wind outflows. I. Static column configuration.}

\author[Kumar \& Komissarov]{
Krishna Kumar\textsuperscript{\orcidlink{0009-0004-0222-788X},1}\thanks{E-mail: K.Kumar@leeds.ac.uk}
Serguei S. Komissarov\textsuperscript{\orcidlink{0000-0003-4545-9774},1}\thanks{E-mail: S.S.Komissarov@leeds.ac.uk}
\\
$^{1}$Department of Applied Mathematics, University of Leeds,
Woodhouse, Leeds, LS2 9JT, United Kingdom\\
}

\date{Accepted XXX. Received YYY; in original form ZZZ}

\pubyear{\the\year}

\begin{document}
\label{firstpage}
\pagerange{\pageref{firstpage}--\pageref{lastpage}}
\maketitle

\begin{abstract}
We describe the results of computer simulations investigating the stability of an axisymmetric static equilibrium combining a highly magnetised ($\sigma=10^3$) central column and a mildly  ($\sigma=0.1$) to weakly ($\sigma=0.001$) magnetised outer shell with a helical and mostly force-free magnetic field of constant pitch.  The measured linear growth rates of the instability are in agreement with the maximum growth rate for a uniform density column with a force-free magnetic field of constant pitch obtained by \citet{Appl-00}, being only a little bit higher. At the non-linear phase, the interface separating the central column and the outer shell develops the characteristic corkscrew shape, with the space inside and between the coils of the corkscrew filled by the shell plasma. In high-resolution runs, a secondary instability is seen as well.  It has the form of helical protuberances of low-$\sigma$ plasma intruding inside the corkscrew coils. These appear to be aligned with the local magnetic field lines. We tentatively identify the secondary instability as the magnetic Rayleigh-Taylor instability.  Overall, the results indicate that the current-driven kink instability may provide the baryon-loading of AGN jets suggested by the observations.  
\end{abstract}




\section{Introduction}
\label{sec: AGN Model}

It is now widely accepted that relativistic jets of Active Galactic Nuclei (AGN) are emerging from supermassive black holes and powered by their rotational energy via the Blandford-Znajek mechanism \citep{10.1093/mnras/179.3.433}.  The key advantage of this model over the accretion disk model is that the disk outflows are heavily-loaded with disk plasma and their terminal speed is only a fraction of the speed of light. This is in conflict with the observations showing that AGN jets can attain Lorentz factors as high as few tens (citations).  In contrast, the outflows emerging from the black hole magnetosphere can have much higher magnetisation than the disk winds and hence may eventually reach much higher speeds. However, the potential terminal speed of the Blandford-Znajek jets is so high that this also creates a problem that needs to be resolved.

The magnetic field of the black hole magnetosphere acts as a shield protecting it from mixing with the plasma of the disk wind. Moreover,  the plasma inside the magnetosphere is either pulled into the black hole when it is sufficiently close to the event horizon, or pushed out by its rotating magnetic field.  This inflow-outflow nature of the plasma dynamics in the black hole magnetosphere implies that it is continuously losing its plasma, which must be replenished somehow. This can be achieved via in situ production of electron-positron pairs. The minimum number density of charged leptons in the magnetosphere, only just sufficient to provide the electric charge density demanded by the electric field of the force-free (FFDE) magnetosphere, is the Goldreich-Julian number density $n_{GJ}=\vdiv{E}/4\pi e$, where $\vv{E}$ is the electric field and $e$ is the electron's electric charge.  The efficiency of pair-production mechanisms is described by the multiplicity factor $\mu =n/n_{\tiny GJ}$, where $n$ is the actual number density of produced leptons.

Only two pair production mechanisms are found relevant for magnetospheres of AGN black holes.  One involves pair cascades in the potential gaps opening in the magnetosphere due to the deficit of electric charges \citep{BIP-92, HO-98}.  The unscreened electric field of the gap accelerates available leptons to energies exceeding the electron rest mass. They scatter the background soft photons produced by the accretion disk via the inverse Compton mechanism, delivering gamma rays capable to produce a pair via the $\gamma$-$\gamma$ interaction with other background soft photons. The produced leptons get accelerated too, and the cycle repeats again and again until the electric field is no longer strong enough to accelerate leptons.  Recent studies show that this process is highly non-stationary \citep[e.g. ][]{Crinq-20}. The multiplicity of produced electron-positron plasma is relatively low, with $\mu<100$ \citep{Levinson-11, Crinq-20, CY-20}, but this could be affected by the limitations of these simulations.         

 The second mechanism involves gamma rays produced outside of the magnetosphere. The mainstream model of this type considers gamma rays emitted by radiatively-inefficient accretion disc \citep{MGDS-11, WRG-21}. Due to their low radiative efficiency, the temperature of electrons near the inner edge can be very high, and they can produce sufficiently energetic gamma rays via multiple inverse Compton scattering.  The pairs in the magnetosphere are produced via the $\gamma$-$\gamma$ interaction of these gamma rays with softer photons emitted by the disk.   The multiplicity of produced pairs strongly depends on the accretion rate and may reach $\mu=100$ \citep{WRG-21}.  
    
Recently, another promising version of the latter mechanism, where the pair-producing gamma rays are radiated during reconnection events in the magnetospheric current sheet (or rather magnetospheric accretion layer),  has emerged following the discovery of such events in ideal GRMD simulations of magnetically-arrested accretion discs \citep{Ripperda22}. Sufficiently large reconnection layers (of the size  $\approx r_g$) may produce plasma with multiplicity ranging from the very large value $\mu=10^{10}$ in their immediate vicinity down to $\mu=10^4$ on the opposite to the reconnection layer side of the black hole   \citep{Ripperda22, KTNH-22, CUD-23, OTK-26}.   The intermittent nature of the observed reconnection events may result in a lower time-averaged value of the multiplicity.            

The predicted multiplicities can be used to estimate the plasma magnetisation $\sigma=B^2/4\pi n m_e c^2$ at the base of AGN jets. In the model of magnetically-arrested accretion, the strength of its poloidal component can be estimated as (see Appendix \ref{app:magnetisation})   
\beq
B_r = 7\times 10^7 \,A_2 \fracp{ \dot{m}}{\eta_{-1}m_{b,9}}^{1/2} \fracp{r}{r_g}^{-2} \mbox{Gauss} \,,
\eeq
where $\dot{m}=\dot{M}/\dot{M}_E$, $\dot{M}_E$ is the Eddington mass accretion rate, $m_b=M_b/M_\odot$, $r_g$ is the gravitational radius, and $A$ is the dimensionless parameter determined via computer simulations \citep{TNM11}  and we apply the common astrophysical use of numerical indices $a_k=a/10^k$.  The corresponding plasma magnetisation is 
 \beq
 \sigma = 10^{12} \frac{A_2}{a\mu_5} \fracp{\dot{m}\,m_{b,9}}{\eta_{-1}}^{1/2} \fracp{r}{r_g}^{-2} 
 \label{eq:sigm-est}
 \eeq 
 (see Appendix \ref{app:magnetisation}). Both estimates were derived assuming a radial poloidal magnetic field, which is suitable for the immediate vicinity of the black hole (up to a few $r_g$).  

Equation \eqref{eq:sigm-est} shows that even for the highest multiplicities predicted by the reconnection model, the plasma magnetisation at the base of the AGN jet exceeds unity by many orders of magnitude.  This creates the conditions for jet acceleration to very large Lorentz factors via the magnetic collimation-acceleration mechanism \citep{Vlahakis_2004,kbvk-07,kvkb-09}, where the Lorentz factor of an axisymmetric jet with a helical magnetic field increases in response to an increase in its collimation.  This is in agreement with the VLBI observations of the best-studied AGN jet in the M87 galaxy, which show a parabolic jet shape and its acceleration up to the distance of 100 pc \citep{JBL-99, Nakamura_2013, Hada-13}.    There  is some observational evidence of simultaneous increase in AGN jet collimation and acceleration up to the de-projected distance of $\approx 100\,$pc in the MOJAVE data \citep{HLK-15}, 

The total energy flux of the jet can be estimated as $L_j=\rho_j\Gamma_j^2 c^3(1+\sigma_j) S_j$, where $\rho_j$, $\Gamma_j$, $\sigma_j$ and $S_j$ are the rest-mass density of jet plasma particles, its Lorentz factor, local magnetisation and cross-section area, respectively (the contribution of plasma thermal energy is ignored for the sake of simplicity. Similarly, the jet rest mass-energy flux is $\dot{M}_j = \rho_j\Gamma_j c^3$. In the absence of additional mass loading, both fluxes are conserved, leading to the result
\beq
\Gamma_j(1+\sigma_j) = \Gamma_{0}(1+\sigma_{0}) \,,
\eeq        
where $\Gamma_0\approx 1$ and $\sigma_0$ are the values at the jet base.  As Poynting flux is converted into the jet kinetic energy, $\sigma_j$ drops, allowing the terminal Lorentz factor to reach values of the order of $\sigma_0$.  In fact, the collimation-acceleration becomes very slow after   $\sigma_j$ reaching unity \citep{Lyubarsky_2009,kbvk-07,kvkb-09}, leading the terminal Lorentz factor  
\beq
\Gamma_t \approx \sigma_0/2\,. 
\label{eq:Gammat}
\eeq  
Using the magnetospheric $\sigma$ given by equation \eqref{eq:sigm-est} as $\sigma_0$, one obtains $\Gamma_t$ that is many orders of magnitude above the observational estimates of the flow Lorentz factor in AGN jets, which give $\Gamma_j\lesssim 50$, clustering around $\Gamma\approx 10$  \citep[e.g. ][]{Lister-09}.   

The magnetic collimation-acceleration is a rather slow process, so one may expect much lower Lorentz factors of the jets on the pc-scales, where VLBI observations reveal an increase of the apparent speed up to the de-projected distance of about $l_j=100\,$pc \citep{HLK-15}. For well-collimated (quasi-cylindrical) Poynting-dominated jets, the Lorentz factor of bulk motion is expected to grow like 
\beq
\Gamma_j = \frac{r_j}{r_{lc}} \,,
\label{eq:Gamma}
\eeq
where $r_{lc}$ is the light cylinder radius of the black hole \citep[e.g.][]{Beskin-10,kvkb-09}. $\Gamma_j\propto r_j$ is also consistent with MOJAVE observations \citep{PKLS-09}. The half-opening angle of  MOJAVE jets is about $\theta_j=1^o$, which corresponds to the jet radius $r_j=\theta_i l_j \approx 1\,$pc.  For a maximally-rotating black hole ($a=1$), $r_{lc}=4r_g=4\times 10^{14}$cm for a black hole of the mass $M_b=10^9M_\odot$.  Substituting these estimates into equation \eqref{eq:Gamma}  yields $\Gamma_j=10^4$.  Though much smaller than the terminal values given by equation  \eqref{eq:Gammat}, this is still well above the values deduced from the observations \citep{Lister-09, PKLS-09}.  A more accurate modelling of Poynting-dominated jets, taking into account the curvature of magnetic field lines, by \citet{LG-M87} yields lower values than those suggested by equation \eqref{eq:Gamma}. Yet the application of this model to the M87 jet still gives unacceptably high $\Gamma\approx 100$ at the location of the HST-1 knot, forcing the authors to conclude the magnetisation of the M87 jet is lower than expected.    

In the numerical investigations of the magnetic collimation-acceleration mechanism by \citep{kbvk-07,kvkb-09}, the initial magnetisation of AGN and GRB jets was assumed to be $\sigma<20$ and $\sigma<1000$ respectively just to ensure that the asymptotic values of $\Gamma$ agree with the observational constraints.   Such low initial values of $\sigma$ imply the existence of additional mass-loading mechanisms which vastly outperform the pair production in the black hole magnetospheres. Several candidates have already been discussed in application to AGN jets. They include a catastrophic pair-cascades in the jets via inverse-Compton scattering of background photons by ultra-relativistic electrons accelerated at shock waves or reconnection cites in the jet while still inside AGN \citep{BL-95, Lev-96, SMMP-97},  mass injection by stellar winds of stars that occupy the same volume as the jets \citep{ssk-94},  turbulent mass entrainment of external gas \citep{bicknell-84}, and episodic mass-loading associated with magnetic flux eruptions of magnetically-arrested discs \citep{CSRA-26}.         

Inside AGN, the jets must be confined and collimated by the magnetised wind emerging from the accretion discs \citep{GVA-09, PF-10, GL-16}.  The self-collimation of non-relativistic magnetised winds via the hoop stress of the rotationally generated azimuthal component of the magnetic field is much more efficient compared to relativistic flows, where the strong electric field largely compensates the hoop-stress action. The planar geometry of the footpoints distribution of the disk's magnetic field also promotes the field-line alignment with the rotational axis. At larger distances, where the wind speed exceeds the fast magnetosonic speed of its plasma, the jet collimation can also be facilitated via shock interaction with the wind \citep{GL-16}.   Disk winds are a common factor of all recent GRMHD simulations of disk accretion onto black holes \citep[e.g. ][]{TNM11,LTQ-20,LMT-22,Chatterjee_2022,Ripperda22,SBD-24}. These winds can produce a strong torque on the accretion disk and extract a substantial fraction of its rotational energy.  In the section originating close to the black hole, they can reach mildly-relativistic speeds $v_w\approx 0.2-0.5c$, similar to the speeds of ultra-fast outflows detected in some AGN \citep[see ][and references therein]{TCR-10, TTM-14}. Their magnetisation is low, $\sigma \ll 1$, and their main mass carriers are protons.   It has long been speculated that the interaction between the disk wind and the Poynting-dominated BZ jet from the black hole ultimately determines the properties of the jet-like outflows emerging from AGN \citep[e.g. ][]{BL-95}. 

Such interaction may be driven by fluid instabilities, of which the current-driven (CD) instabilities seem to be most important, given the high-strength wind's and jet's magnetic field. The CD kink mode is considered particularly promising, given the strong restructuring of the flow it leads to at the nonlinear phase. The helical geometry of the perturbation alone already suggests the possibility of interpenetration of the BZ jet and fast inner disk wind, leading to a slower resultant outflow.  The great potential role of current-driven kink instability for converting Poynting flux into kinetic energy, via triggering magnetic reconnection and dissipation, was pointed out by \citet{GS-06}, who used a simplified one-dimensional steady-state model to show the possibility of full conversion within $1000 r_g$ from the black hole.  This is an attractive alternative to the collimation-acceleration mechanism as it incorporates magnetic dissipation as a source of energy for the observed jet emission.  However, to achieve the terminal Lorentz factors consistent with the observations, this model must either assume low initial magnetisation of the jet (like this is done in \citet{kbvk-07}), or include some prescription for the jet mass-loading.  It is quite possible that this mass-loading is also triggered by the kink instability.        

The kink instability of magnetised jets has been already a subject of extensive theoretical analysis \citep[e.g.][]{AC-92, IP-94, IP-96, Begelman-98, Lyubarski-99, Appl-00, Narayan_2009, 10.1093/mnras/stx807} and computer simulations \citep[e.g.][]{Mizuno_2009, Mizuno_2011, MLNH-12, mizuno2014, OBB-12, anjiri2014, BT-16, TB-16, Singh_2016, BSDP-19, bodo2020, 10.1093/mnras/stab3492, MMM-24}.   However, none of these studies included the magnetised wind of the accretion disk in their models. 

To fill this gap, we initiated a program of 3D computer simulations where we plan to explore the interaction between Poynting-dominated jets and kinetic-energy-dominated magnetised wind driven by the kink instability. Here we report the first outcome of this program, where we focus on a simpler problem involving a static high-$\sigma$ cylinder (corresponding to the BZ jet) surrounded by a much denser low-$\sigma$ static cylindrical shell (corresponding to the wind from the accretion disk). The simulations show that at its nonlinear phase, the kink-mode instability leads to significant interpenetration of the high-$\sigma$ and low-$\sigma$ plasmas. Moreover, it triggers a secondary instability of Rayleigh-Taylor type, which drives strong mixing between the two components.  

Section \ref{sec:Model} describes the initial equilibrium configuration. Section \ref{sec:setup} describes the numerical method and the setup of computer simulations. Their results are presented in section \ref{sec:results}.  The limitations of this study and topics of future investigations are outlined in section \ref{discuss}.  Appendix \ref{app:magnetisation} describes basic results concerning the mass-loading of black hole magnetospheres via pair production. Appendix \ref{sec:converge} describes the dependence of solutions on the numerical resolution and the size of the computational domain.  

In the rest of the paper, we employ the Lorentz-Heaviside, resulting in the disappearance of the speed of light and the geometric factor $4\pi$.       

\section{Equilibrium solution}
\label{sec:Model}

\begin{figure}
    \centering
    \includegraphics[width=0.85\linewidth]{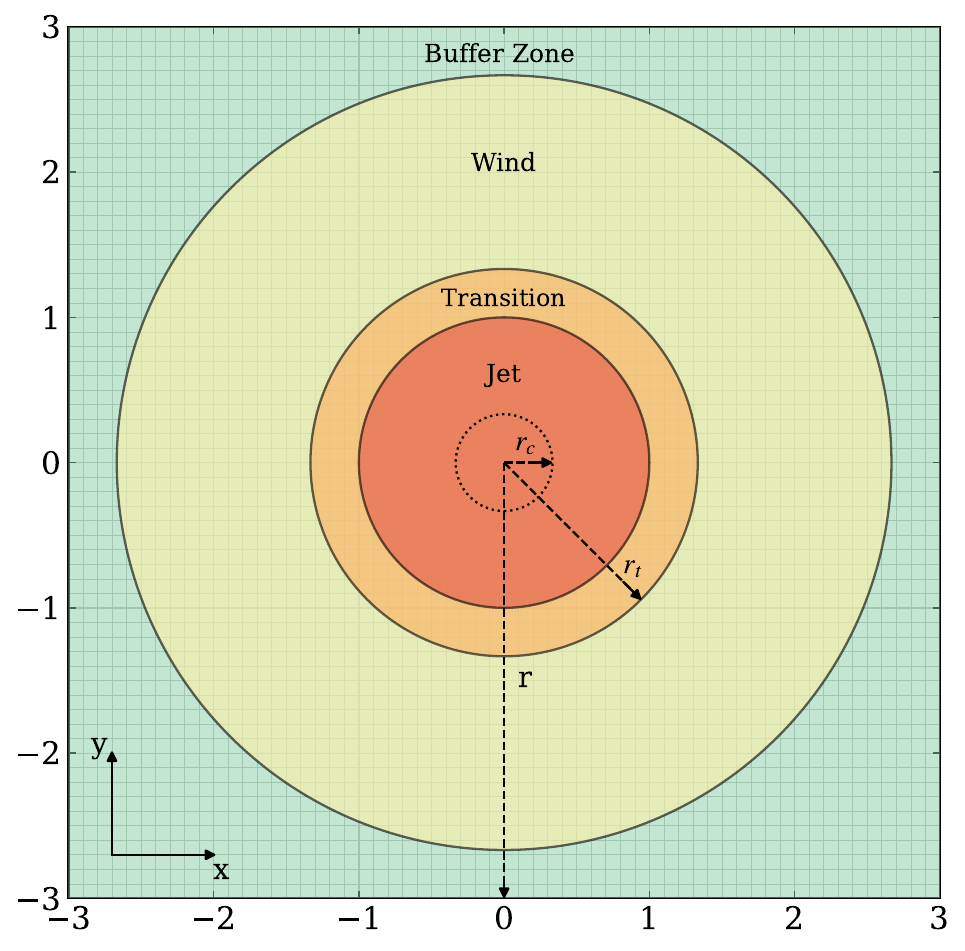}
    \caption{Schematic structure of the equilibrium model showing the high-$\sigma$ inner cylinder (red) and low-$\sigma$ outer shell (yellow), corresponding to the BZ jet and disk wind, respectively. Also shown are the smooth transition layer (orange) between the two components and the outer buffer zone (green) introduced for numerical reasons.}
    \label{fig:model}
\end{figure}

The axisymmetric equilibrium solution used in this study describes a high-$\sigma$ inner cylinder inside a low-$\sigma$ cylindrical shell (see figure \ref{fig:model}).  In the rest of the paper, for the sake of brevity,  we will refer to these two components as jet and wind.   In cylindrical coordinates centred on the symmetry axis, it satisfies the force balance equation 
\begin{equation}
\frac{d}{dr}(p+p_m) + \frac{{{B}_\phi}^2}{r} = 0,
\label{eq:balance}
\end{equation}
where $p$ is the gas pressure and $p_m=(B_z^2+B_\phi^2)/2$ is the magnetic pressure, and $B_r=0$. In the jet, the magnetic field is force-free and hence  
\begin{equation}
\frac{d\bar{p}_m}{dr} + \frac{{\bar{B}_\phi}^2}{r} = 0\,.
\label{eq:ff-balance}
\end{equation}
In this study, we employ the helical force-free magnetic field 
\begin{align}
\bar{B}_z &= B_0 \left(1 + \fracp{r}{r_c}^2\right)^{-1} , \label{eq:ff-Bz}\\
\bar{B}_\phi &= \bar{B}_z \fracp{r}{r_c} ,
\label{eq:ff-Bphi}
\end{align}
with constant pitch parameter $P = r \bar{B}_z/\bar{B}_\phi=r_c$ \citep{Appl-00}.  We will refer to $r_c$ as the jet core radius. At this radius, the azimuthal component of the magnetic field reaches its maximum value.  According to the Kruskal-Shafranov criterion \citep{Shafranov1956, Kruskal1958, Bateman1978},  this solution is unstable to kink modes with wavelength  $ \lambda > 2\pi r_c$.

\begin{table}
\centering
\caption{Parameters of the simulations with mildly magnetised outer shell ($\sigma_w=0.1$). $r_c$ is the magnetic core radius, which equals to the uniform pitch of magnetic field lines; $L_x$, $L_y$, and $L_z$ are the lengths of the computational domain in the x, y, and z directions, respectively; $n_x$, $n_y$, and $n_z$ are the numbers of grid points in these directions;  $L_z/2\pi r_c$ is the discriminant of the kink mode instability. According to the Kruskal-Shafranov criterion, it has to be above unity for instability. }
\label{tab1}
\begin{tabular}{lccccr}
\hline
Model & $r_c$ &  $L_x\!\times\! L_y \!\times\! L_z$ &  $n_x\!\times\! n_y\!\times\! n_z$ & $L_z/2\pi r_c$ \\
\hline
$M_1$ & $1/6$ & $6 \times 6 \times 3$ & $540 \times 540 \times 270$ & 2.86\\
$M_2$ & $1/4$  & $6 \times 6 \times 3$ & $540 \times 540 \times 270$ & 1.91\\
$M_3$ & $1/3$  & $6 \times 6 \times 3$ & $540 \times 540 \times 270$ & 1.43\\
$M_4$ & $1/2$  & $6 \times 6 \times 4$ & $540 \times 540 \times 360$ & 1.27\\
\hline
\end{tabular}
\end{table}

\begin{figure*}
\centering
\includegraphics[width=0.325\textwidth]{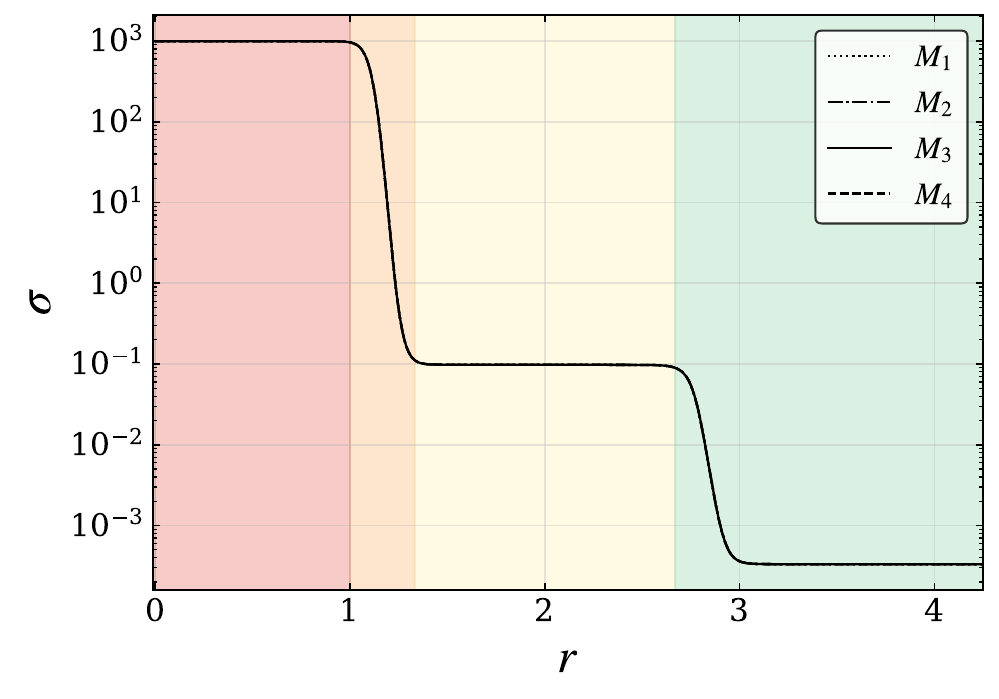}
\includegraphics[width=0.325\textwidth]{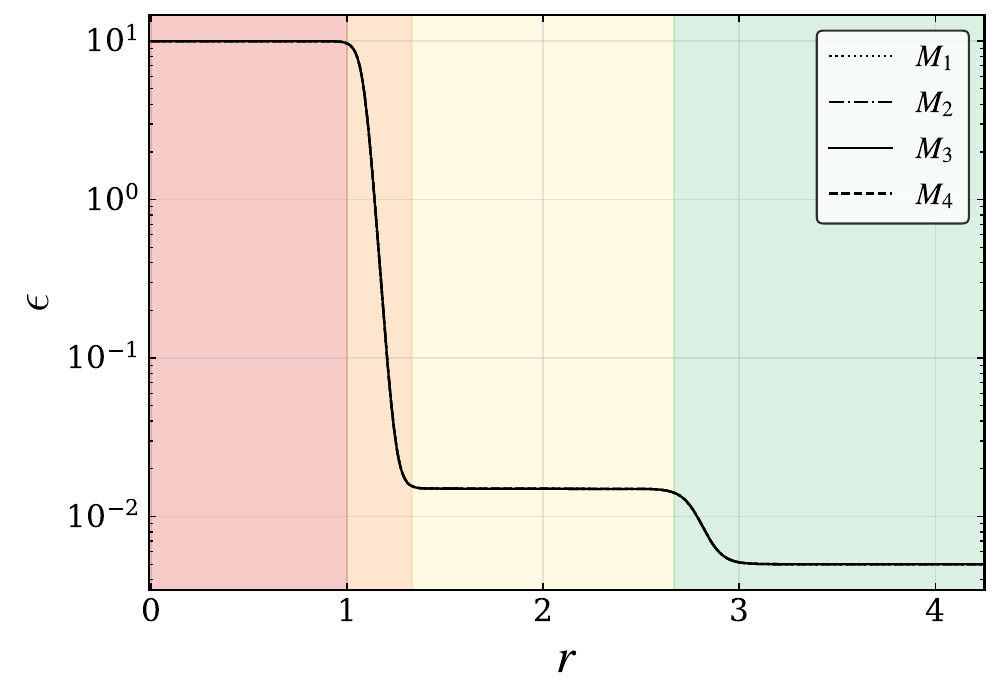}
\includegraphics[width=0.325\textwidth]{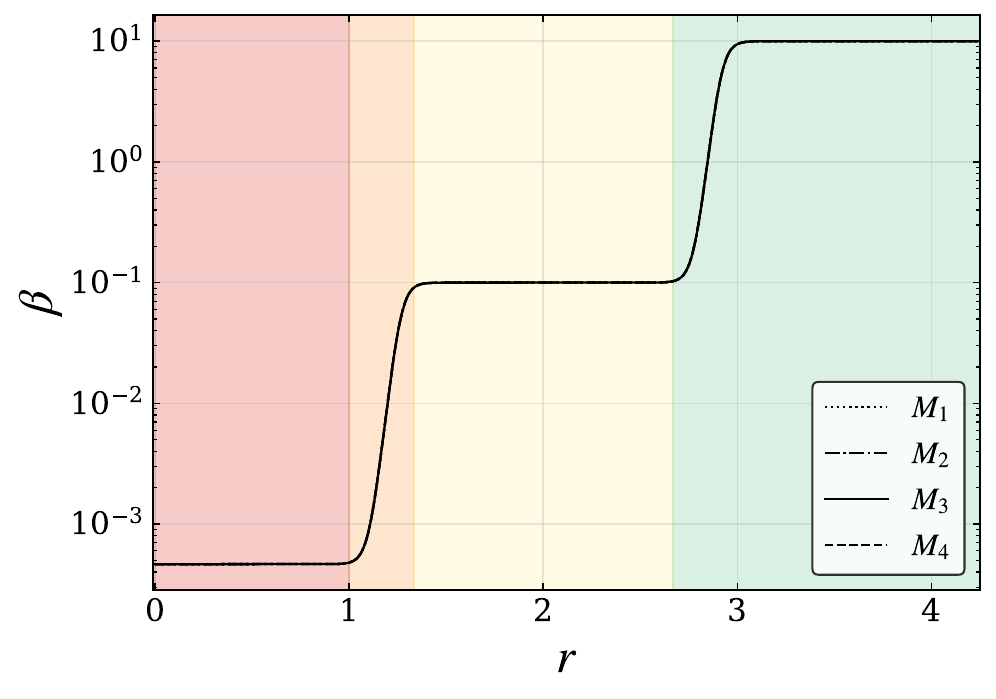}
\includegraphics[width=0.325\textwidth]{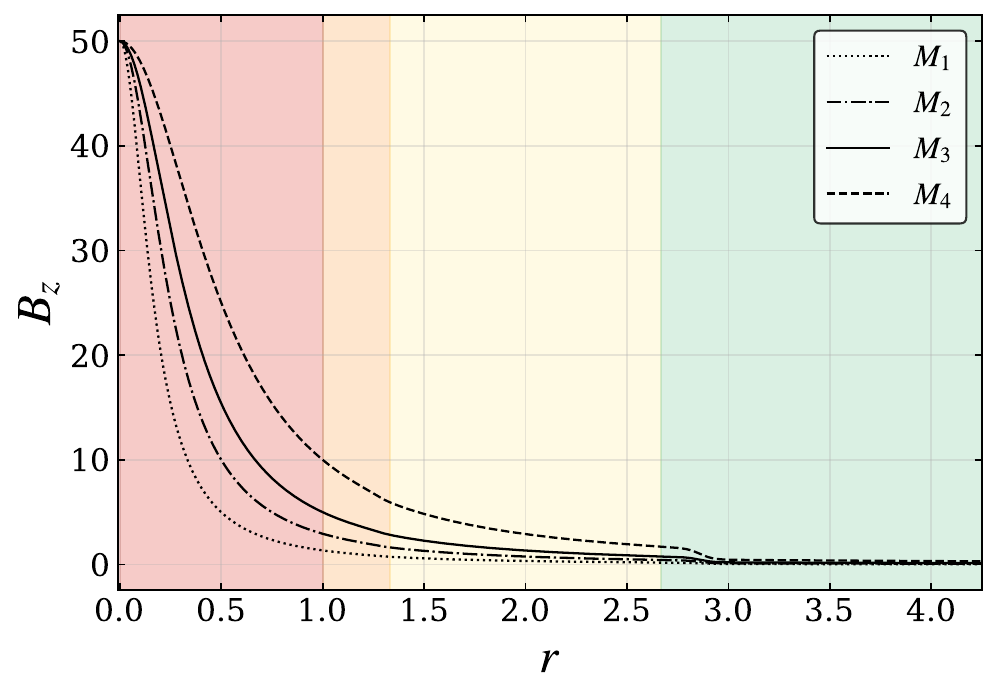}
\includegraphics[width=0.325\textwidth]{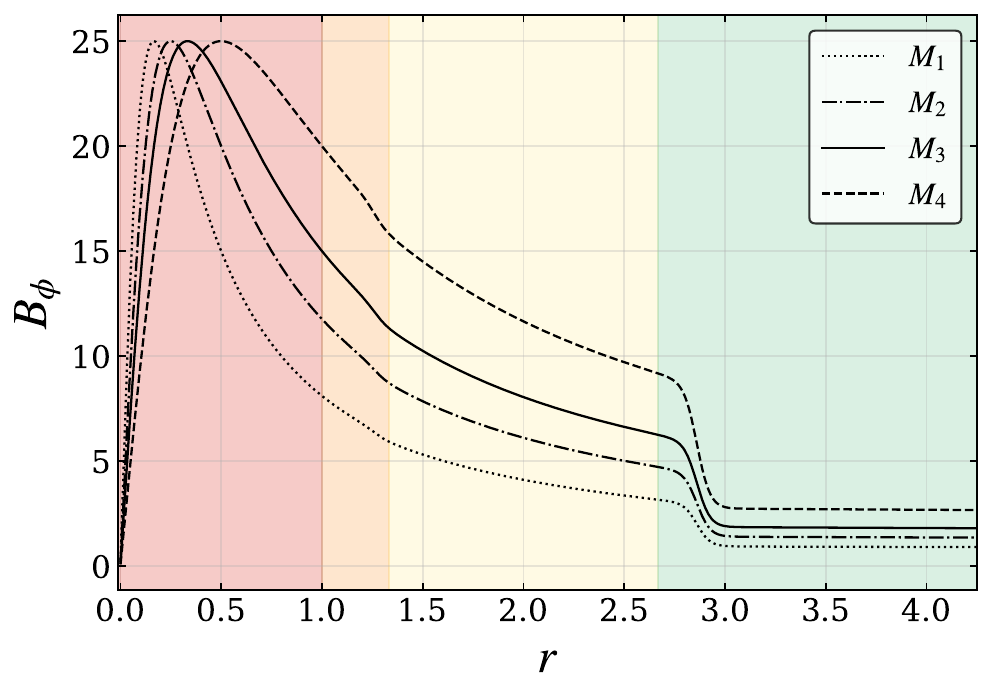}
\includegraphics[width=0.325\textwidth]{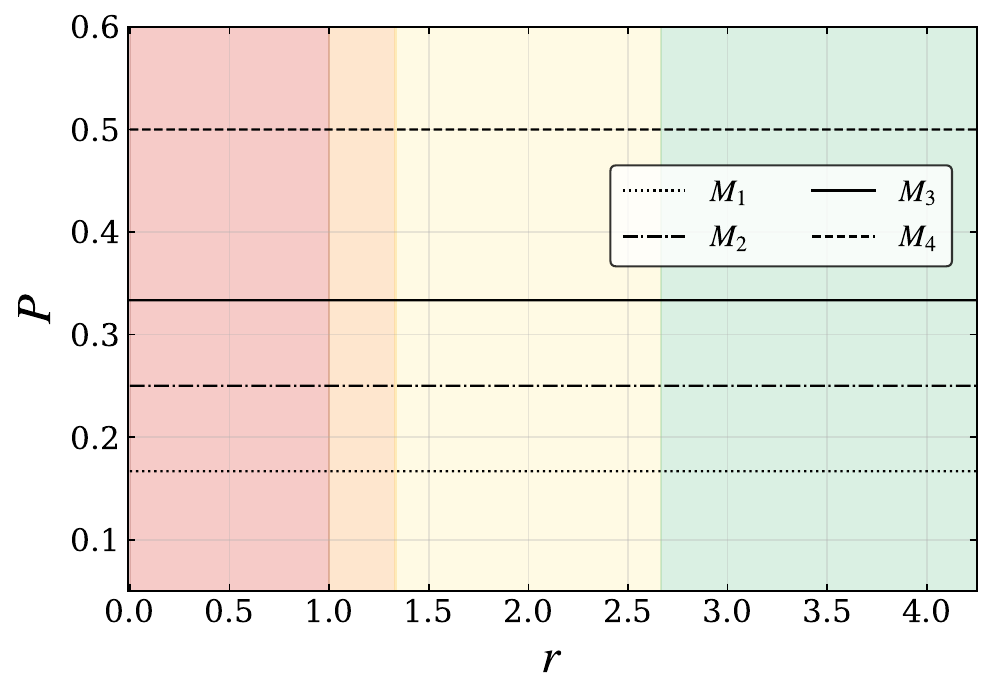}
\includegraphics[width=0.325\textwidth]{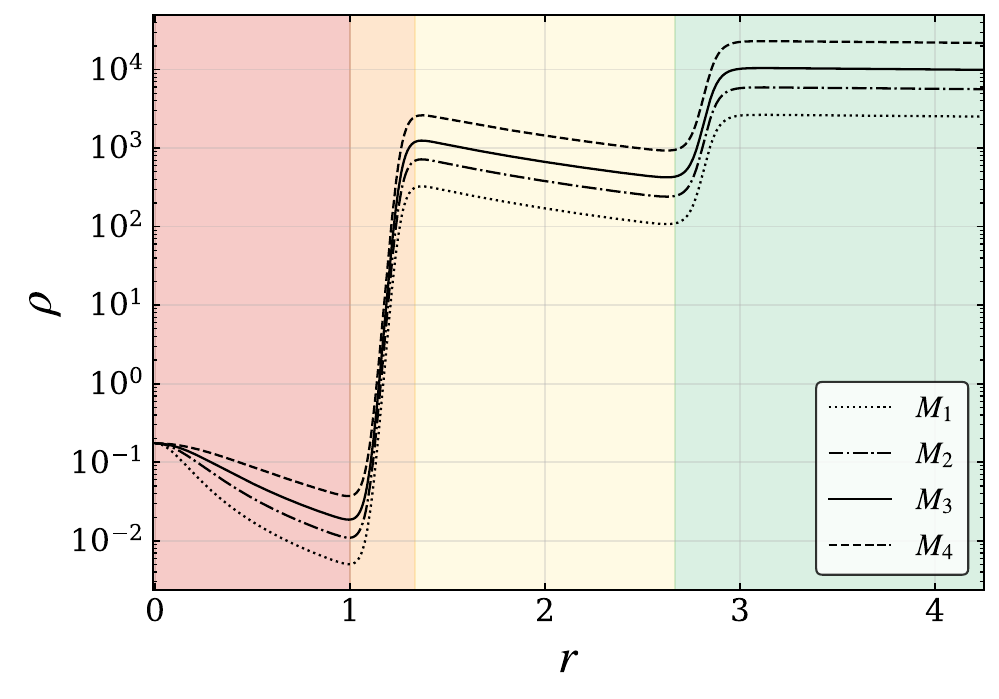}
\includegraphics[width=0.325\textwidth]{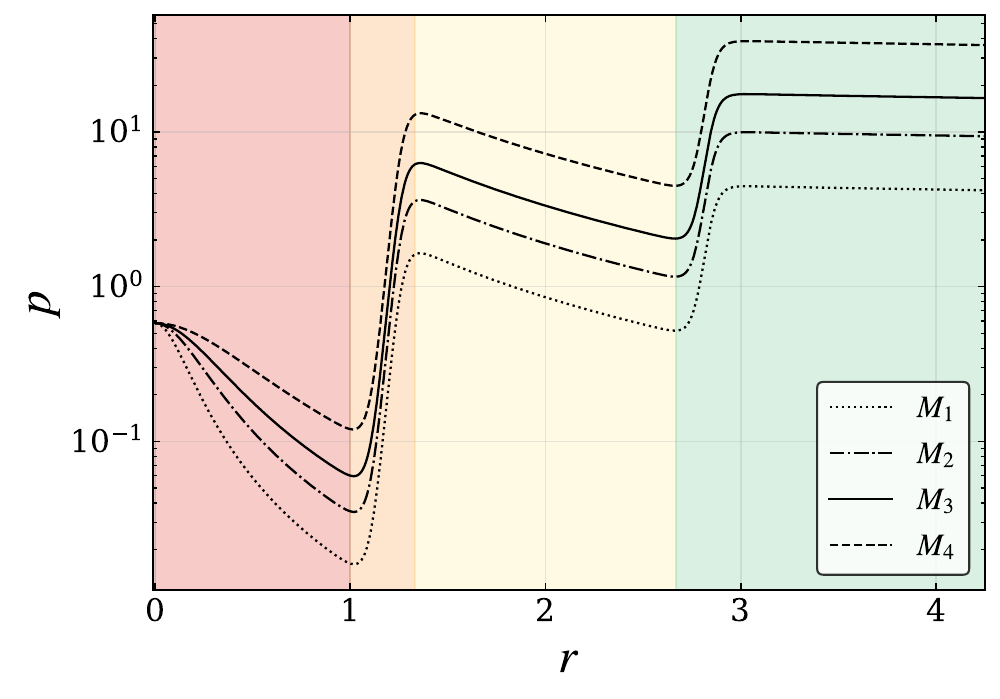}
\includegraphics[width=0.325\textwidth]{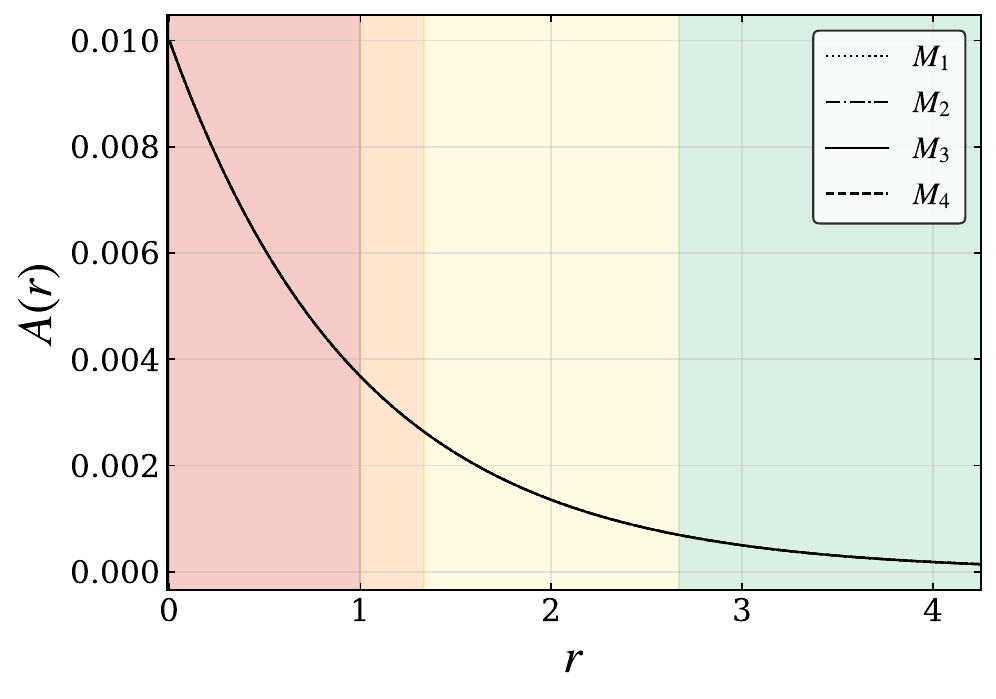}
\caption{Initial equilibrium solutions. The plot shows the distributions of the axial $B_z$ and azimuthal $B_\phi$ components of the magnetic field, rest-mass plasma density $\rho$, thermal pressure $p$, temperature parameter $\epsilon=p/(\gamma-1)\rho$, plasma magnetisation $\beta=p/p_m$, plasma magnetisation $\sigma=B^2/w$, magnetic pitch $P$, and the amplitude of the initial velocity perturbation $A(r)$. The dotted, dashed-dotted, solid, and dashed lines represent the models $M_1$, $M_2$, $M_3$,  and $M_4$, respectively.  The background colouring shows the structural components of the solution as in figure \ref{fig:model}.}
\label{fig:equilibrium}
\end{figure*}

In the wind, the gas pressure is no longer negligible, and so using a uniform pressure distribution for the whole domain is not reasonable. As a result, the force-free magnetic field is no longer justified. This is especially true for the transition layer between the jet and wind where a significant gradient of gas pressure is expected. To describe the deviation from the force-free magnetic field, we introduce the rescaling factor $f(r)$, such that  
\begin{equation}
B_r = 0, B_z = \sqrt{f} \bar{B}_z, B_\phi = \sqrt{f} \bar{B}_\phi \,.
\label{eq4a}
\end{equation}
This way the magnetic field retains the pitch of the force-free field (\ref{eq:ff-Bz}-\ref{eq:ff-Bphi}). Substituting these into the balance equation \eqref{eq:balance}, one obtains the relation between the scaling factor and the gas pressure 
\begin{equation}
    \frac{dp}{dr} = - \bar{p}_m \frac{df}{dr},
    \label{eq:p-f}
\end{equation}
where $ \bar{p}_m$ is the magnetic pressure of the force-free field. In magnetised plasma, the relative dynamical importance of gas pressure is quantified via the magnetisation parameter $\beta=p/p_m$. In terms of $\beta(r)$, equation \eqref{eq:p-f} reads   
\begin{equation}
 [(1 + \beta)\bar{p}_m ]\frac{df}{dr} + f \frac{d }{dr}\left(\beta \bar{p}_m\right) = 0.
\label{eq:beta-f}
\end{equation}
Given the function $\beta(r)$, this equation can be easily integrated to find the required scaling function $f(r)$ and then $p(r)=\beta(r) p_m(r)$. 

To set a meaningful density distribution, it is convenient to start with prescribing the gas temperature parameter $\epsilon=p/(\gamma-1)\rho$ as a function of $r$, and then to compute $\rho(r)$ based on the distributions $p(r)$ and $\epsilon(r)$. The relativistic magnetisation parameter $\sigma$ is related to $\beta$ and $\epsilon$ as 
\beq
\sigma=\frac{2(\gamma-1)}{\beta(\gamma+\epsilon^{-1})}\,,
\label{eq:sbe}
\eeq
where $\gamma$ is the ratio of specific heats. 

The distributions of $\beta(r)$ and $\epsilon(r)$ in the jet and wind are chosen to be uniform and connected via a tanh transition layer, e.g. 
 \begin{equation}
\beta(r) = 
\frac12 (\beta_{\rm w}-\beta_{\rm j})
        \tanh\!\left(\frac{r-r_t}{\Delta r_t}\right)
        + \frac12 (\beta_{\rm w}+\beta_{\rm j})  \,,
\label{eq:transition}
\end{equation}
where $r_t$ is the location of the layer, $\Delta r_t$ is its characteristic thickness, and the suffixes $j$ and $w$  refer to the jet and wind, respectively.  

\begin{figure*}
    \centering
    \includegraphics[width=0.325\linewidth]{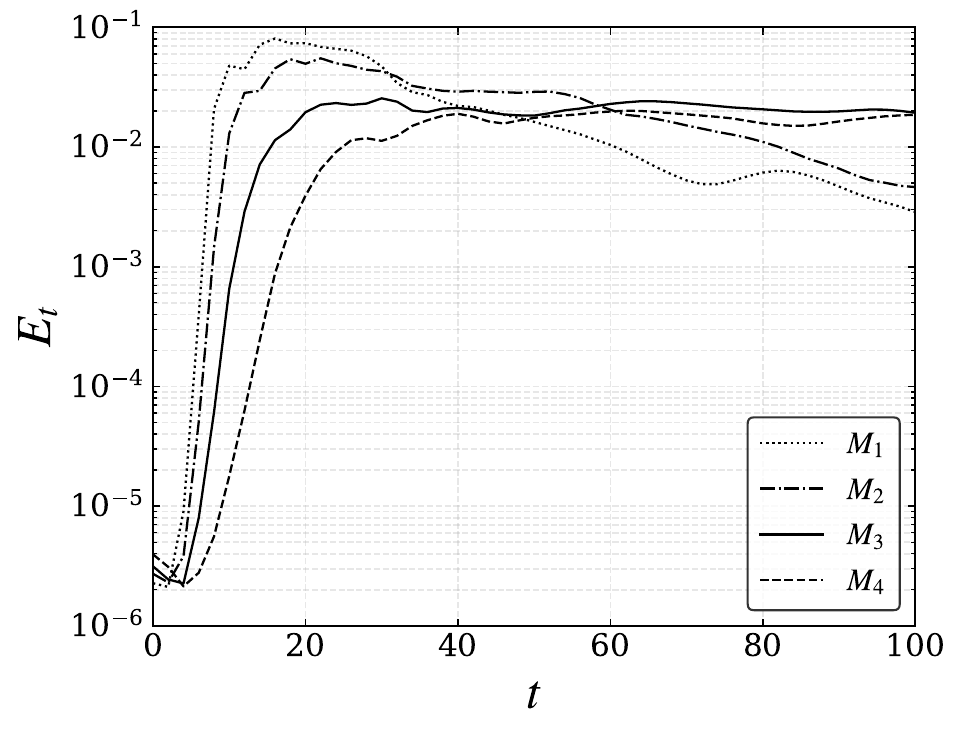}
    \includegraphics[width=0.325\linewidth]{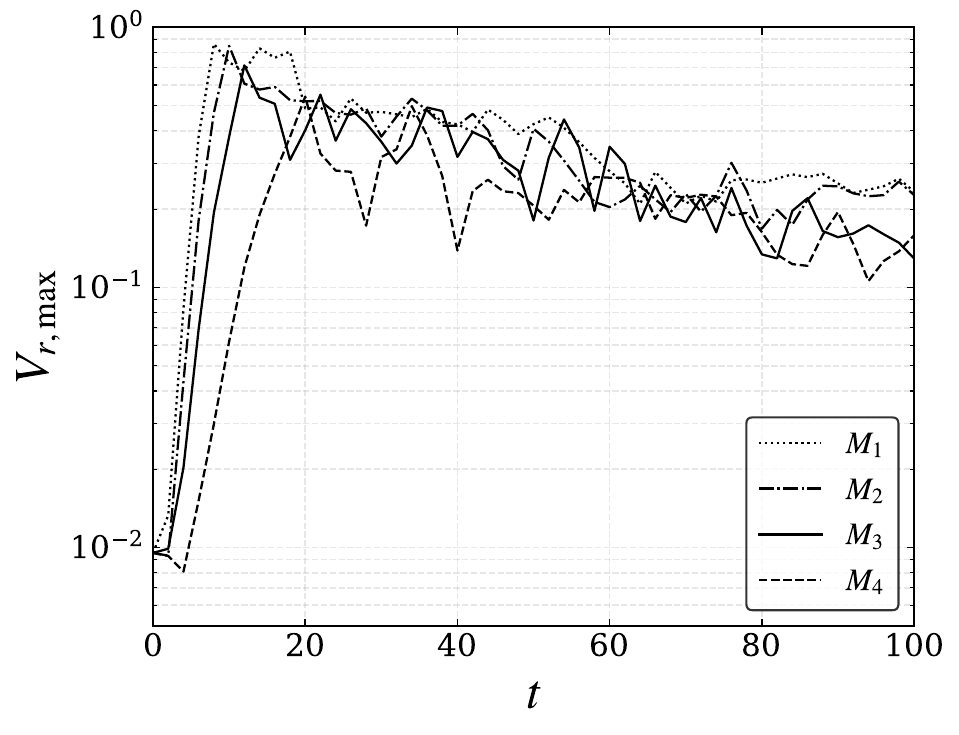}
     \includegraphics[width=0.325\linewidth]{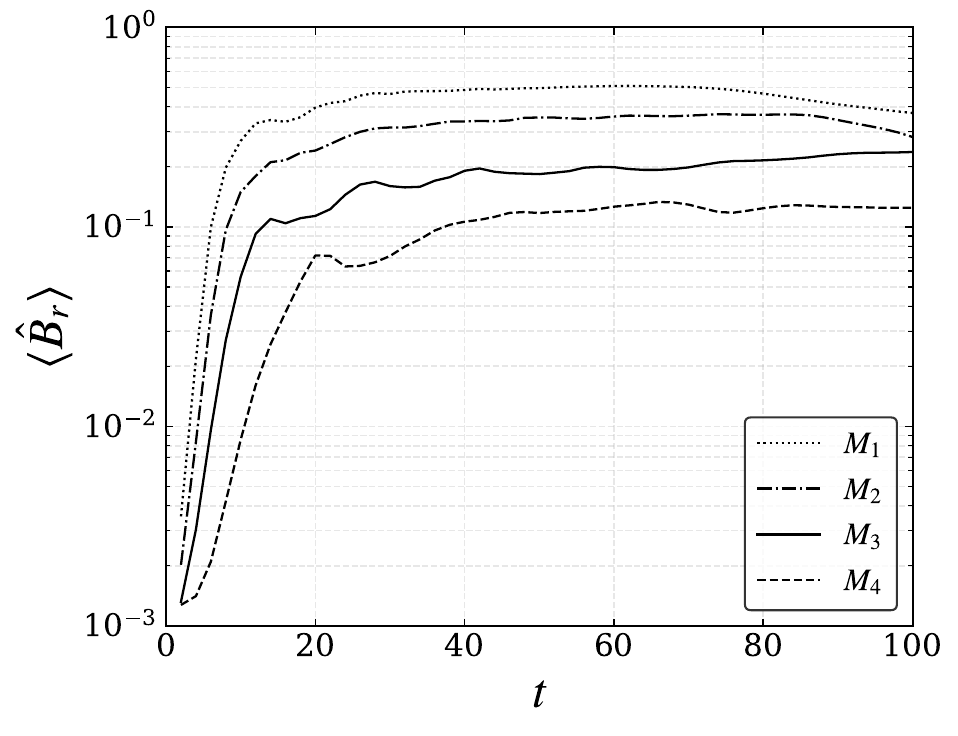}     
 \caption{Temporal evolution of the $E_t, V_{r, max}$ and $\langle \hat{B}_r \rangle$ for the entire jet-wind region. The dotted, dashed dotted, solid, and dashed lines represent models $M_1, M_2, M_3$, and $M_4$,  respectively.}
    \label{fig:growth}
\end{figure*}

\begin{figure*}
    \centering 
    \includegraphics[width=0.325\linewidth]{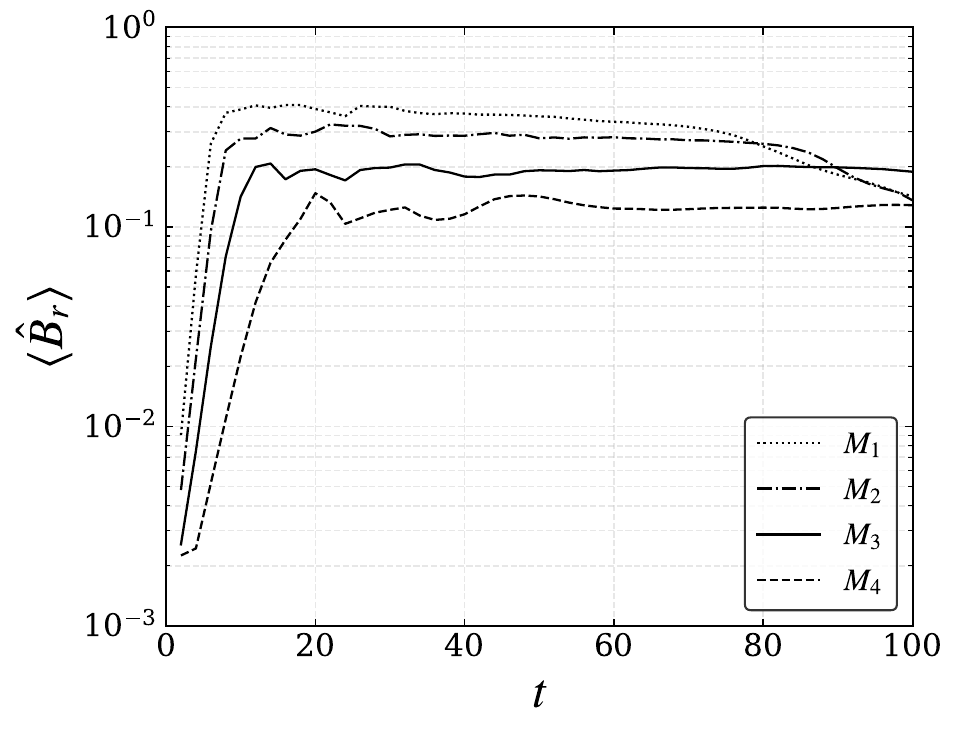}
    \includegraphics[width=0.325\linewidth]{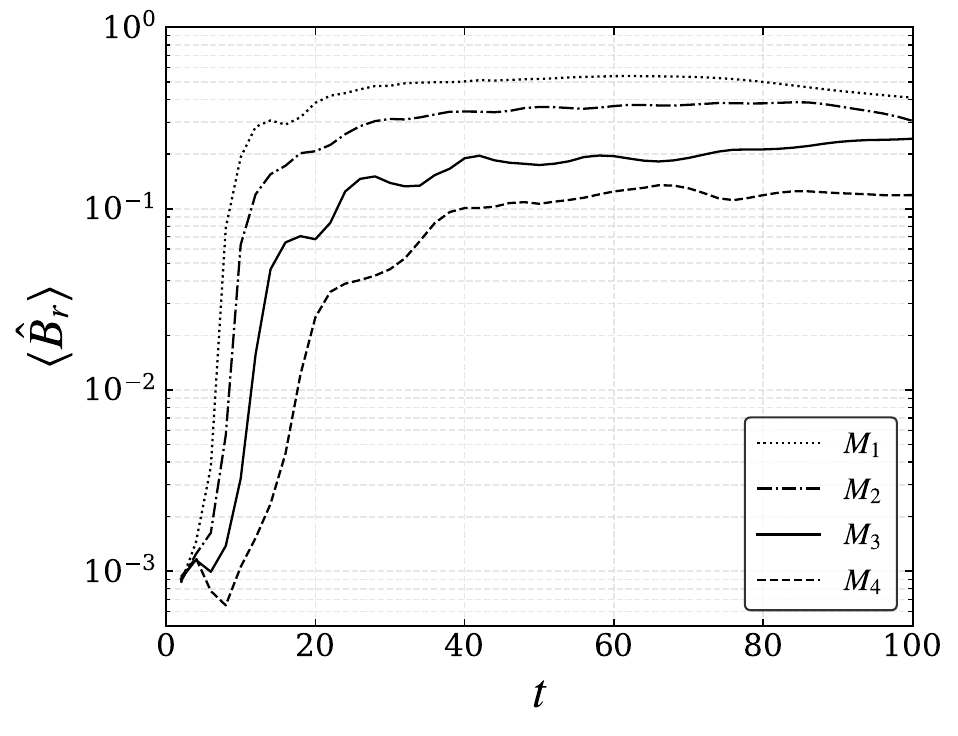}
   \includegraphics[width=0.325\linewidth]{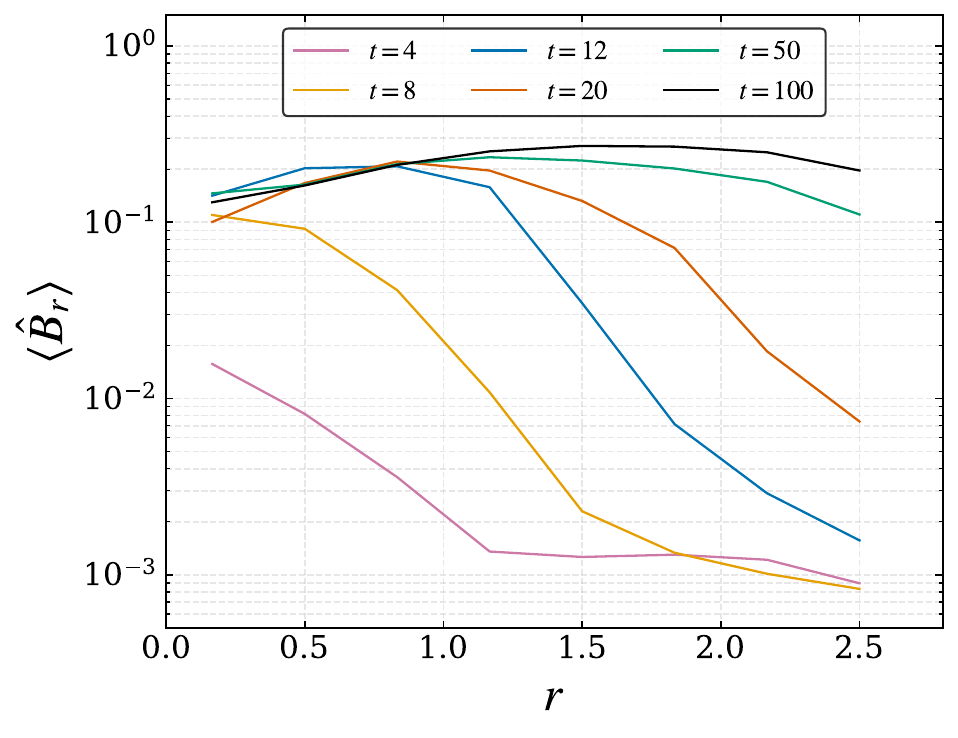}  
    \caption{{\it Left panel:}   $\langle \hat{B}_r \rangle(t)$ for the jet region. {\it Middle panel:}   $\langle \hat{B}_r \rangle(t)$ for the wind region. {\it Right panel:}  Temporal evolution of $\langle \hat{B}_r \rangle(r)$ in the model $M_3$.}
    \label{fig:Br-growth}
\end{figure*}

\begin{table}
\centering
\caption{Growth rate of the kink instability based on the $\langle \hat{B}_r \rangle$ parameter in the jet, wind, and the combined jet-wind regions. The last column shows the maximum growth rate predicted by equation \eqref{eq:G-appl}.}
\label{tab:2}
\begin{tabular}{lcccc}
\hline
Model  &
jet $\Gamma_g$ &
wind $\Gamma_g$ &
total $\Gamma_g$ & 
max $\Gamma_g$ \\
\hline
$M_1$ & $0.84 \pm 0.05$ & $0.67 \pm 0.09$ & $0.68 \pm 0.09$ & 0.80\\
$M_2$ & $0.66 \pm 0.05$ & $0.51 \pm 0.07$ & $0.55 \pm 0.06$ & 0.53\\
$M_3$ & $0.51 \pm 0.03$ & $0.41 \pm 0.05$ & $0.44 \pm 0.03$ & 0.40\\
$M_4$ & $0.34 \pm 0.01$ & $0.28 \pm 0.03$ & $0.29 \pm 0.01$ & 0.27\\
\hline
\end{tabular}
\end{table}

\section{SIMULATION SETUP}
\label{sec:setup}
The simulations were performed using the novel  3D RMHD code PERT, which utilised the splitting numerical scheme of \citet{Phillips_2024}. It is parallelised using MPI, which allows it to run on CPU clusters. Its memory distribution model is adopted from the JANUS code \citet{10.1093/mnras/stt2247}. The computational grid is uniform Cartesian, with the z-axis aligned with the symmetry axis of the equilibrium solution.  For most of the simulations, the computational domain is $(-3,3)$ in the x and y directions and $(0,3)$ in the z direction, with $540\times540\times270$ grid points. A few support simulations were carried out with lower resolution. 

For the jet, the selected  parameters are $\sigma_j=10^3$ and $\epsilon_j=10$, for which equation \eqref{eq:sbe} yeilds $\beta_j\approx 5\times 10^{-4}$. For the mildly magnetised wind, we set $\sigma_w=0.1$ and $\beta_w=0.1$, for which equation \eqref{eq:sbe} yeilds $\epsilon_w\approx1.5\times 10^{-2}$. The values of $\sigma_w$ and $\beta_w$ are aligned with the wind parameters in the GRMHD simulations of MAD accretion disks by \citet{ARW-24}. They are derived from the figures 4 and 6 in this paper and are characteristic of the wind near the disk only. However, our main purpose here is just to combine a high-$\sigma$ jet with a low-$\sigma$ wind, and so the exact values of these parameters are not particularly important for the purpose.  

For the weakly magnetised wind, $\epsilon_w=1$ and $\sigma_w=10^{-3}$. The corresponding value of the magnetisation parameter $\beta$ is $3\times10^2$. 

The radius of the transition layer $r_t=1.15$ and its thickness $\Delta r=0.05$. 
The amplitude parameter of the magnetic field $B_0=50$. The equation of state describes an ideal gas with the ratio of specific heats $\gamma = 4/3$, both for the wind and jet plasmas.  

At the z-boundaries,  we employ periodic boundary conditions. The x and y-boundaries are a bit problematic because the Cartesian box does not accommodate the cylindrical symmetry of the problem. To mitigate this, the boundary region with $r>2.7$ is filled with cold $\epsilon \approx 5\times10^{-3}$ and dense ($\sigma\approx 10^{-3}$) plasma. This creates a buffer which prevents the highly magnetised plasma from crossing the x and y boundaries, where the zero-gradient boundary conditions are used\footnote{The solid-wall BCs would probably be more appropriate there.}.    
 
Following \citet{Mizuno_2009}, the equilibrium is perturbed via introducing the radial velocity 
\begin{equation}
v_r = A(r)
\cos(m\phi) \sin\left( kz \right),
\label{eq:perturbation}
\end{equation}
where $A(r) =  10^{-2} \exp(-r)$, $m=1$, and $k=2\pi/L_z$.  Note that the periodic boundary conditions allow only the wavelengths $\lambda_n=L_z/n$, where $n\in{\cal N}$. With the exception of the model $M_1$, for which $n=2$ mode can also grow, in all other models only $n=1$ modes are unstable.

\begin{figure*}
    \centering
    \includegraphics[width=0.24\textwidth,trim=0cm 0cm 0cm 2cm,clip]{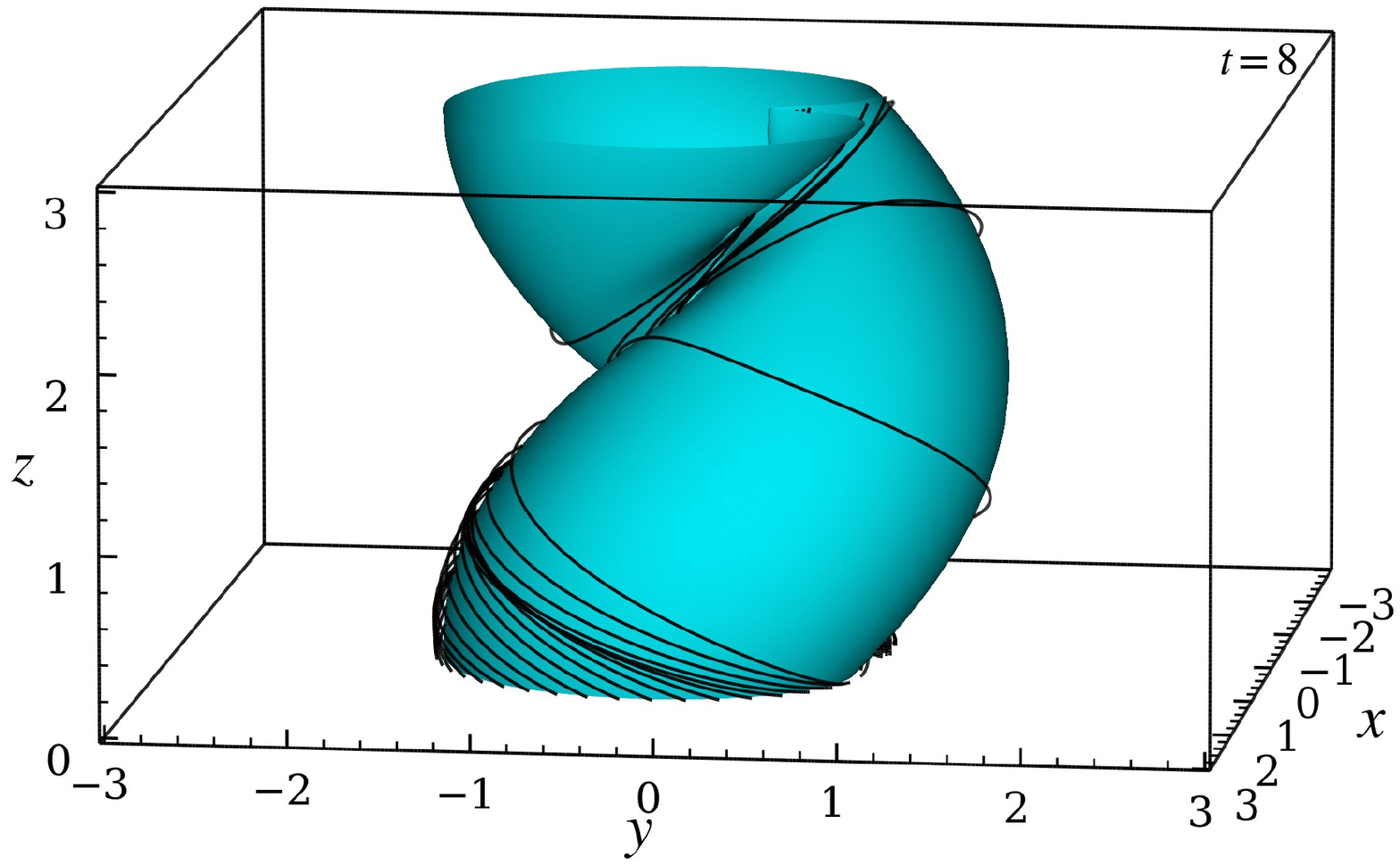}
    \includegraphics[width=0.24\textwidth,trim=0cm 0cm 0cm 2cm,clip]{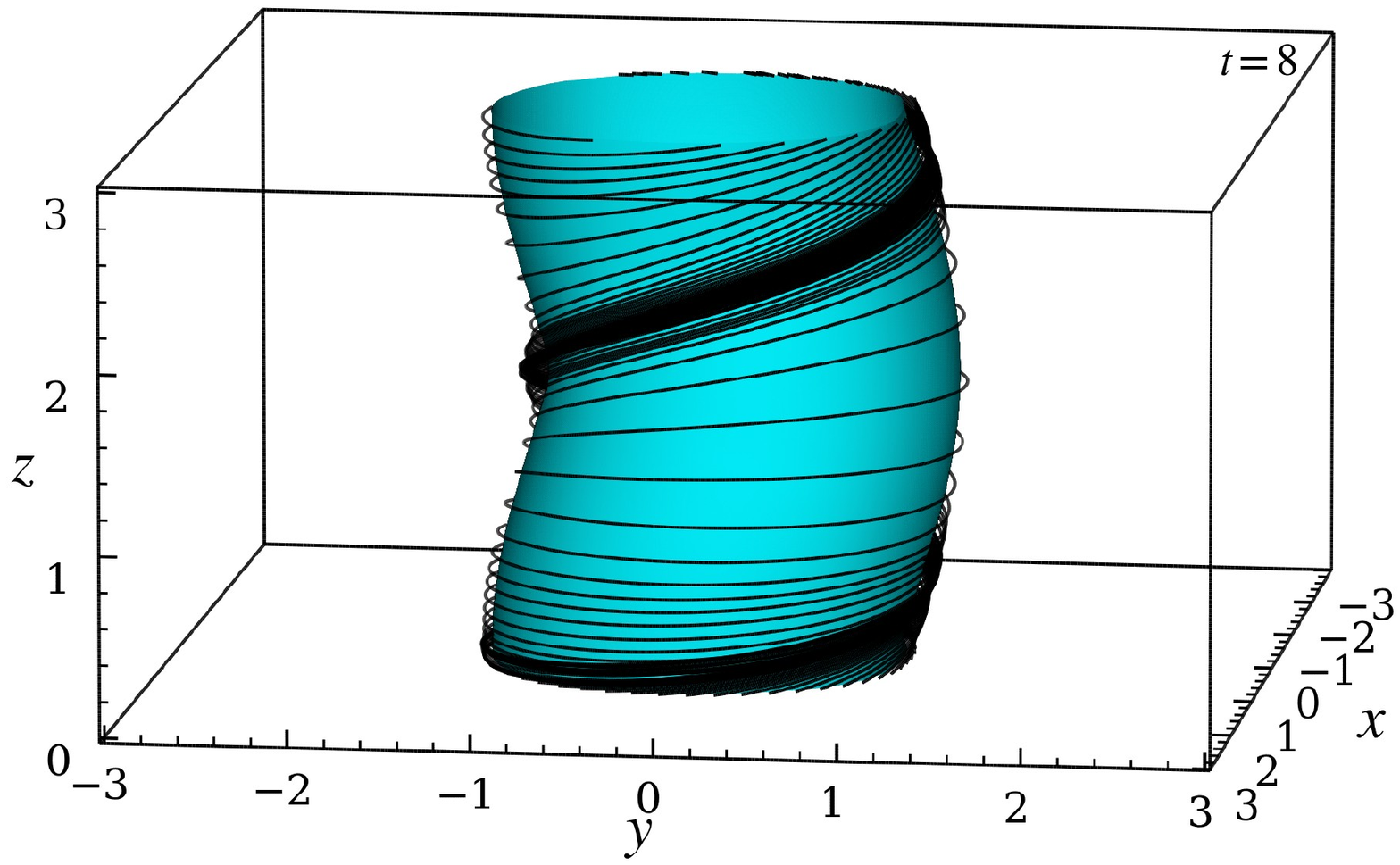}
    \includegraphics[width=0.24\textwidth,trim=0cm 0cm 0cm 2cm,clip]{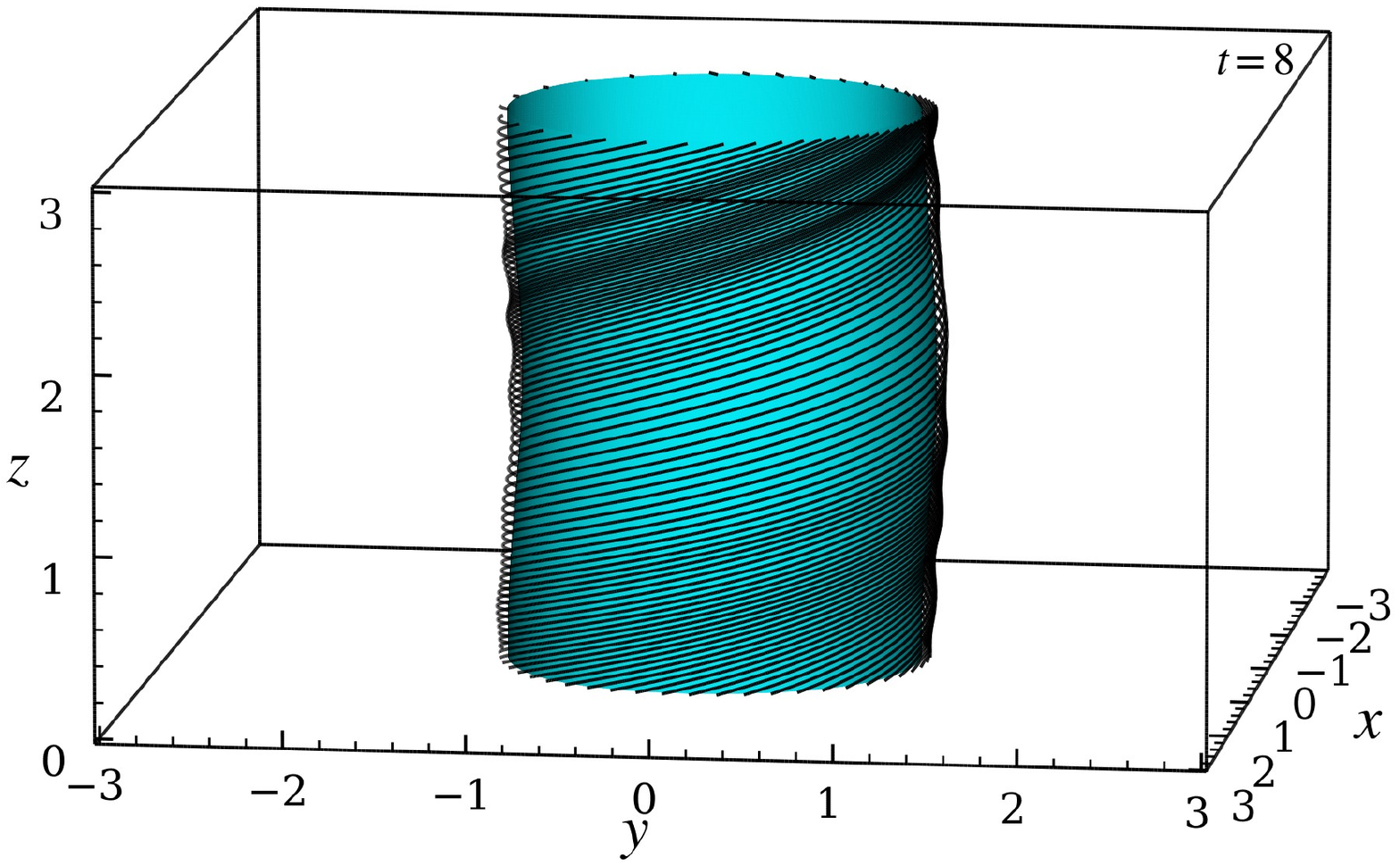}
    \includegraphics[width=0.24\textwidth,trim=0cm 0.75cm 0cm 1.0cm,clip]{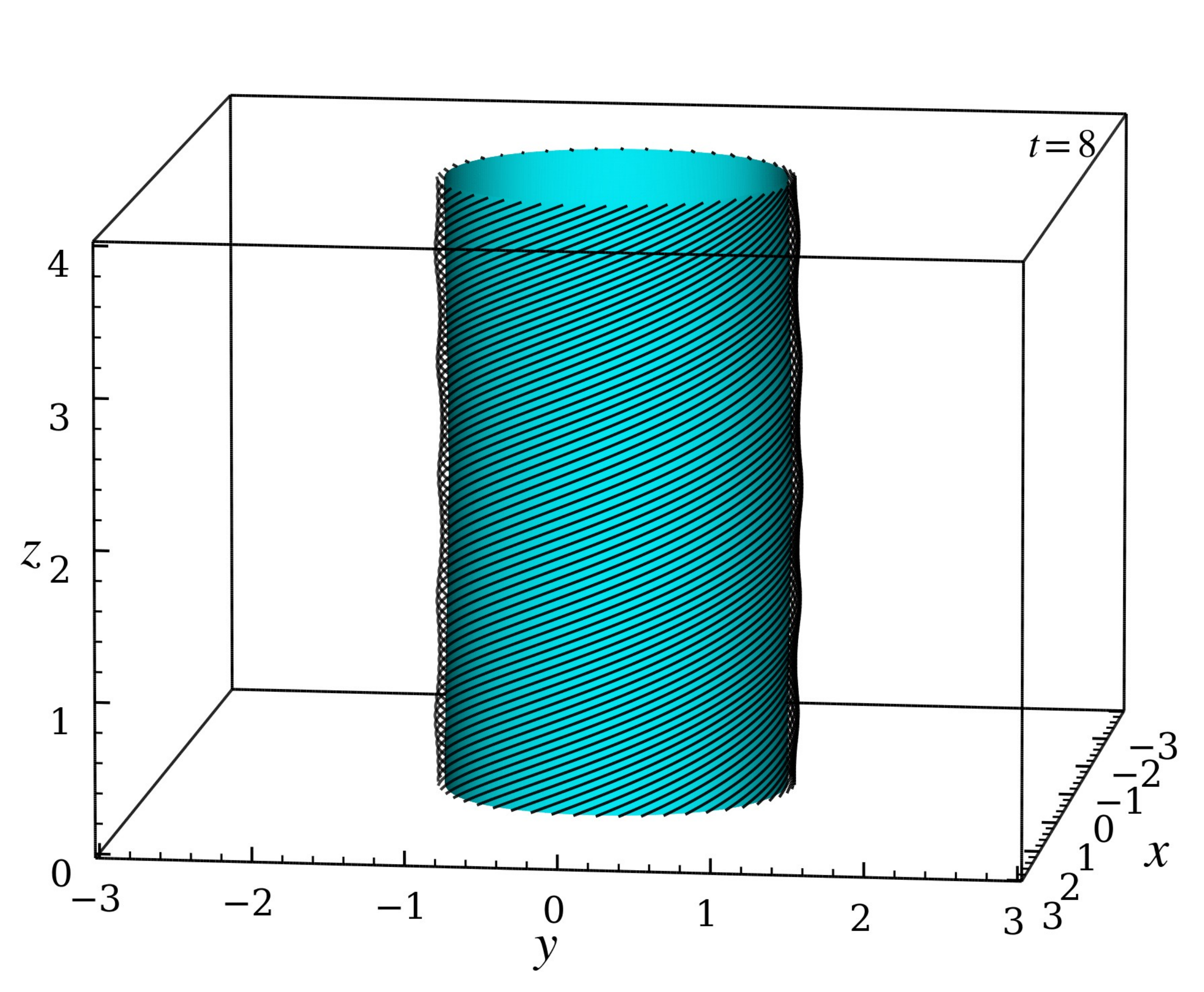}
    
    \includegraphics[width=0.24\textwidth,trim=0cm 0cm 0cm 2cm,clip]{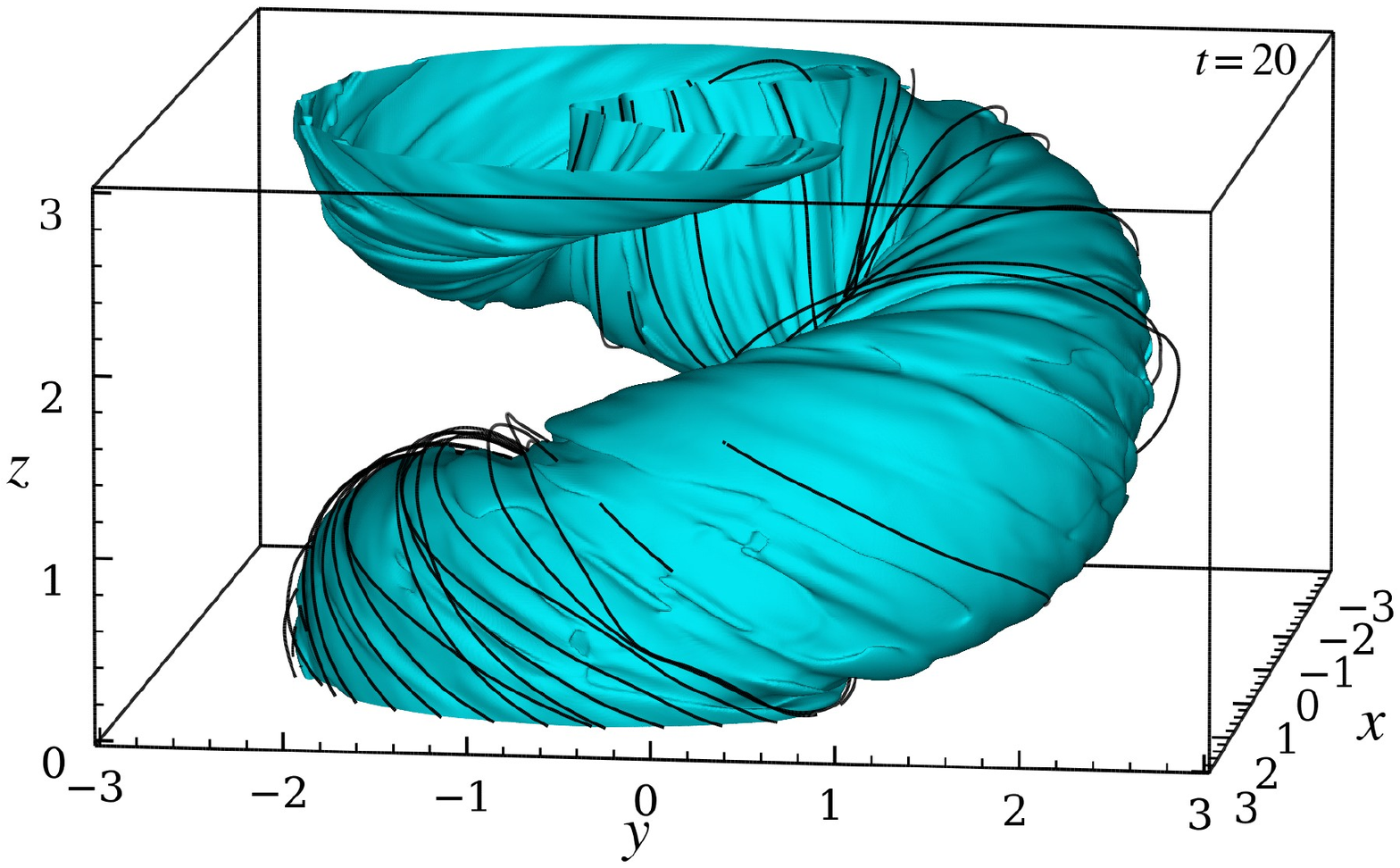}
    \includegraphics[width=0.24\textwidth,trim=0cm 0cm 0cm 2cm,clip]{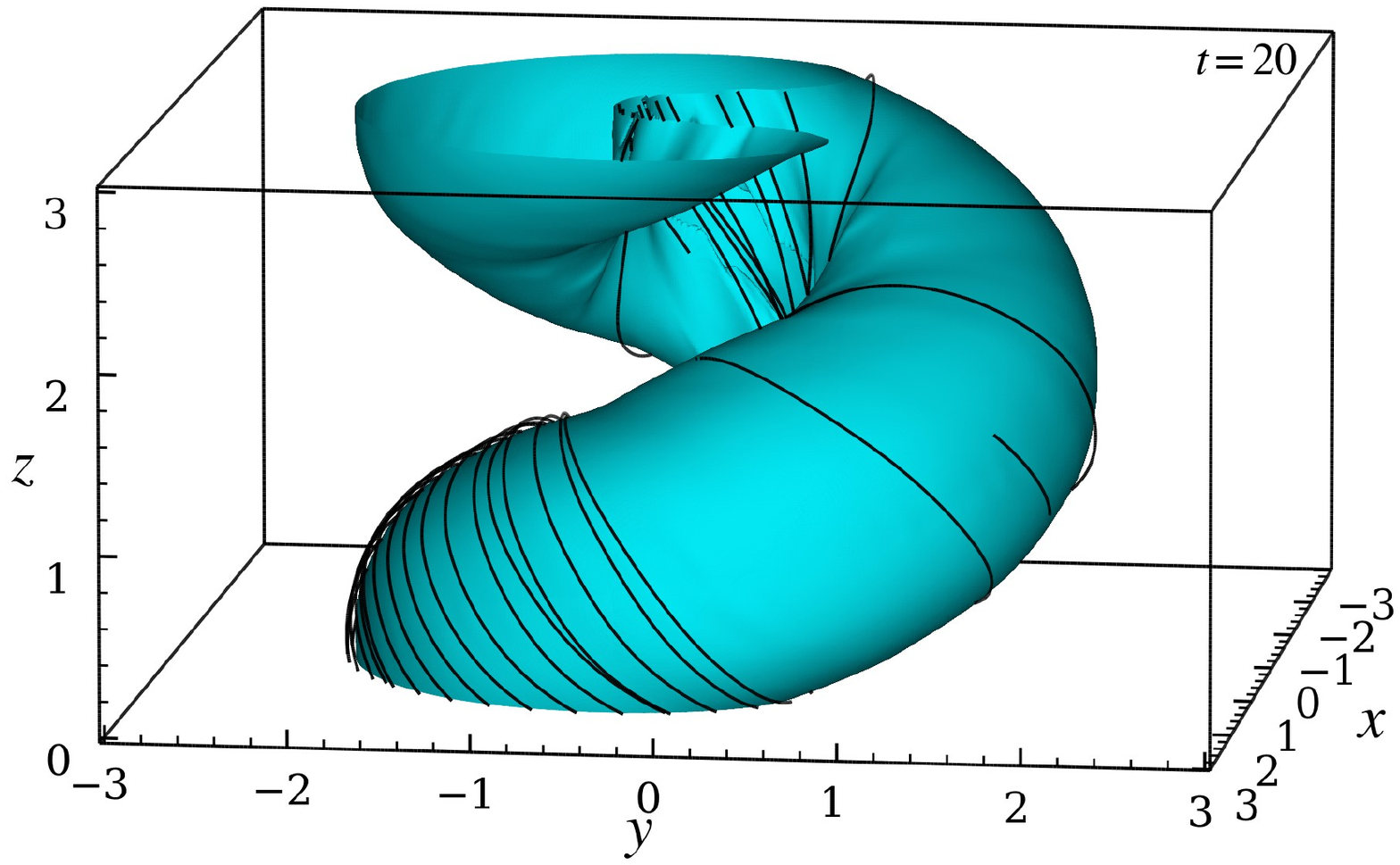}
    \includegraphics[width=0.24\textwidth,trim=0cm 0cm 0cm 2cm,clip]{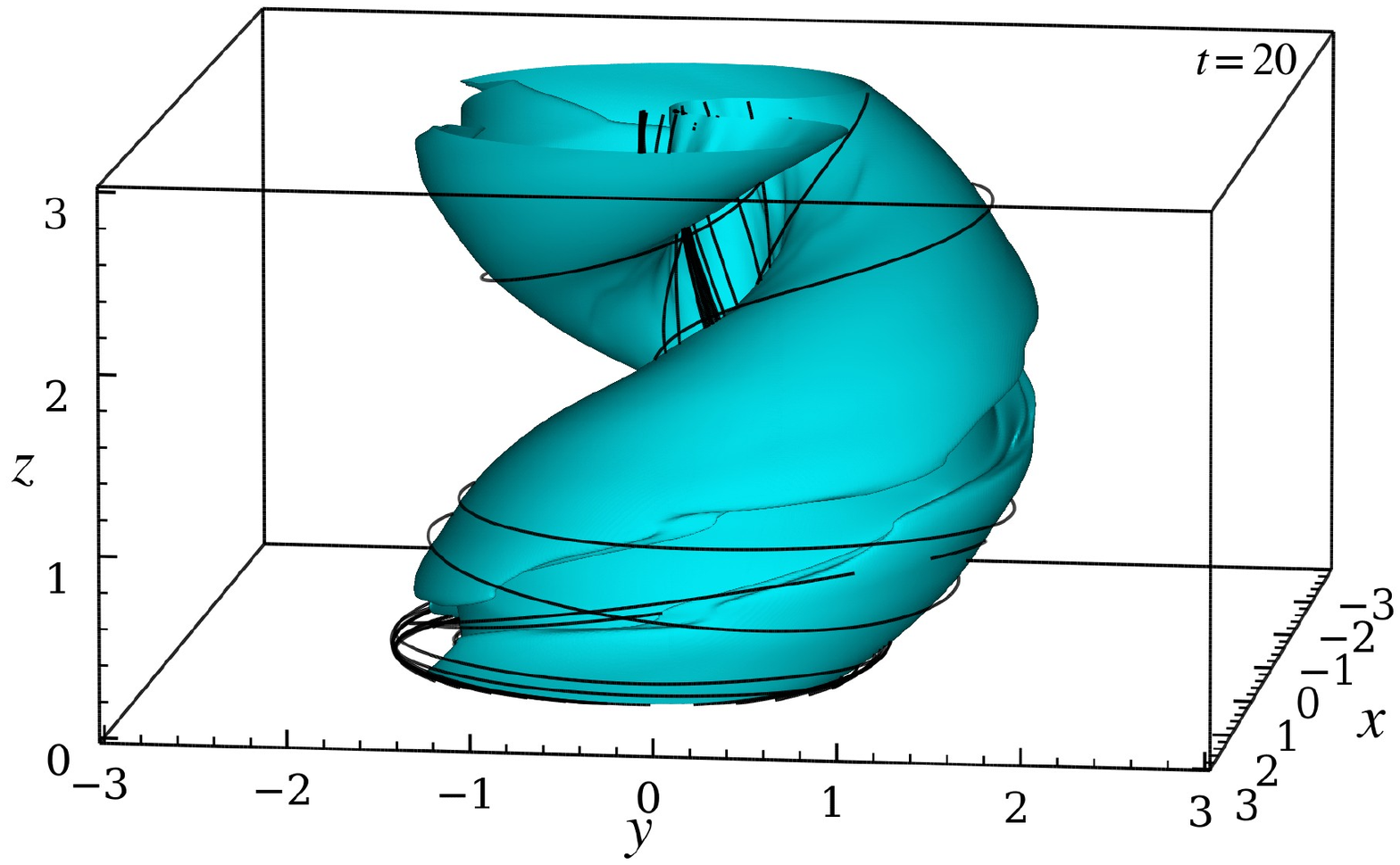}
    \includegraphics[width=0.24\textwidth,trim=0cm 0.75cm 0cm 1.0cm,clip]{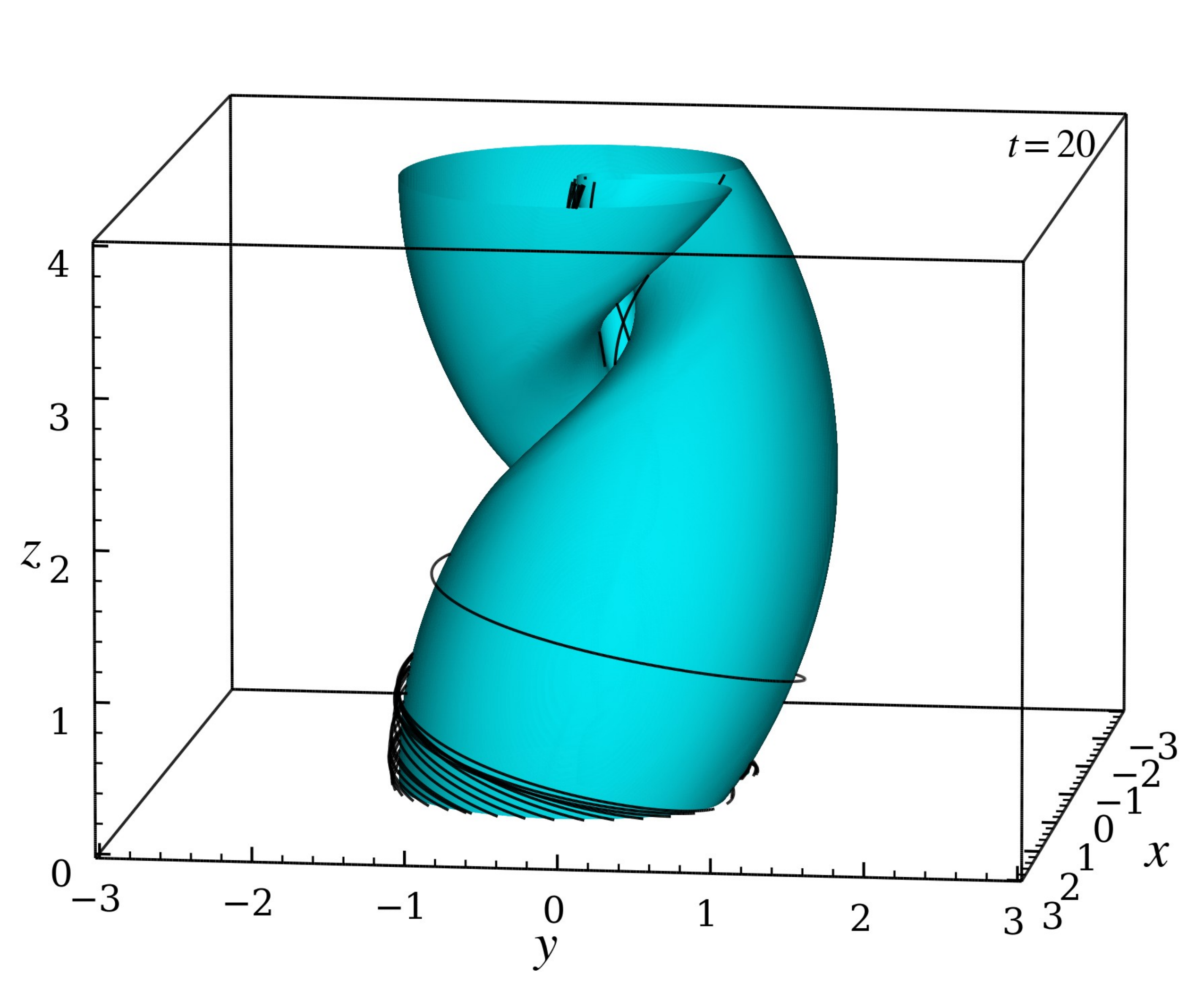}
    
    \includegraphics[width=0.24\textwidth,trim=0cm 0cm 0cm 2cm,clip]{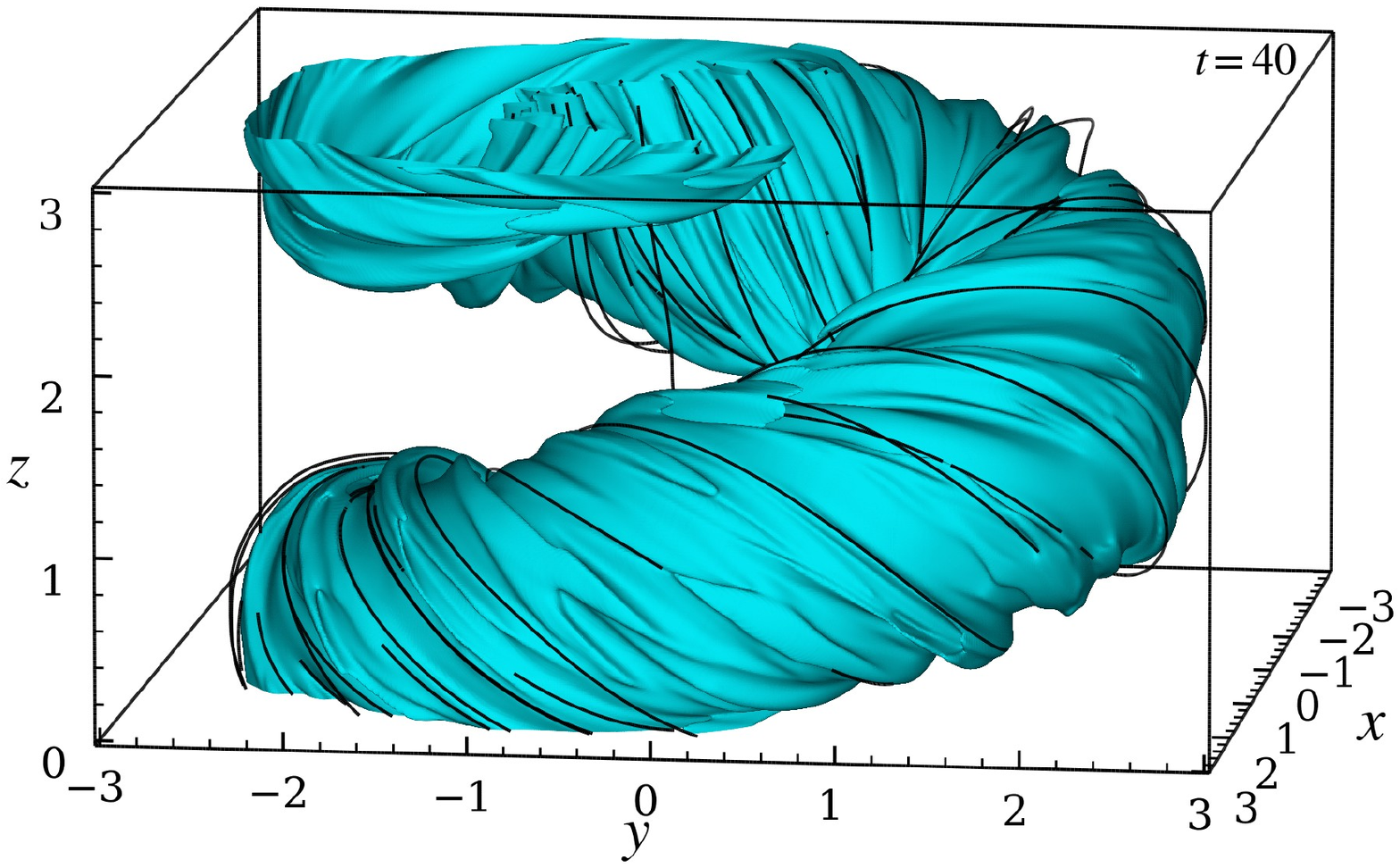}
    \includegraphics[width=0.24\textwidth,trim=0cm 0cm 0cm 2cm,clip]{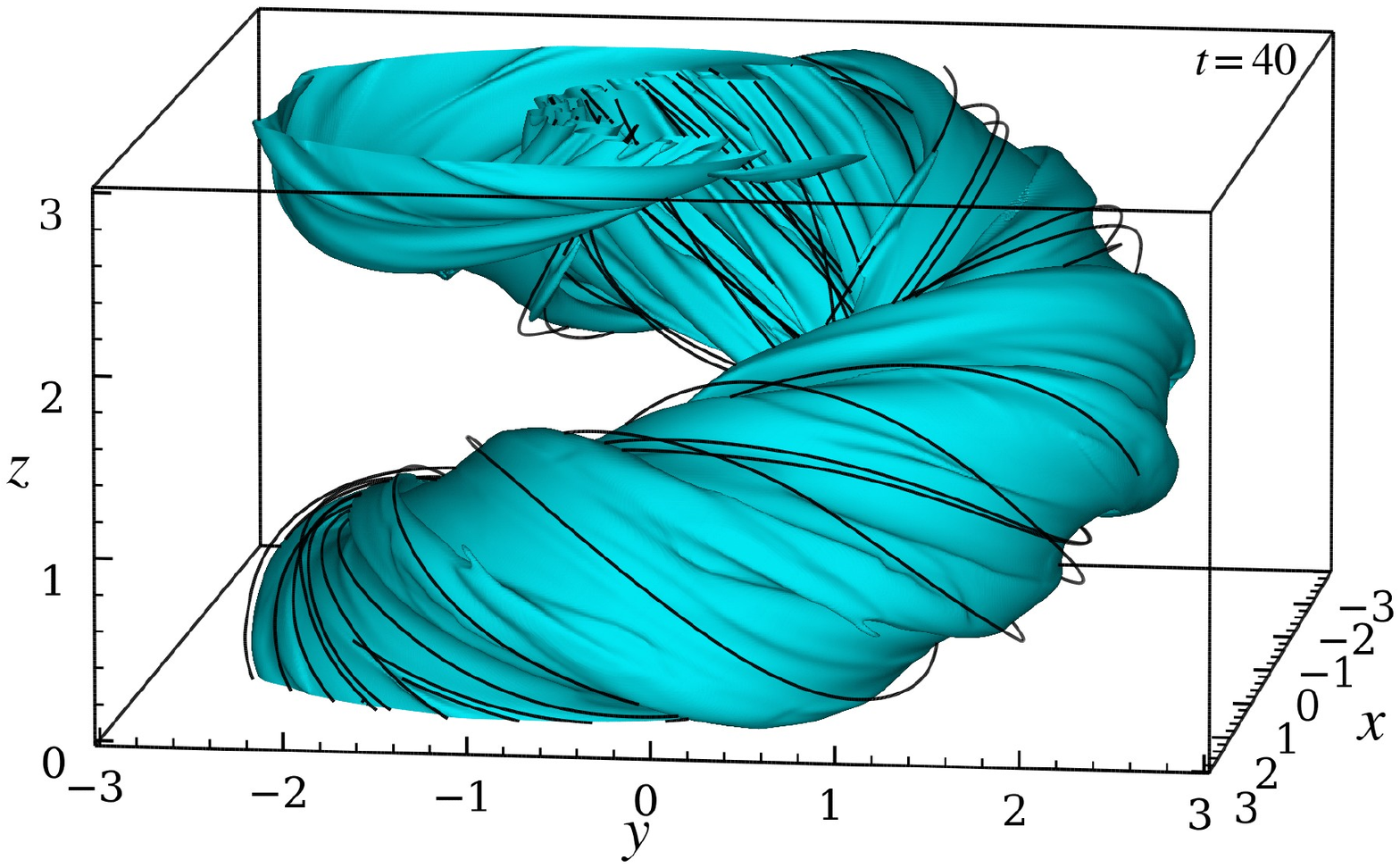}
    \includegraphics[width=0.24\textwidth,trim=0cm 0cm 0cm 2cm,clip]{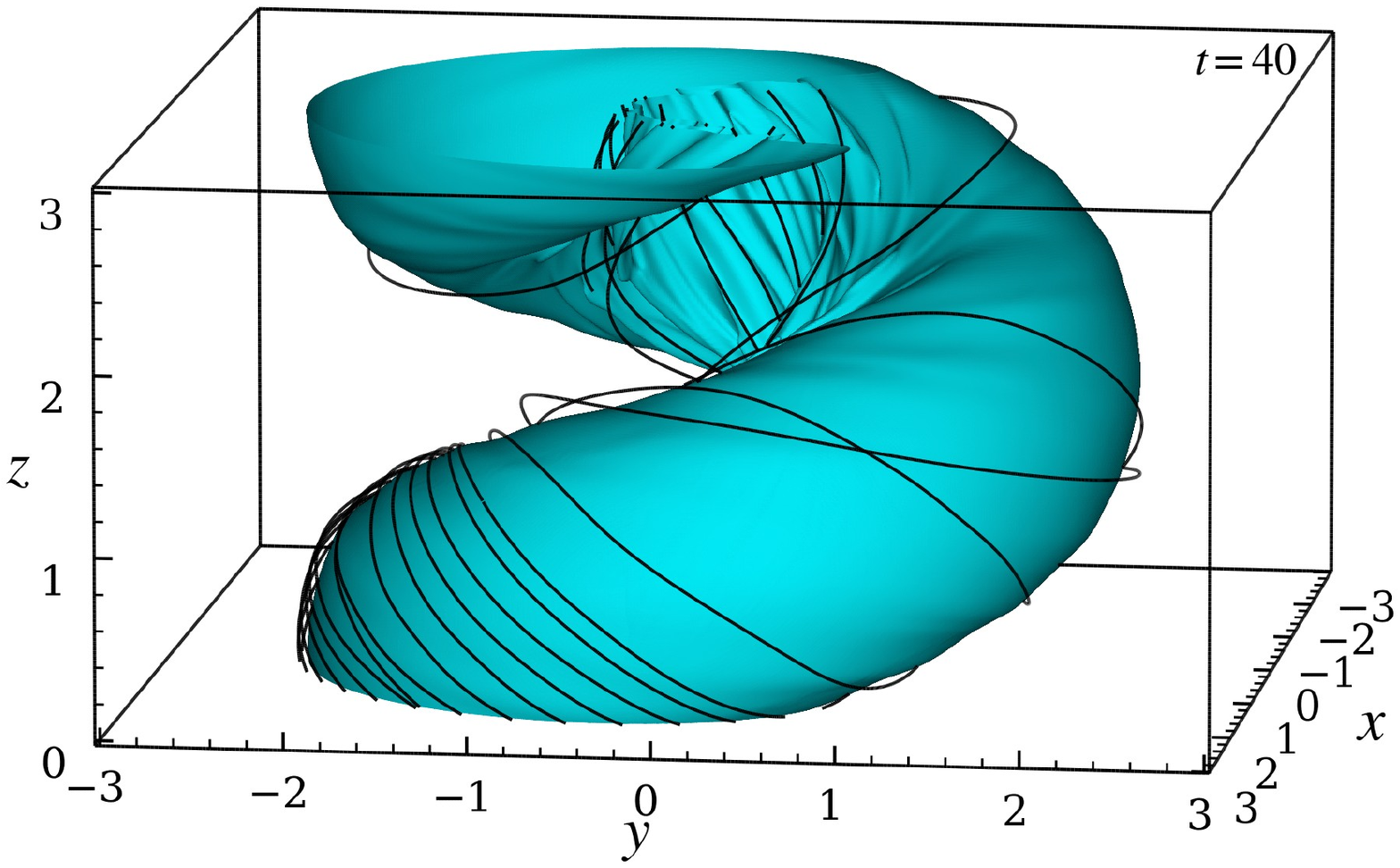}
    \includegraphics[width=0.24\textwidth,trim=0cm 0.75cm 0cm 1.0cm,clip]{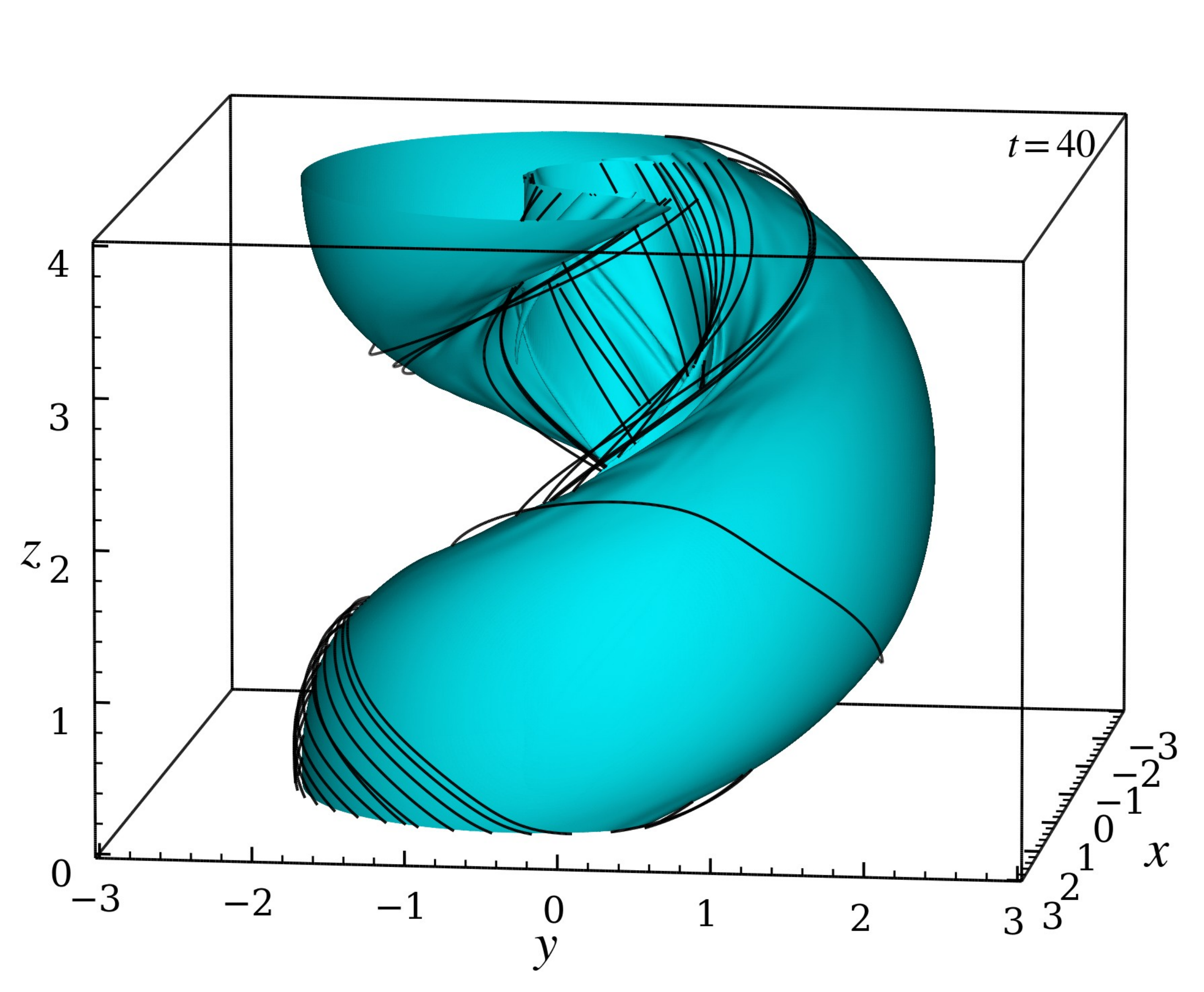}

    \includegraphics[width=0.24\textwidth,trim=0cm 0cm 0cm 2cm,clip]{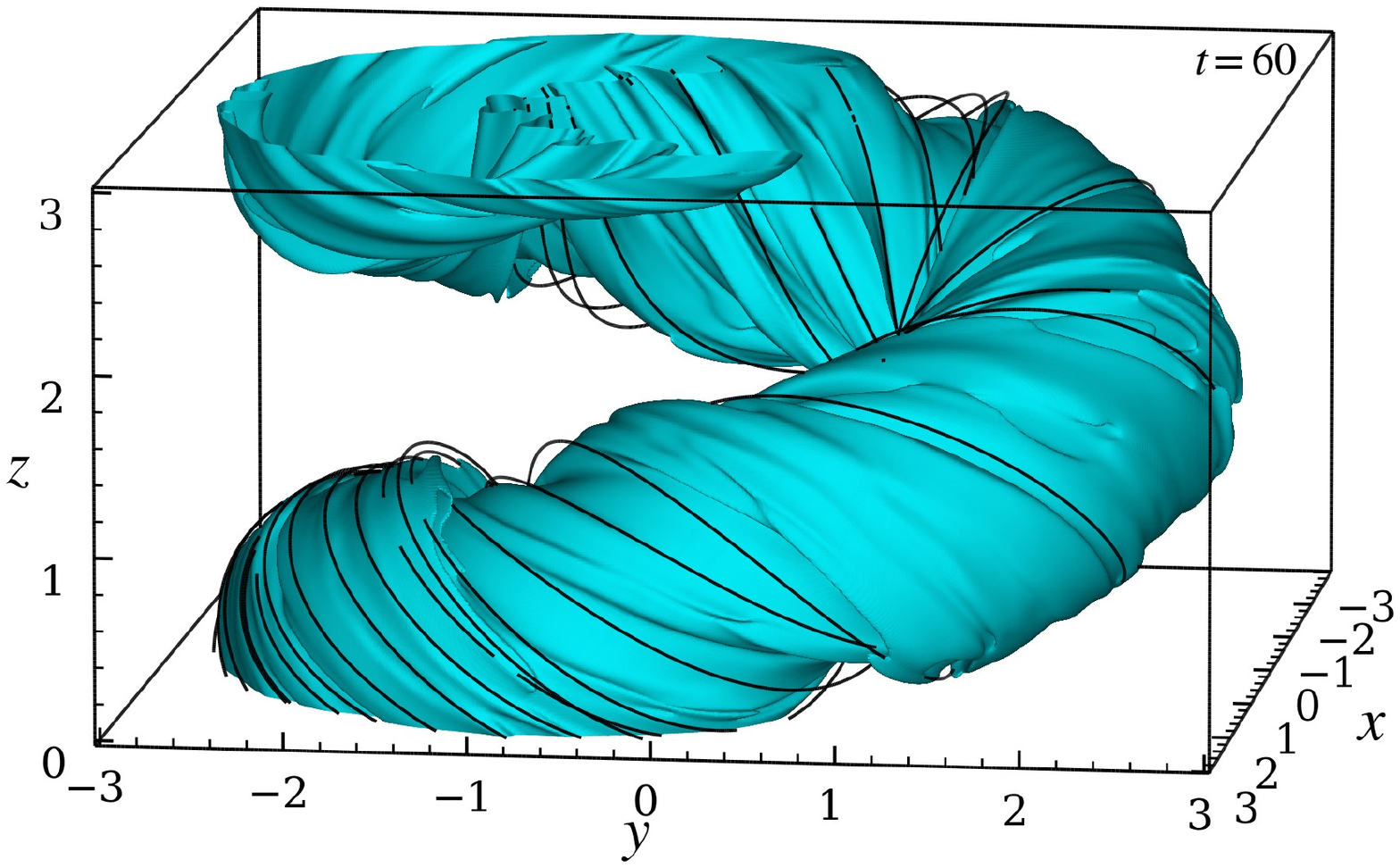}
    \includegraphics[width=0.24\textwidth,trim=0cm 0cm 0cm 2cm,clip]{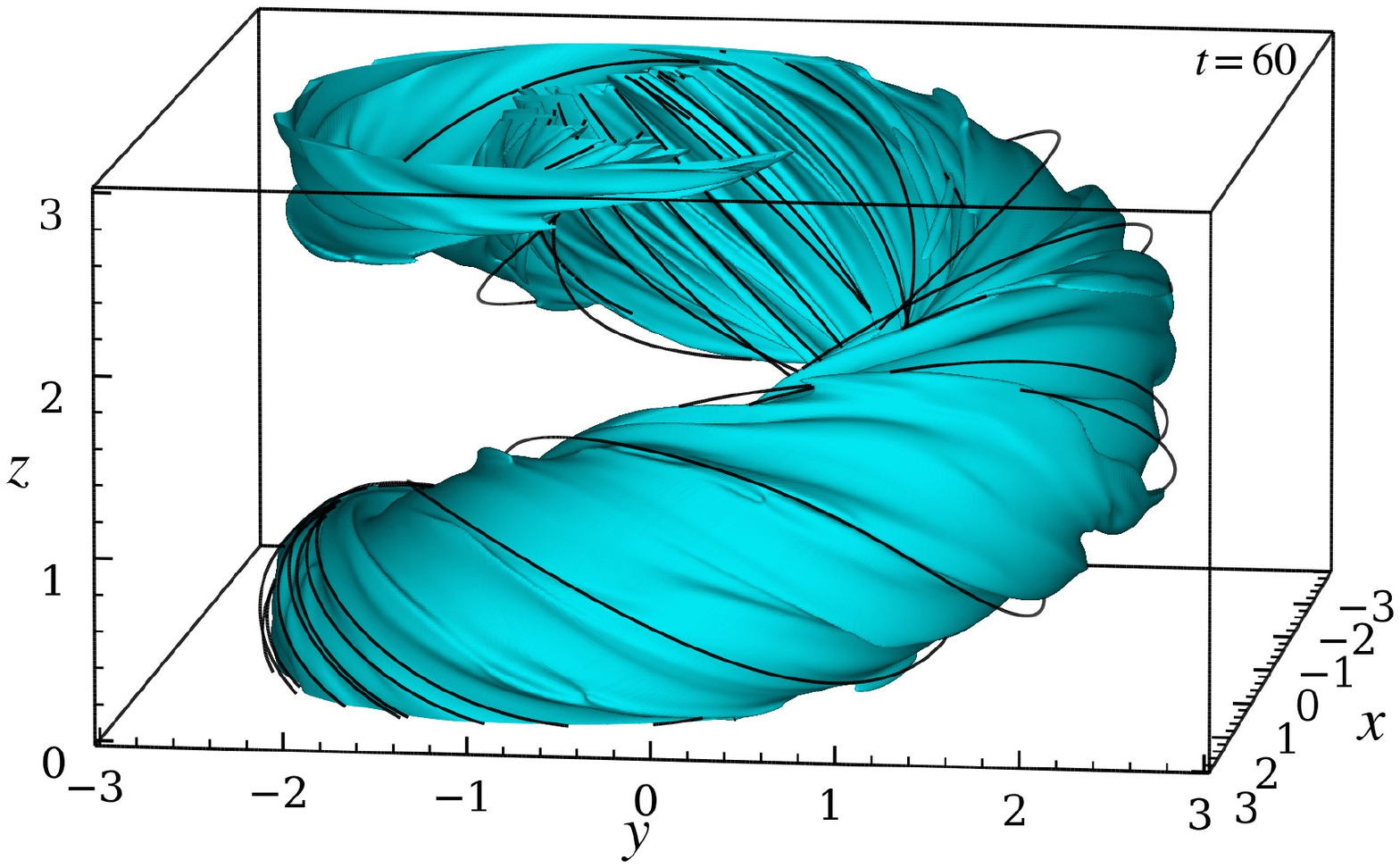}
    \includegraphics[width=0.24\textwidth,trim=0cm 0cm 0cm 2cm,clip]{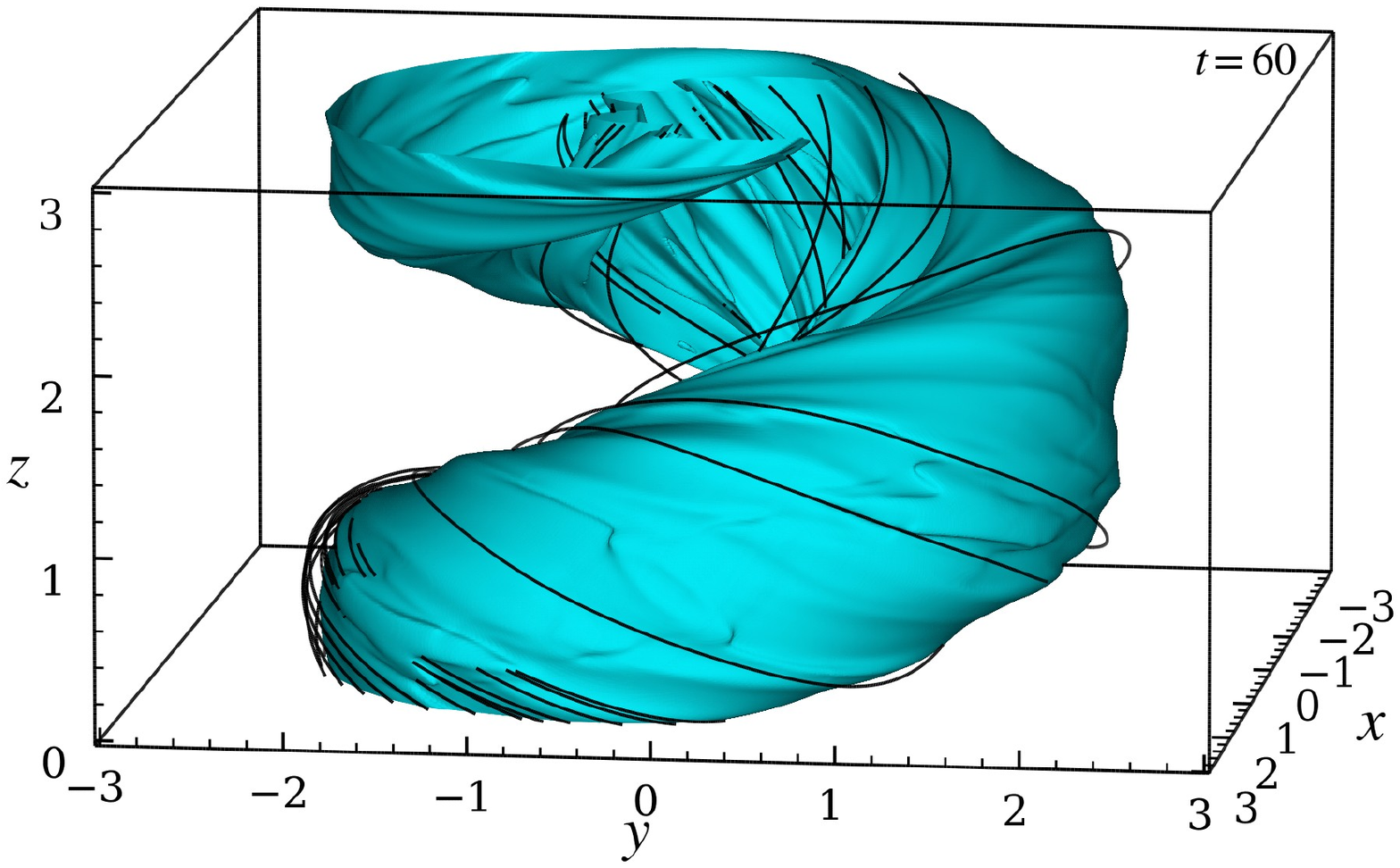}
    \includegraphics[width=0.24\textwidth,trim=0cm 0.75cm 0cm 1.0cm,clip]{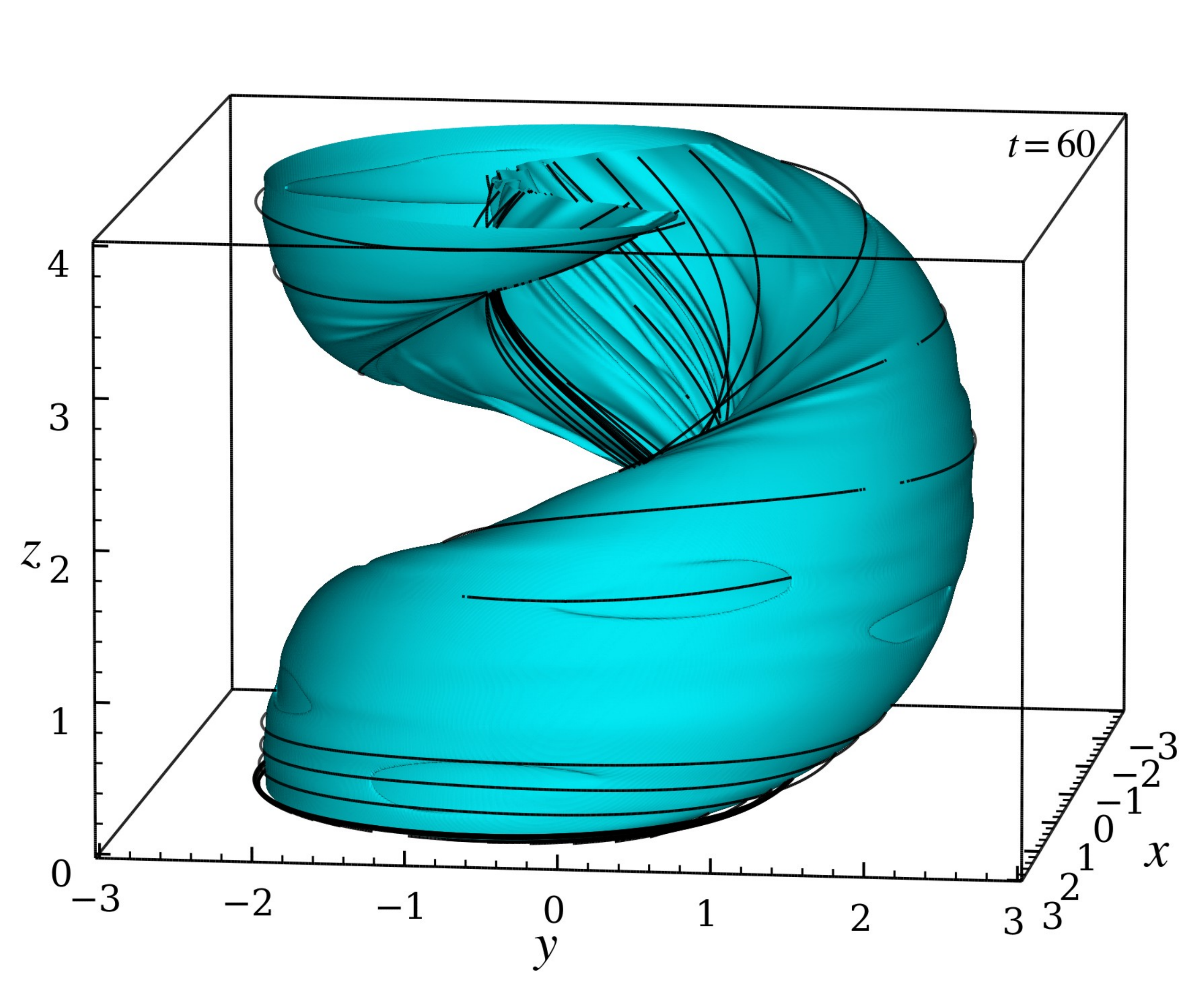}
    
    \includegraphics[width=0.24\textwidth,trim=0cm 0cm 0cm 2cm,clip]{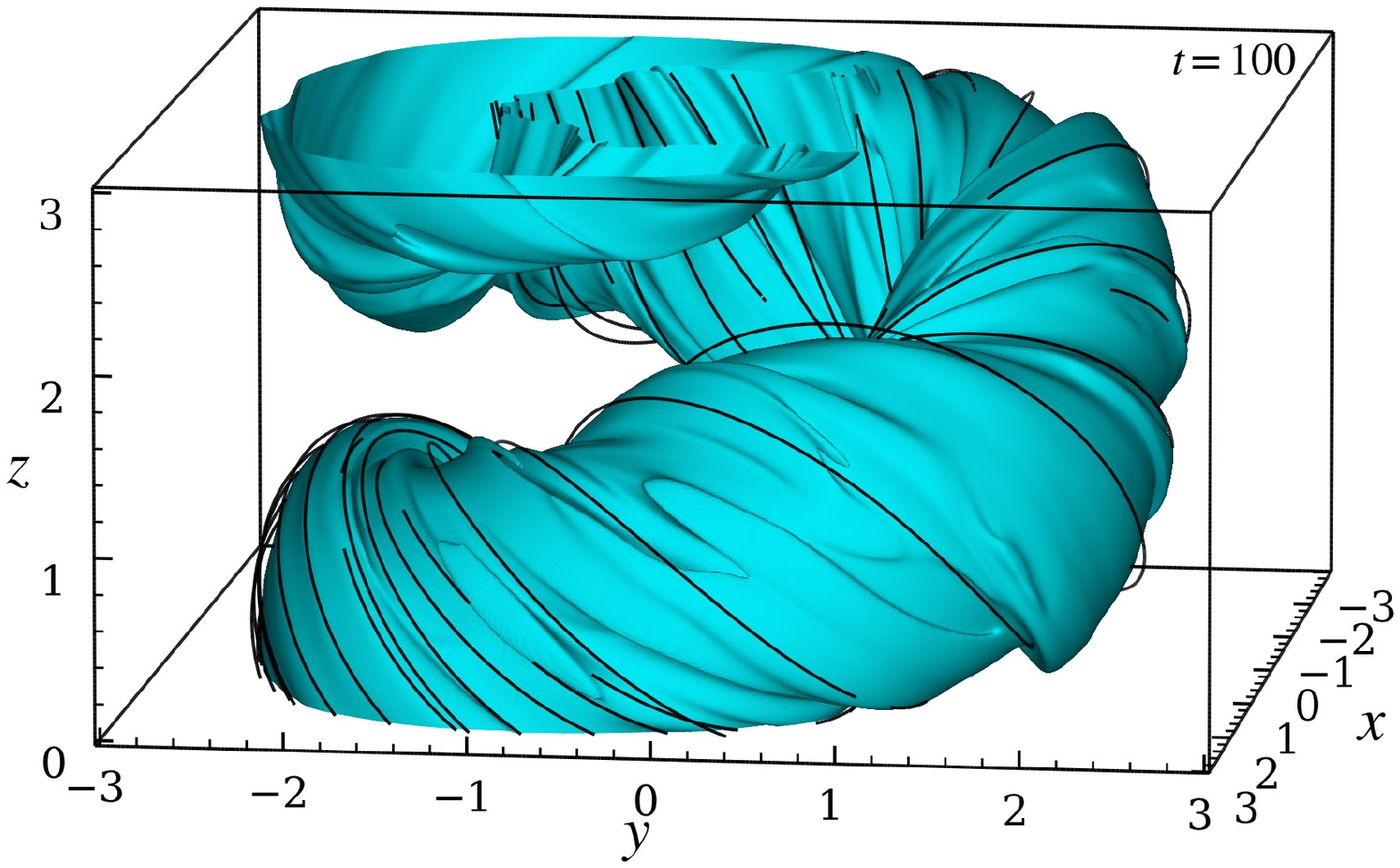}
    \includegraphics[width=0.24\textwidth,trim=0cm 0cm 0cm 2cm,clip]{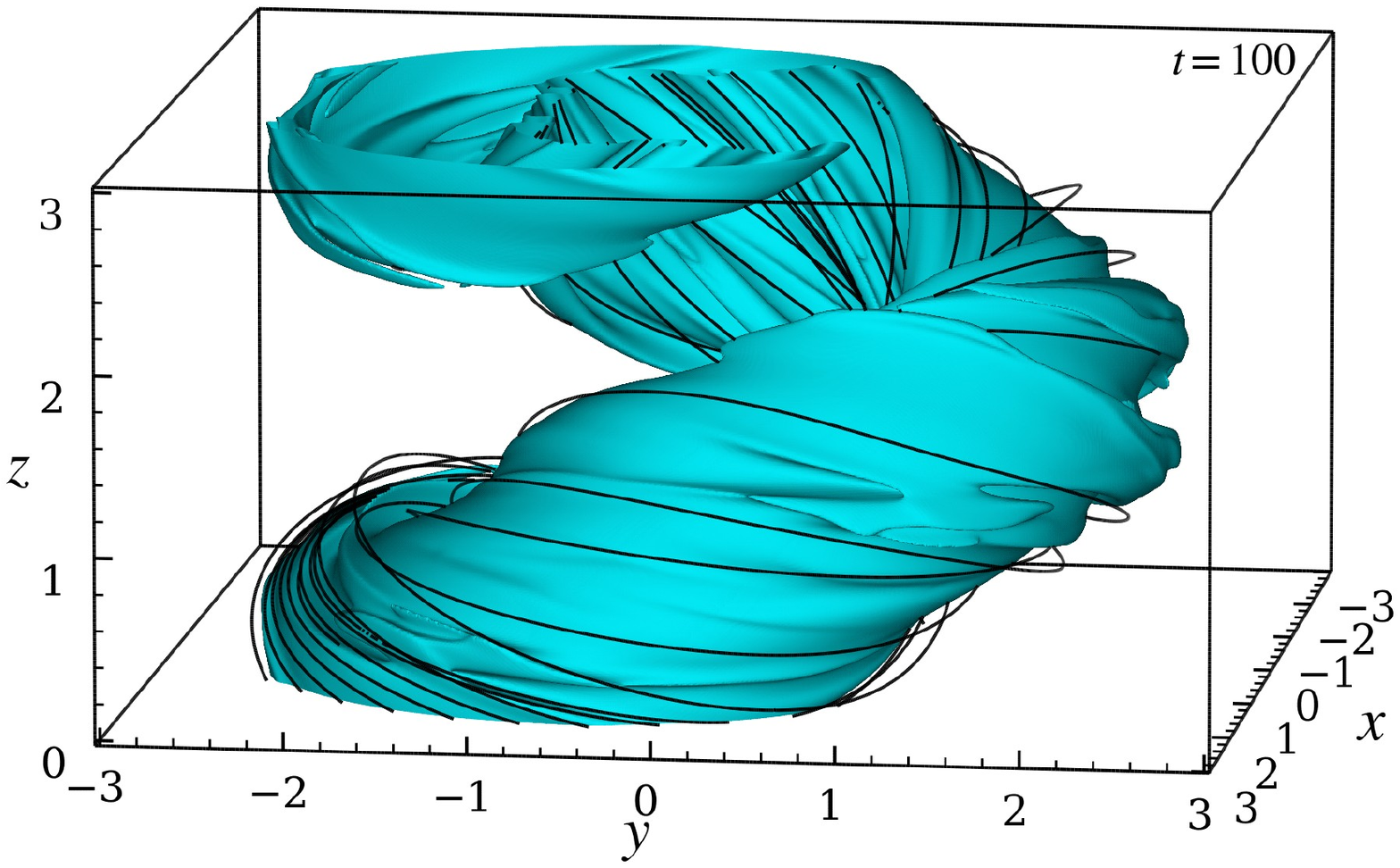}
    \includegraphics[width=0.24\textwidth,trim=0cm 0cm 0cm 2cm,clip]{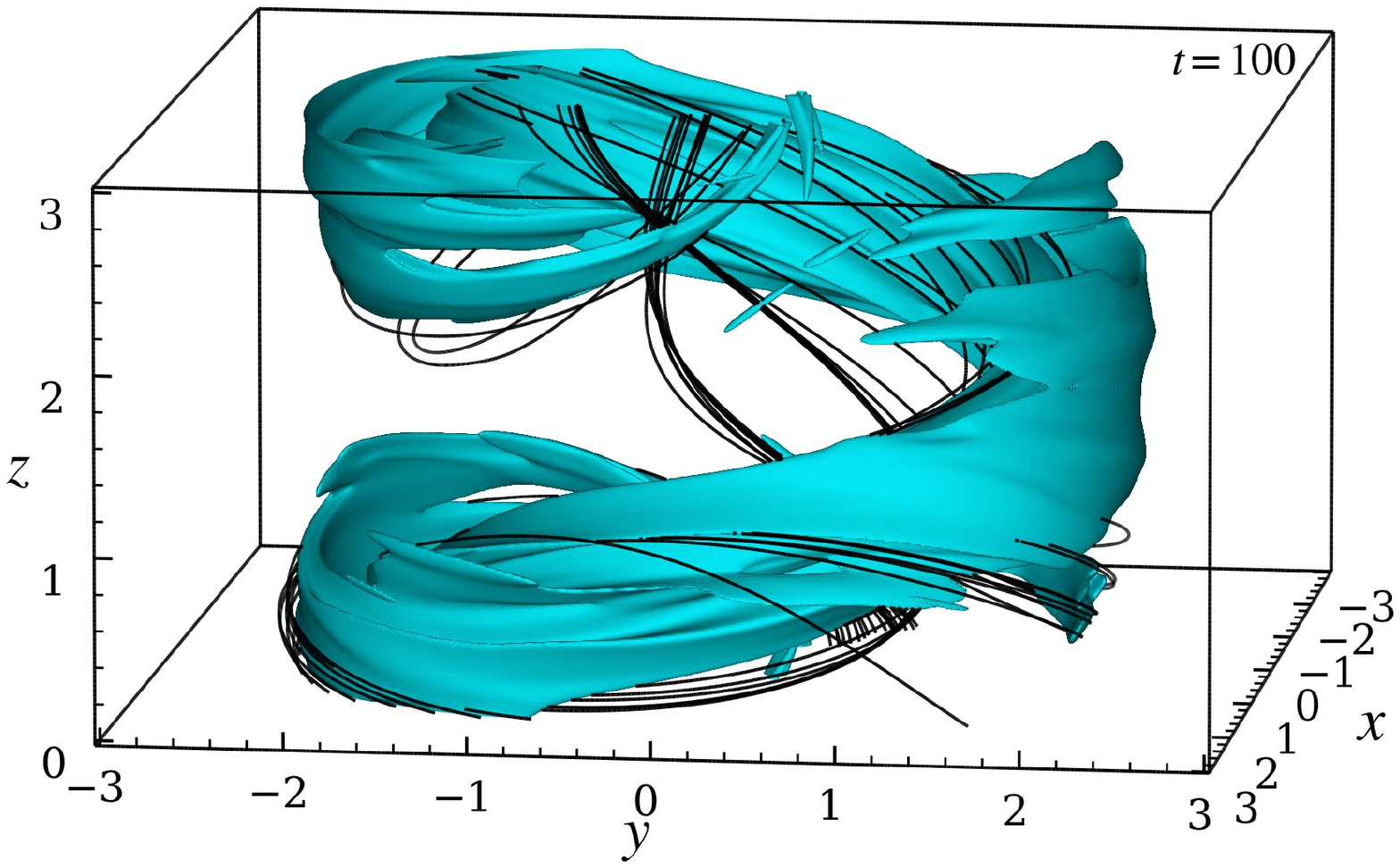}
    \includegraphics[width=0.24\textwidth,trim=0cm 0.75cm 0cm 1.0cm,clip]{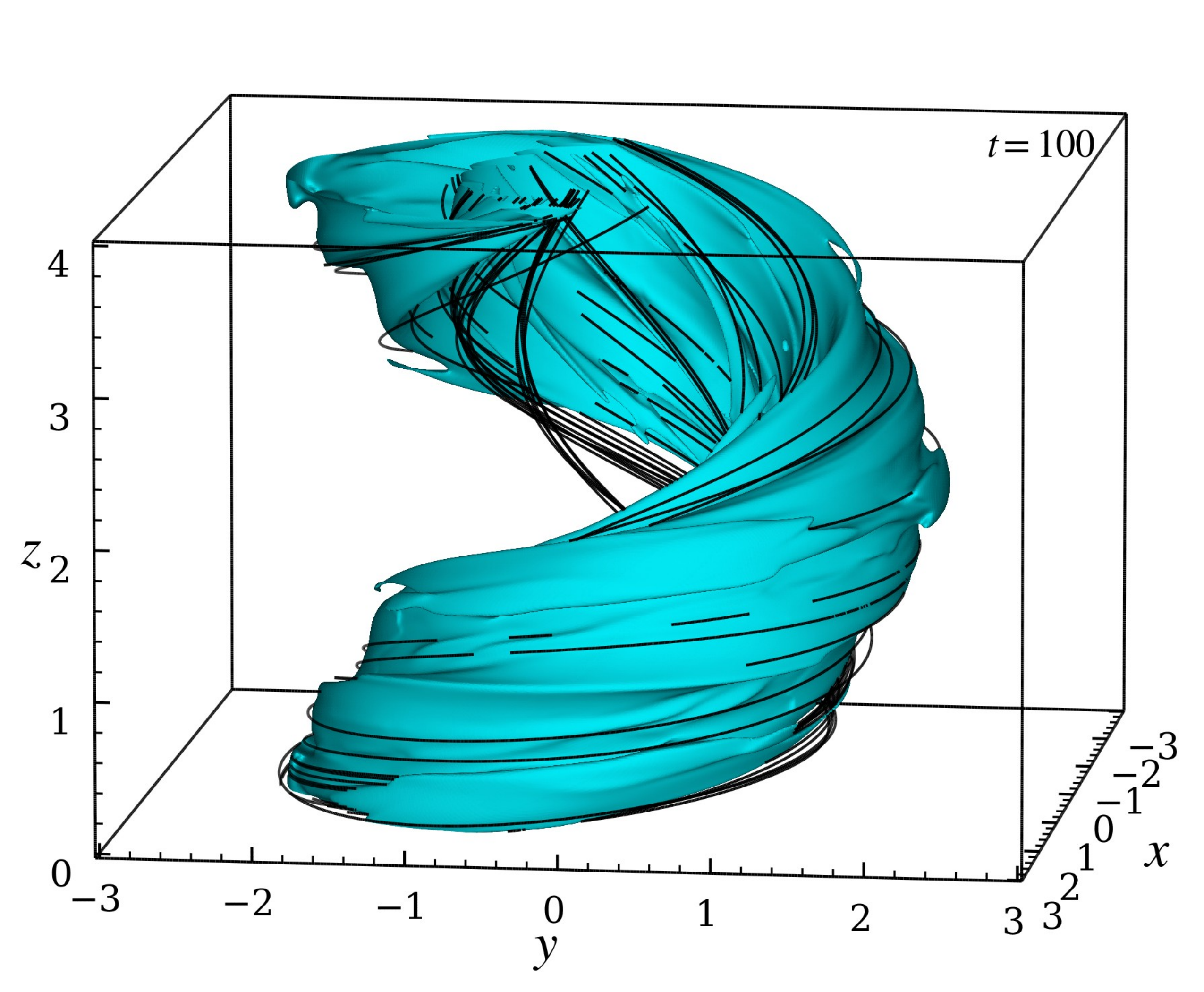}

 \caption{Evolution of the density iso-surface with $\rho = 10^{-3/4}$, which separates the jet from the wind. The images show the solutions for models $M_1, M_2, M_3, $ and $M_4$ ( from left to right ) at $t = 8$, 20, 40, 60 and $100$ ( from top to bottom ). The lines are the magnetic field lines originating from the footpoints located just outside of the iso-surface in the $z=0$ plane, by about 4 cells.}
    \label{fig:isosurf}
\end{figure*}

\begin{figure*}
    \centering
    \includegraphics[width=0.195\linewidth]{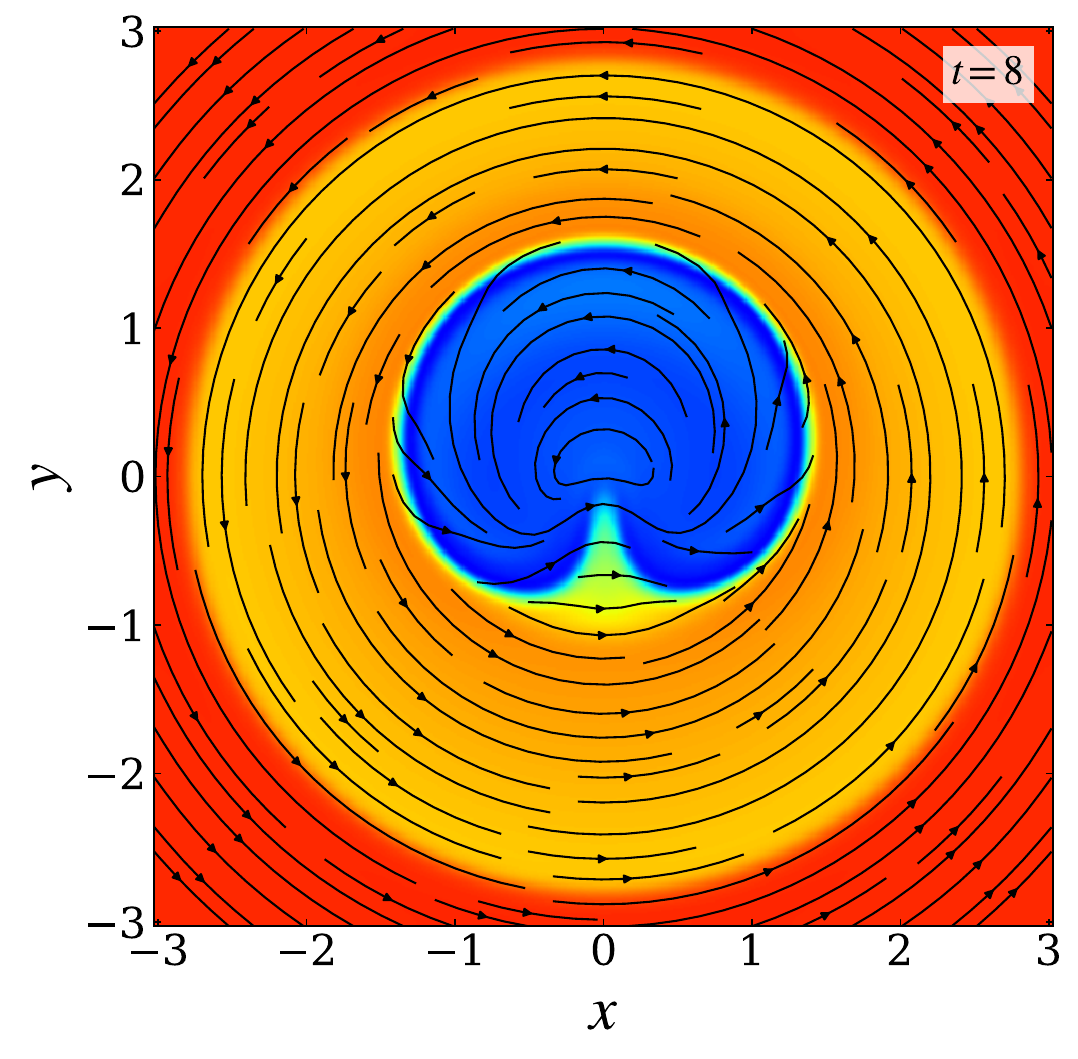}
    \includegraphics[width=0.195\linewidth]{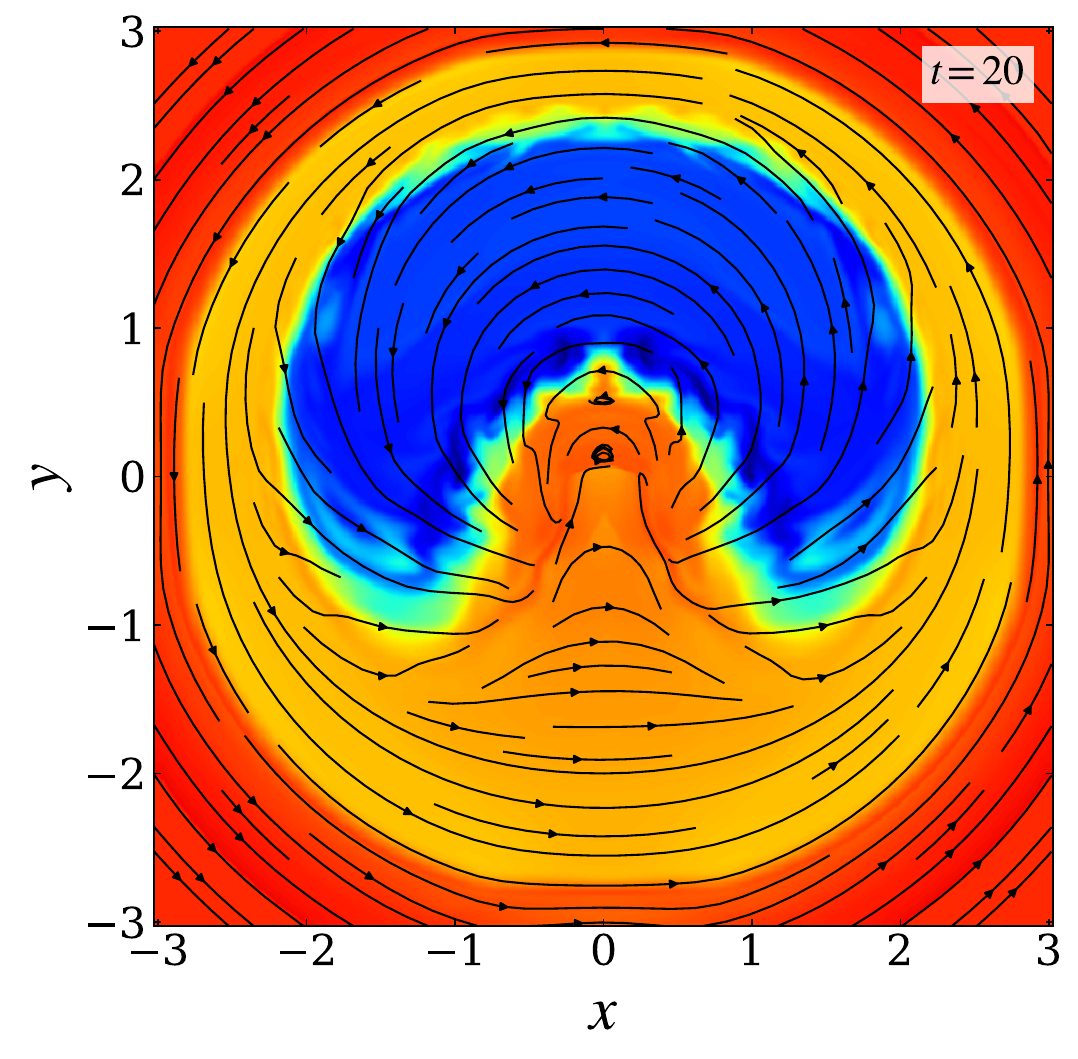}
    \includegraphics[width=0.195\linewidth]{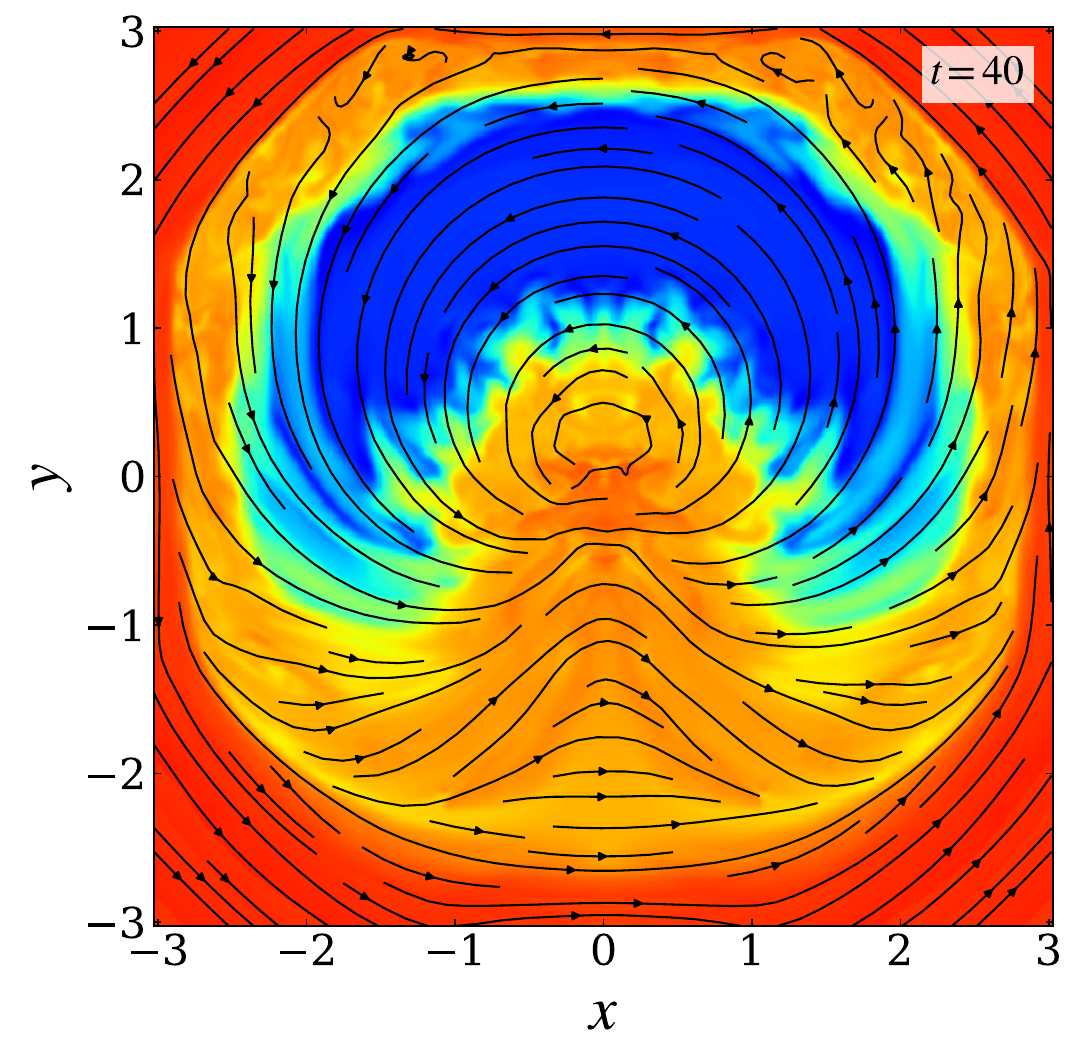}
    \includegraphics[width=0.195\linewidth]{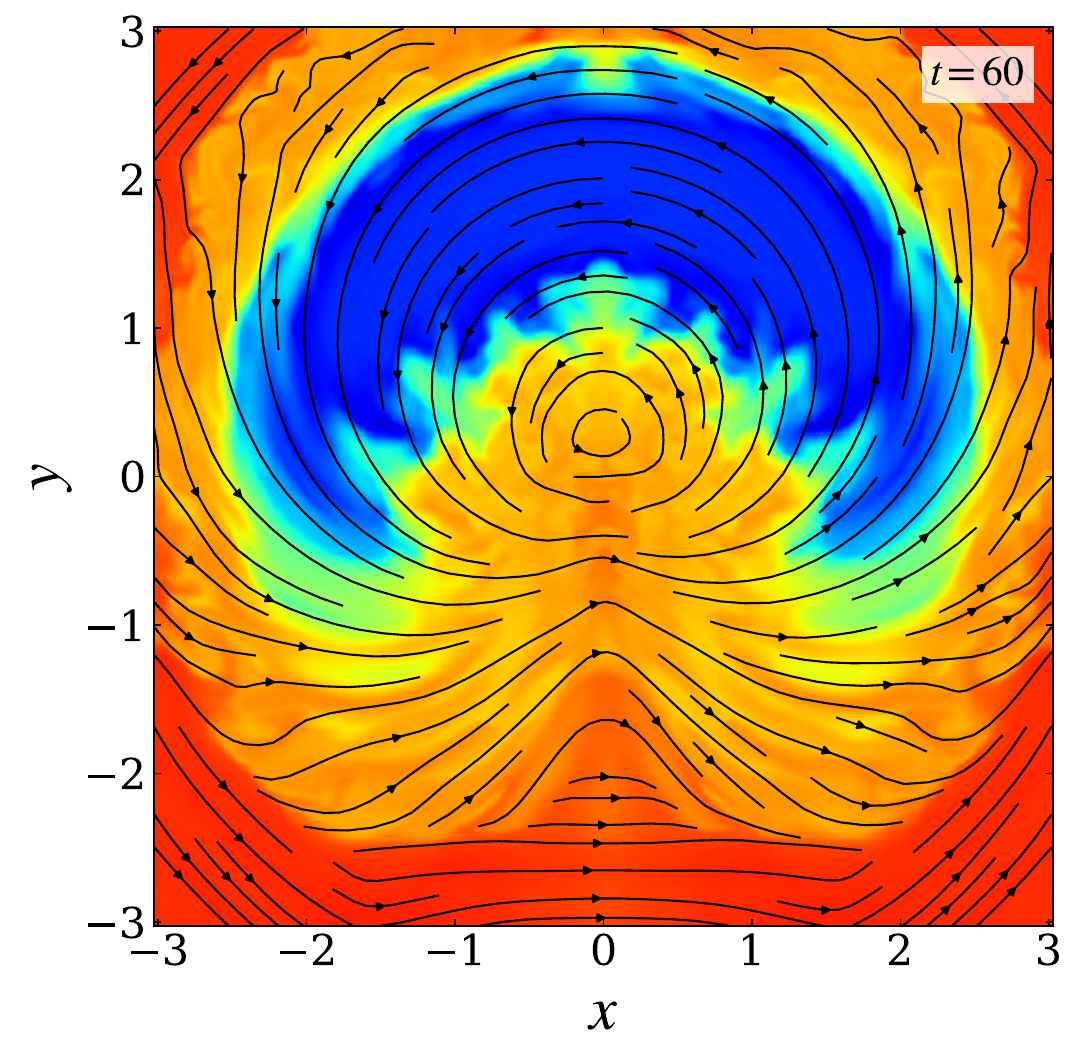}
    \includegraphics[width=0.195\linewidth]{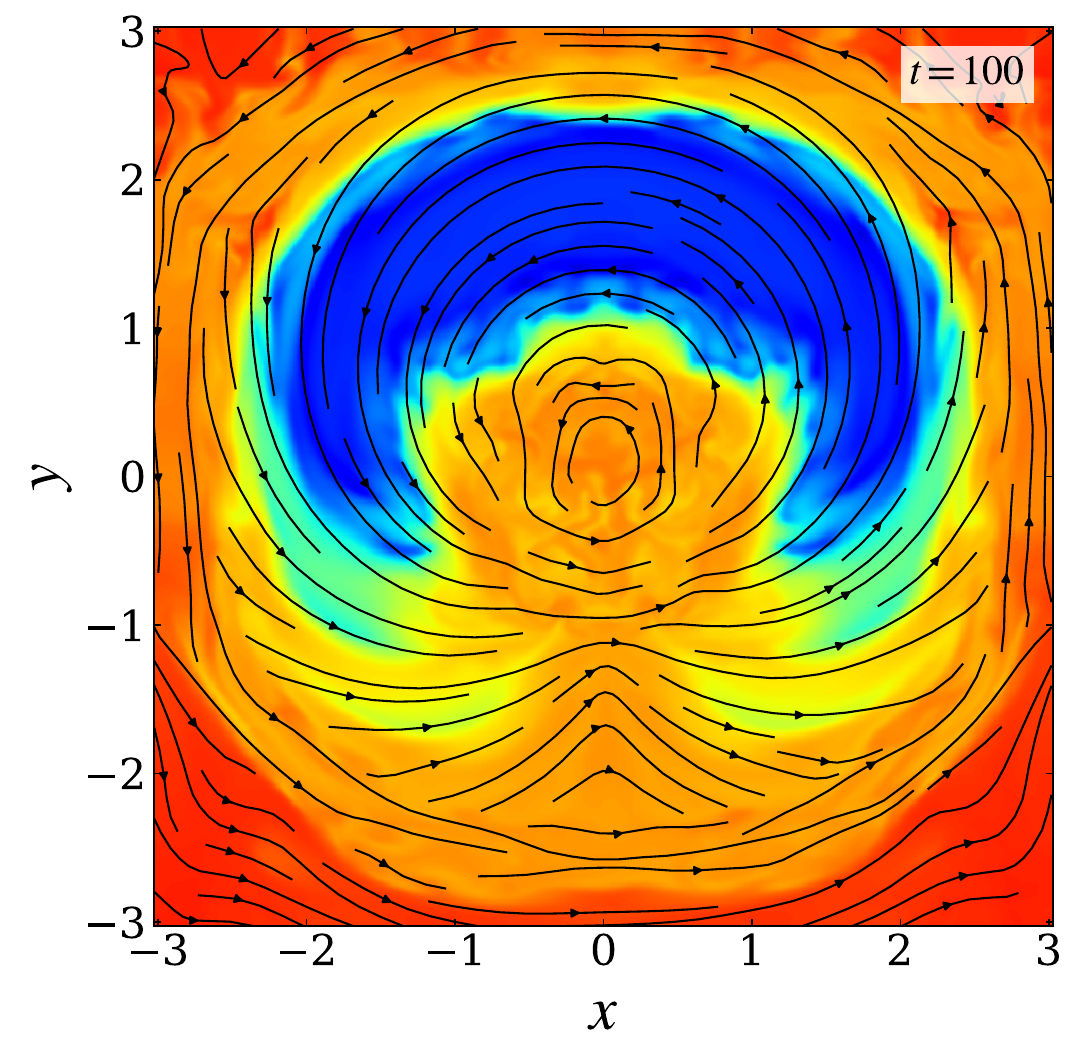}
    
     \includegraphics[width=0.195\linewidth]{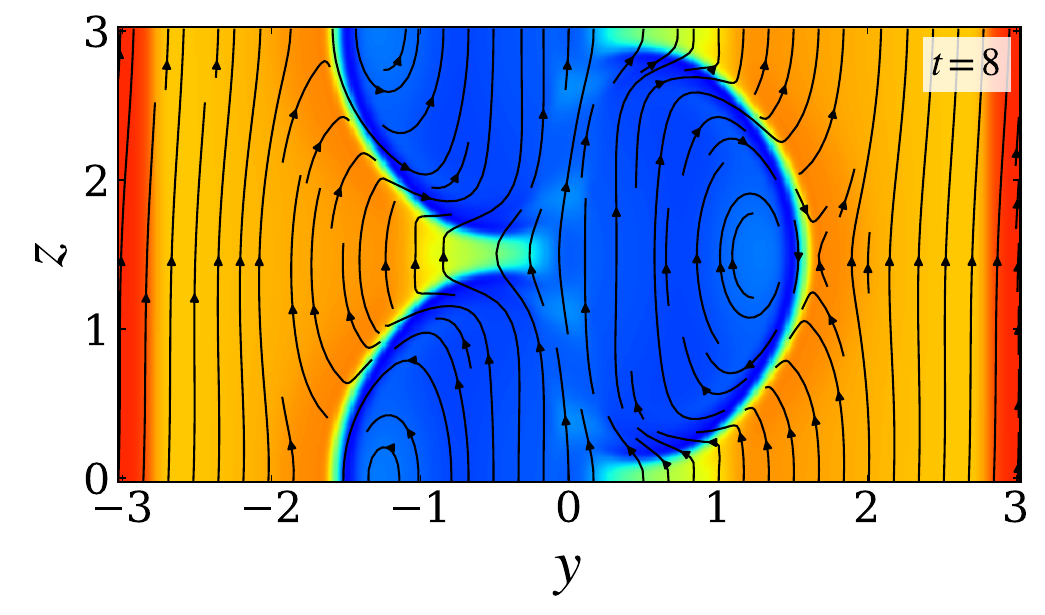}
    \includegraphics[width=0.195\linewidth]{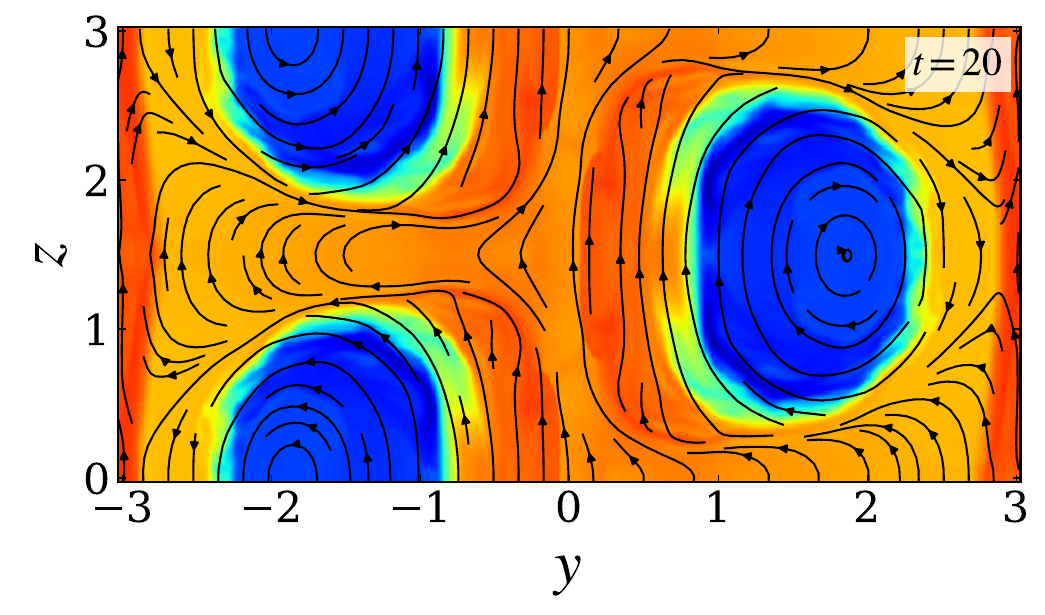}
    \includegraphics[width=0.195\linewidth]{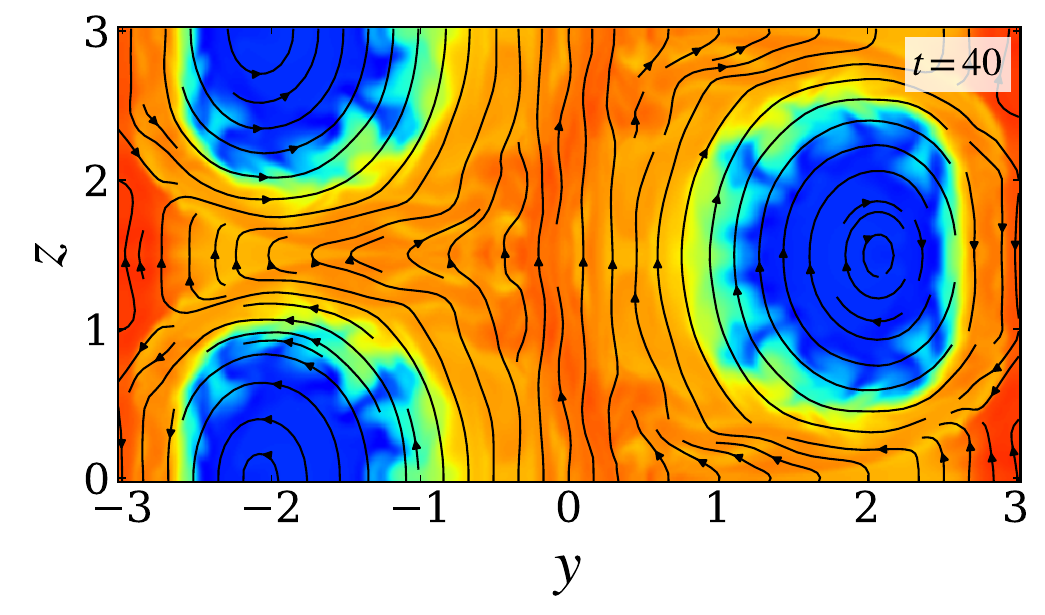}
    \includegraphics[width=0.195\linewidth]{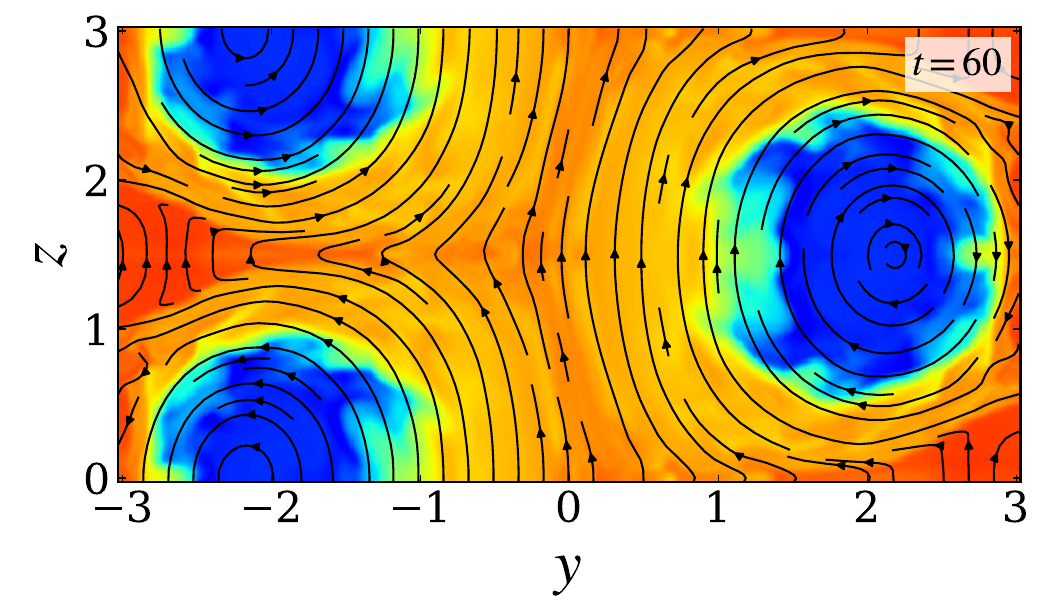}
    \includegraphics[width=0.195\linewidth]{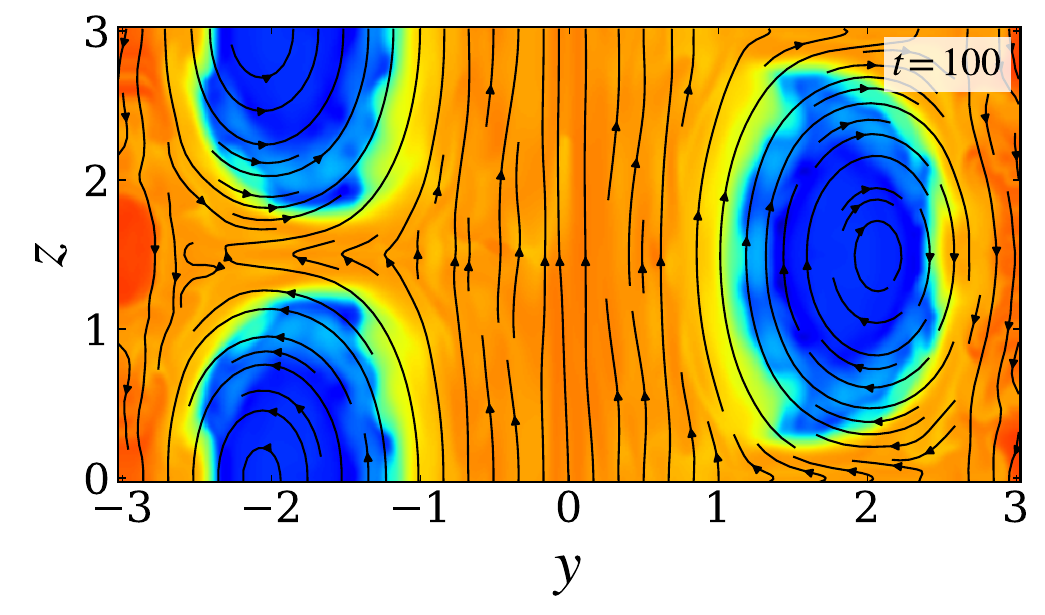}
     
    \includegraphics[width=0.195\linewidth]{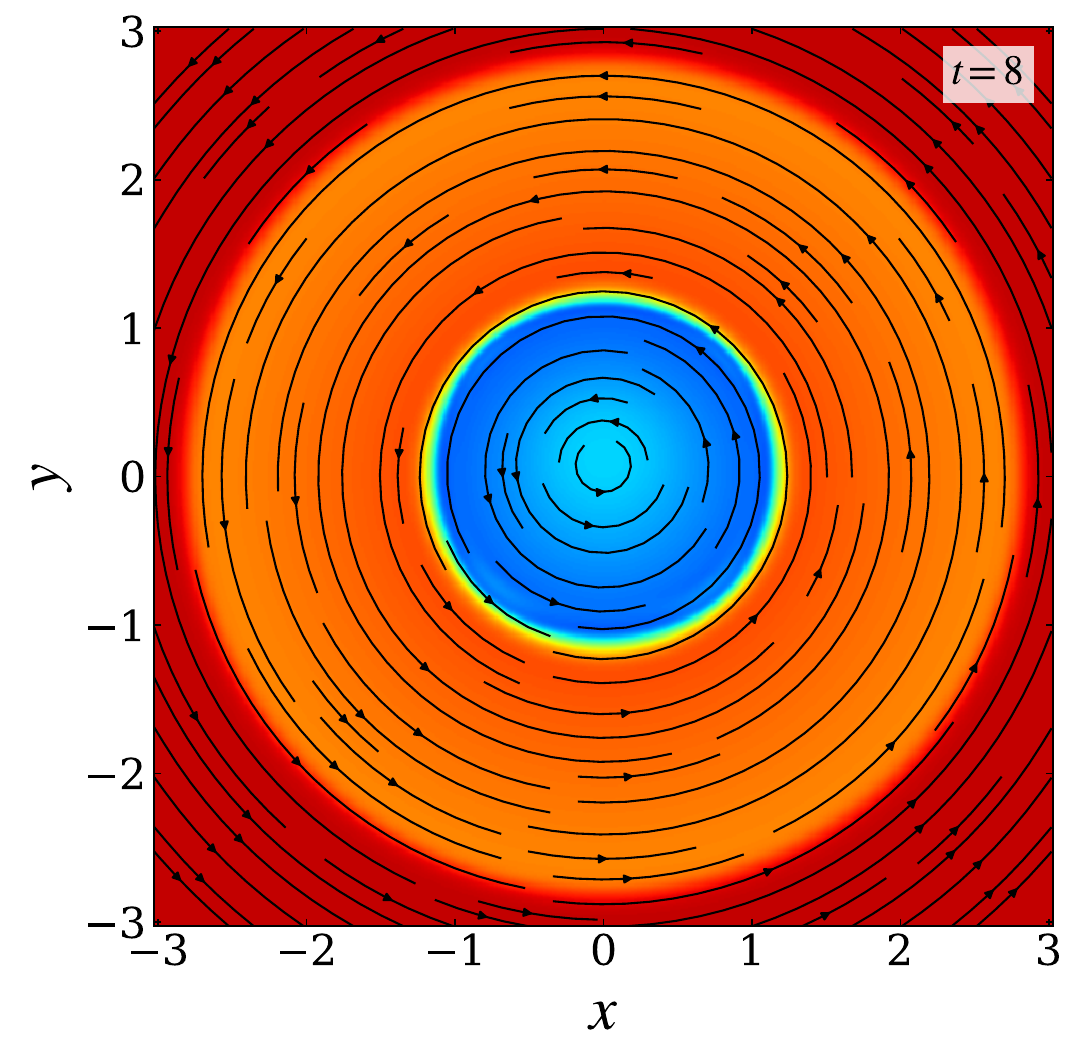}
    \includegraphics[width=0.195\linewidth]{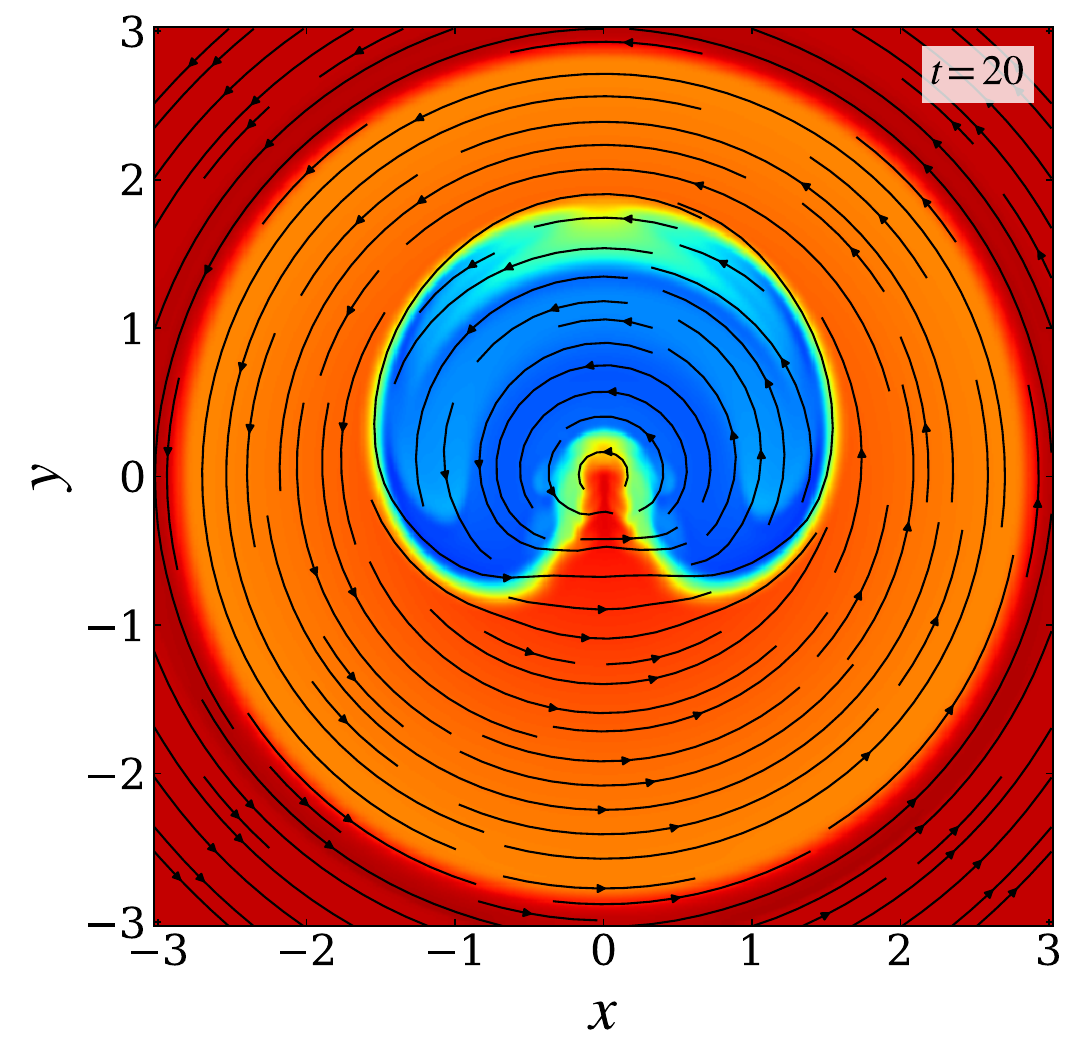}
    \includegraphics[width=0.195\linewidth]{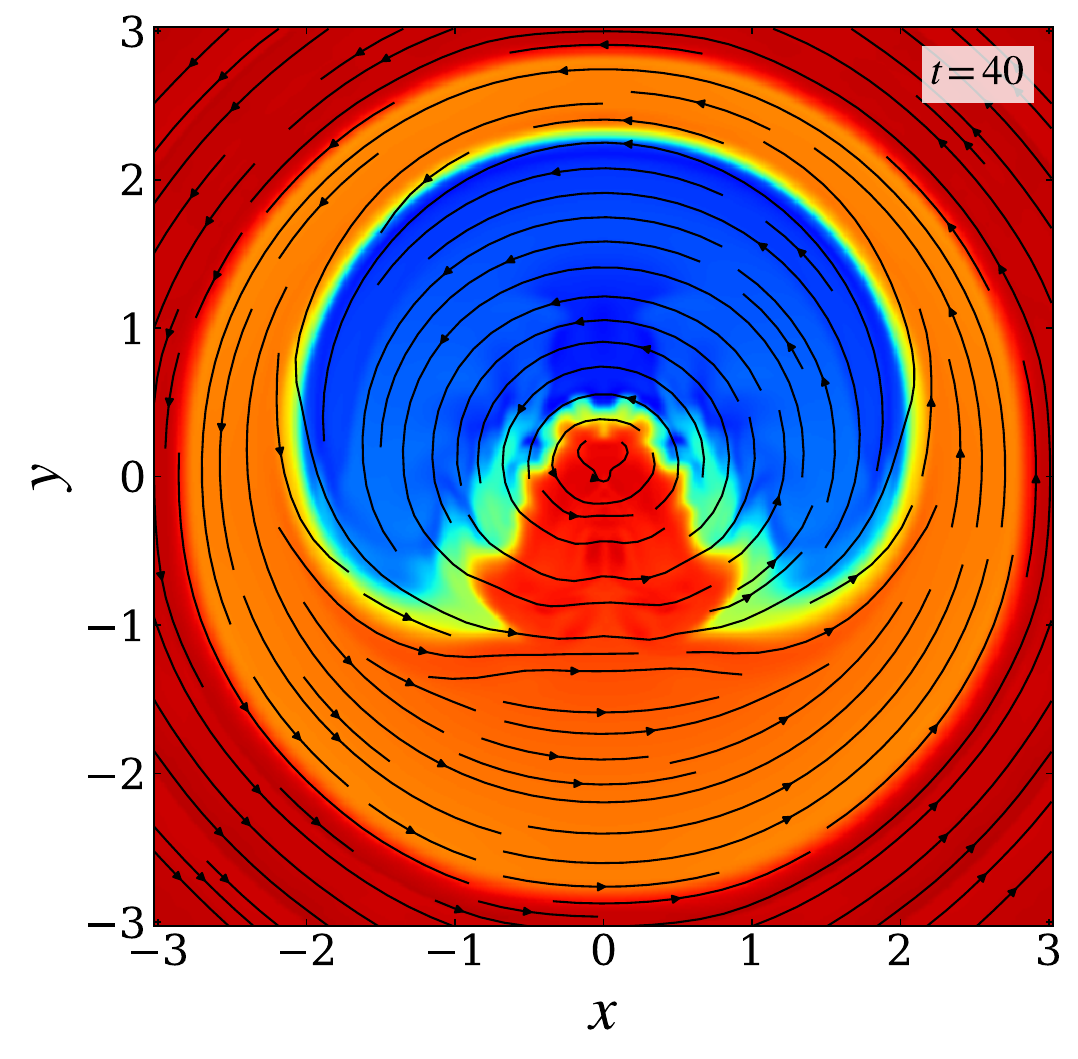}
    \includegraphics[width=0.195\linewidth]{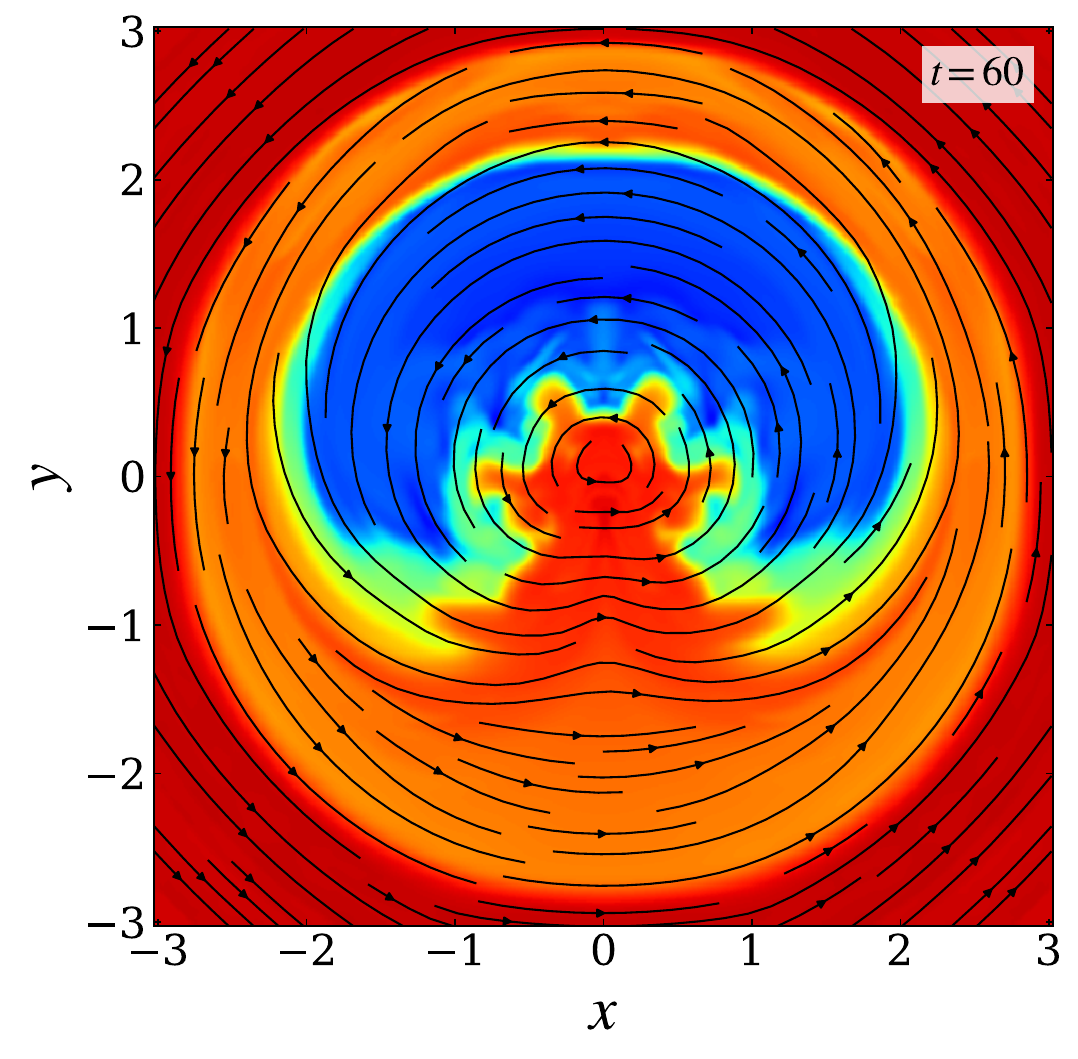}
    \includegraphics[width=0.195\linewidth]{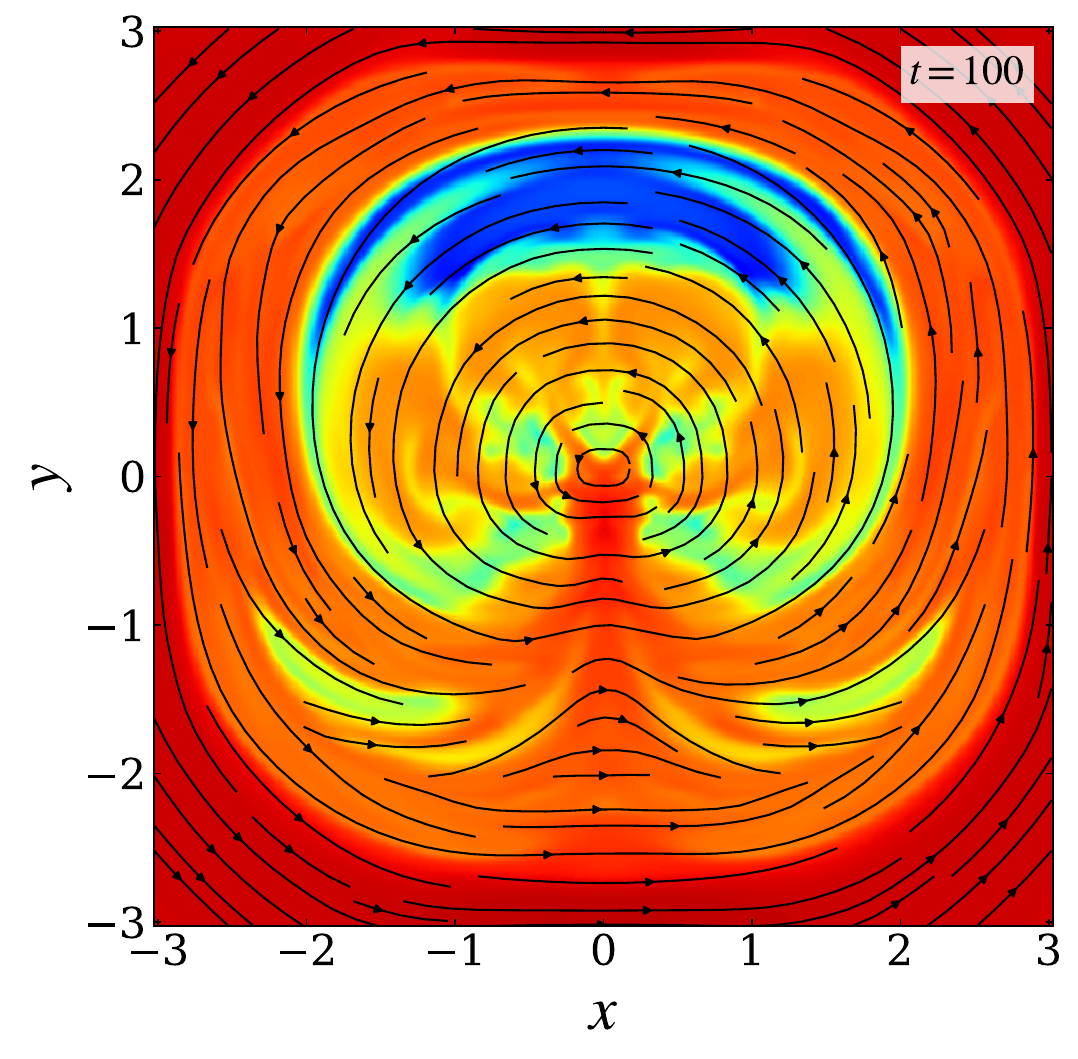}
    
     \includegraphics[width=0.195\linewidth]{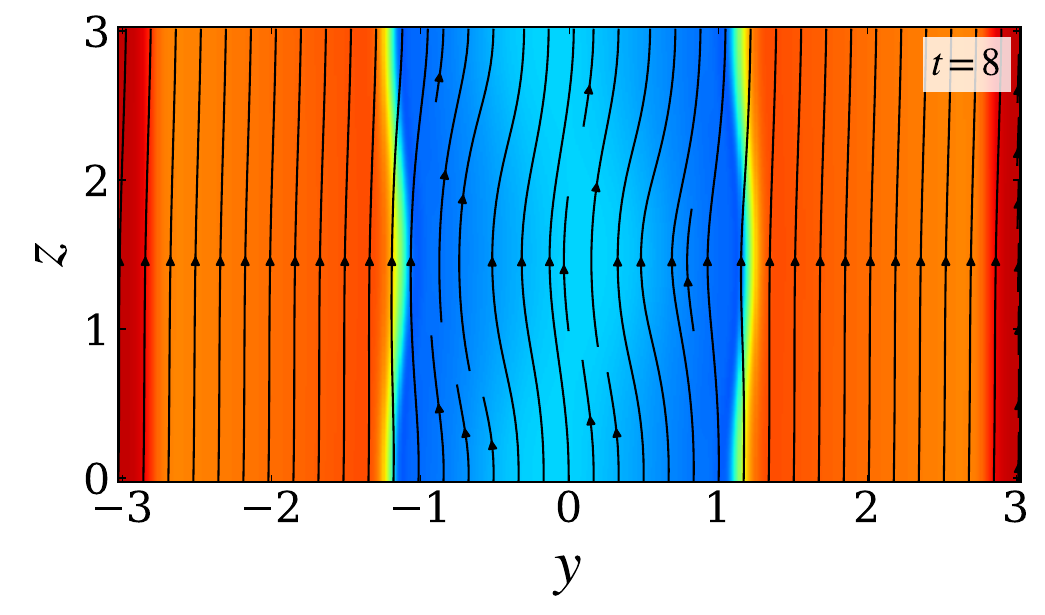}
    \includegraphics[width=0.195\linewidth]{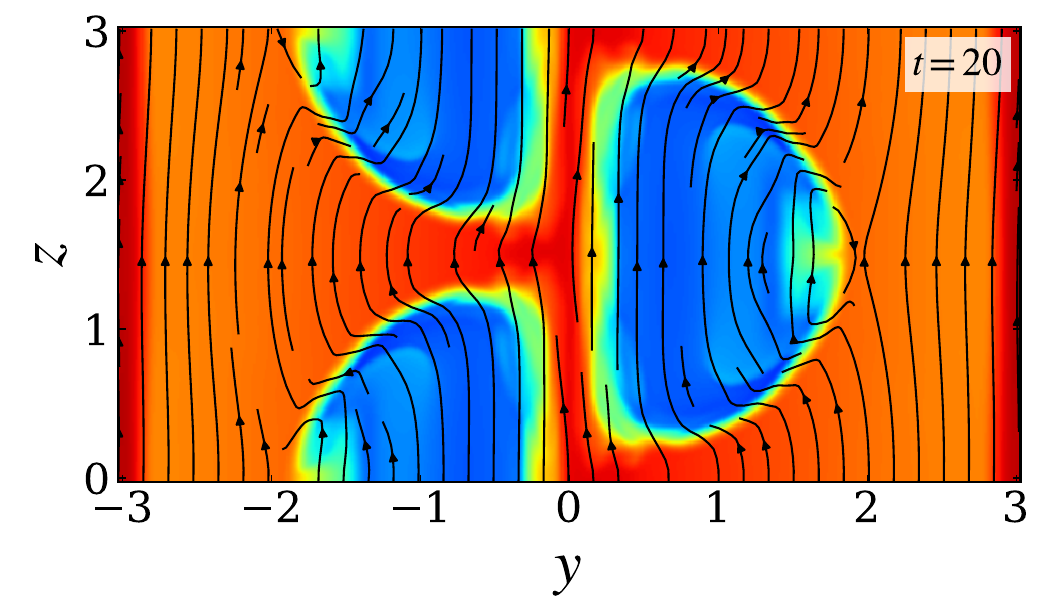}
    \includegraphics[width=0.195\linewidth]{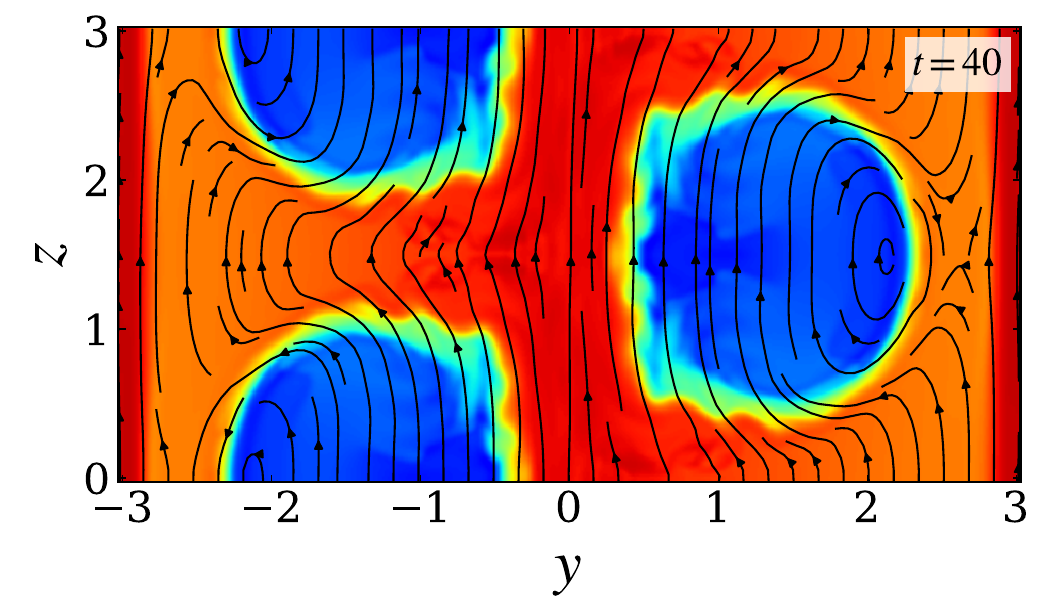}
    \includegraphics[width=0.195\linewidth]{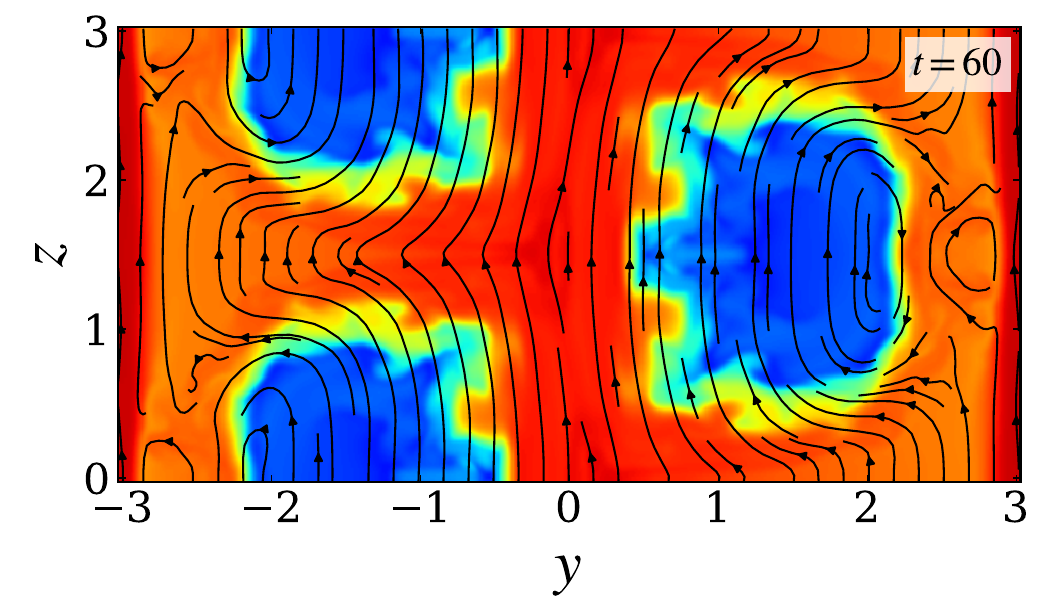}
    \includegraphics[width=0.195\linewidth]{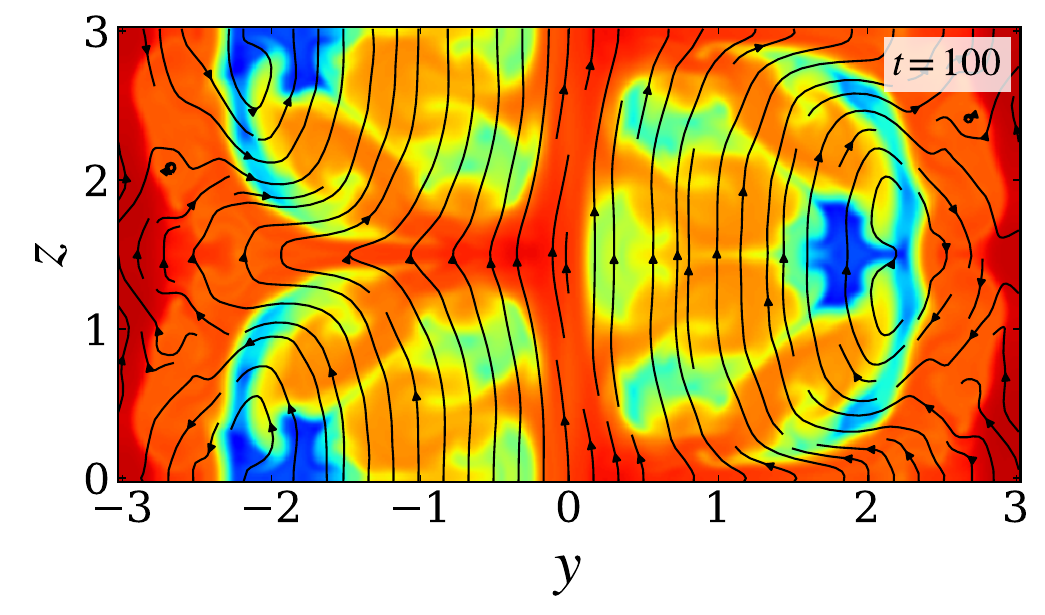}
    
    \includegraphics[width=0.925\linewidth]{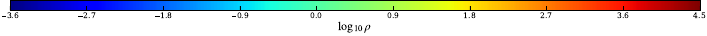}
    
    \caption{Evolution of the density distribution. The top two rows show the $M_1$ model in the horizontal ( $z=1.5$ ) and vertical ($x=0$) mid-planes of the computational domain.  The second two rows show the same for the $M_3$ model. In this and other similar plots of mid-plane slices, the lines show the tangent lines to the magnetic field vectors in the corresponding mid-plane.  They are neither the real magnetic field lines, which are three-dimensional structures, nor their projections on this plane.}
    \label{fig:rho-slices}
\end{figure*}

\begin{figure}
    \centering
    \includegraphics[width=0.7\linewidth]{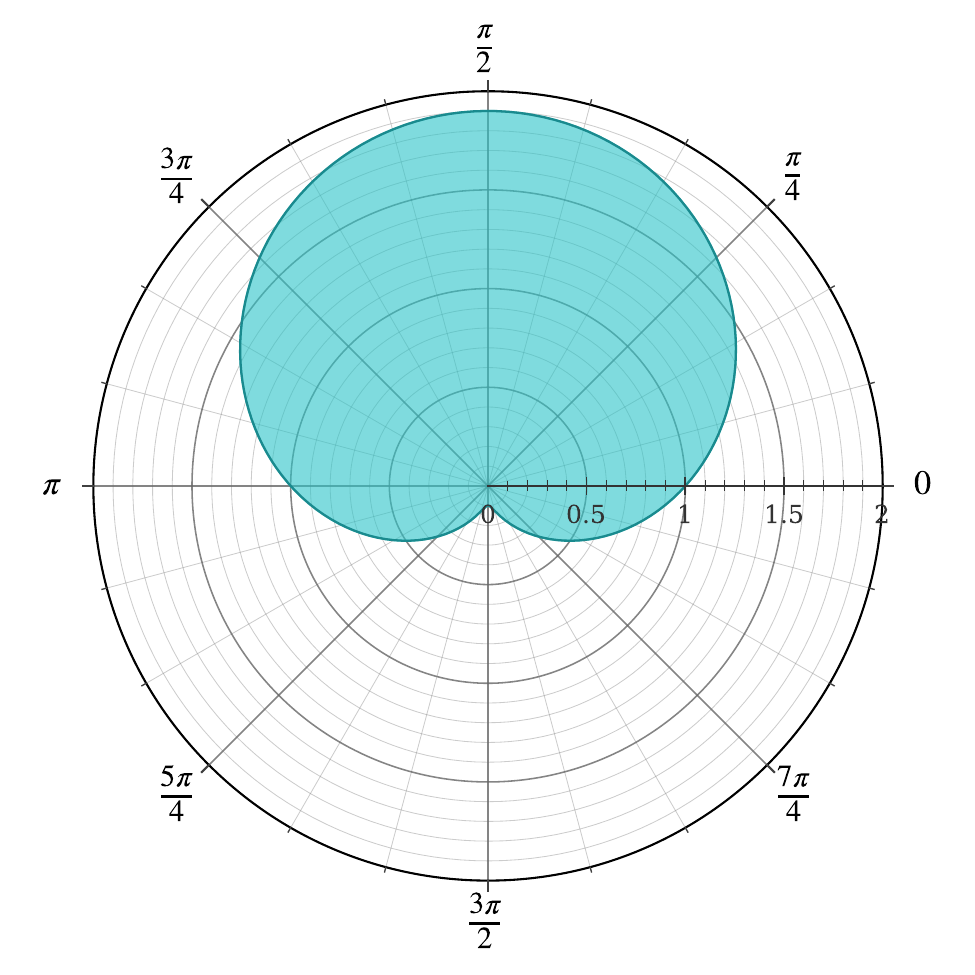}    
    \caption{Shape of the circle $r=1$ subject to the kink perturbation $\Delta r=0.9\cos(\phi-\pi/2)$.}
    \label{fig:nl-kink}
\end{figure}

\section{RESULTS}
\label{sec:results}

\subsection{Mildly magnetised wind} 
\label{sec:mmw}

This case was studied most rigorously, with the main suite of high-resolution simulations consisting of four models, which differ by the size of the force-free core $r_c$ of the initial equilibrium solution. They are denoted as  $M_1, M_2, M_3$ and $M_4$, and their core radius $r_c=P$ equals to 1/6, 1/4, 1/3, and 1/2 respectively (see table \ref{tab1} and figure \ref{fig:equilibrium}).   For all models, the Kruskal-Shafranov instability criterion  $L_z>2\pi r_c$, where $L_z$ is the size of the computational domain in the z direction, is satisfied both in the jet and the wind zones (see table \ref{tab1}).

\subsubsection{Linear phase}
\label{sec:linear}

Following  \citet{Mizuno_2009}, the growth of kink instability is investigated by measuring (i) $E_t$, the kinetic energy of transverse motion normalised to the initial value of the magnetic energy, (ii) $V_{r, max}$, the maximum value of the radial velocity, and (iii) $\langle \hat{B}_r\rangle$, the rms value of $B_r/||\vv{B}||$. 
The temporal evolution of these parameters obtained for the whole domain is shown in figure \ref{fig:growth}. For all three parameters, the data indicate a reduction of the growth rate with the magnetic core radius, in general agreement with the findings by \citet{Mizuno_2009}. 

The highly inhomogeneous nature of the equilibrium solution, however, suggests that the instability may not develop at a single rate for the whole domain. The most suitable parameter for exploring this possibility seems to be $\langle \hat{B}_r\rangle$ as it directly reflects the distortion of the initial axisymmetric magnetic configuration.   The left and middle panels of figure \ref{fig:Br-growth} show the evolution of this parameter inside the volumes corresponding in the initial solution to the jet and wind, respectively. The plots indicate the phase of exponential growth in both these regions. However, one can also see that in the jet region this phase gets established very quickly, whereas in the wind region it takes much longer to set in. As the result, there is a significant lag between the times for the instability to reach the same amplitude.   To illustrate this further, the right panel of figure \ref{fig:Br-growth} shows the evolution of $\langle \hat{B}_r \rangle$ measured in eight cylindrical shells of equal width $ \Delta r = 1/3$ and increasing radius for the model $M_3$.  One can see that by the time $t=8$, when $\langle \hat{B}_r \rangle$ has reached its saturation value near the z-axis, it is still very low in the wind zone.  Moreover, it has hardly changed compared to the value induced by the initial perturbation.  However, starting from $t\approx 8$, the value of $\langle \hat{B}_r \rangle$ in the wind zone begins to increase, and its spatial evolution is reminiscent of an outgoing wave.

Table \ref{tab:2} shows the linear growth rates based on the $\langle \hat{B}_r \rangle(t)$ data obtained for the jet, wind, and the whole domain using ordinary least-squares (OLS) regression. One can see that the growth rates are higher in the jet than in the wind, but not by much, thus reaffirming the conclusion that the main reason for the delayed development of the instability in the wind is its longer setting-in period.  Figure \ref{fig:Br-growth} also shows that when the growth of the kink mode in the jet saturates, its growth rate in the wind slows down. This is particularly pronounced in the $M_4$ model, where $\langle \hat{B}_r \rangle(t)$ stops growing in the jet and its growth in the wind slows down at about the same time, $t\approx 20$.   

Linear stability analysis of a static column with the force-free magnetic field given by equations (\ref{eq:ff-Bz},\ref{eq:ff-Bphi}) produced the maximum growth rate 
\beq
\Gamma_g^{max} = 0.133 \frac{c_a}{r_c} \,,
\label{eq:G-appl}
\eeq  
where $c_a$ is the Alfv\'en speed given by the equation $c_a^2=B^2/(w+B^2)=\sigma/(1+\sigma)$ \citep{Appl-00}. This growth rate is reached at $\lambda_{m}=8.43 r_c$, which is only slightly higher than the marginal-stability value of $2\pi r_c$. The last column of table \ref{tab:2} shows the values of $\Gamma_g^{max}$  predicted by this equation for $c_a=1$ (the jet value). These are systematically lower, though not much lower, than those found in our simulations.  The equilibrium solution explored in \citep{Appl-00} is different from ours by extending to infinity in the radial direction, with an asymptotic boundary condition corresponding to outgoing sound waves.  So some difference is expected. 
 
 \begin{figure*}
    \centering

    \includegraphics[width=0.35\linewidth]{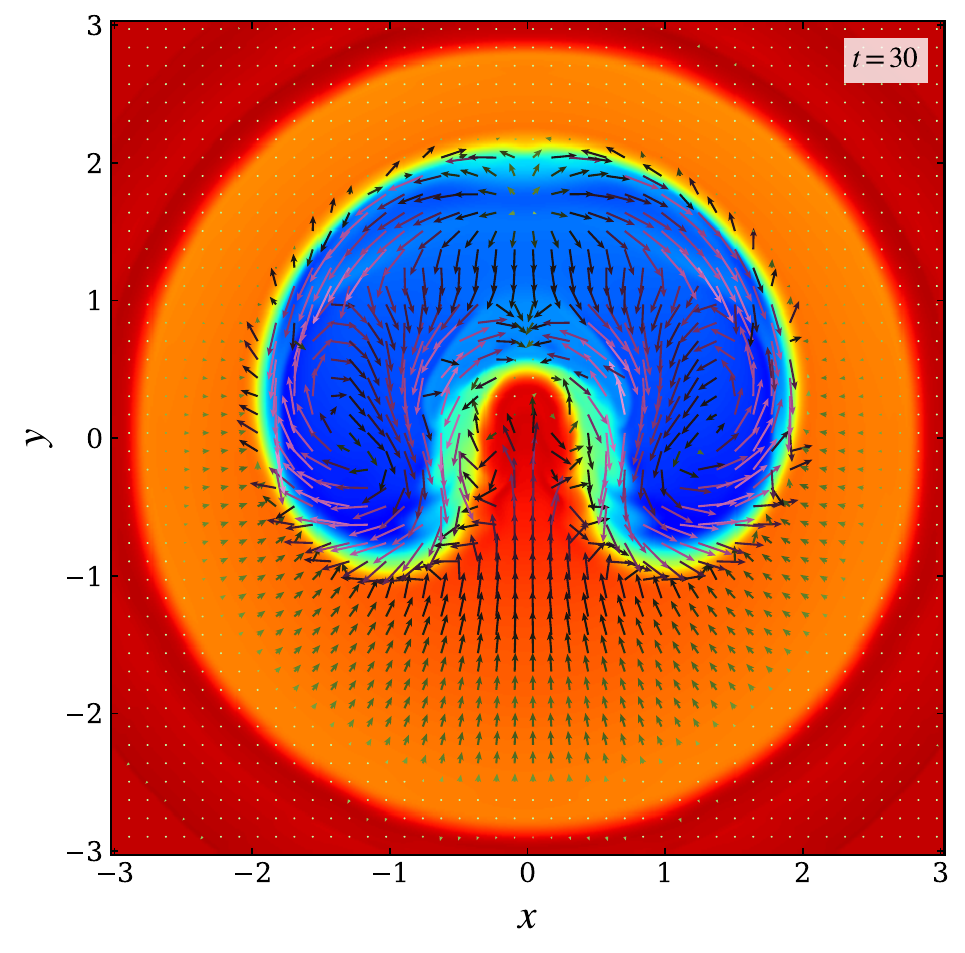}
    \includegraphics[width=0.35\linewidth]{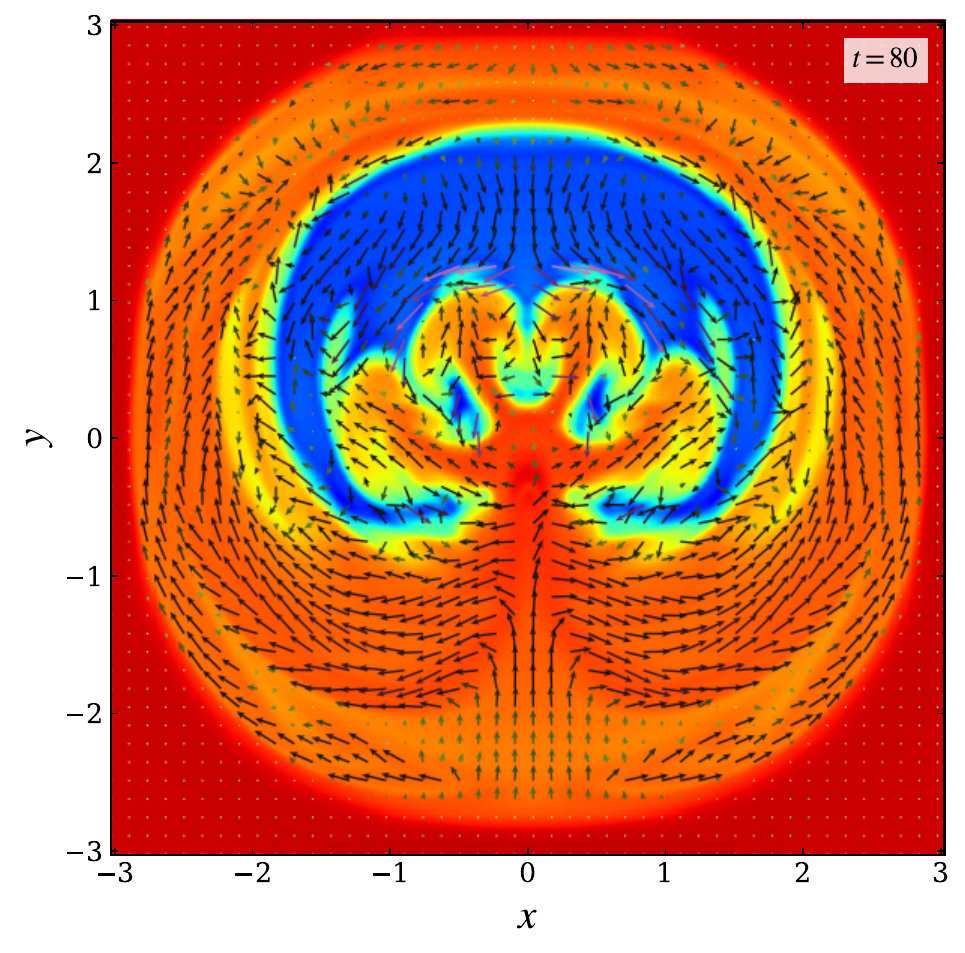}
    \includegraphics[width=0.7\linewidth]{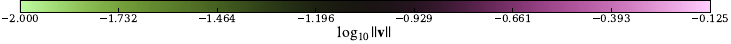}
     \includegraphics[width=0.69\linewidth]{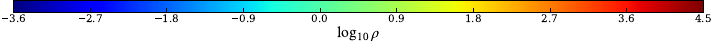}

 \caption{The velocity field in the $M_3$ model before and after the development of the secondary instability. The left panel shows the solution at $t = 30$ and the right one at $t=80$ in the horizontal mid-plane ( $z=1.5$ ).  The length of the velocity arrows and their colour scale as $\log_{10}||\vv{v}||$.  The colour maps show the rest mass density $\rho$ on a logarithmic scale too.}
    \label{fig:M3-v}
\end{figure*}

\begin{figure*}
    \centering        
    \includegraphics[width=0.242\linewidth]{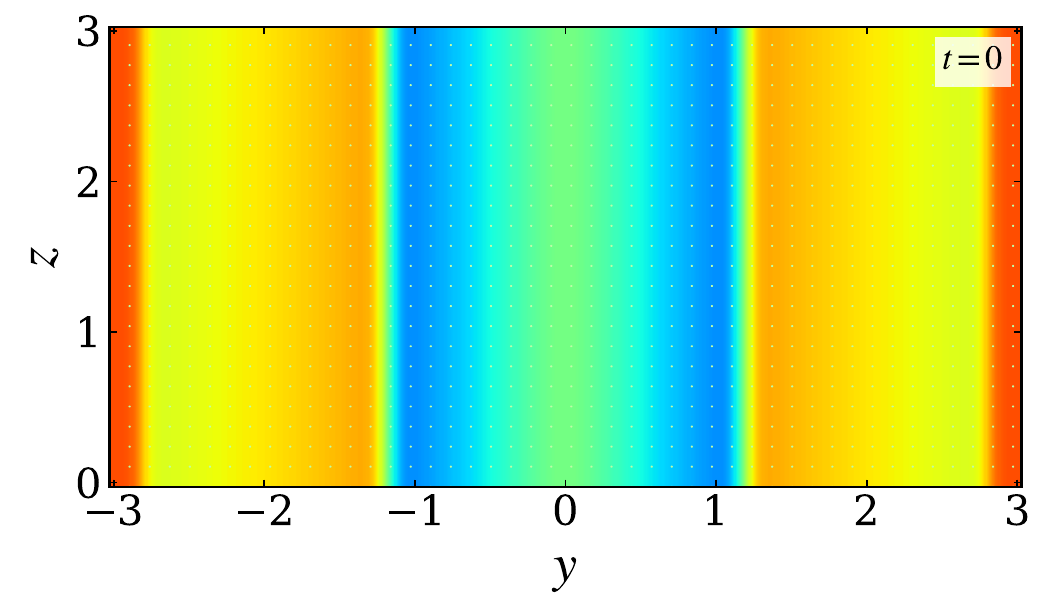}
    \includegraphics[width=0.242\linewidth]{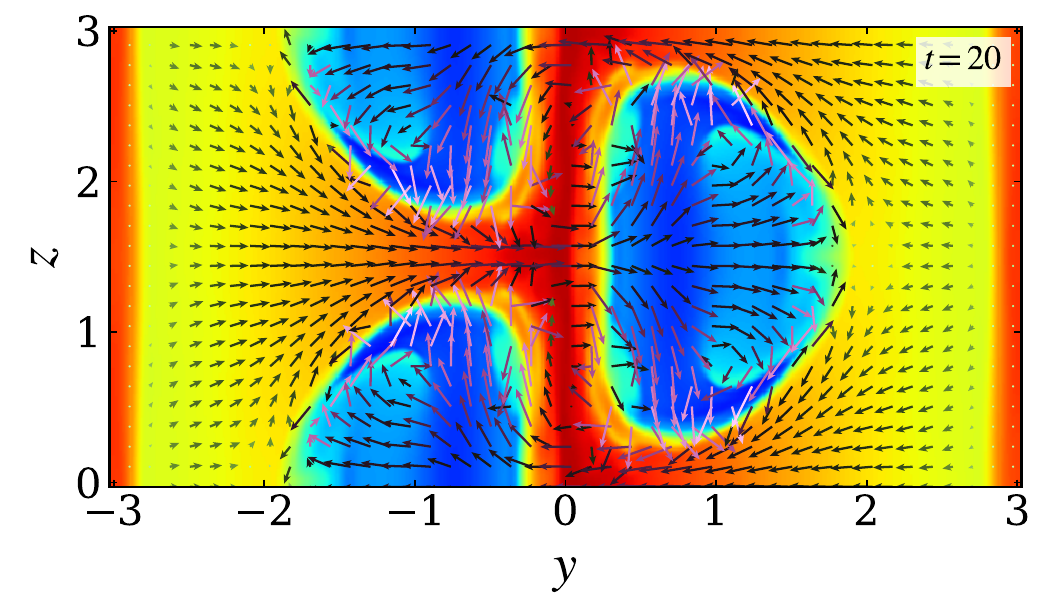}
    \includegraphics[width=0.242\linewidth]{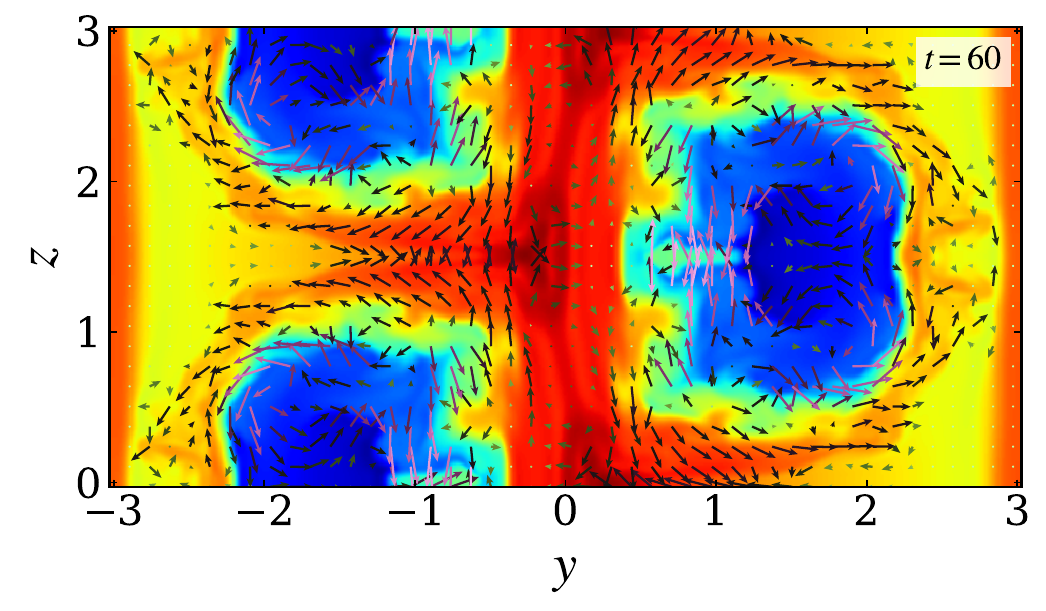}
    \includegraphics[width=0.242\linewidth]{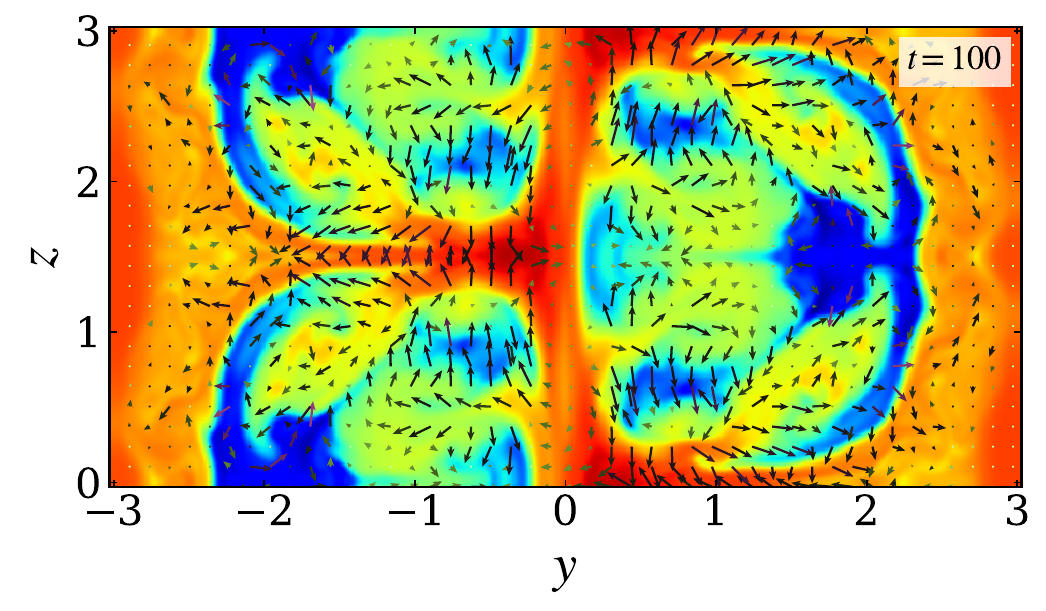}
    \includegraphics[width=0.925\linewidth]{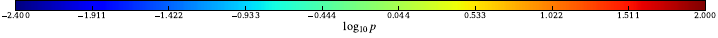}
    
    \includegraphics[width=0.242\linewidth]{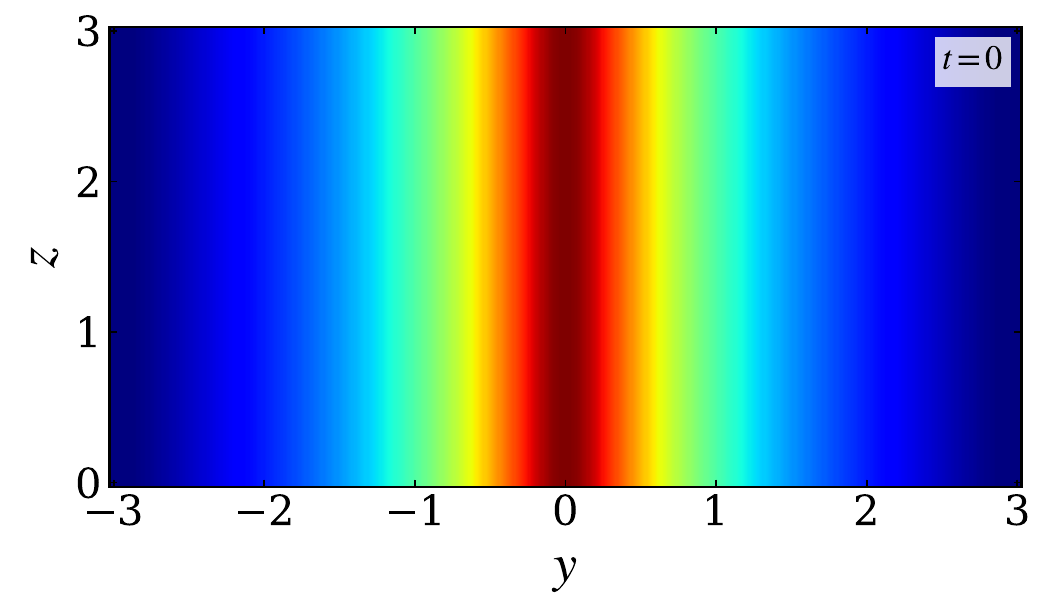}
    \includegraphics[width=0.242\linewidth]{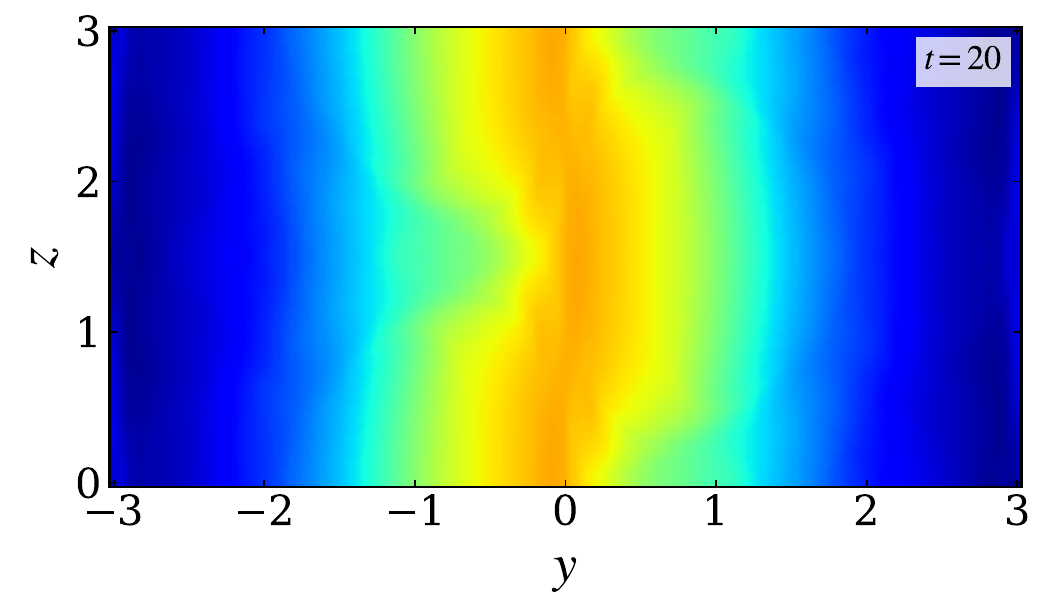}
    \includegraphics[width=0.242\linewidth]{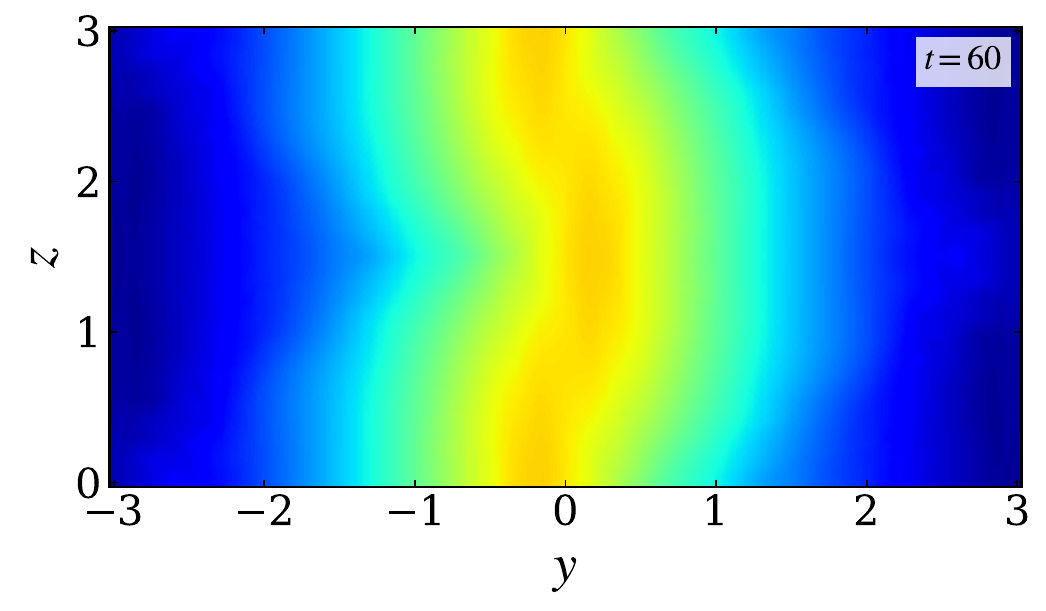}
    \includegraphics[width=0.242\linewidth]{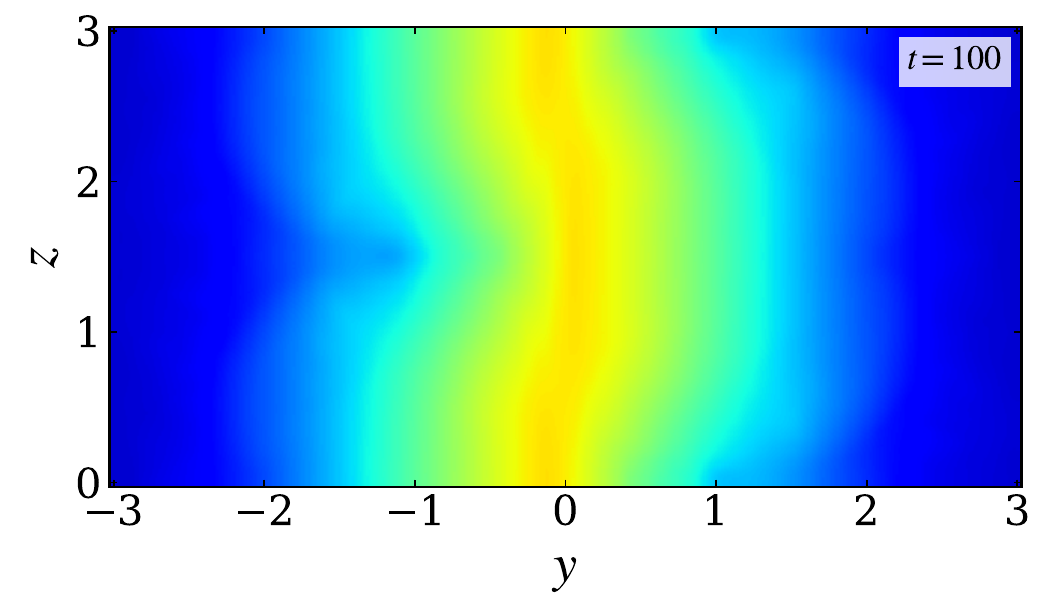}
    \includegraphics[width=0.925\linewidth]{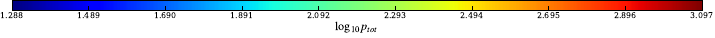}
    
    \includegraphics[width=0.242\linewidth]{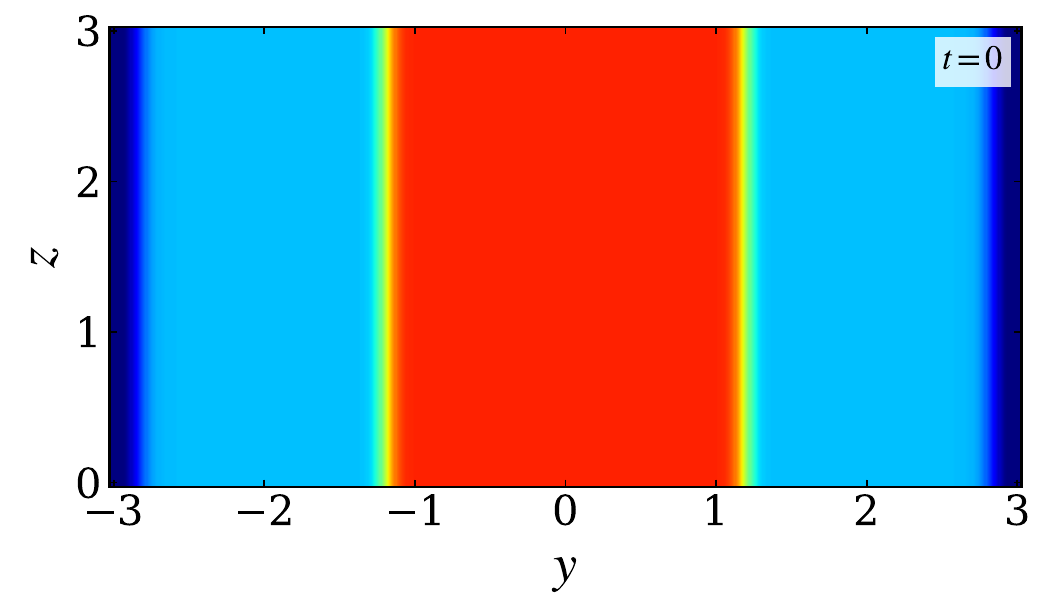}
    \includegraphics[width=0.242\linewidth]{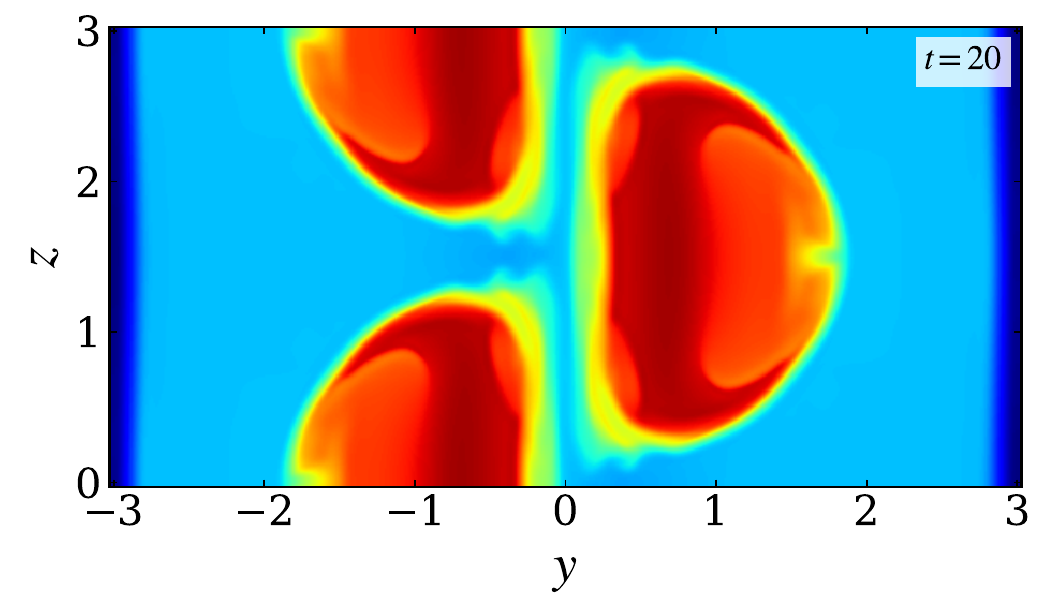}
    \includegraphics[width=0.242\linewidth]{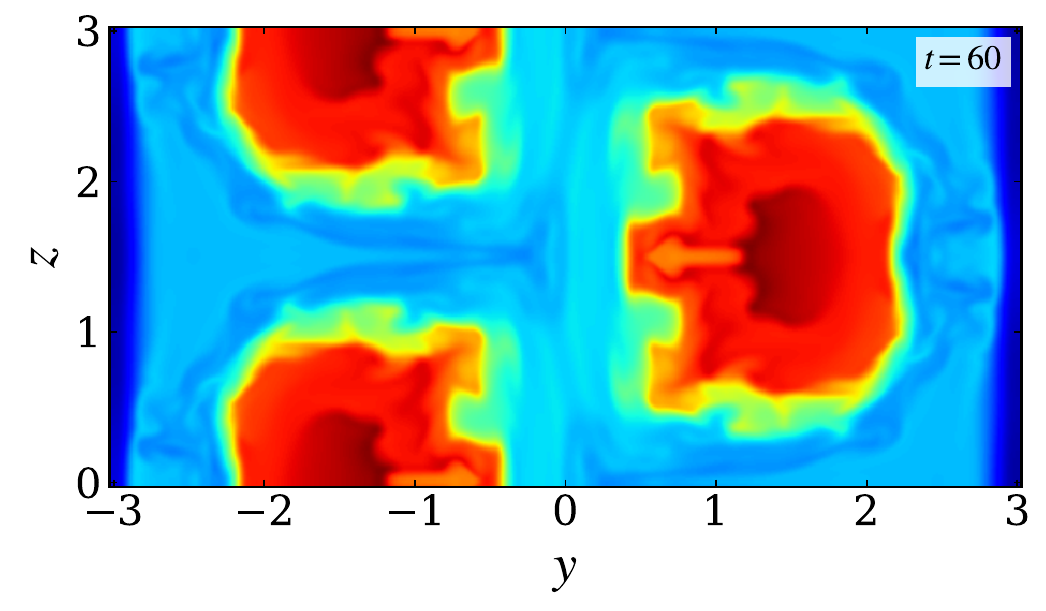}
    \includegraphics[width=0.242\linewidth]{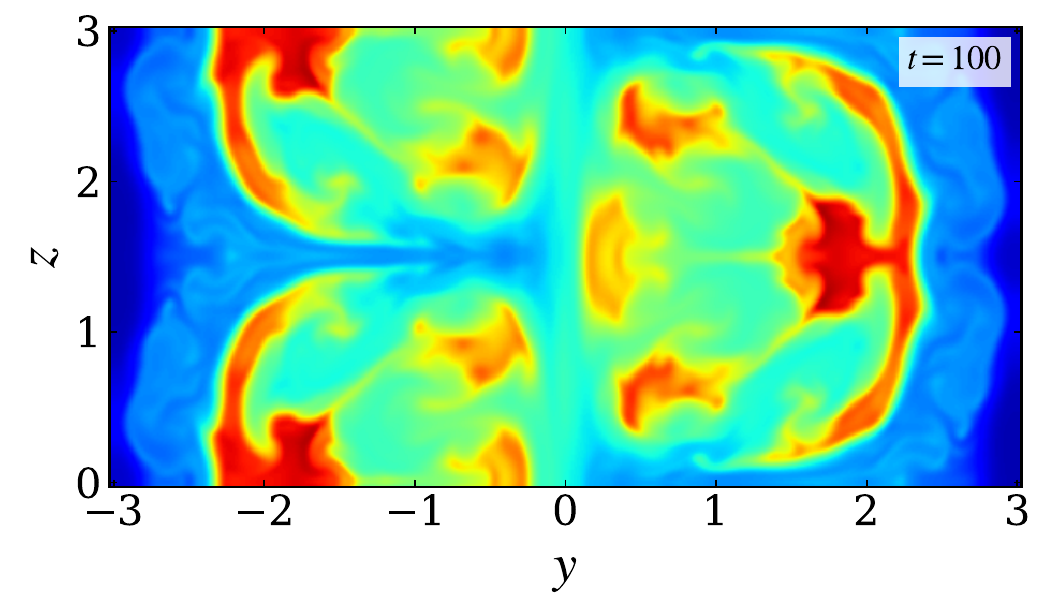}
     \includegraphics[width=0.925\linewidth]{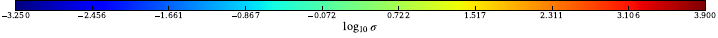}

    \caption{Evolution gas pressure $p$ (top row), total pressure $p_{tot} =p +p_m$ (middle row), and the magnetisation $\sigma$ (bottom row) in the mid-plane $x=0$ for the $M_3$ model.}
    \label{fig:M3-others}
\end{figure*}

\subsubsection{Nonlinear phase}
\label{sec:nonlinear}

Fig.~\ref{fig:isosurf} shows the evolution of the density iso-surface with $\rho = 10^{-3/4})$, which separates the jet and wind in the initial solutions, and the magnetic field lines at the non-linear phase of the kink instability.  This value of $\rho$ corresponds to the separation layer between the jet and the wind zones in the initial solution (see figure \ref{fig:equilibrium}).   The footpoints of plotted magnetic field lines are located in the $z=0$ plane slightly outside of the iso-surface, by about 4 computational cells.  Like in the previous studies, the iso-surface develops the characteristic of kink instability corkscrew shape.  As a result, some of the magnetic foot-points end up inside the jet and become invisible in the plots, whereas others end up outside of it and hence describe the magnetic field of the wind.  The radial size of the corkscrew increases with time until reaching the final relaxed state. This takes longer in models with larger core radius $r_c$.   

The corkscrew shape signifies interpenetration of the jet and wind on the scale set by the wavelength of the kink mode.  Early on, the separation surface is quite smooth. However, soon after its overall structure relaxes, it develops conspicuous ripples, which appear to be aligned with local magnetic field lines.  These ripples, not seen in the simulations of  \citet{Mizuno_2009}, suggest the development of a secondary instability on a smaller scale.

Figure \ref{fig:rho-slices} gives further information on the origin of the ripples by showing the evolution of the density distribution in the horizontal ($z=1.5$)  and vertical ($x=0$) mid-planes of the computational domain.  All four models exhibit qualitatively the same behaviour.  At first, as the jet cross-section moves away from the initial symmetry axis (the point $x=y=0$ in the plots), it retains a smooth boundary but changes its shape from the original disk into something resembling lobes of kidney beans. In figure  \ref{fig:rho-slices}, this phase is represented by the $t=8$ images of the model $M_1$ and $t=20$ images of the model $M_3$. The kidney-bean structure is seen in every horizontal cross section of the solutions and hence is a characteristic of their global structure.  Such structure has been seen in the simulations by \citet{Mizuno_2009}. Its origin can be explained if one extrapolates the effect of kink perturbation on the cross-section shape of the plasma column from linear to non-linear amplitudes. To illustrate this, figure \ref{fig:nl-kink} shows the line $r=r_0 + \delta r \cos(\phi-\pi/2)$ for $r_0=1$ and $\delta r=0.9r_0$. The kidney-bean shape is manifest there.      

As time progresses, the central protuberance of wind plasma responsible for the kidney-bean morphology grows in size and then spreads out via several distinct streams (fingers).  In figure  \ref{fig:rho-slices}, these streams become prominent at $t=40$ for the model $M_1$, and at $t=60$ for the model $M_3$. The development of fingers coincides in time with the appearance of ripples of the density iso-surface seen in figure \ref{fig:isosurf}.  In models $M_1$ and $M_2$,  the fingers eventually stop growing in size and diffuse, leaving most of the ''kidney'' lobe intact - compare the $M_1$ images at $t=60$  and $t=100$ in figure \ref{fig:isosurf}.  However, in models $M_3$ and $M_4$, they are seen to penetrate and spread throughout the whole lobe via plumes and arc-like features -- see the $M_3$ images at $t=100$ in figure \ref{fig:isosurf}.  

Figure \ref{fig:M3-v} illustrates the difference in the velocity field before ($t=30$) and after ($t=80$) the development of secondary instability in the $M_3$ model. At $t=30$, the wind plasma streams towards the symmetry axis ($x=y=0$) of the initial solution. This stream seems to produce the central protuberance, which turns the kidney-bean shape of the jet cross-section into a travel-pillow shape.  This stream is driven by the tension of the wind's magnetic field, which gets stretched by the kink of the jet column. In the jet, the velocity field is more complicated, with a pronounced vortical component.  At $t=80$, the velocity field of the wind shows (i) the flow along the magnetic field lines winded round the jet,  driven by the gas pressure, and (ii) deep penetration of the jet via four sub-streams originated from the original protuberance. This structure is repeated in other cross-sections of the domain, but its orientation changes following the corkscrew structure. So the 3D geometry of the penetrating streams is helical.                             

To provide further details, figure \ref{fig:M3-others} shows the evolution of a few more fluid parameters in the model $M_3$. It is consistent with the above description of strong regular inflow of wind plasma at the nonlinear phase of the kink mode, its eventual deceleration about the z-axis, and the development of the secondary instability. The initial regular inflow of wind plasma is accompanied by its compression, which is manifest in the gas pressure distribution (top panel).  The distribution of total pressure $p_{tot}=p+p_m$ develops a kink, but its amplitude is few times smaller than in the density distribution. The overall increase of $p_{tot}$ towards the z-axis implies that it is the magnetic tension force that drives the wind plasma towards the axis. The jet magnetisation increases locally by almost ten-fold during the kink phase. Its overall distribution becomes highly inhomogeneous after the development of the secondary instability. The magnetisation of plasma in dense RT-figures increases presumably due to numerical diffusion.

The secondary instability can be related to the growth of  CD modes with $m>1$, triggered by the kink instability \citep{LBA-00}. However, the longitudinal wave number $k$ of these modes grows like $k=m/P$, and so for $m=4$ perturbation one would expect the longitudinal wavelength $\lambda\approx L_z/4$, which is inconsistent with the alignment of ripples with the magnetic field lines.     

The emergence of mushroom-like plumes at the non-linear phase of the secondary instability hints that it could be the magnetic Rayleigh-Taylor instability \citep[e.g.][]{chandra-stability,JNS-95,TH-25}. When the jet develops its corkscrew shape, the wind plasma gets pulled by the magnetic forces into the vacated space inside the jet coil. Then the initially rapid coil expansion slows down, and the interface between the jet and the wind plasma decelerates. In the frame of the interface, the fluids on both sides of the interface are subject to the inertial force equivalent to gravity, with the gravitational acceleration vector $\vv{g}$ pointing radially outwards. On the inner side of the coil, this is equivalent to heavy fluid being placed on top of the light fluid, which is a Rayleigh-Taylor unstable configuration, whereas on the outer side of the coil it is the other way around. This interpretation is consistent with the observed secondary instability developing predominantly on the inner side of the coil and mostly during the late phase of the coil expansion. One may also expect to see this instability on the outer side of the coil at the early time when its expansion is accelerated. Although some evidence of it is seen in the simulations ( for example, in the images of figure \ref{fig:rho-slices} for the model $M_3$ at $t=20$ ), but it never reaches such a large amplitude as on the inner side. The difference may be connected to the different shapes of the interface on the inner and outer sides of the coil.

\begin{figure}
    \centering
    \includegraphics[width=0.8\linewidth]{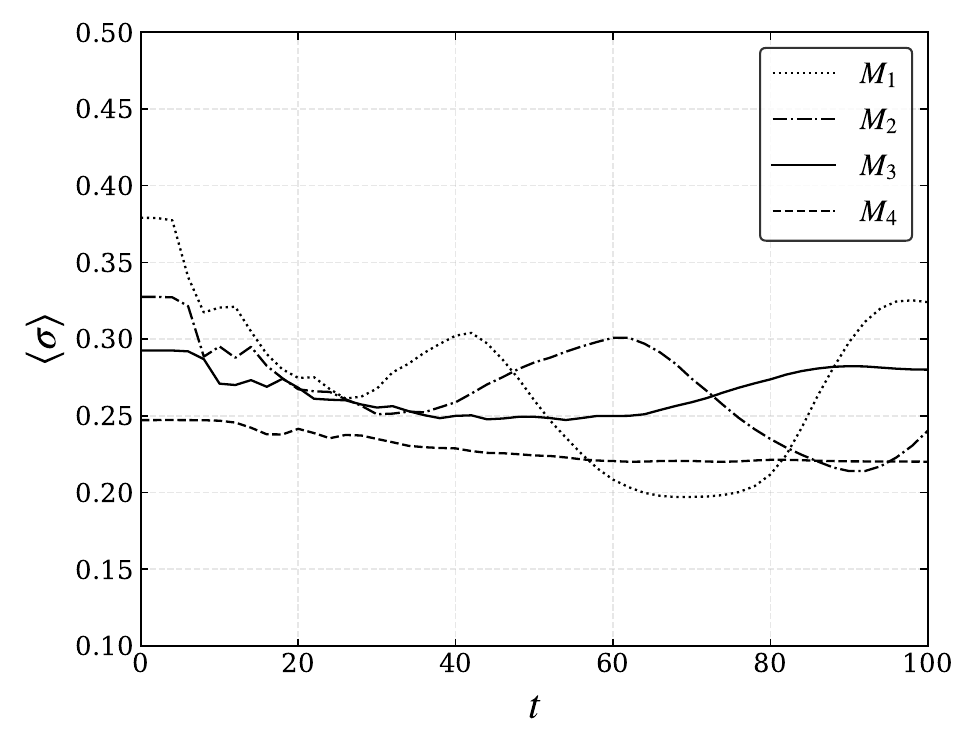}
    
    \caption{Temporal evolution of $\langle\sigma\rangle$ in the entire jet-wind region.}
    \label{fig:sigma2}
\end{figure}

\begin{figure*}
    \centering
    \includegraphics[width=0.195\linewidth]{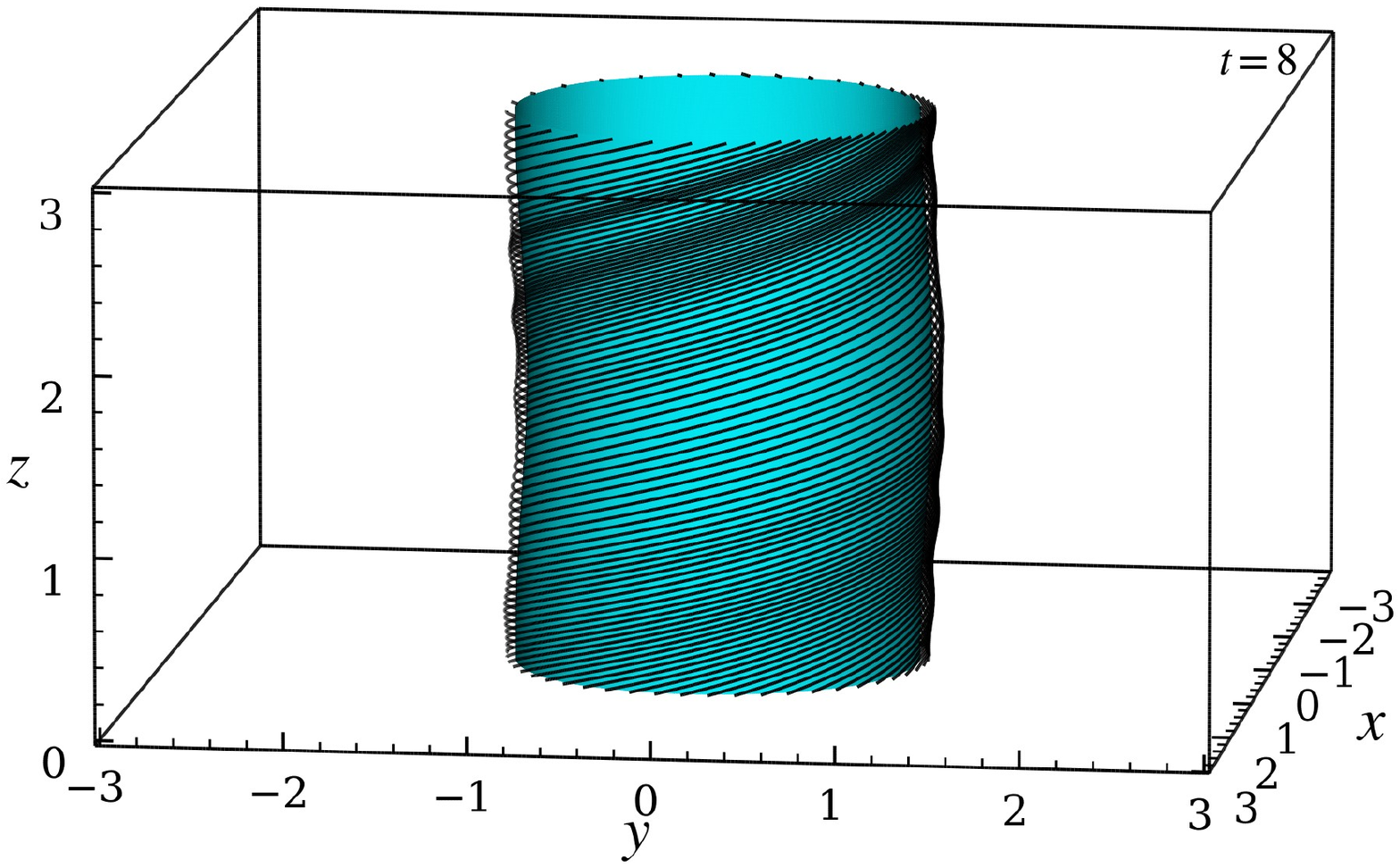}
    \includegraphics[width=0.195\linewidth]{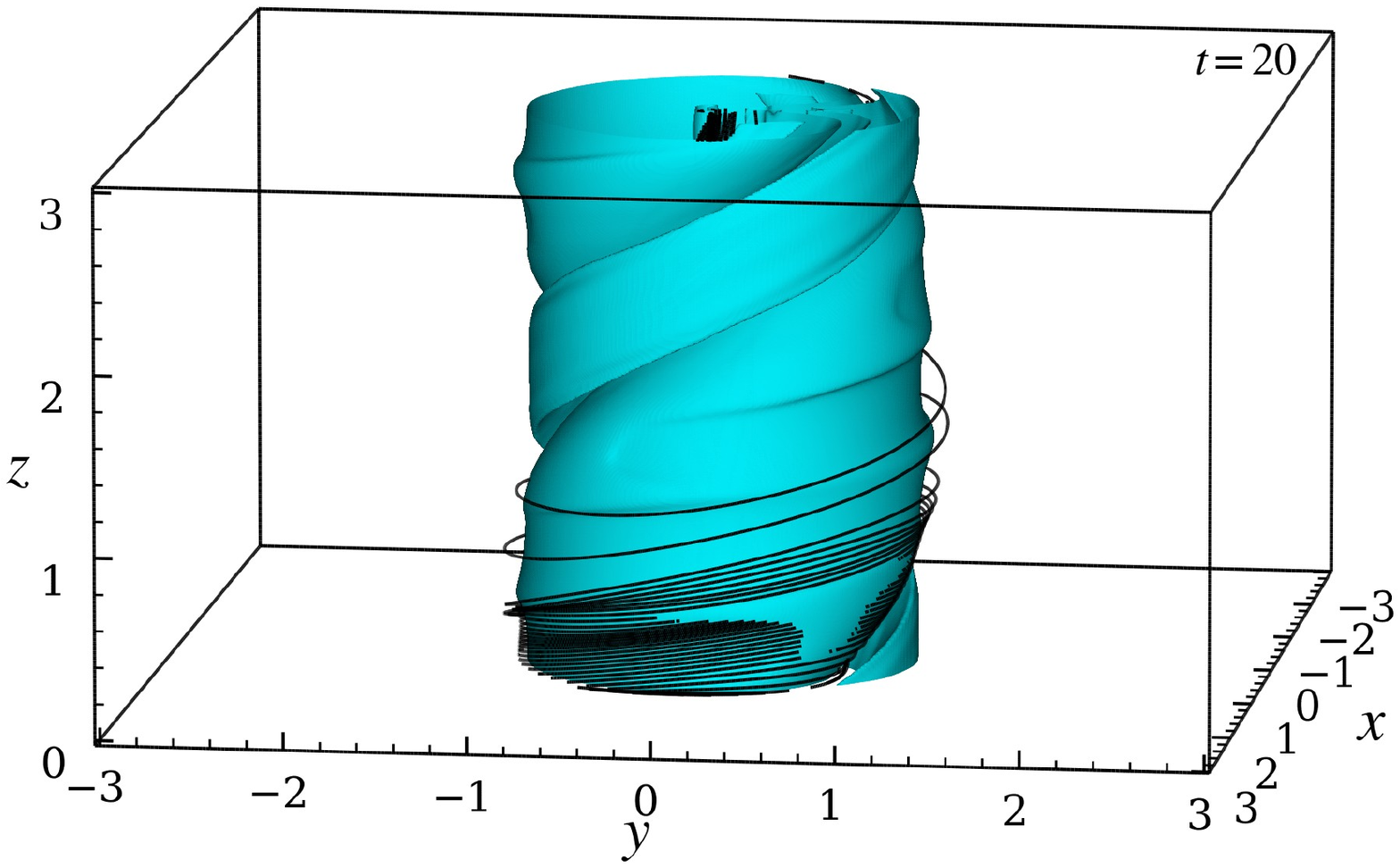}
    \includegraphics[width=0.195\linewidth]{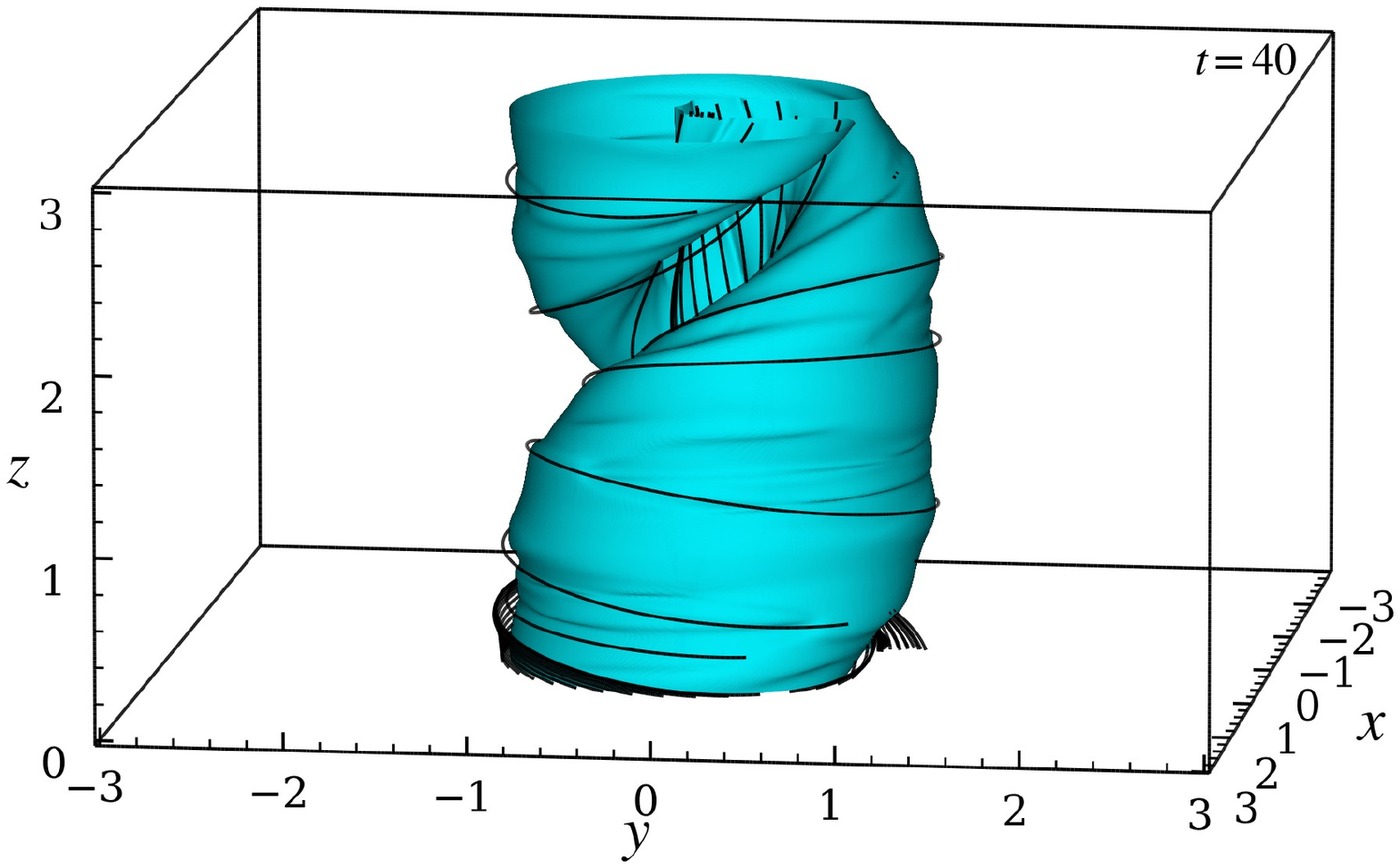}
    \includegraphics[width=0.195\linewidth]{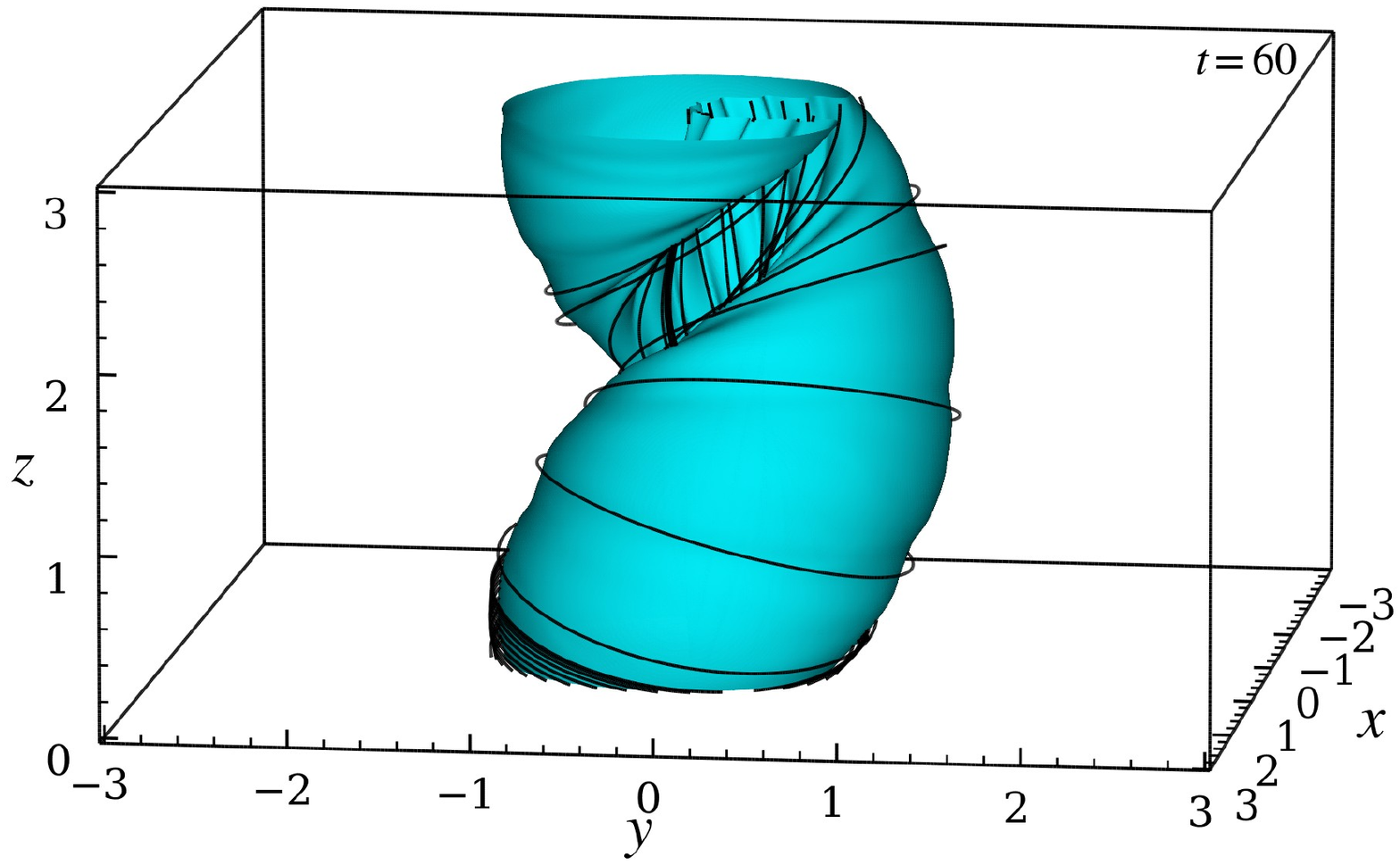}
    \includegraphics[width=0.195\linewidth]{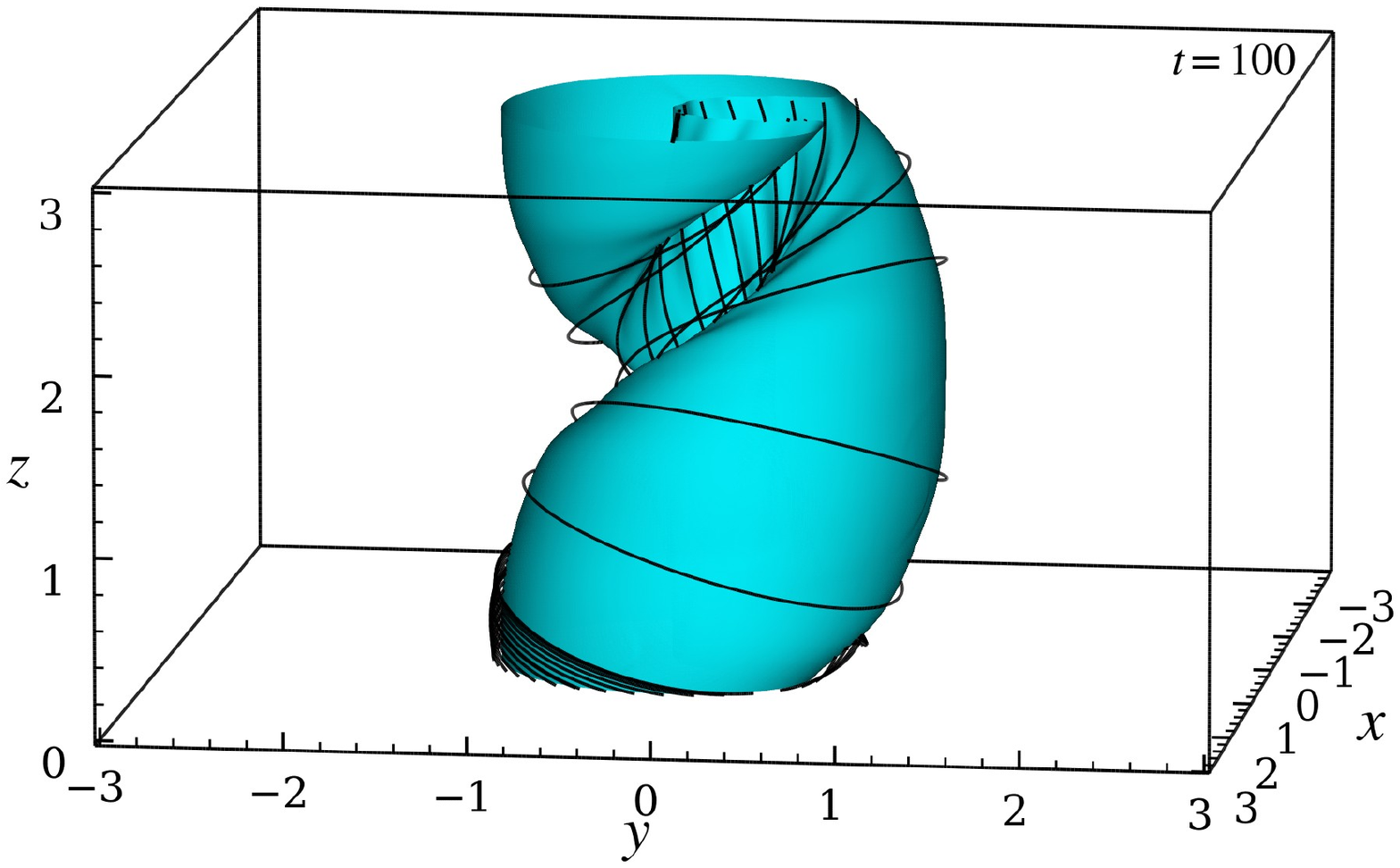}  
    
\caption{Evolution of the density iso-surface separating the jet from the wind in the modified $M_3$ model with weakly magnetised wind  ($\epsilon_w = 1$ and $\sigma_w = 0.001$). The lines show the magnetic field with footpoints located in the plane $z=0$ just outside of the iso-surface.}
    \label{fig:isosurf-M6}
\end{figure*}

\begin{figure*}
    \centering
     \includegraphics[width=0.195\linewidth]{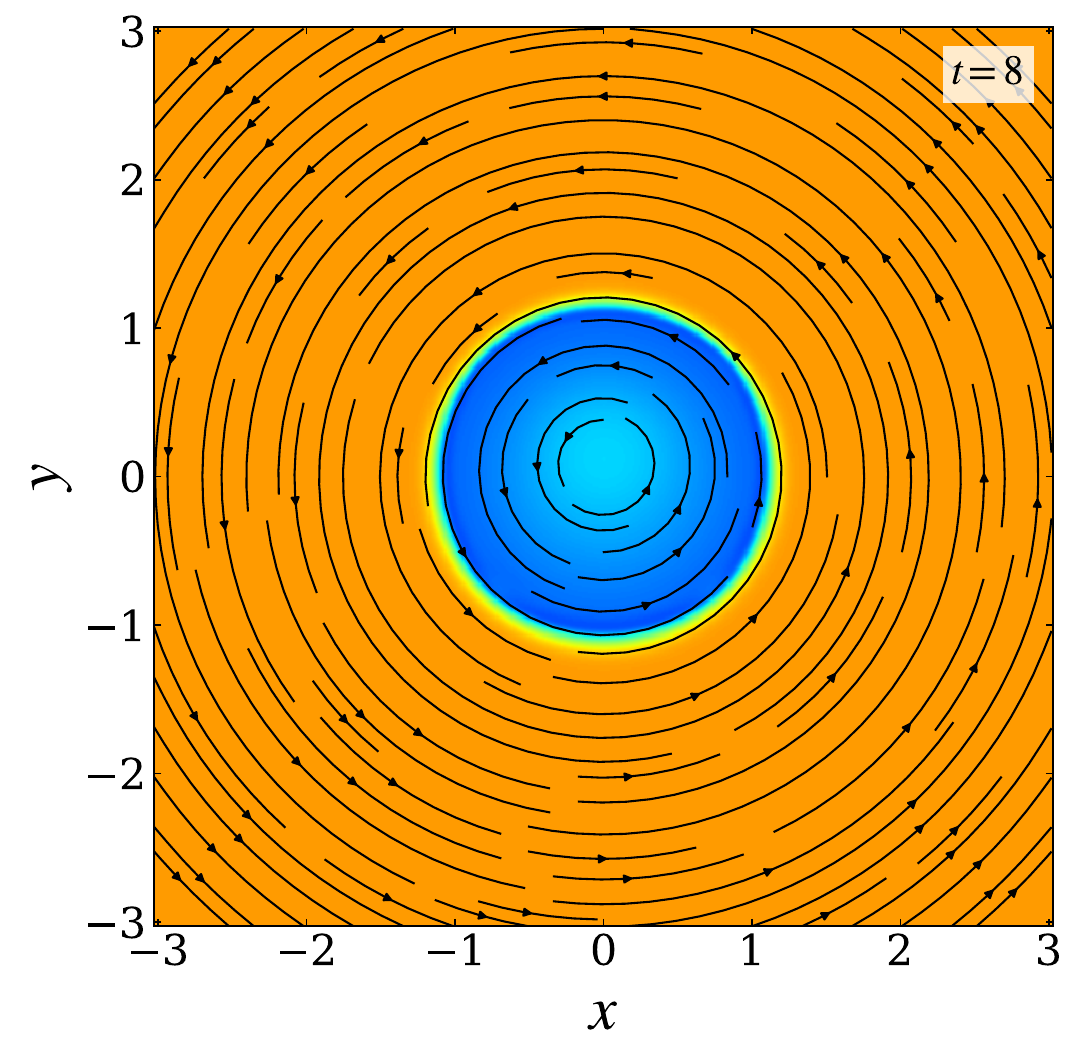}
    \includegraphics[width=0.195\linewidth]{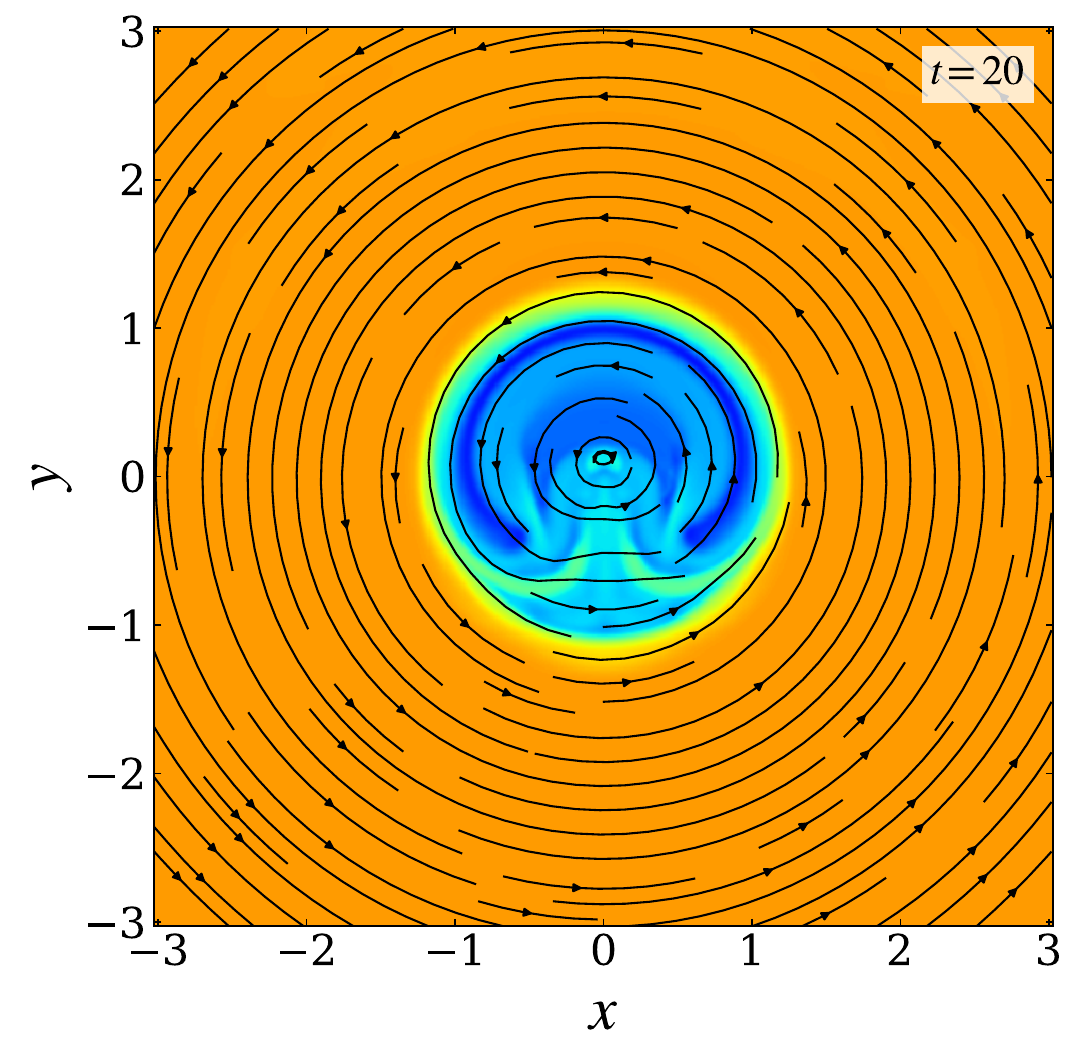}
   \includegraphics[width=0.195\linewidth]{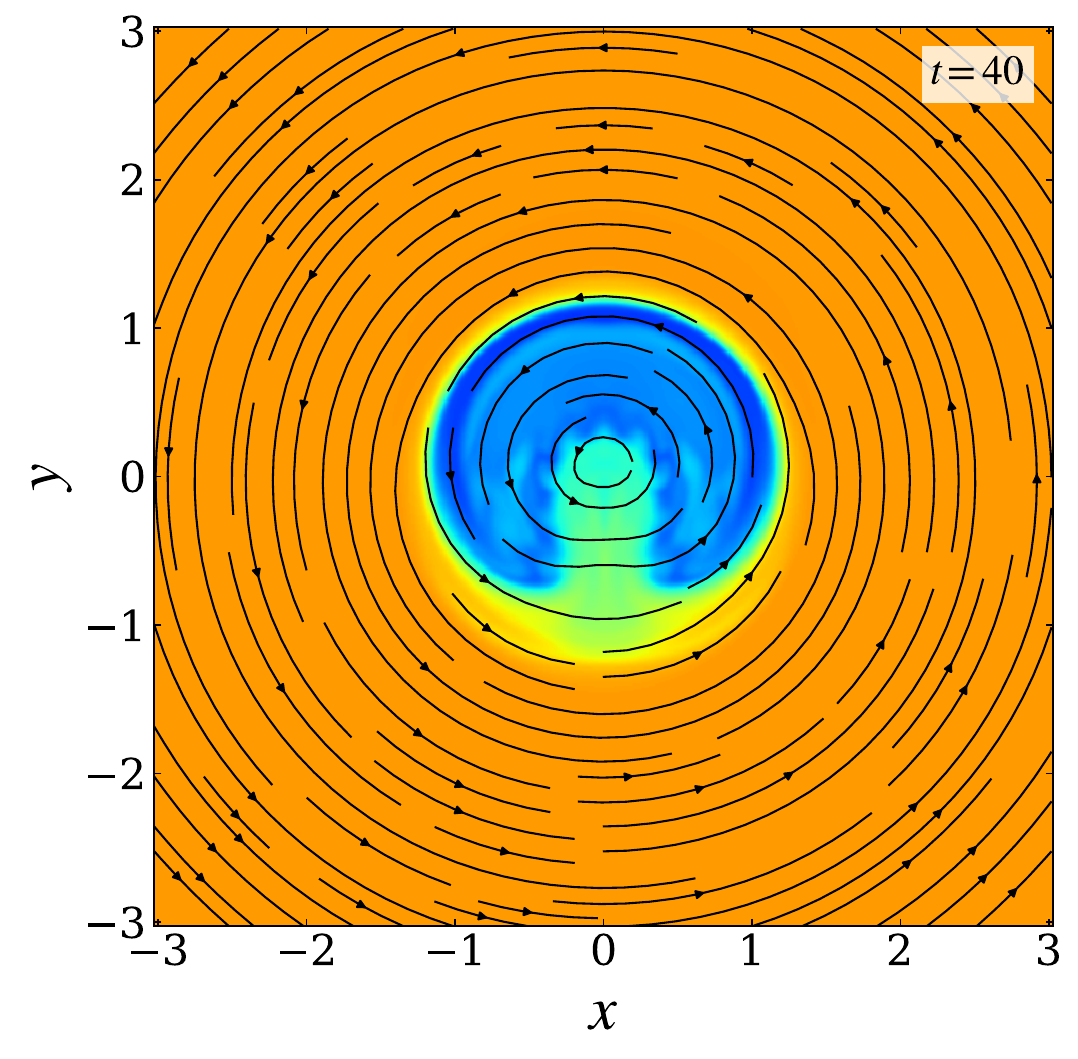}
    \includegraphics[width=0.195\linewidth]{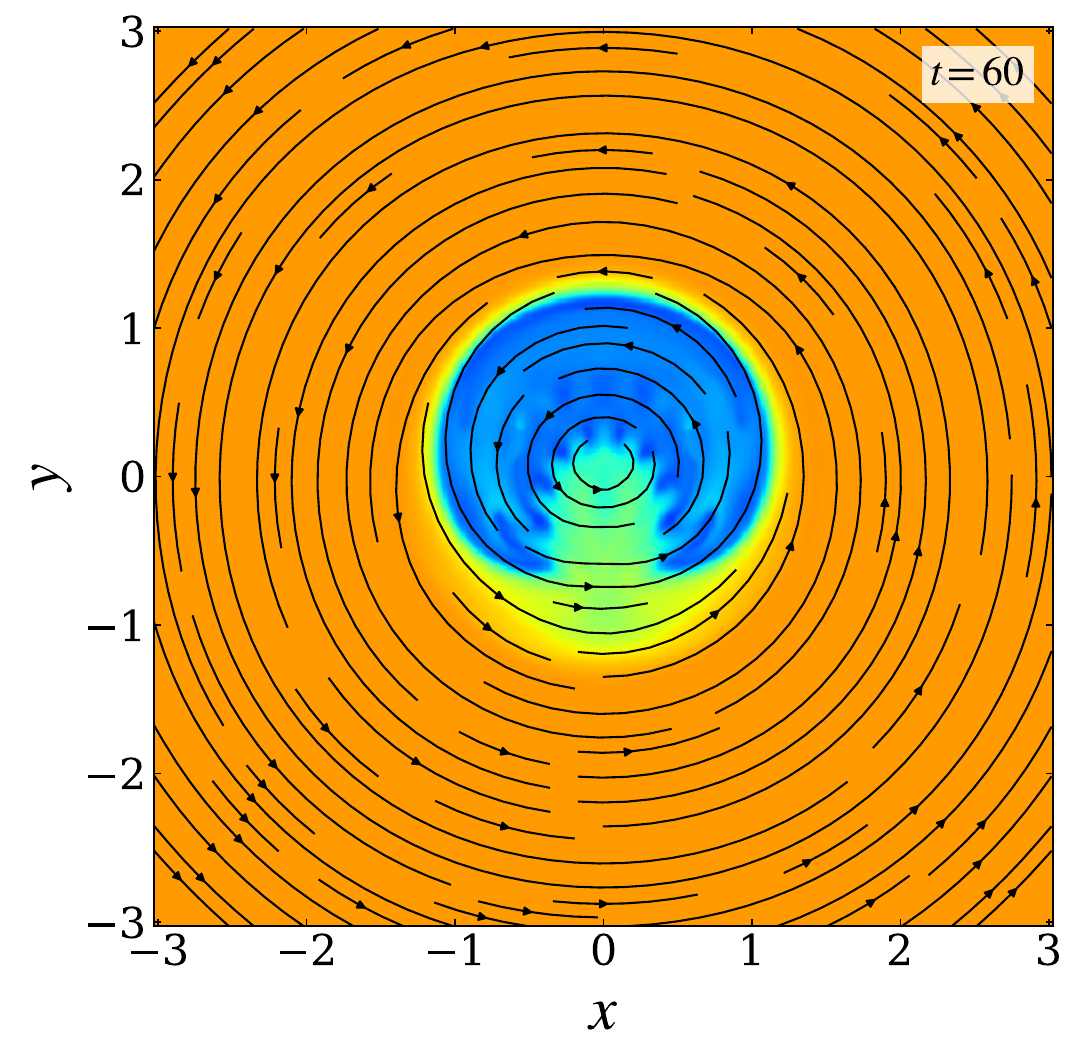}
   \includegraphics[width=0.195\linewidth]{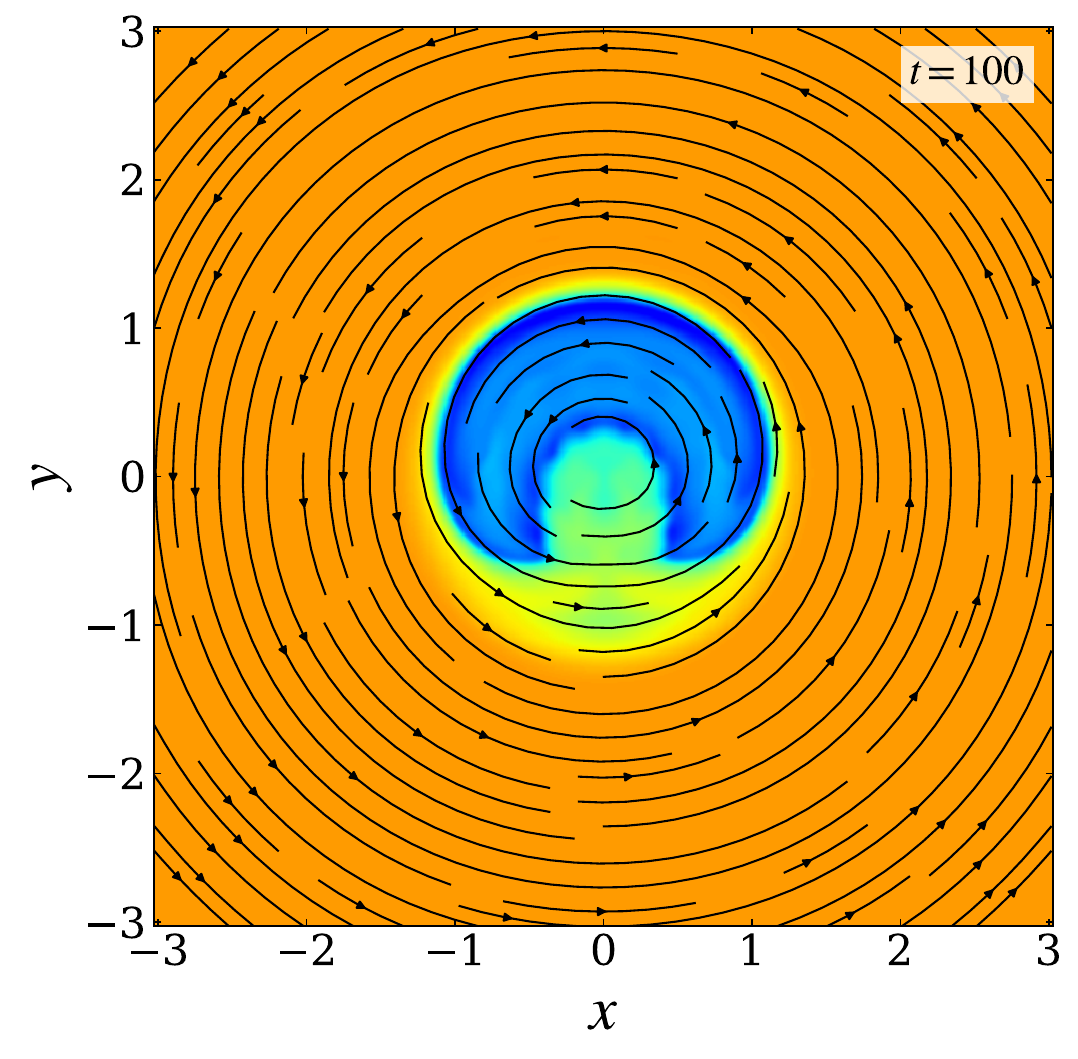}
     \includegraphics[width=0.195\linewidth]{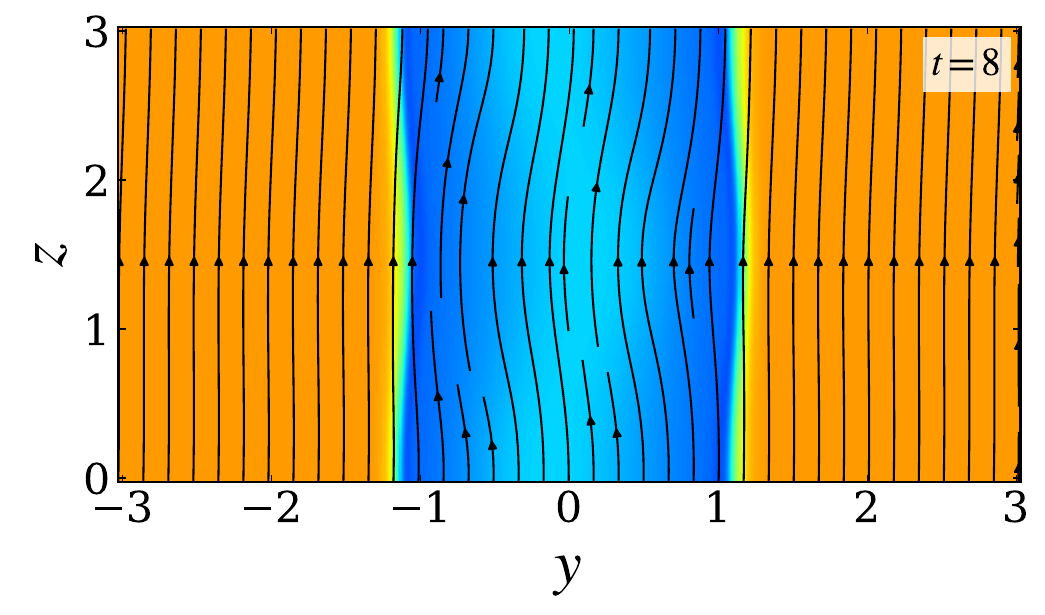}
    \includegraphics[width=0.195\linewidth]{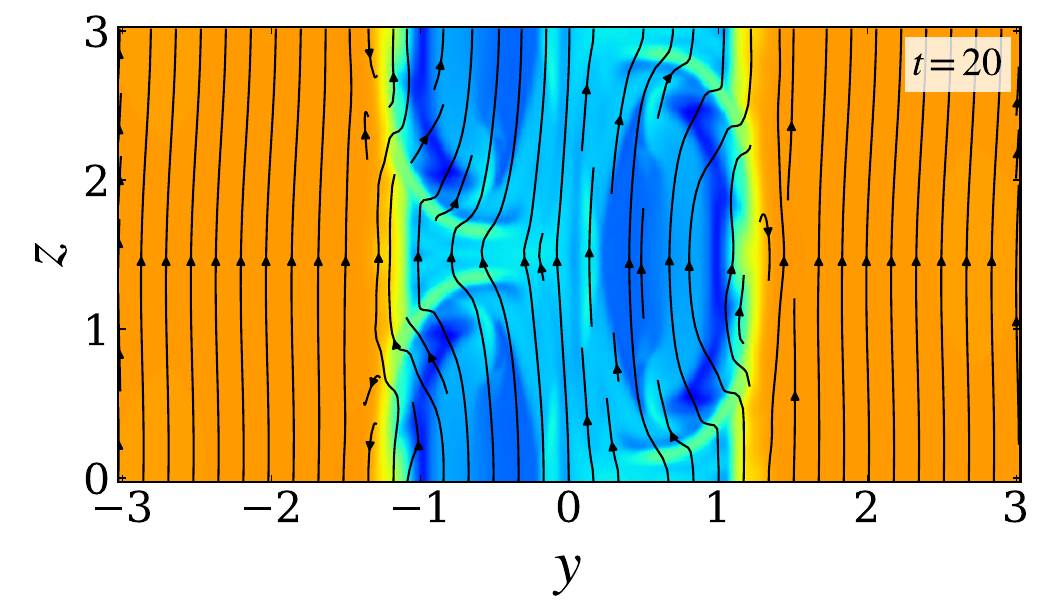}
   \includegraphics[width=0.195\linewidth]{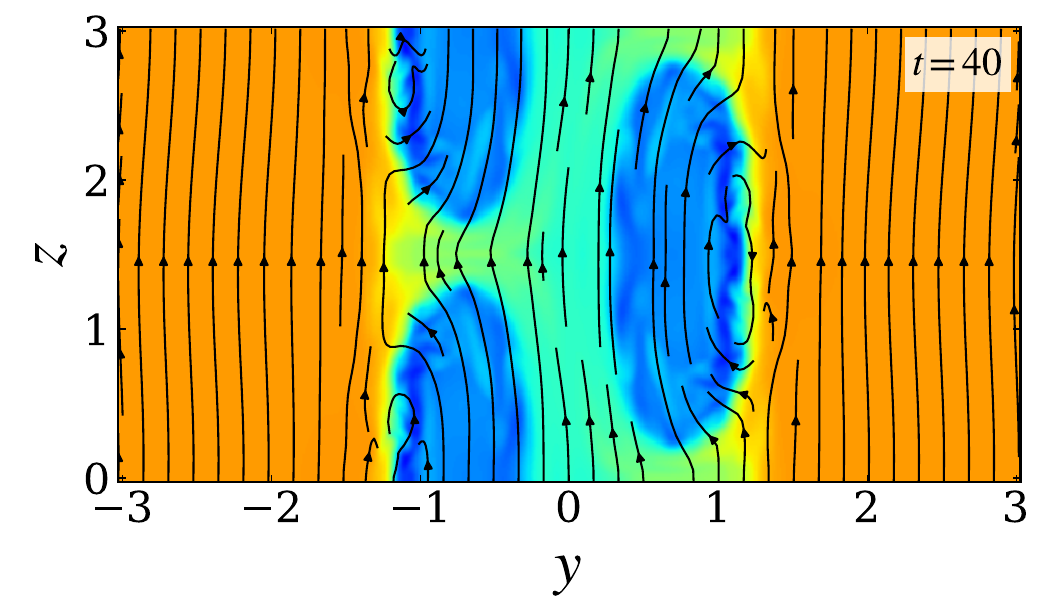}
    \includegraphics[width=0.195\linewidth]{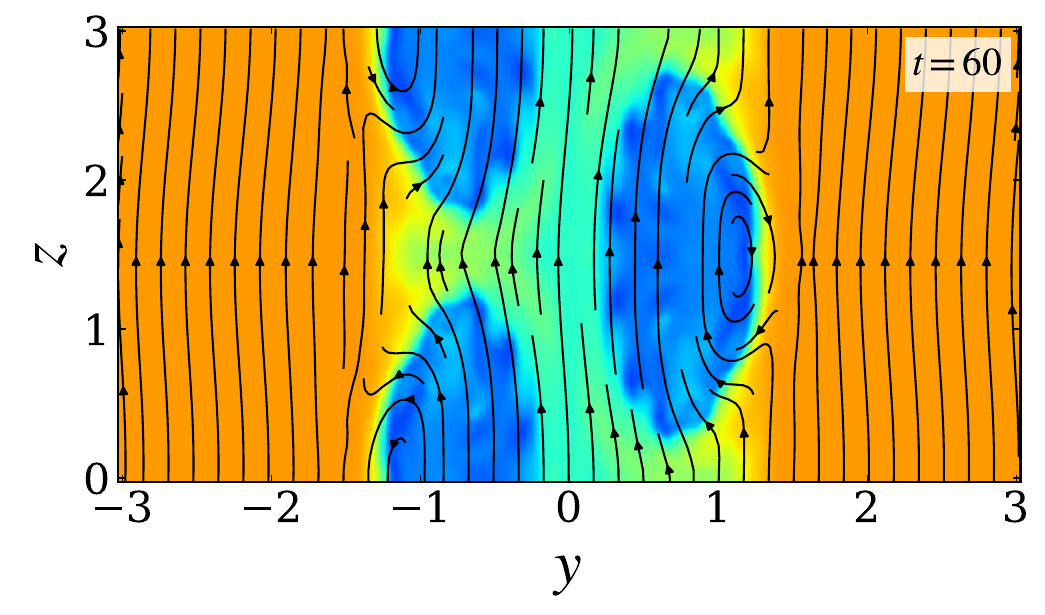}
   \includegraphics[width=0.195\linewidth]{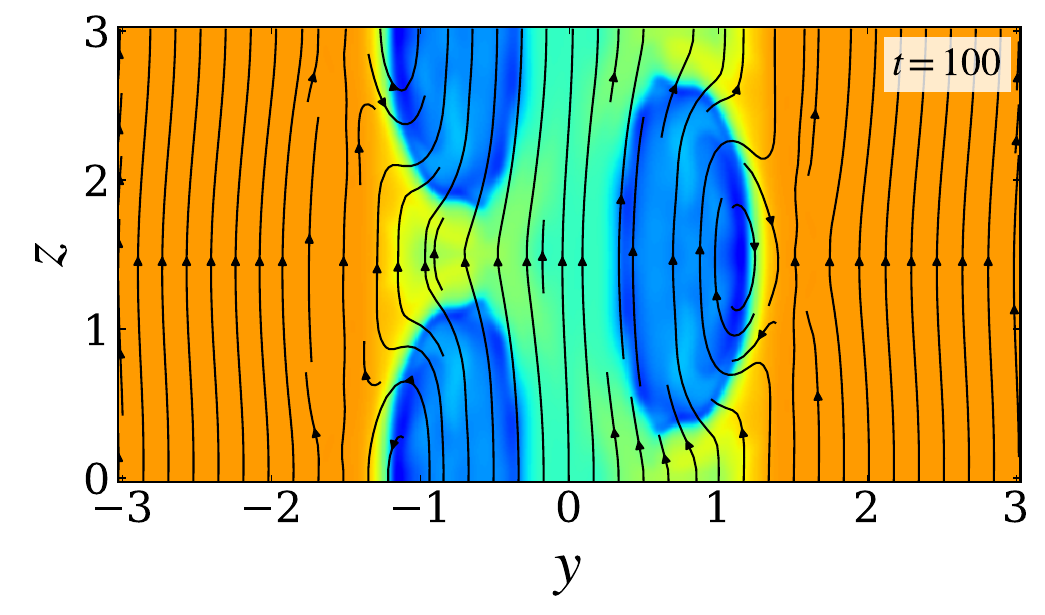}        
    \includegraphics[width=0.925\linewidth]{Figures/CDKI/colorbar_rho_2D.pdf}    
 \caption{Evolution of the density distribution for the modified $M_3$ model with weakly magnetised wind. The top and bottom rows show the horizontal ( $z=1.5$ ) and the vertical ($x=0$) mid-planes of the computational domain, respectively. The lines are tangent to the plane's magnetic field vectors.}
    \label{fig:rho-slices-M6}
\end{figure*}

\begin{figure*}
    \centering
    \includegraphics[width=0.242\linewidth]{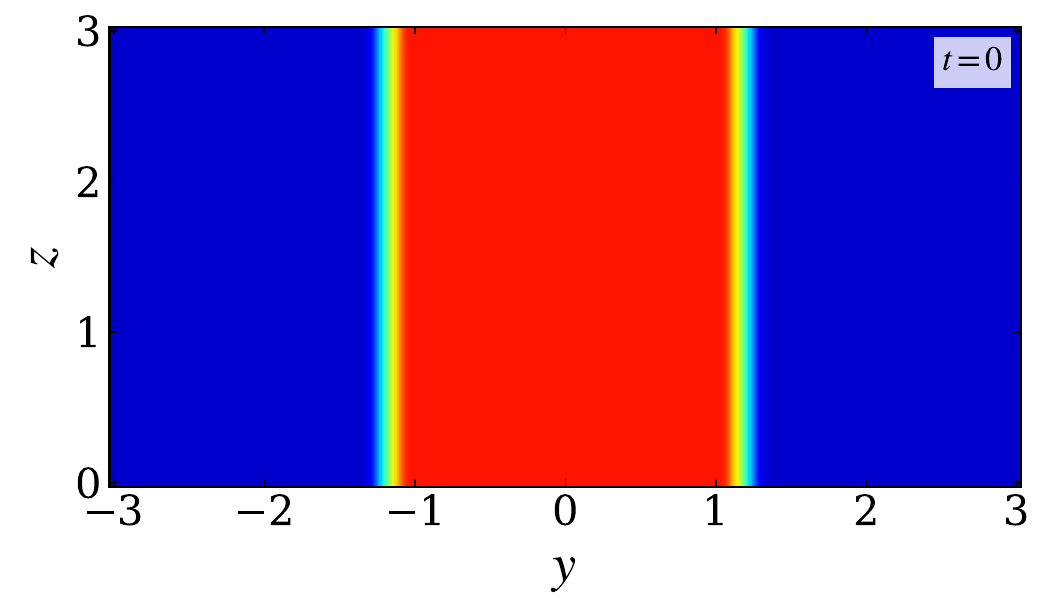}
    \includegraphics[width=0.242\linewidth]{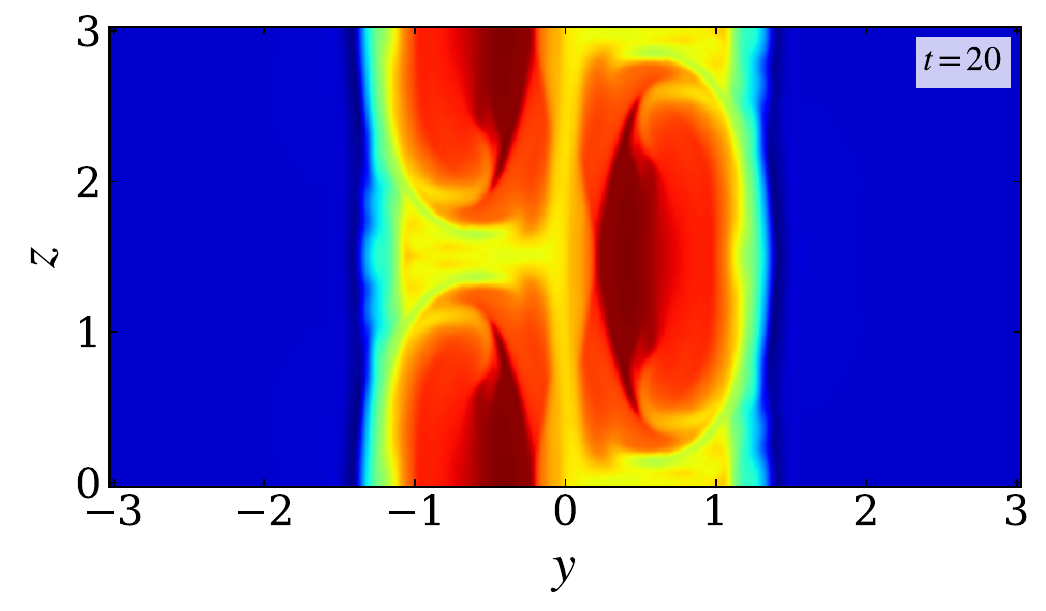}
    \includegraphics[width=0.242\linewidth]{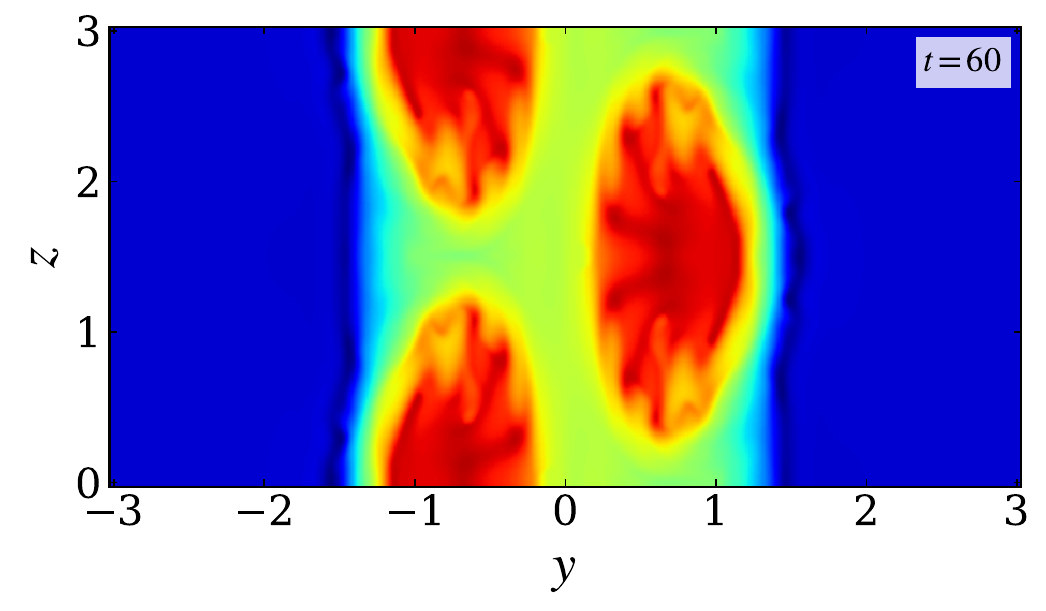}
    \includegraphics[width=0.242\linewidth]{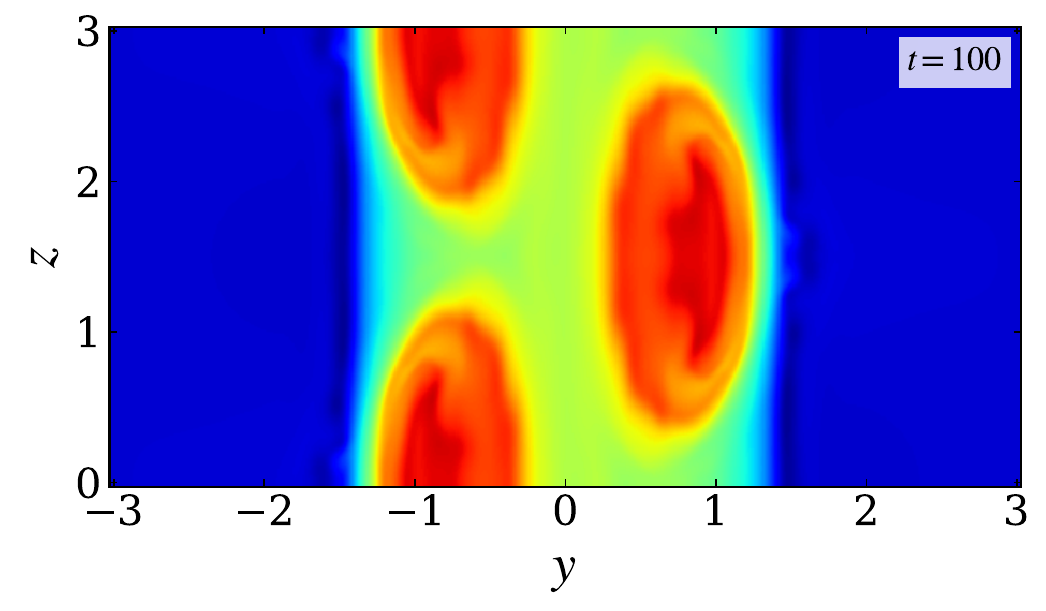}
    \includegraphics[width=0.925\linewidth]{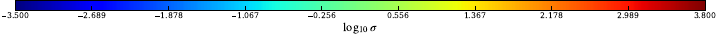}
    
    \caption{Evolution of the $\sigma$ distribution in mid-plane $x=0$ for the modified $M_3$ model with weakly magnetised wind.}
    \label{fig:sigma-slices-M6}
\end{figure*}

Given the highly inhomogeneous structure of the configuration reached by the end of the simulations, it is not clear how to quantify its overall magnetisation in a meaningful way.  One option is to employ $\langle\sigma\rangle = \langle B^2\rangle/\langle w\rangle$, where $\langle B^2\rangle$ and $\langle w\rangle$ are volume-averaged values. Figure  \ref{fig:sigma2} shows the evolution of this parameter, with the integration being carried out over the space occupied by the jet and wind in the initial solution.  This selection of the integration volume looks reasonable as it reflects the size of the interpenetration seen in the simulations.  Obviously, one would not expect this parameter to change by more than a factor of few compared to its initial value, and it does not.  Ultimately, the low value of $\langle\sigma\rangle$ is determined by the large mass of the wind plasma involved in the interpenetration.

\subsection{Weakly magnetised wind}
\label{sec:wmw}

The results presented above indicate that the magnetic field of the wind is an important factor in the nonlinear dynamics of the kink instability in the models explored there. To confirm this, we ran an additional model where the wind magnetic field is much weaker and the jet is confined largely by the wind thermal pressure. As a reference, we used the $M_3$ model with the wind temperature increased to $\epsilon_w=1$ and its magnetisation reduced to $\sigma_w=10^{-3}$. In addition, there was no buffer zone at the boundaries of the computational domain. The size of the domain and resolution are the same as in the $M_3$ model.

Since the magnetic acceleration of non-relativistic flows is much more efficient, the electromagnetic energy of the wind (its Poynting flux) can be converted into kinetic energy much more rapidly than in the jet. This kinetic energy can be dissipated into thermal energy by shock waves, creating hot plasma confining the BZ jet \citep[e.g. ][]{GL-16}.   Thus, this problem may be quite relevant for AGN jets as well. 

Figure \ref{fig:isosurf-M6} shows the evolution of the density iso-surface separating the wind from the jet as well as the magnetic field lines just outside of this surface.  Like in the original $M_3$ model, one can see the coiling of the jet and the development of ripples on its surface. However, on a much lower amplitude now. Moreover, the ripples disappear completely by $t=100$.     

\begin{figure}
    \centering
    \includegraphics[width=0.8\linewidth]{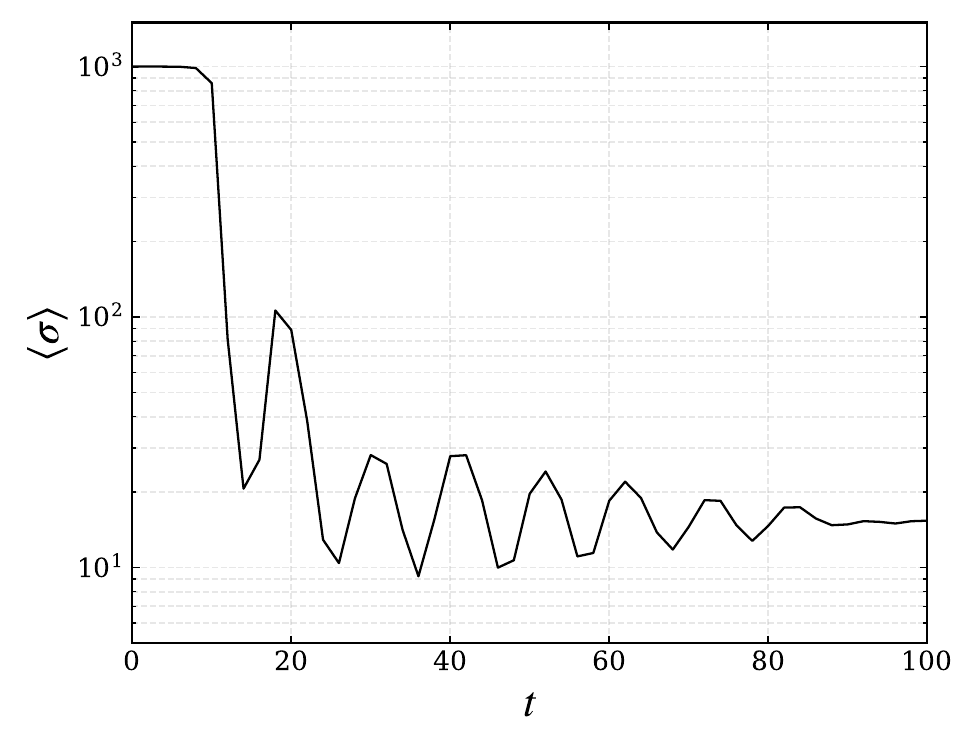}    
    \caption{Evolution of the mean magnetisation $\langle\sigma\rangle$ in the jet for the modified $M_3$ model with weakly magnetised wind.}
    \label{fig:sigma3}
\end{figure}

Figure \ref{fig:rho-slices-M6} shows the evolution of mass density distribution in this model. It exhibits many of the features seen in the models with mildly magnetised winds, but with a reduced amplitude. In particular, one can see the development of the kidney bean morphology for the jet cross-section, the emergence of the central protuberance, and the onset of the secondary instability. However, the interpenetration of the jet and wind plasmas seen in the $M_3$ model seems to have been replaced by a more one-sided process, where the wind plasma penetrates the jet. There is no compression of the wind plasma, attributed in the mildly magnetised case to the hoop stress of the wind's magnetic field. Instead, the density of the wind plasma entering the jet is significantly lower, indicating the action of numerical diffusion.  By the end of the simulation ($t=100$), the solution seems to have reached a relaxed state. Due to the much weaker secondary instability, the mixing of the jet and wind plasma inside the jet is not so efficient as in the $M_3$ model, and the jet coil retains more-or-less regular structure in this relaxed state.      
  
Figure \ref{fig:sigma-slices-M6} shows the evolution of the $\sigma$ distribution in the $x=0$ mid-plane of the domain.  By the end of the simulations, the jet coil remains highly magnetised, but the space between the coils has $\sigma \gtrsim 1$. Figure \ref{fig:sigma3} shows the evolution of $\langle\sigma\rangle$ in the volume occupied by the jet in the initial solution. Soon after the end of the linear phase of the instability, it drops from the initial value of $10^3$ down to $\approx 20$ and then exhibits damped oscillations, finally relaxing to  $\langle\sigma\rangle\approx 10$. This suggests that the initial coiling of the jet, accompanied by the inflow of the wind plasma into the space between coils, is the main factor in the reduction of its mean magnetisation.     
However, the large gradient seen at the boundary of the jet in figure \ref{fig:sigma-slices-M6} indicates that the numerical diffusion makes a big impact on the ultimate magnetisation distribution in this solution too.             
  
\section{Discussion}
\label{discuss}

Our study shows that the development of kink instability of a highly-magnetised light jet surrounded by a dense weakly-magnetised wind of the accretion disk may result in highly efficient mass-loading of the jet.  Moreover, in the static problem explored in this study, the mass-loading is too strong, as the mean $\langle\sigma\rangle$ of the mixed relaxed state is lower than unity.  

Another interesting result of this study is the discovery of secondary instability which enhances the mixing of wind and jet plasma, initially produced via the coiling of the jet.  It is possible that this is only ''the tip of an iceberg'',  and higher resolution simulations will reveal more secondary small-scale instabilities which are dumped in our simulations by numerical viscosity, diffusion, and resistivity.   

Obviously, the static column model is a gross simplification of the much richer astrophysical problem.   

The lack of relative motion between the jet and wind is probably its most drastic simplification.  As the kink instability produces the corkscrew structure of the jet, it begins acting as sort of a piston driving waves into the wind. With its speed being very close to the speed of light and hence exceeding the fast magnetosonic speed of the wind's plasma, these waves are likely to be shocks, pushing the wind plasma away from the jet. This may strongly reduce the jet mass loading below what is seen in the static column simulation. Moreover, since the wind plasma is only weakly magnetised, these shocks may accelerate a fraction of wind electrons to relativistic speeds via the diffusive shock-acceleration mechanism.  This leads quite naturally to enhanced synchrotron that may explain the observed limb-brightening of the ANG jets on sub-pc scale \citep{Walker-M87, CenA-21, NGC315-24}. 

The relativistic motion of the jet may reduce the observed growth rate of the instabilities via the time dilation effect. However, it has been argued that for the kink instability it is not the Lorentz factor of the jet what is important,  but the Lorentz factor corresponding to the group speed of the kink mode \citep{Lyubarski-99}.   With the wind moving at a much slower speed than the jet, the group speed of the kink mode may move at an intermediate speed, thus mitigating the relativistic time dilation effect.   

We also did not take into account the rotation of magnetic field lines. This rotation induces the electric field $E_r=\Omega r B_z/c$ and the radial electrostatic force, which changes the equilibrium solution. As a result, the Kruskal-Shafranov stability criterion no longer applies. Instead, $B_z(r)$ must be a decreasing function for the kink instability to exist \citep{Lyubarski-99}, and its growth rate depends on the rate of this decrease.            

We considered only the case of constant magnetic pitch, whereas the jet and wind magnetic fields may have rather different pitches.       
 
The observed radius of AGN jets grows with the distance from the black hole \citep[e.g. ][]{Nakamura_2013, Hada-13}. It is known that such radial expansion can strongly reduce the growth rate of jet instabilities \citep[e.g.][]{Hardee-82, PK-15}. 
              
Given all these limitations of our current study,  much more has to be done before a definitive conclusion on the role of CD kink instability in mass-loading of AGN jets can be made.

\section{Conclusion}
\label{sec:conclusion}

As the first step in a series of numerical simulations aimed at studying the current-driven kink instability of the black-hole jet collimated by the wind of the accretion disk, we investigated the stability of a simplified axisymmetric static equilibrium combining a highly magnetised ($\sigma=10^3$) central column and a moderately ($\sigma=10^{-1}$) of weakly ($\sigma=10^{-3}$) magnetised outer shell in a periodic box. The stability condition for this equilibrium is given by the Kruskal-Shafranov criterion on the magnetic pitch. The study is limited to the case of constant pitch only. Inside the column and in the outer shell, the magnetic field is force-free, but the gas pressure gradient supports the equilibrium in the transition layer between the two. The measured linear growth rates of the instability are in agreement with the maximum growth rate for a uniform density column with a force-free magnetic field of constant pitch obtained by \citet{Appl-00}, being only a little bit higher. At the non-linear phase, the interface separating the central column and the outer shell develops the characteristic corkscrew shape, with the space inside and between the coils of the corkscrew filled by the shell plasma. In high-resolution runs, a secondary instability is seen as well.  It has the form of helical protuberances of the low-$\sigma$ plasma intruding inside the corkscrew coils. They appear to be aligned with the local magnetic field lines. We tentatively identify the secondary instability as the magnetic Rayleigh-Taylor instability.                 
Overall, the results indicate that the CD kink instability may provide the baryon-loading of AGN jets, which is strongly hinted by their observed properties.

\section*{Acknowledgements}
The simulations were carried out on the CPU cluster of the AIRE High-Performance Computing facility of Leeds University. K.Kumar is supported by the Science and Technology Facilities Council (STFC) via a postgraduate research studentship.

\section*{Data Availability}

The 3D simulation data underlying this article can be provided upon reasonable request.



\bibliographystyle{mnras}
\bibliography{references, added} 

@BOOK{chandra-stability,
   author = {{Chandrasekhar}, S.},
    title = "{Hydrodynamic and hydromagnetic stability}",
publisher = {International Series of Monographs on Physics, Oxford: Clarendon, 1961},
     year = 1961,
   adsurl = {http://adsabs.harvard.edu/abs/1961hhs..book.....C}
}

@ARTICLE{JNS-95,
       author = {{Jun}, Byung-Il and {Norman}, Michael L. and {Stone}, James M.},
        title = "{A Numerical Study of Rayleigh-Taylor Instability in Magnetic Fluids}",
      journal = {\apj},
         year = 1995,
        month = nov,
       volume = {453},
        pages = {332},
          doi = {10.1086/176393},
       adsurl = {https://ui.adsabs.harvard.edu/abs/1995ApJ...453..332J}
}

@ARTICLE{TH-25,
       author = {{Teja Kalluri}, Manohar and {Hillier}, Andrew},
        title = "{Comparing the magnetic Rayleigh-Taylor instability dynamics in two- and three-dimensions}",
      journal = {arXiv e-prints},
         year = 2025,
        month = oct,
          eid = {arXiv:2510.27053},
        pages = {arXiv:2510.27053},
          doi = {10.48550/arXiv.2510.27053},
archivePrefix = {arXiv},
       eprint = {2510.27053},
 primaryClass = {physics.flu-dyn},
       adsurl = {https://ui.adsabs.harvard.edu/abs/2025arXiv251027053T}
}

@ARTICLE{TNM11,
       author = {{Tchekhovskoy}, Alexander and {Narayan}, Ramesh and {McKinney}, Jonathan C.},
        title = "{Efficient generation of jets from magnetically arrested accretion on a rapidly spinning black hole}",
      journal = {\mnras},
         year = 2011,
        month = nov,
       volume = {418},
       number = {1},
        pages = {L79-L83},
          doi = {10.1111/j.1745-3933.2011.01147.x},
archivePrefix = {arXiv},
       eprint = {1108.0412},
 primaryClass = {astro-ph.HE},
       adsurl = {https://ui.adsabs.harvard.edu/abs/2011MNRAS.418L..79T}
}

@ARTICLE{SSK01,
       author = {{Komissarov}, S.~S.},
        title = "{Direct numerical simulations of the Blandford-Znajek effect}",
      journal = {\mnras},
         year = 2001,
        month = sep,
       volume = {326},
       number = {3},
        pages = {L41-L44},
          doi = {10.1046/j.1365-8711.2001.04863.x},
       adsurl = {https://ui.adsabs.harvard.edu/abs/2001MNRAS.326L..41K}
}

@ARTICLE{TNM10,
       author = {{Tchekhovskoy}, Alexander and {Narayan}, Ramesh and {McKinney}, Jonathan C.},
        title = "{Black Hole Spin and The Radio Loud/Quiet Dichotomy of Active Galactic Nuclei}",
      journal = {\apj},
         year = 2010,
        month = mar,
       volume = {711},
       number = {1},
        pages = {50-63},
          doi = {10.1088/0004-637X/711/1/50},
archivePrefix = {arXiv},
       eprint = {0911.2228},
 primaryClass = {astro-ph.HE},
       adsurl = {https://ui.adsabs.harvard.edu/abs/2010ApJ...711...50T}
}

@ARTICLE{Ripperda22,
       author = {{Ripperda}, B. and {Liska}, M. and {Chatterjee}, K. and {Musoke}, G. and {Philippov}, A.~A. and {Markoff}, S.~B. and {Tchekhovskoy}, A. and {Younsi}, Z.},
        title = "{Black Hole Flares: Ejection of Accreted Magnetic Flux through 3D Plasmoid-mediated Reconnection}",
      journal = {\apjl},
         year = 2022,
        month = jan,
       volume = {924},
       number = {2},
          eid = {L32},
        pages = {L32},
          doi = {10.3847/2041-8213/ac46a1},
archivePrefix = {arXiv},
       eprint = {2109.15115},
 primaryClass = {astro-ph.HE},
       adsurl = {https://ui.adsabs.harvard.edu/abs/2022ApJ...924L..32R}
}

@ARTICLE{BIP-92,
       author = {{Beskin}, V.~S. and {Istomin}, Y.~N. and {Parev}, V.~I.},
        title = "{Filling the Magnetosphere of a Supermassive Black-Hole with Plasma}",
      journal = {\sovast},
         year = 1992,
        month = dec,
       volume = {36},
        pages = {642},
       adsurl = {https://ui.adsabs.harvard.edu/abs/1992SvA....36..642B}
}

@ARTICLE{MGDS-11,
       author = {{Mo{\'s}cibrodzka}, M. and {Gammie}, C.~F. and {Dolence}, J.~C. and {Shiokawa}, H.},
        title = "{Pair Production in Low-luminosity Galactic Nuclei}",
      journal = {\apj},
         year = 2011,
        month = jul,
       volume = {735},
       number = {1},
          eid = {9},
        pages = {9},
          doi = {10.1088/0004-637X/735/1/9},
archivePrefix = {arXiv},
       eprint = {1104.2042},
 primaryClass = {astro-ph.HE},
       adsurl = {https://ui.adsabs.harvard.edu/abs/2011ApJ...735....9M}
}

@ARTICLE{WRG-21,
       author = {{Wong}, George N. and {Ryan}, Benjamin R. and {Gammie}, Charles F.},
        title = "{Pair Drizzle around Sub-Eddington Supermassive Black Holes}",
      journal = {\apj},
         year = 2021,
        month = feb,
       volume = {907},
       number = {2},
          eid = {73},
        pages = {73},
          doi = {10.3847/1538-4357/abd0f9},
archivePrefix = {arXiv},
       eprint = {2012.04658},
 primaryClass = {astro-ph.HE},
       adsurl = {https://ui.adsabs.harvard.edu/abs/2021ApJ...907...73W}
}

@ARTICLE{HO-98,
       author = {{Hirotani}, Kouichi and {Okamoto}, Isao},
        title = "{Pair Plasma Production in a Force-free Magnetosphere around a Supermassive Black Hole}",
      journal = {\apj},
         year = 1998,
        month = apr,
       volume = {497},
       number = {2},
        pages = {563-572},
          doi = {10.1086/305479},
       adsurl = {https://ui.adsabs.harvard.edu/abs/1998ApJ...497..563H}
}

@article{Crinq-20,
  title = {Multidimensional Simulations of Ergospheric Pair Discharges around Black Holes},
  author = {Crinquand, Benjamin and Cerutti, Beno\^{\i}t and Philippov, Alexander and Parfrey, Kyle and Dubus, Guillaume},
  journal = {Phys. Rev. Lett.},
  volume = {124},
  issue = {14},
  pages = {145101},
  numpages = {6},
  year = {2020},
  month = {Apr},
  publisher = {American Physical Society},
  doi = {10.1103/PhysRevLett.124.145101},
  url = {https://link.aps.org/doi/10.1103/PhysRevLett.124.145101}
  }

@ARTICLE{CY-20,
       author = {{Chen}, Alexander Y. and {Yuan}, Yajie},
        title = "{Physics of Pair Producing Gaps in Black Hole Magnetospheres. II. General Relativity}",
      journal = {\apj},
         year = 2020,
        month = jun,
       volume = {895},
       number = {2},
          eid = {121},
        pages = {121},
          doi = {10.3847/1538-4357/ab8c46},
archivePrefix = {arXiv},
       eprint = {1908.06919},
 primaryClass = {astro-ph.HE},
       adsurl = {https://ui.adsabs.harvard.edu/abs/2020ApJ...895..121C}
}

@article{Levinson-11,
doi = {10.1088/0004-637X/730/2/123},
url = {https://doi.org/10.1088/0004-637X/730/2/123},
year = {2011},
month = {mar},
publisher = {The American Astronomical Society},
volume = {730},
number = {2},
pages = {123},
author = {Levinson, Amir and Rieger, Frank},
title = {VARIABLE TeV EMISSION AS A MANIFESTATION OF JET FORMATION IN M87?},
journal = {The Astrophysical Journal},
}

@article{KTNH-22,
doi = {10.3847/2041-8213/ac8d5a},
url = {https://doi.org/10.3847/2041-8213/ac8d5a},
year = {2022},
month = {sep},
publisher = {The American Astronomical Society},
volume = {937},
number = {2},
pages = {L34},
author = {Kimura, Shigeo S. and Toma, Kenji and Noda, Hirofumi and Hada, Kazuhiro},
title = {Magnetic Reconnection in Black Hole Magnetospheres: Lepton Loading into Jets, Superluminal Radio Blobs, and Multiwavelength Flares},
journal = {The Astrophysical Journal Letters},
}

@ARTICLE{CUD-23,
       author = {{Chen}, Alexander Y. and {Uzdensky}, Dmitri and {Dexter}, Jason},
        title = "{Synchrotron Pair Production Equilibrium in Relativistic Magnetic Reconnection}",
      journal = {\apj},
         year = 2023,
        month = feb,
       volume = {944},
       number = {2},
          eid = {173},
        pages = {173},
          doi = {10.3847/1538-4357/acb68a},
archivePrefix = {arXiv},
       eprint = {2209.03249},
 primaryClass = {astro-ph.HE},
       adsurl = {https://ui.adsabs.harvard.edu/abs/2023ApJ...944..173C}
}

@ARTICLE{OTK-26,
       author = {{Oikawa}, Rin and {Toma}, Kenji and {Kimura}, Shigeo S.},
        title = "{Nonuniform Particle Injection into Black Hole Jets by Radiative Magnetic Reconnection}",
      journal = {\apj},
         year = 2026,
        month = sep,
       volume = {1008},
       number = {2},
          eid = {185},
        pages = {185},
          doi = {10.3847/1538-4357/ae969b},
archivePrefix = {arXiv},
       eprint = {2605.01002},
 primaryClass = {astro-ph.HE},
       adsurl = {https://ui.adsabs.harvard.edu/abs/2026ApJ..1008..185O}
}

@ARTICLE{kvkb-09,
   author = {{Komissarov}, S.~S. and {Vlahakis}, N. and {K{\"o}nigl}, A. and 
	{Barkov}, M.~V.},
    title = "{Magnetic acceleration of ultrarelativistic jets in gamma-ray burst sources}",
  journal = {\mnras},
archivePrefix = "arXiv",
   eprint = {0811.1467},
     year = 2009,
    month = apr,
   volume = 394,
    pages = {1182-1212},
      doi = {10.1111/j.1365-2966.2009.14410.x},
   adsurl = {http://adsabs.harvard.edu/abs/2009MNRAS.394.1182K}
}

@ARTICLE{kbvk-07,
   author = {{Komissarov}, S.~S. and {Barkov}, M.~V. and {Vlahakis}, N. and 
	{K{\"o}nigl}, A.},
    title = "{Magnetic acceleration of relativistic active galactic nucleus jets}",
  journal = {\mnras},
   eprint = {arXiv:astro-ph/0703146},
     year = 2007,
    month = sep,
   volume = 380,
    pages = {51-70},
      doi = {10.1111/j.1365-2966.2007.12050.x},
   adsurl = {http://adsabs.harvard.edu/abs/2007MNRAS.380...51K}
}

@ARTICLE{Lister-09,
       author = {{Lister}, M.~L. and {Cohen}, M.~H. and {Homan}, D.~C. and {Kadler}, M. and {Kellermann}, K.~I. and {Kovalev}, Y.~Y. and {Ros}, E. and {Savolainen}, T. and {Zensus}, J.~A.},
        title = "{MOJAVE: Monitoring of Jets in Active Galactic Nuclei with VLBA Experiments. VI. Kinematics Analysis of a Complete Sample of Blazar Jets}",
      journal = {\aj},
         year = 2009,
        month = dec,
       volume = {138},
       number = {6},
        pages = {1874-1892},
          doi = {10.1088/0004-6256/138/6/1874},
archivePrefix = {arXiv},
       eprint = {0909.5100},
 primaryClass = {astro-ph.CO},
       adsurl = {https://ui.adsabs.harvard.edu/abs/2009AJ....138.1874L}
}

@ARTICLE{HLK-15,
   author = {{Homan}, D.~C. and {Lister}, M.~L. and {Kovalev}, Y.~Y. and 
	{Pushkarev}, A.~B. and {Savolainen}, T. and {Kellermann}, K.~I. and 
	{Richards}, J.~L. and {Ros}, E.},
    title = "{MOJAVE. XII. Acceleration and Collimation of Blazar Jets on Parsec Scales}",
  journal = {\apj},
archivePrefix = "arXiv",
   eprint = {1410.8502},
 primaryClass = "astro-ph.HE",
     year = 2015,
    month = jan,
   volume = 798,
      eid = {134},
    pages = {134},
      doi = {10.1088/0004-637X/798/2/134},
   adsurl = {http://adsabs.harvard.edu/abs/2015ApJ...798..134H}
}

@ARTICLE{JBL-99,
       author = {{Junor}, William and {Biretta}, John A. and {Livio}, Mario},
        title = "{Formation of the radio jet in M87 at 100 Schwarzschild radii from the central black hole}",
      journal = {\nat},
         year = 1999,
        month = oct,
       volume = {401},
       number = {6756},
        pages = {891-892},
          doi = {10.1038/44780},
       adsurl = {https://ui.adsabs.harvard.edu/abs/1999Natur.401..891J}
}

@ARTICLE{Hada-13,
       author = {{Hada}, Kazuhiro and {Kino}, Motoki and {Doi}, Akihiro and {Nagai}, Hiroshi and {Honma}, Mareki and {Hagiwara}, Yoshiaki and {Giroletti}, Marcello and {Giovannini}, Gabriele and {Kawaguchi}, Noriyuki},
        title = "{The Innermost Collimation Structure of the M87 Jet Down to \raisebox{-0.5ex}\textasciitilde10 Schwarzschild Radii}",
      journal = {\apj},
         year = 2013,
        month = sep,
       volume = {775},
       number = {1},
          eid = {70},
        pages = {70},
          doi = {10.1088/0004-637X/775/1/70},
archivePrefix = {arXiv},
       eprint = {1308.1411},
 primaryClass = {astro-ph.CO},
       adsurl = {https://ui.adsabs.harvard.edu/abs/2013ApJ...775...70H}
}

@ARTICLE{PKLS-09,
       author = {{Pushkarev}, A.~B. and {Kovalev}, Y.~Y. and {Lister}, M.~L. and {Savolainen}, T.},
        title = "{Jet opening angles and gamma-ray brightness of AGN}",
      journal = {\aap},
         year = 2009,
        month = nov,
       volume = {507},
       number = {2},
        pages = {L33-L36},
          doi = {10.1051/0004-6361/200913422},
archivePrefix = {arXiv},
       eprint = {0910.1813},
 primaryClass = {astro-ph.CO},
       adsurl = {https://ui.adsabs.harvard.edu/abs/2009A&A...507L..33P}
}

@ARTICLE{Beskin-10,
       author = {{Beskin}, Vasily S.},
        title = "{Magnetohydrodynamic models of astrophysical jets}",
      journal = {Physics Uspekhi},
         year = 2010,
        month = dec,
       volume = {53},
       number = {12},
        pages = {1199-1233},
          doi = {10.3367/UFNe.0180.201012b.1241},
archivePrefix = {arXiv},
       eprint = {1103.3375},
 primaryClass = {astro-ph.HE},
       adsurl = {https://ui.adsabs.harvard.edu/abs/2010PhyU...53.1199B}
}

@ARTICLE{BL-95,
       author = {{Blandford}, R.~D. and {Levinson}, A.},
        title = "{Pair Cascades in Extragalactic Jets. I. Gamma Rays}",
      journal = {\apj},
         year = 1995,
        month = mar,
       volume = {441},
        pages = {79},
          doi = {10.1086/175338},
       adsurl = {https://ui.adsabs.harvard.edu/abs/1995ApJ...441...79B}
}

@ARTICLE{Lev-96,
       author = {{Levinson}, Amir},
        title = "{Constraints on the Dynamics of Electron-Positron Jets in Active Galactic Nuclei}",
      journal = {\apj},
         year = 1996,
        month = aug,
       volume = {467},
        pages = {546},
          doi = {10.1086/177632},
       adsurl = {https://ui.adsabs.harvard.edu/abs/1996ApJ...467..546L}
}

@ARTICLE{SMMP-97,
       author = {{Sikora}, Marek and {Madejski}, Greg and {Moderski}, Rafa{\L} and {Poutanen}, Juri},
        title = "{Learning about Active Galactic Nucleus Jets from Spectral Properties of Blazars}",
      journal = {\apj},
         year = 1997,
        month = jul,
       volume = {484},
       number = {1},
        pages = {108-117},
          doi = {10.1086/304305},
       adsurl = {https://ui.adsabs.harvard.edu/abs/1997ApJ...484..108S}
}

@ARTICLE{ssk-94,
   author = {{Komissarov}, S.~S.},
    title = "{Mass-Loaded Relativistic Jets}",
  journal = {\mnras},
     year = 1994,
    month = jul,
   volume = 269,
    pages = {394},
   adsurl = {http://adsabs.harvard.edu/abs/1994MNRAS.269..394K}
}

@ARTICLE{bicknell-84,
   author = {{Bicknell}, G.~V.},
    title = "{A model for the surface brightness of a turbulent low Mach number jet. I - Theoretical development and application to 3C 31}",
  journal = {\apj},
     year = 1984,
    month = nov,
   volume = 286,
    pages = {68-87},
      doi = {10.1086/162577},
   adsurl = {http://adsabs.harvard.edu/abs/1984ApJ...286...68B}
}

@ARTICLE{CSRA-26,
       author = {{Chow}, Anthony and {Sironi}, Lorenzo and {Ripperda}, Bart and {Levinson}, Amir},
        title = "{The baryon content of magnetically arrested black hole discs and jets}",
      journal = {\mnras},
         year = 2026,
        month = oct,
       volume = {552},
       number = {1},
          eid = {stag1631},
        pages = {stag1631},
          doi = {10.1093/mnras/stag1631},
archivePrefix = {arXiv},
       eprint = {2603.10122},
       adsurl = {https://ui.adsabs.harvard.edu/abs/2026MNRAS.552g1631C}
}

@ARTICLE{LTQ-20,
       author = {{Liska}, M. and {Tchekhovskoy}, A. and {Quataert}, E.},
        title = "{Large-scale poloidal magnetic field dynamo leads to powerful jets in GRMHD simulations of black hole accretion with toroidal field}",
      journal = {\mnras},
         year = 2020,
        month = may,
       volume = {494},
       number = {3},
        pages = {3656-3662},
          doi = {10.1093/mnras/staa955},
archivePrefix = {arXiv},
       eprint = {1809.04608},
 primaryClass = {astro-ph.HE},
       adsurl = {https://ui.adsabs.harvard.edu/abs/2020MNRAS.494.3656L}
}

@article{Chatterjee_2022,
doi = {10.3847/1538-4357/ac9d97},
url = {https://doi.org/10.3847/1538-4357/ac9d97},
year = {2022},
month = {dec},
publisher = {The American Astronomical Society},
volume = {941},
number = {1},
pages = {30},
author = {Chatterjee, K. and Narayan, R.},
title = {Flux Eruption Events Drive Angular Momentum Transport in Magnetically Arrested Accretion Flows},
journal = {The Astrophysical Journal}
}

@ARTICLE{LMT-22,
       author = {{Liska}, M.~T.~P. and {Musoke}, G. and {Tchekhovskoy}, A. and {Porth}, O. and {Beloborodov}, A.~M.},
        title = "{Formation of Magnetically Truncated Accretion Disks in 3D Radiation-transport Two-temperature GRMHD Simulations}",
      journal = {\apjl},
         year = 2022,
        month = aug,
       volume = {935},
       number = {1},
          eid = {L1},
        pages = {L1},
          doi = {10.3847/2041-8213/ac84db},
archivePrefix = {arXiv},
       eprint = {2201.03526},
 primaryClass = {astro-ph.HE},
       adsurl = {https://ui.adsabs.harvard.edu/abs/2022ApJ...935L...1L}
}

@ARTICLE{SBD-24,
       author = {{Scepi}, Nicolas and {Begelman}, Mitchell C. and {Dexter}, Jason},
        title = "{Magnetic support, wind-driven accretion, coronal heating, and fast outflows in a thin magnetically arrested disc}",
      journal = {\mnras},
         year = 2024,
        month = jan,
       volume = {527},
       number = {1},
        pages = {1424-1443},
          doi = {10.1093/mnras/stad3299},
archivePrefix = {arXiv},
       eprint = {2302.10226},
 primaryClass = {astro-ph.HE},
       adsurl = {https://ui.adsabs.harvard.edu/abs/2024MNRAS.527.1424S}
}

@ARTICLE{TCR-10,
       author = {{Tombesi}, F. and {Cappi}, M. and {Reeves}, J.~N. and {Palumbo}, G.~G.~C. and {Yaqoob}, T. and {Braito}, V. and {Dadina}, M.},
        title = "{Evidence for ultra-fast outflows in radio-quiet AGNs. I. Detection and statistical incidence of Fe K-shell absorption lines}",
      journal = {\aap},
         year = 2010,
        month = oct,
       volume = {521},
          eid = {A57},
        pages = {A57},
          doi = {10.1051/0004-6361/200913440},
archivePrefix = {arXiv},
       eprint = {1006.2858},
 primaryClass = {astro-ph.HE},
       adsurl = {https://ui.adsabs.harvard.edu/abs/2010A&A...521A..57T}
}

@ARTICLE{TTM-14,
       author = {{Tombesi}, F. and {Tazaki}, F. and {Mushotzky}, R.~F. and {Ueda}, Y. and {Cappi}, M. and {Gofford}, J. and {Reeves}, J.~N. and {Guainazzi}, M.},
        title = "{Ultrafast outflows in radio-loud active galactic nuclei}",
      journal = {\mnras},
         year = 2014,
        month = sep,
       volume = {443},
       number = {3},
        pages = {2154-2182},
          doi = {10.1093/mnras/stu1297},
archivePrefix = {arXiv},
       eprint = {1406.7252},
 primaryClass = {astro-ph.HE},
       adsurl = {https://ui.adsabs.harvard.edu/abs/2014MNRAS.443.2154T}
}

@ARTICLE{Lyubarski-99,
       author = {{Lyubarskii}, Yu. E.},
        title = "{Kink instability of relativistic force-free jets}",
      journal = {\mnras},
         year = 1999,
        month = oct,
       volume = {308},
       number = {4},
        pages = {1006-1010},
          doi = {10.1046/j.1365-8711.1999.02763.x},
       adsurl = {https://ui.adsabs.harvard.edu/abs/1999MNRAS.308.1006L}
}

@ARTICLE{IP-94,
       author = {{Istomin}, Y.~N. and {Pariev}, V.~I.},
        title = "{Stability of a relativistic rotating electron-positron jet.}",
      journal = {\mnras},
         year = 1994,
        month = apr,
       volume = {267},
        pages = {629-636},
          doi = {10.1093/mnras/267.3.629},
       adsurl = {https://ui.adsabs.harvard.edu/abs/1994MNRAS.267..629I}
}

@ARTICLE{AC-92,
       author = {{Appl}, S. and {Camenzind}, M.},
        title = "{The stability of current-carrying jets.}",
      journal = {\aap},
         year = 1992,
        month = mar,
       volume = {256},
        pages = {354-370},
       adsurl = {https://ui.adsabs.harvard.edu/abs/1992A&A...256..354A}
}

@ARTICLE{IP-96,
       author = {{Istomin}, Ya. N. and {Pariev}, V.~I.},
        title = "{Stability of a relativistic rotating electron-positron jet: non-axisymmetric perturbations}",
      journal = {\mnras},
         year = 1996,
        month = jul,
       volume = {281},
       number = {1},
        pages = {1-26},
          doi = {10.1093/mnras/281.1.1},
archivePrefix = {arXiv},
       eprint = {astro-ph/9512145},
 primaryClass = {astro-ph},
       adsurl = {https://ui.adsabs.harvard.edu/abs/1996MNRAS.281....1I}
}

@ARTICLE{Begelman-98,
       author = {{Begelman}, Mitchell C.},
        title = "{Instability of Toroidal Magnetic Field in Jets and Plerions}",
      journal = {\apj},
         year = 1998,
        month = jan,
       volume = {493},
       number = {1},
        pages = {291-300},
          doi = {10.1086/305119},
archivePrefix = {arXiv},
       eprint = {astro-ph/9708142},
 primaryClass = {astro-ph},
       adsurl = {https://ui.adsabs.harvard.edu/abs/1998ApJ...493..291B}
}

@ARTICLE{OBB-12,
       author = {{O'Neill}, Sean M. and {Beckwith}, Kris and {Begelman}, Mitchell C.},
        title = "{Local simulations of instabilities in relativistic jets - I. Morphology and energetics of the current-driven instability}",
      journal = {\mnras},
         year = 2012,
        month = may,
       volume = {422},
       number = {2},
        pages = {1436-1452},
          doi = {10.1111/j.1365-2966.2012.20721.x},
archivePrefix = {arXiv},
       eprint = {1201.2681},
 primaryClass = {astro-ph.HE},
       adsurl = {https://ui.adsabs.harvard.edu/abs/2012MNRAS.422.1436O}
}

@ARTICLE{MMM-24,
       author = {{Musso}, M. and {Bodo}, G. and {Mamatsashvili}, G. and {Rossi}, P. and {Mignone}, A.},
        title = "{Evolution of current- and pressure-driven instabilities in relativistic jets}",
      journal = {\mnras},
         year = 2024,
        month = aug,
       volume = {532},
       number = {4},
        pages = {4810-4825},
          doi = {10.1093/mnras/stae1788},
archivePrefix = {arXiv},
       eprint = {2407.15477},
 primaryClass = {astro-ph.HE},
       adsurl = {https://ui.adsabs.harvard.edu/abs/2024MNRAS.532.4810M}
}

@ARTICLE{GS-06,
       author = {{Giannios}, D. and {Spruit}, H.~C.},
        title = "{The role of kink instability in Poynting-flux dominated jets}",
      journal = {\aap},
         year = 2006,
        month = may,
       volume = {450},
       number = {3},
        pages = {887-898},
          doi = {10.1051/0004-6361:20054107},
archivePrefix = {arXiv},
       eprint = {astro-ph/0601172},
 primaryClass = {astro-ph},
       adsurl = {https://ui.adsabs.harvard.edu/abs/2006A&A...450..887G}
}

@ARTICLE{BSDP-19,
       author = {{Bromberg}, Omer and {Singh}, Chandra B. and {Davelaar}, Jordy and {Philippov}, Alexander A.},
        title = "{Kink Instability: Evolution and Energy Dissipation in Relativistic Force-free Nonrotating Jets}",
      journal = {\apj},
         year = 2019,
        month = oct,
       volume = {884},
       number = {1},
          eid = {39},
        pages = {39},
          doi = {10.3847/1538-4357/ab3fa5},
archivePrefix = {arXiv},
       eprint = {1908.08620},
 primaryClass = {astro-ph.HE},
       adsurl = {https://ui.adsabs.harvard.edu/abs/2019ApJ...884...39B}
}

@ARTICLE{ARW-24,
       author = {{Anantua}, Richard and {Ricarte}, Angelo and {Wong}, George and {Emami}, Razieh and {Blandford}, Roger and {Oramas}, Lani and {West}, Hayley and {Duran}, Joaquin and {Curd}, Brandon},
        title = "{On the comparison of AGN with GRMHD simulations - II. M87}",
      journal = {\mnras},
         year = 2024,
        month = feb,
       volume = {528},
       number = {1},
        pages = {735-756},
          doi = {10.1093/mnras/stad3998},
archivePrefix = {arXiv},
       eprint = {2309.05602},
 primaryClass = {astro-ph.HE},
       adsurl = {https://ui.adsabs.harvard.edu/abs/2024MNRAS.528..735A}
}

@ARTICLE{NGC315-24,
       author = {{Park}, Jongho and {Zhao}, Guang-Yao and {Nakamura}, Masanori and {Mizuno}, Yosuke and {Pu}, Hung-Yi and {Asada}, Keiichi and {Takahashi}, Kazuya and {Toma}, Kenji and {Kino}, Motoki and {Cho}, Ilje and {Hada}, Kazuhiro and {Edwards}, Phil G. and {Ro}, Hyunwook and {Kam}, Minchul and {Yi}, Kunwoo and {Lee}, Yunjeong and {Koyama}, Shoko and {Byun}, Do-Young and {Phillips}, Chris and {Reynolds}, Cormac and {Hodgson}, Jeffrey A. and {Lee}, Sang-Sung},
        title = "{Discovery of Limb Brightening in the Parsec-scale Jet of NGC 315 through Global Very Long Baseline Interferometry Observations and Its Implications for Jet Models}",
      journal = {\apjl},
         year = 2024,
        month = oct,
       volume = {973},
       number = {2},
          eid = {L45},
        pages = {L45},
          doi = {10.3847/2041-8213/ad7137},
archivePrefix = {arXiv},
       eprint = {2408.09069},
 primaryClass = {astro-ph.HE},
       adsurl = {https://ui.adsabs.harvard.edu/abs/2024ApJ...973L..45P}
}

@ARTICLE{CenA-21,
       author = {{Janssen}, Michael and {Falcke}, Heino and {Kadler}, Matthias and {Ros}, Eduardo and {Wielgus}, Maciek and {Akiyama}, Kazunori and {Balokovi{\'c}}, Mislav and {Blackburn}, Lindy and {Bouman}, Katherine L. and {Chael}, Andrew and {Chan}, Chi-kwan and {Chatterjee}, Koushik and {Davelaar}, Jordy and {Edwards}, Philip G. and {Fromm}, Christian M. and {G{\'o}mez}, Jos{\'e} L. and {Goddi}, Ciriaco and {Issaoun}, Sara and {Johnson}, Michael D. and {Kim}, Junhan and {Koay},},
        title = "{Event Horizon Telescope observations of the jet launching and collimation in Centaurus A}",
      journal = {Nature Astronomy},
         year = 2021,
        month = jul,
       volume = {5},
        pages = {1017-1028},
          doi = {10.1038/s41550-021-01417-w},
archivePrefix = {arXiv},
       eprint = {2111.03356},
 primaryClass = {astro-ph.GA},
       adsurl = {https://ui.adsabs.harvard.edu/abs/2021NatAs...5.1017J}
}

@ARTICLE{Walker-M87,
       author = {{Walker}, R. Craig and {Hardee}, Philip E. and {Davies}, Frederick B. and {Ly}, Chun and {Junor}, William},
        title = "{The Structure and Dynamics of the Subparsec Jet in M87 Based on 50 VLBA Observations over 17 Years at 43 GHz}",
      journal = {\apj},
         year = 2018,
        month = mar,
       volume = {855},
       number = {2},
          eid = {128},
        pages = {128},
          doi = {10.3847/1538-4357/aaafcc},
archivePrefix = {arXiv},
       eprint = {1802.06166},
 primaryClass = {astro-ph.HE},
       adsurl = {https://ui.adsabs.harvard.edu/abs/2018ApJ...855..128W}
}

@ARTICLE{GVA-09,
       author = {{Gracia}, J. and {Vlahakis}, N. and {Agudo}, I. and {Tsinganos}, K. and {Bogovalov}, S.~V.},
        title = "{Synthetic Synchrotron Emission Maps from MHD Models for the Jet of M87}",
      journal = {\apj},
         year = 2009,
        month = apr,
       volume = {695},
       number = {1},
        pages = {503-510},
          doi = {10.1088/0004-637X/695/1/503},
archivePrefix = {arXiv},
       eprint = {0901.2634},
 primaryClass = {astro-ph.GA},
       adsurl = {https://ui.adsabs.harvard.edu/abs/2009ApJ...695..503G}
}

@ARTICLE{GL-16,
       author = {{Globus}, N. and {Levinson}, A.},
        title = "{The collimation of magnetic jets by disc winds}",
      journal = {\mnras},
         year = 2016,
        month = sep,
       volume = {461},
       number = {3},
        pages = {2605-2615},
          doi = {10.1093/mnras/stw1474},
archivePrefix = {arXiv},
       eprint = {1604.07408},
 primaryClass = {astro-ph.HE},
       adsurl = {https://ui.adsabs.harvard.edu/abs/2016MNRAS.461.2605G}
}

@ARTICLE{PF-10,
       author = {{Porth}, Oliver and {Fendt}, Christian},
        title = "{Acceleration and Collimation of Relativistic Magnetohydrodynamic Disk Winds}",
      journal = {\apj},
         year = 2010,
        month = feb,
       volume = {709},
       number = {2},
        pages = {1100-1118},
          doi = {10.1088/0004-637X/709/2/1100},
archivePrefix = {arXiv},
       eprint = {0911.3001},
 primaryClass = {astro-ph.HE},
       adsurl = {https://ui.adsabs.harvard.edu/abs/2010ApJ...709.1100P}
}

@ARTICLE{MLNH-12,
       author = {{Mizuno}, Yosuke and {Lyubarsky}, Yuri and {Nishikawa}, Ken-Ichi and {Hardee}, Philip E.},
        title = "{Three-dimensional Relativistic Magnetohydrodynamic Simulations of Current-driven Instability. III. Rotating Relativistic Jets}",
      journal = {\apj},
         year = 2012,
        month = sep,
       volume = {757},
       number = {1},
          eid = {16},
        pages = {16},
          doi = {10.1088/0004-637X/757/1/16},
archivePrefix = {arXiv},
       eprint = {1207.4949},
 primaryClass = {astro-ph.HE},
       adsurl = {https://ui.adsabs.harvard.edu/abs/2012ApJ...757...16M}
}

@ARTICLE{LG-M87,
       author = {{Levinson}, A. and {Globus}, N.},
        title = "{Reconfinement of highly magnetized jets: implications for HST-1 in M87}",
      journal = {\mnras},
         year = 2017,
        month = feb,
       volume = {465},
       number = {2},
        pages = {1608-1612},
          doi = {10.1093/mnras/stw2902},
archivePrefix = {arXiv},
       eprint = {1609.01091},
 primaryClass = {astro-ph.HE},
       adsurl = {https://ui.adsabs.harvard.edu/abs/2017MNRAS.465.1608L}
}

@ARTICLE{Appl-00,
       author = {{Appl}, S. and {Lery}, T. and {Baty}, H.},
        title = "{Current-driven instabilities in astrophysical jets. Linear analysis}",
      journal = {\aap},
         year = 2000,
        month = mar,
       volume = {355},
        pages = {818-828},
       adsurl = {https://ui.adsabs.harvard.edu/abs/2000A&A...355..818A}
}

@ARTICLE{LBA-00,
       author = {{Lery}, T. and {Baty}, H. and {Appl}, S.},
        title = "{Current-driven instabilities in astrophysical jets. Non linear development}",
      journal = {\aap},
         year = 2000,
        month = mar,
       volume = {355},
        pages = {1201-1208},
       adsurl = {https://ui.adsabs.harvard.edu/abs/2000A&A...355.1201L}
}

@ARTICLE{Hardee-82,
       author = {{Hardee}, P.~E.},
        title = "{Helical and pinching instability of supersonic expanding jets in extragalactic radio sources}",
      journal = {\apj},
         year = 1982,
        month = jun,
       volume = {257},
        pages = {509-526},
          doi = {10.1086/160008},
       adsurl = {https://ui.adsabs.harvard.edu/abs/1982ApJ...257..509H}
}

@ARTICLE{PK-15,
       author = {{Porth}, Oliver and {Komissarov}, Serguei S.},
        title = "{Causality and stability of cosmic jets}",
      journal = {\mnras},
         year = 2015,
        month = sep,
       volume = {452},
       number = {2},
        pages = {1089-1104},
          doi = {10.1093/mnras/stv1295},
archivePrefix = {arXiv},
       eprint = {1408.3318},
 primaryClass = {astro-ph.HE},
       adsurl = {https://ui.adsabs.harvard.edu/abs/2015MNRAS.452.1089P}
}

@ARTICLE{BT-16,
       author = {{Bromberg}, Omer and {Tchekhovskoy}, Alexander},
        title = "{Relativistic MHD simulations of core-collapse GRB jets: 3D instabilities and magnetic dissipation}",
      journal = {\mnras},
         year = 2016,
        month = feb,
       volume = {456},
       number = {2},
        pages = {1739-1760},
          doi = {10.1093/mnras/stv2591},
archivePrefix = {arXiv},
       eprint = {1508.02721},
 primaryClass = {astro-ph.HE},
       adsurl = {https://ui.adsabs.harvard.edu/abs/2016MNRAS.456.1739B}
}

@ARTICLE{TB-16,
       author = {{Tchekhovskoy}, Alexander and {Bromberg}, Omer},
        title = "{Three-dimensional relativistic MHD simulations of active galactic nuclei jets: magnetic kink instability and Fanaroff-Riley dichotomy}",
      journal = {\mnras},
         year = 2016,
        month = sep,
       volume = {461},
       number = {1},
        pages = {L46-L50},
          doi = {10.1093/mnrasl/slw064},
archivePrefix = {arXiv},
       eprint = {1512.04526},
 primaryClass = {astro-ph.HE},
       adsurl = {https://ui.adsabs.harvard.edu/abs/2016MNRAS.461L..46T}
}

@article{Kruskal1958,
    author = {Kruskal, M. and Tuck, J. L.},
    title = {The instability of a pinched fluid with a longitudinal magnetic field},
    journal = {Proceedings of the Royal Society of London. A. Mathematical and Physical Sciences},
    volume = {245},
    number = {1241},
    pages = {222-237},
    year = {1958},
    month = {06},
    issn = {0080-4630},
    doi = {10.1098/rspa.1958.0079},
    url = {https://doi.org/10.1098/rspa.1958.0079},
    eprint = {https://royalsocietypublishing.org/rspa/article-pdf/245/1241/222/50920/rspa.1958.0079.pdf},
}

@article{Shafranov1956,
  author = {Shafranov, V. D.},
  title = {The stability of a cylindrical gaseous conductor in a magnetic field},
  journal = {The Soviet Journal of Atomic Energy},
  year = {1956},
  month = {10},
  day = {01},
  volume = {1},
  number = {5},
  pages = {709--713},
  issn = {1573-8205},
  doi = {10.1007/BF01480907},
  url = {https://doi.org/10.1007/BF01480907}
}

@article{10.1093/mnras/179.3.433,
    author = {Blandford, R. D. and Znajek, R. L.},
    title = {Electromagnetic extraction of energy from Kerr black holes},
    journal = {Monthly Notices of the Royal Astronomical Society},
    volume = {179},
    number = {3},
    pages = {433-456},
    year = {1977},
    month = {07},
    issn = {0035-8711},
    doi = {10.1093/mnras/179.3.433},
    url = {https://doi.org/10.1093/mnras/179.3.433},
    eprint = {https://academic.oup.com/mnras/article-pdf/179/3/433/9333653/mnras179-0433.pdf},
}

@article{Nakamura_2013,
doi = {10.1088/0004-637X/775/2/118},
url = {https://doi.org/10.1088/0004-637X/775/2/118},
year = {2013},
month = {sep},
publisher = {The American Astronomical Society},
volume = {775},
number = {2},
pages = {118},
author = {Nakamura, Masanori and Asada, Keiichi},
title = {THE PARABOLIC JET STRUCTURE IN M87 AS A MAGNETOHYDRODYNAMIC NOZZLE},
journal = {The Astrophysical Journal},
}

@article{Vlahakis_2004,
doi = {10.1086/382670},
url = {https://doi.org/10.1086/382670},
year = {2004},
month = {apr},
publisher = {},
volume = {605},
number = {2},
pages = {656},
author = {Vlahakis, Nektarios and Königl, Arieh},
title = {Magnetic Driving of Relativistic Outflows in Active Galactic Nuclei. I. Interpretation of Parsec-Scale Accelerations},
journal = {The Astrophysical Journal},
}

@article{Lyubarsky_2009,
doi = {10.1088/0004-637X/698/2/1570},
url = {https://doi.org/10.1088/0004-637X/698/2/1570},
year = {2009},
month = {jun},
publisher = {The American Astronomical Society},
volume = {698},
number = {2},
pages = {1570},
author = {Lyubarsky, Yuri},
title = {ASYMPTOTIC STRUCTURE OF POYNTING-DOMINATED JETS},
journal = {The Astrophysical Journal},
}

@article{bodo2020,
    author = {Bodo, G and Tavecchio, F and Sironi, L},
    title = {Kink-driven magnetic reconnection in relativistic jets: consequences for X-ray polarimetry of BL Lacs},
    journal = {Monthly Notices of the Royal Astronomical Society},
    volume = {501},
    number = {2},
    pages = {2836-2847},
    year = {2020},
    month = {11},
    issn = {0035-8711},
    doi = {10.1093/mnras/staa3620},
    url = {https://doi.org/10.1093/mnras/staa3620},
    eprint = {https://academic.oup.com/mnras/article-pdf/501/2/2836/35559260/staa3620.pdf},
}

@article{10.1093/mnras/stx807,
    author = {Sobacchi, E. and Lyubarsky, Y. E. and Sormani, M. C.},
    title = {Kink instability of force-free jets: a parameter space study},
    journal = {Monthly Notices of the Royal Astronomical Society},
    volume = {468},
    number = {4},
    pages = {4635-4641},
    year = {2017},
    month = {04},
    issn = {0035-8711},
    doi = {10.1093/mnras/stx807},
    url = {https://doi.org/10.1093/mnras/stx807},
    eprint = {https://academic.oup.com/mnras/article-pdf/468/4/4635/14161709/stx807.pdf},
}

@article{anjiri2014,
    author = {Anjiri, M. and Mignone, A. and Bodo, G. and Rossi, P.},
    title = {Linear and non-linear evolution of current-carrying highly magnetized jets},
    journal = {Monthly Notices of the Royal Astronomical Society},
    volume = {442},
    number = {3},
    pages = {2228-2239},
    year = {2014},
    month = {06},
    issn = {0035-8711},
    doi = {10.1093/mnras/stu1004},
    url = {https://doi.org/10.1093/mnras/stu1004},
    eprint = {https://academic.oup.com/mnras/article-pdf/442/3/2228/3477932/stu1004.pdf},
}

@article{10.1093/mnras/stab3492,
    author = {Bodo, G and Mamatsashvili, G and Rossi, P and Mignone, A},
    title = {Current-driven kink instabilities in relativistic jets: dissipation properties},
    journal = {Monthly Notices of the Royal Astronomical Society},
    volume = {510},
    number = {2},
    pages = {2391-2406},
    year = {2022},
    month = {02},
    issn = {0035-8711},
    doi = {10.1093/mnras/stab3492},
    url = {https://doi.org/10.1093/mnras/stab3492},
    eprint = {https://academic.oup.com/mnras/article-pdf/510/2/2391/42076598/stab3492.pdf},
}

@article{mizuno2014,
doi = {10.1088/0004-637X/784/2/167},
url = {https://doi.org/10.1088/0004-637X/784/2/167},
year = {2014},
month = {mar},
publisher = {The American Astronomical Society},
volume = {784},
number = {2},
pages = {167},
author = {Mizuno, Yosuke and Hardee, Philip E. and Nishikawa, Ken-Ichi},
title = {SPATIAL GROWTH OF THE CURRENT-DRIVEN INSTABILITY IN RELATIVISTIC JETS},
journal = {The Astrophysical Journal},
}

@article{Mizuno_2011,
doi = {10.1088/0004-637X/734/1/19},
url = {https://doi.org/10.1088/0004-637X/734/1/19},
year = {2011},
month = {may},
publisher = {The American Astronomical Society},
volume = {734},
number = {1},
pages = {19},
author = {Mizuno, Yosuke and Hardee, Philip E. and Nishikawa, Ken-Ichi},
title = {THREE-DIMENSIONAL RELATIVISTIC MAGNETOHYDRODYNAMIC SIMULATIONS OF CURRENT-DRIVEN INSTABILITY WITH A SUB-ALFVÉNIC JET: TEMPORAL PROPERTIES},
journal = {The Astrophysical Journal}
}

@article{Mizuno_2009,
doi = {10.1088/0004-637X/700/1/684},
url = {https://doi.org/10.1088/0004-637X/700/1/684},
year = {2009},
month = {jul},
publisher = {The American Astronomical Society},
volume = {700},
number = {1},
pages = {684},
author = {Mizuno, Yosuke and Lyubarsky, Yuri and Nishikawa, Ken-Ichi and Hardee, Philip E.},
title = {THREE-DIMENSIONAL RELATIVISTIC MAGNETOHYDRODYNAMIC SIMULATIONS OF CURRENT-DRIVEN INSTABILITY. I. INSTABILITY OF A STATIC COLUMN},
journal = {The Astrophysical Journal}
}

@article{Phillips_2024,
    author = {Komissarov, Serguei S and Phillips, David},
    title = {A splitting method for numerical relativistic magnetohydrodynamics},
    journal = {Monthly Notices of the Royal Astronomical Society},
    volume = {536},
    number = {2},
    pages = {1268-1302},
    year = {2024},
    month = {11},
    issn = {0035-8711},
    doi = {10.1093/mnras/stae2620},
    url = {https://doi.org/10.1093/mnras/stae2620},
}

@article{Narayan_2009,
doi = {10.1088/0004-637X/697/2/1681},
url = {https://doi.org/10.1088/0004-637X/697/2/1681},
year = {2009},
month = {may},
publisher = {The American Astronomical Society},
volume = {697},
number = {2},
pages = {1681},
author = {Narayan, Ramesh and Li, Jason and Tchekhovskoy, Alexander},
title = {STABILITY OF RELATIVISTIC FORCE-FREE JETS},
journal = {The Astrophysical Journal}
}

@article{Singh_2016,
 doi = {10.3847/0004-637X/824/1/48},
 url = {https://doi.org/10.3847/0004-637X/824/1/48},
 year = {2016},
 month = {jun},
 publisher = {The American Astronomical Society},
 volume = {824},
 number = {1},
 pages = {48},
 author = {Singh, Chandra B. and Mizuno, Yosuke and Pino, Elisabete M. de Gouveia Dal},
 title = {SPATIAL GROWTH OF CURRENT-DRIVEN INSTABILITY IN RELATIVISTIC ROTATING JETS AND THE SEARCH FOR MAGNETIC RECONNECTION},
 journal = {The Astrophysical Journal}
 }

@BOOK{Bateman1978,
       author = {{Bateman}, G.},
        title = "{MHD instabilities}",
        year = 1978,
        publisher = {MIT Press},
        address   = {Cambridge, MA},
       adsurl = {https://ui.adsabs.harvard.edu/abs/1978mit..book.....B}}

@article{10.1093/mnras/stt2247,
    author = {Barkov, Maxim and Komissarov, Serguei S. and Korolev, Vitaly and Zankovich, Andrey},
    title = {A multidimensional numerical scheme for two-fluid relativistic magnetohydrodynamics},
    journal = {Monthly Notices of the Royal Astronomical Society},
    volume = {438},
    number = {1},
    pages = {704-716},
    year = {2014},
    month = {02},
    issn = {0035-8711},
    doi = {10.1093/mnras/stt2247},
    url = {https://doi.org/10.1093/mnras/stt2247},
    eprint = {https://academic.oup.com/mnras/article-pdf/438/1/704/18467454/stt2247.pdf},
}



\appendix

\section{Magnetisation of black hole magnetospheres}
\label{app:magnetisation}

For the purpose of order of magnitude estimating the black hole magnetic field and its plasma magnetisation, one may ignore the many complications introduced by the non-Euclidean geometry of the space around black holes.  The only general relativistic results used below are the existence of the gravitational radius, the constraint on the rotation rate of black holes, and its relation to the angular velocity of the black hole magnetosphere.  The main benefit of this approximation is the simplicity of the outcome.

He we assume that the disk accretion onto the black hole occurs in the magnetically-arrested regime.  
The starting point is the relation between the magnetic flux penetrating the black hole horizon $\Phi_b$ and the accretion rate $\dot{M}$, and the black hole mass $M_b$ suggested by GRMHD simulations,  
\beq
\Phi_b=A_\Phi  (\dot{M}c)^{1/2} r_g\,,
\label{eq:Phi1}
\eeq
$r_g=GM_b/c^2$ is the gravitational radius for a black hole of mass $M_b$, and $A_\Phi \simeq 100$ is a constant numerical factor found in the simulations \citep{TNM11}.  The accretion rate is normally measured in  units of the Eddington mass rate 
\beq
\dot{M}_{E} = \frac{L_{E}}{\eta c^2} \,, 
\eeq 
where    
\beq
L_{E} = L_{E,\odot} \frac{M}{M_\odot} \mbox{erg}/s\,,
\eeq
is the Eddington luminosity, $\eta$ is the radiative efficiency of accretion disk, and $L_{E,\odot}=1.3\times10^{38}\mbox{erg}/\mbox{s}$ is the Eddington luminosity for a solar-mass central object. 

Up to few $r_g$, the poloidal magnetic field in the magnetosphere is almost radial  \citep[e.g. ][]{Ripperda22}, and so its strength can be estimated as 
\beq
B_r=\Phi_b/S_h(r)\,,
\eeq
where $S_h(r)$ is the surface area of the hemisphere of radius $r$, for which we use the Euclidean formula $S_h=2\pi r^2$. This yields
\beq
B_r = C_B \,A_\Phi \fracp{\dot{m}}{\eta m_{b}}^{1/2} \fracp{r}{r_g}^{-2} \mbox{Gauss} \,,
\eeq
where
$$
C_B=\frac{1}{4\pi}\fracp{L_{E,\odot}}{c}^{1/2}\frac{c^2}{GM_\odot}=7\times10^7\mbox{Gauss}\,,
$$
$\dot{m}=\dot{M}/\dot{M}_E$ and $m_b=M_b/M_\odot$.

In a steady-state force-free magnetosphere rotating with the angular velocity $\Omega$, the electric field 
\beq
\vv{E} = -\frac{\vpr{v}{B}}{c} = \frac{\Omega\varpi B_r}{c}\ort{\theta}\,.
\eeq               
Here $\{r,\theta,\phi\}$ are Boyer-Lindquist coordinates of black-hole space, treated as normal spherical coordinates of Euclidean space, and $\varpi=r\sin\theta$ is the cylindrical radius.  The electric charge density required to sustain this electric field is
\beq
q=\vdiv{\vv{E}}/4\pi = \frac{1}{4\pi}\frac{\Omega B_r}{c} \sin\theta\,.
\eeq
The corresponding number density of leptons is the Goldreich-Julian number density $n_{GJ}=|q/e|$, where $e$ is the electron charge. As a rule, the pair-production in potential gaps overproduces pairs, leading to the number density $n=\mu n_{GJ}$, where $\mu$ is called the multiplicity factor.

For the split-monopole (monopole) force-free magnetosphere of a rotating black hole, 
\beq
\Omega \approx \frac{\Omega_b}{2}\,,
\eeq
\citep[e.g.][]{10.1093/mnras/179.3.433,SSK01,TNM10} where 
\beq
\Omega_b = \frac{c}{r_g} \frac{a}{r_+^2+a^2} \approx \frac{c}{r_g} a
\eeq
is the angular velocity of the black hole, $r_+=1+\sqrt{1-a^2}$ is the radius of the outer event horizon in units of $r_g$,   and $a \in(-1,1)$ is the standard dimensionless spin parameter describing the constraints imposed on the rotation rate by General Relativity.

Once the number density $n$ of leptons is known, the magnetisation parameter of the magnetosphere can be estimated via 
\beq
\sigma=\frac{B_r^2}{4\pi n m_e c^2}\,,
\eeq
 where $m_e$ is the electron mass. Straightforward calculation yields 
 \beq
 \sigma = C_\sigma\, \frac{A_\Phi}{a\mu} \fracp{\dot{m}\,m_b}{\eta}^{1/2} \fracp{r}{r_g}^{-2} \,,
 \eeq
 where 
 \beq
 C_\sigma=\frac{1}{4\pi} \fracp{L_{E,\odot}}{c}^{1/2} \fracp{e}{m_e c^2} = 3\times 10^9\,.
 \eeq

\section{Test runs}
\label{sec:test-runs}
   
\begin{figure*}
    \centering
     \includegraphics[width=0.325\linewidth]{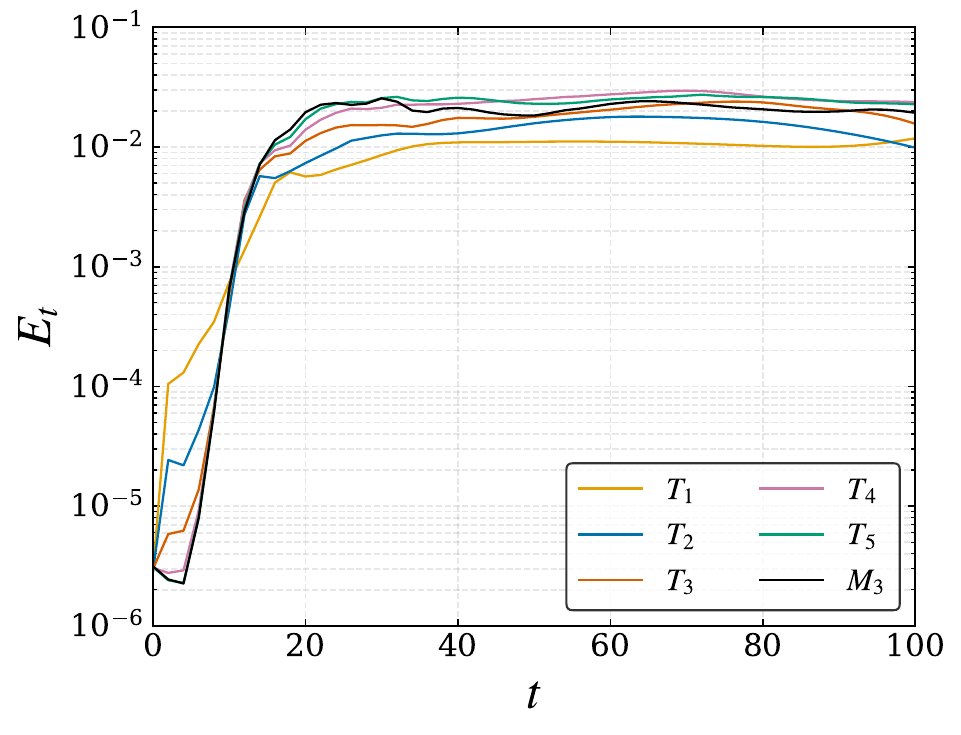}
      \includegraphics[width=0.325\linewidth]{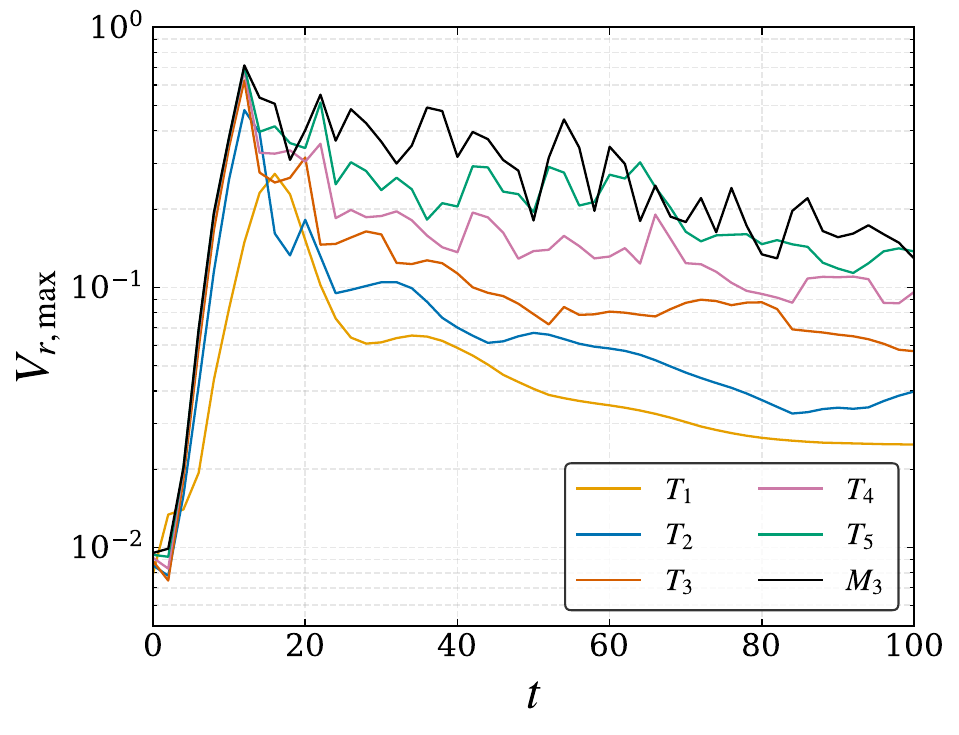}
       \includegraphics[width=0.325\linewidth]{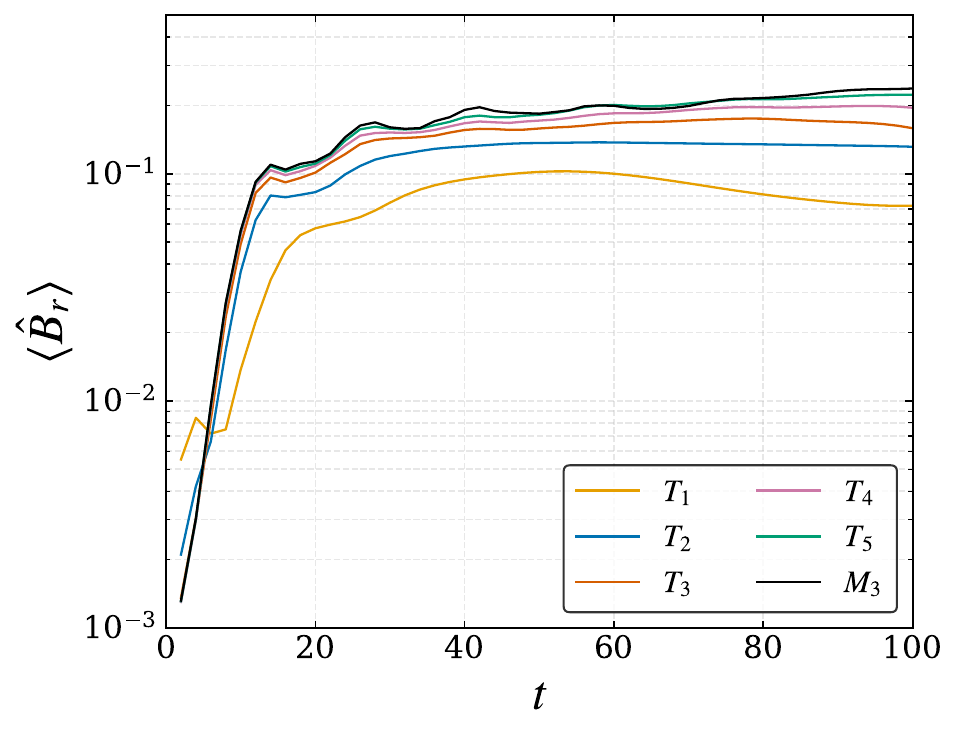}    
    \caption{Temporal evolution of the perturbation-amplitude parameters $E_t$, $V_{r,max}$ and $\langle \hat{B}_r\rangle$ for the entire jet-wind region.}
    \label{fig:CTests-params}
\end{figure*}

\begin{figure*}
    \centering
    \includegraphics[width=0.195\linewidth]{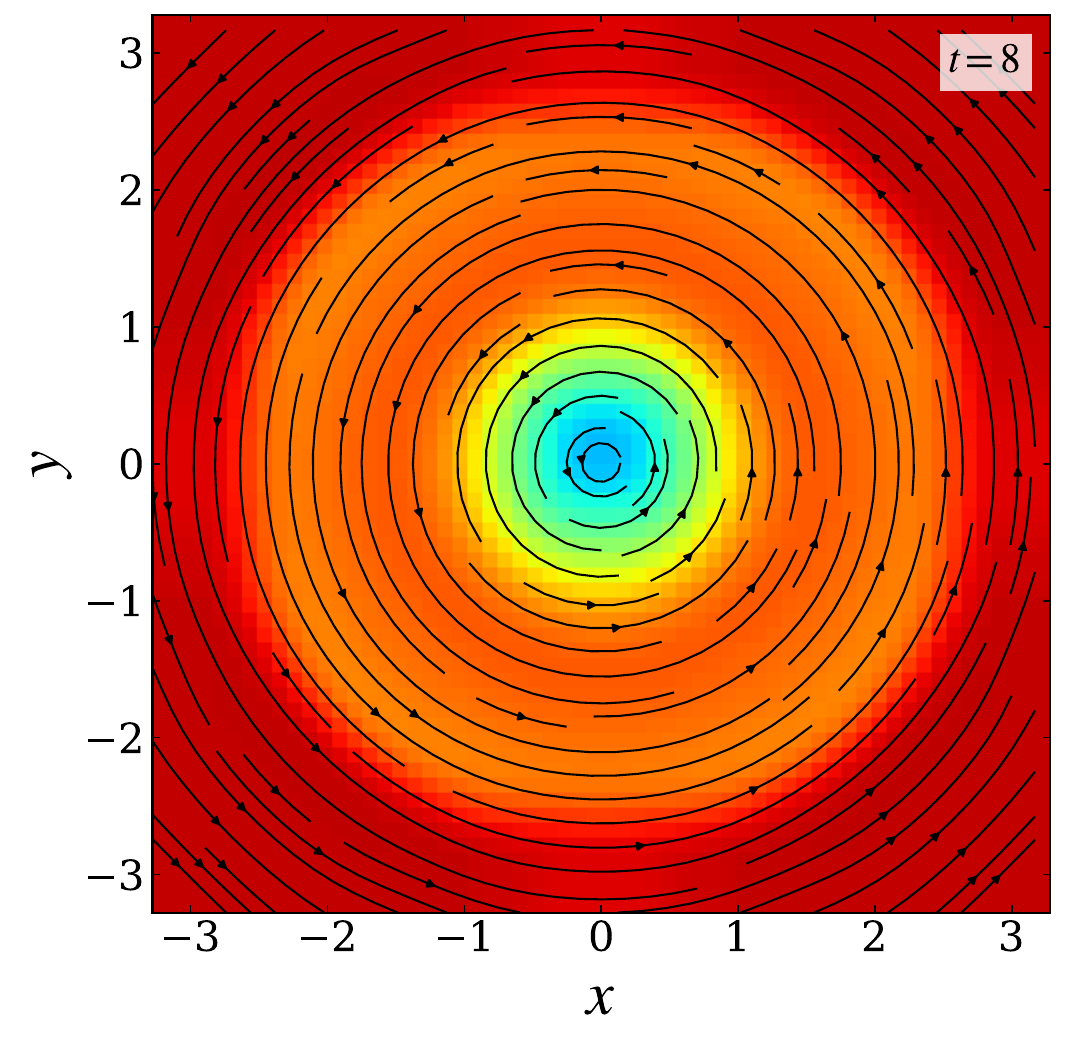}
    \includegraphics[width=0.195\linewidth]{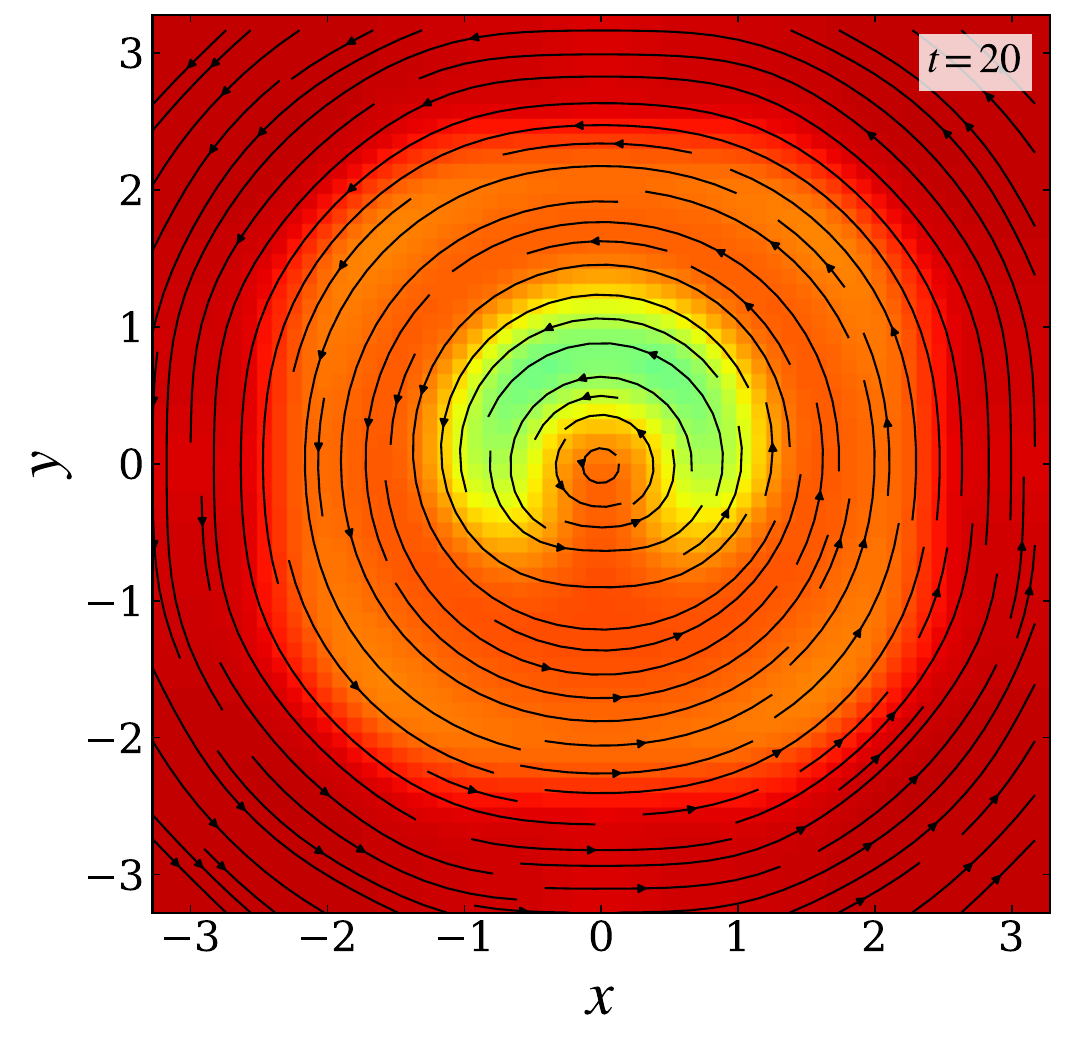}
    \includegraphics[width=0.195\linewidth]{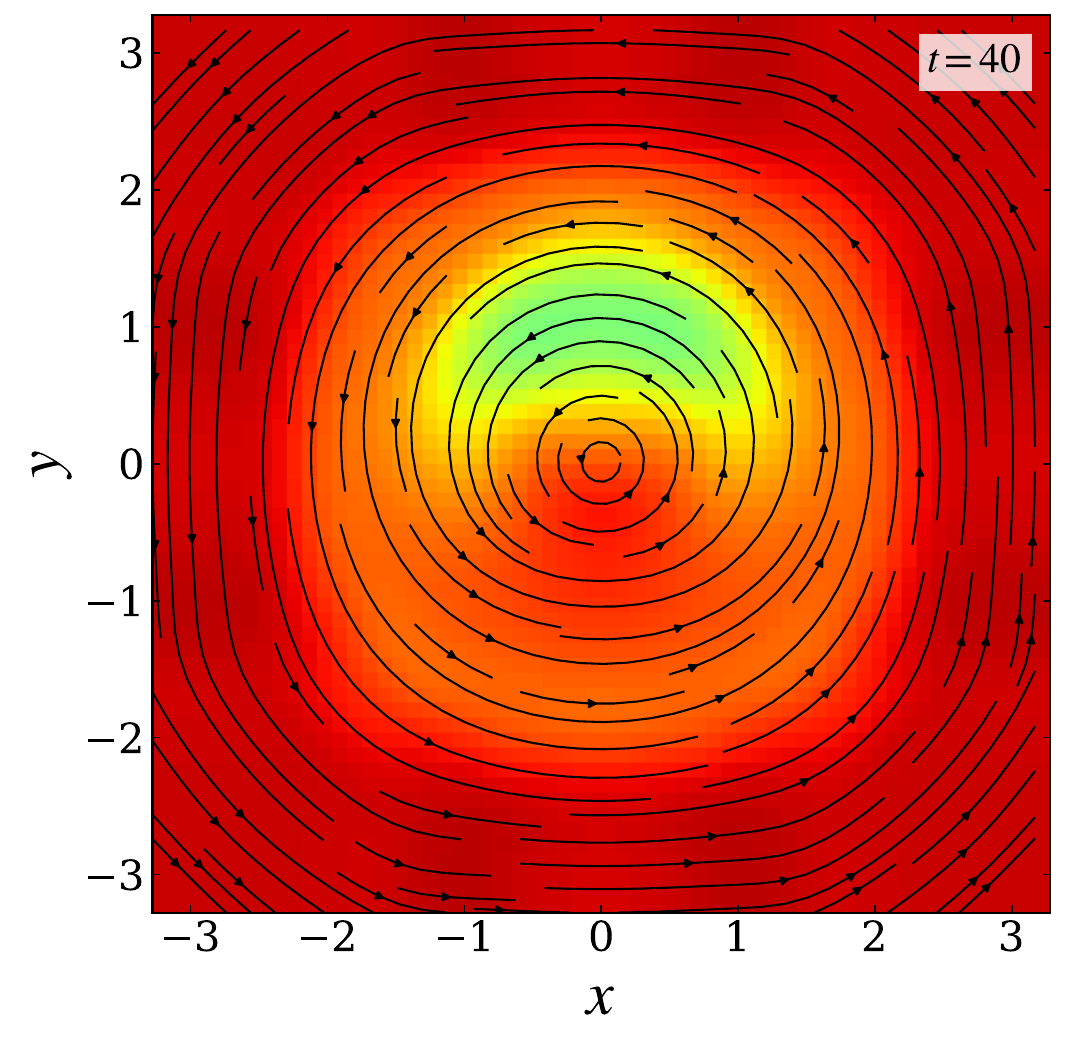}
    \includegraphics[width=0.195\linewidth]{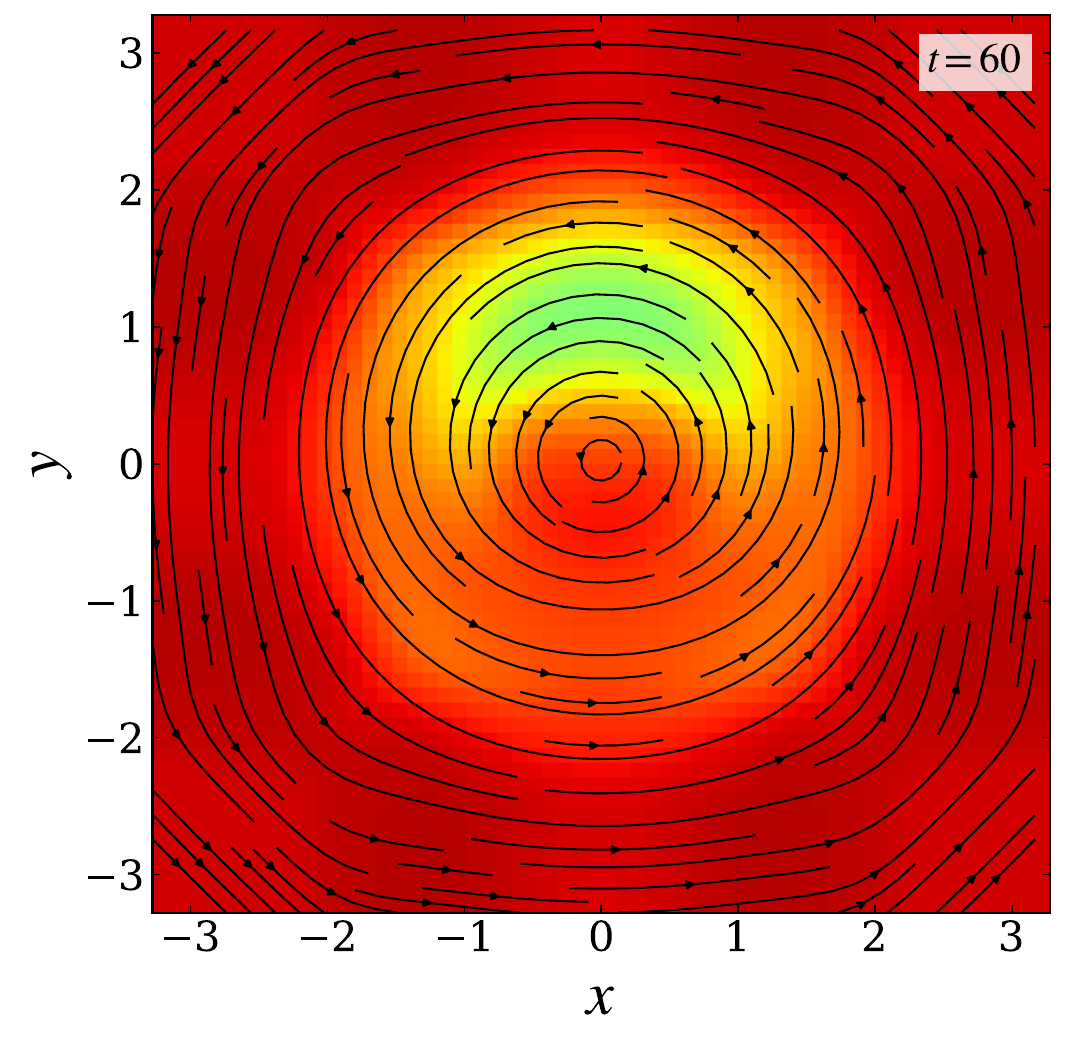}
    \includegraphics[width=0.195\linewidth]{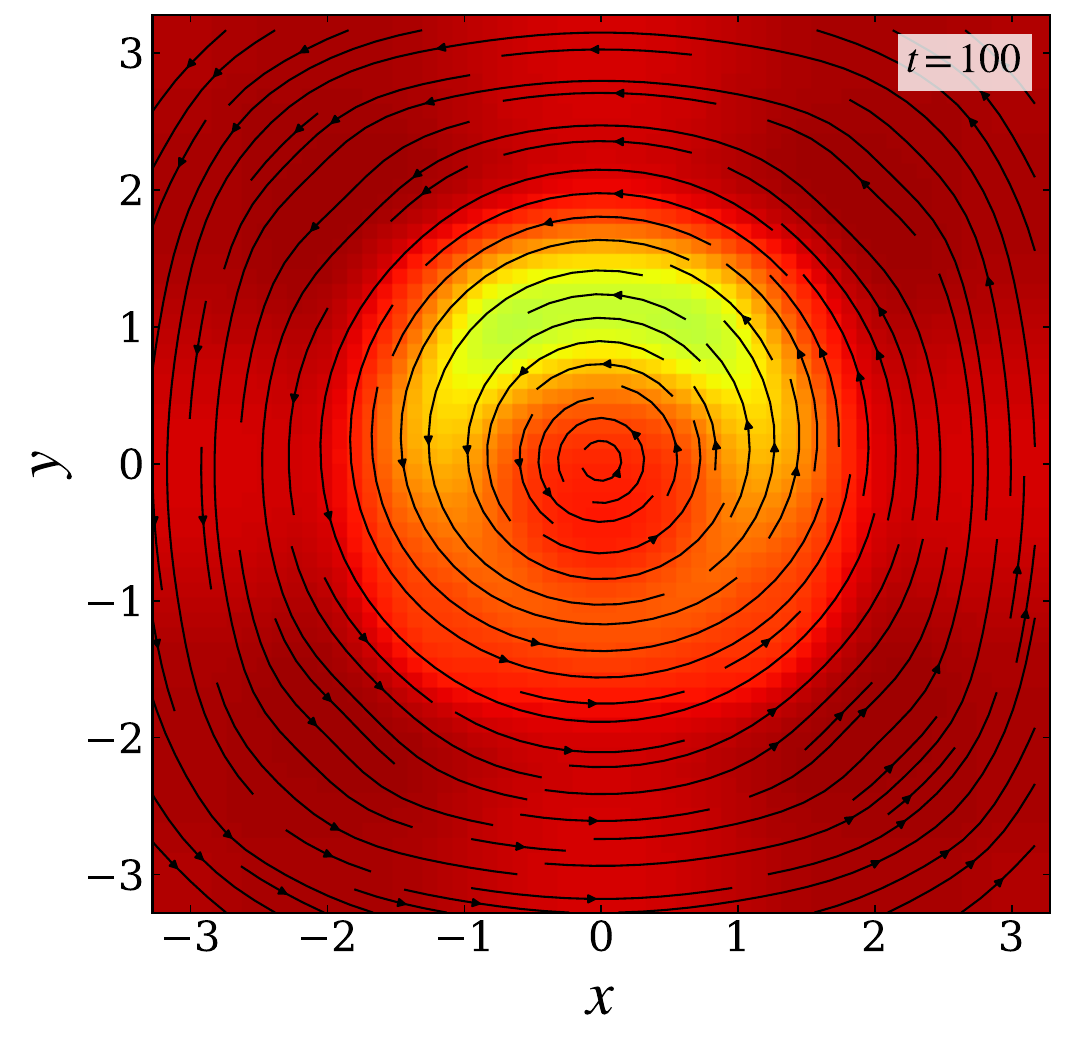}
    \includegraphics[width=0.195\linewidth]{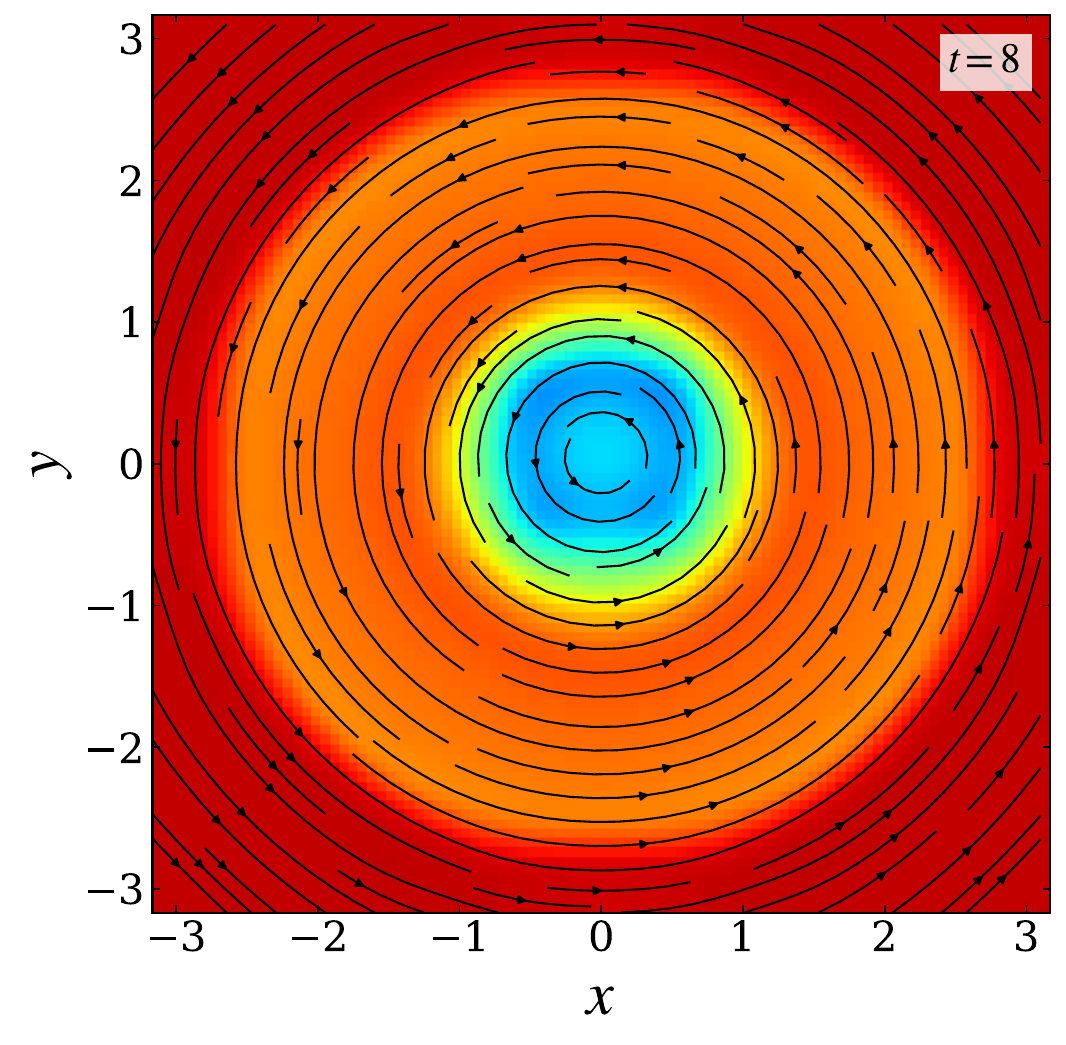}
    \includegraphics[width=0.195\linewidth]{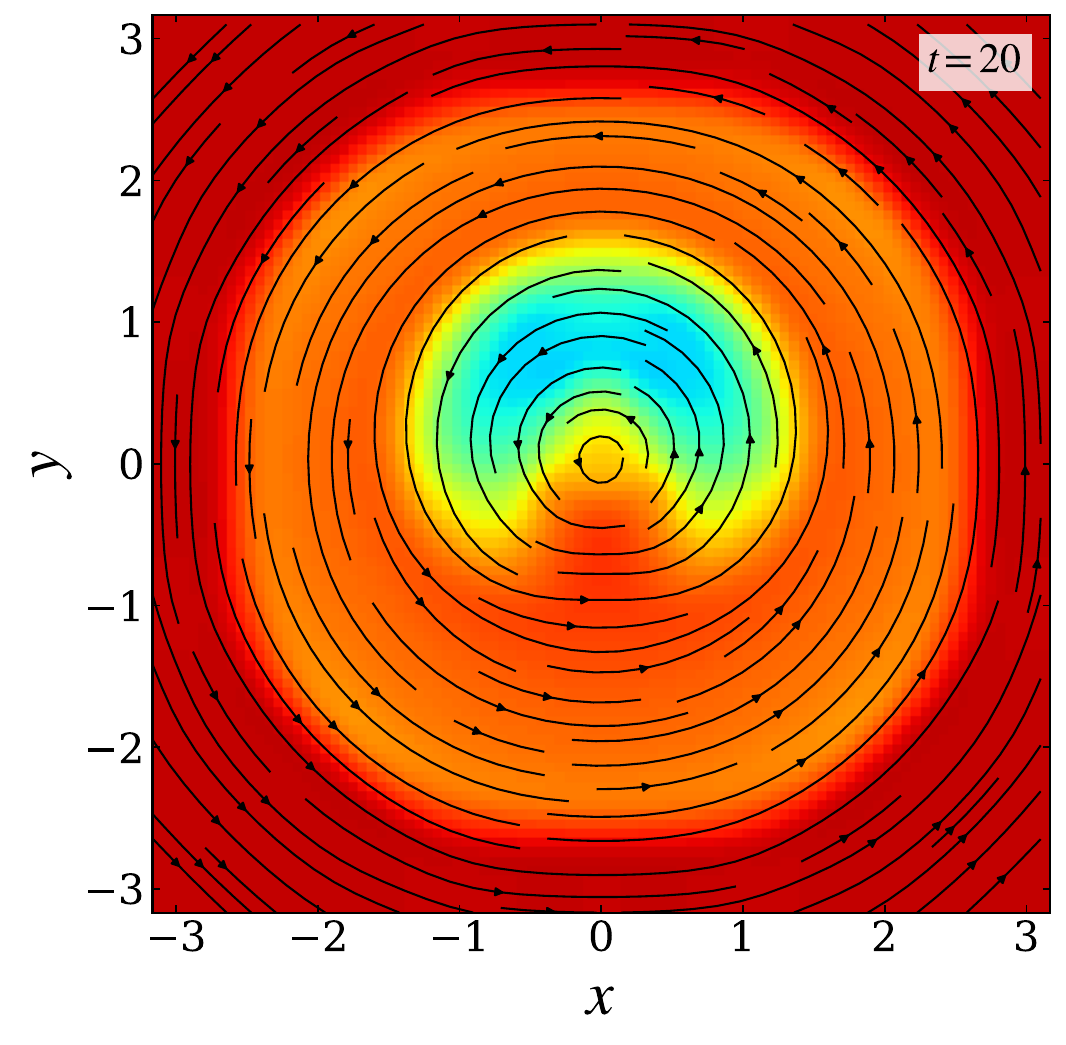}
    \includegraphics[width=0.195\linewidth]{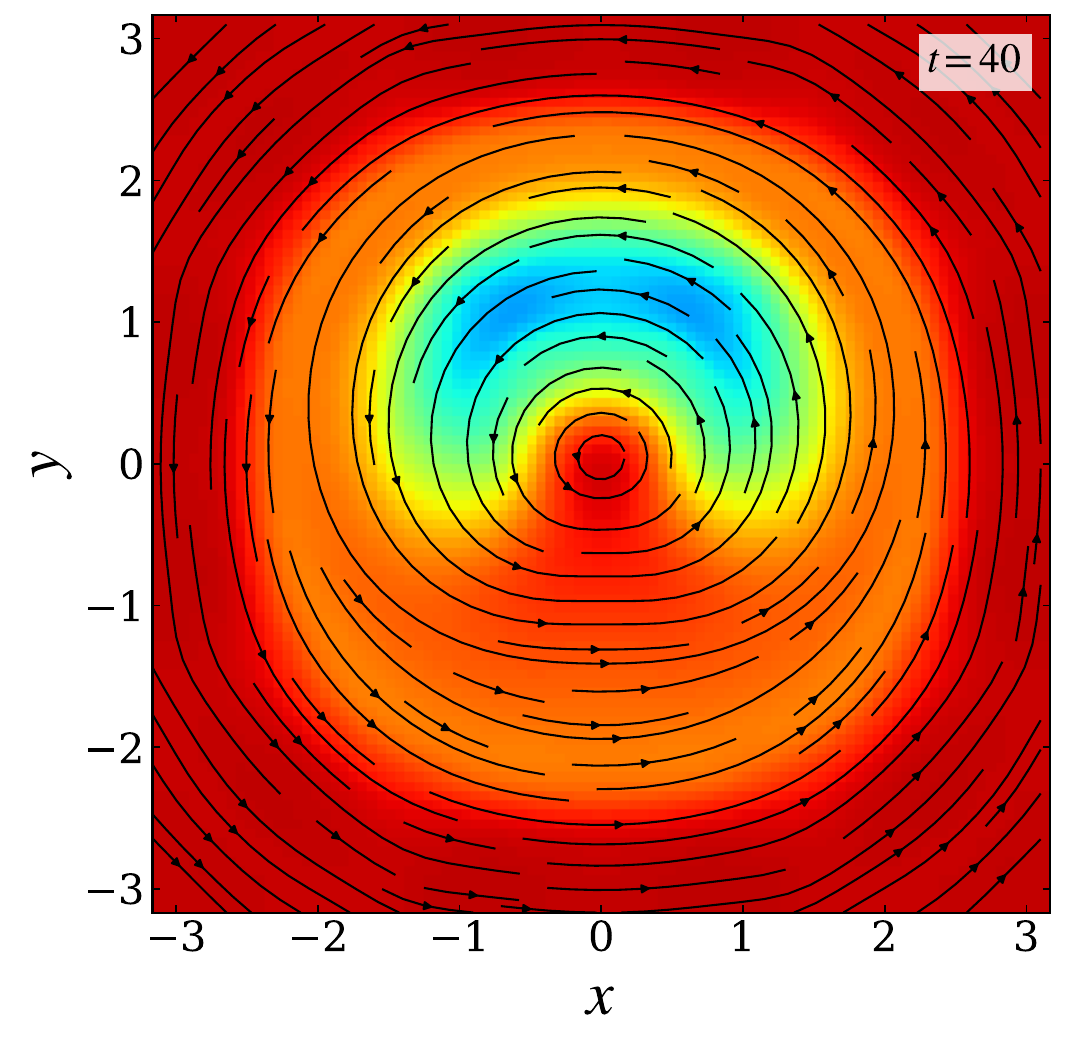}
    \includegraphics[width=0.195\linewidth]{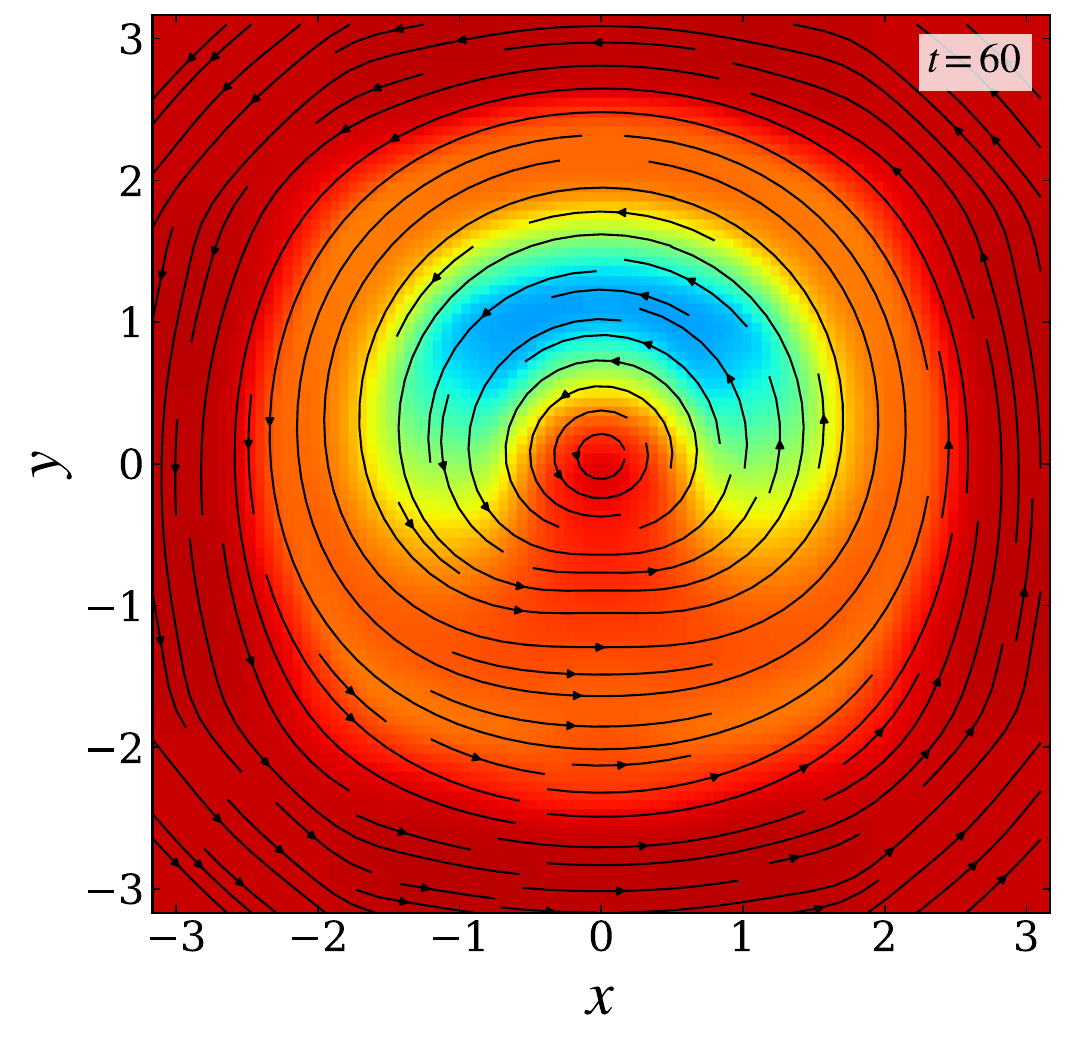}
    \includegraphics[width=0.195\linewidth]{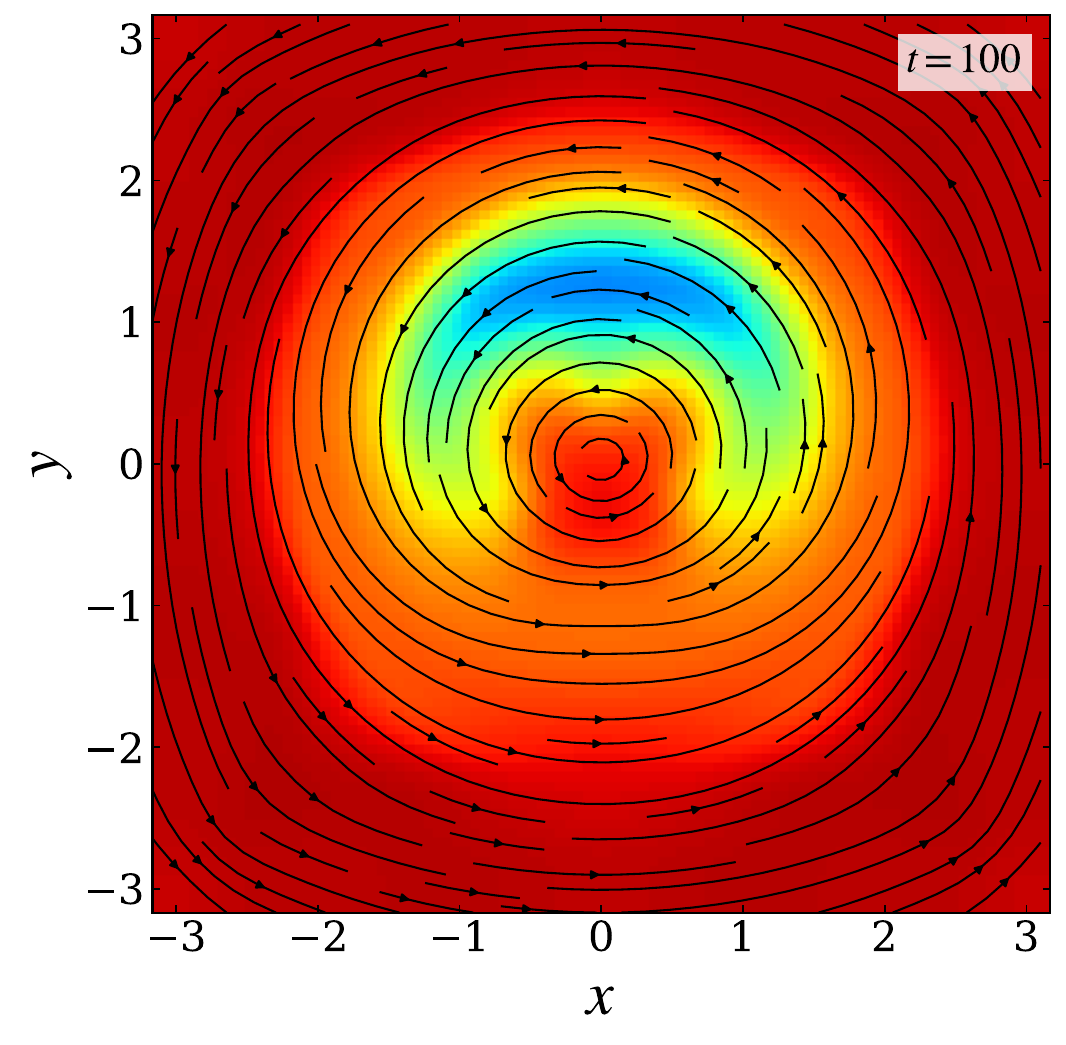}
    \includegraphics[width=0.195\linewidth]{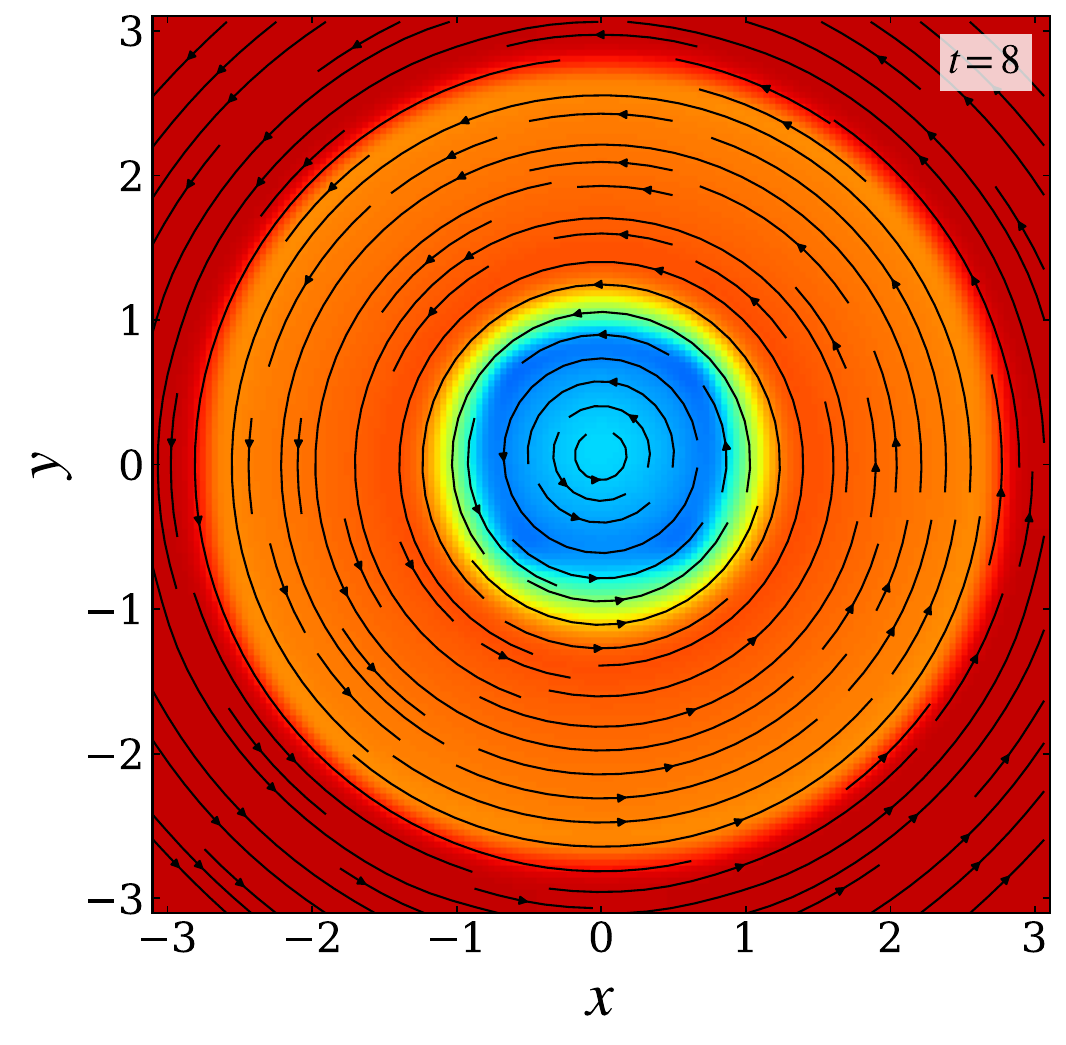}
    \includegraphics[width=0.195\linewidth]{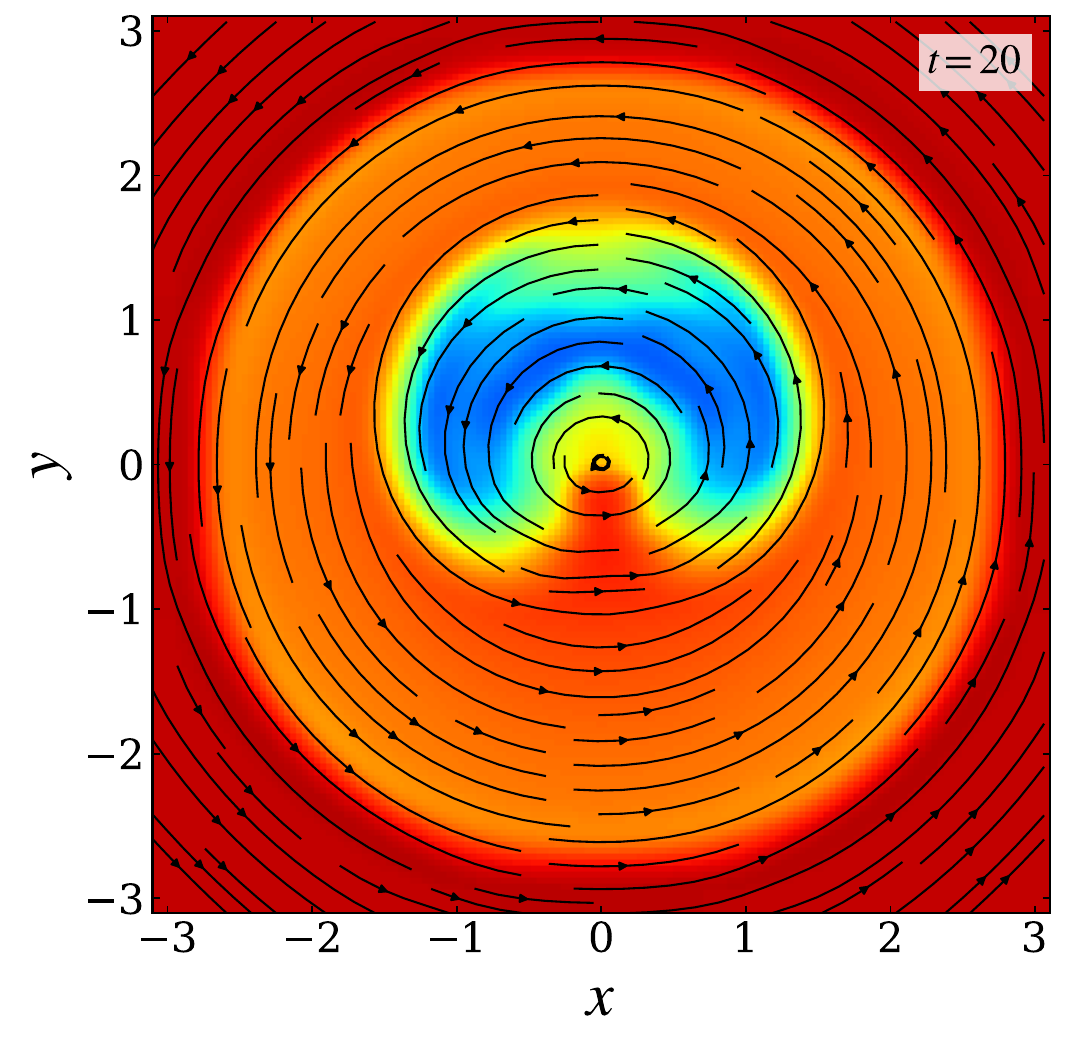}
    \includegraphics[width=0.195\linewidth]{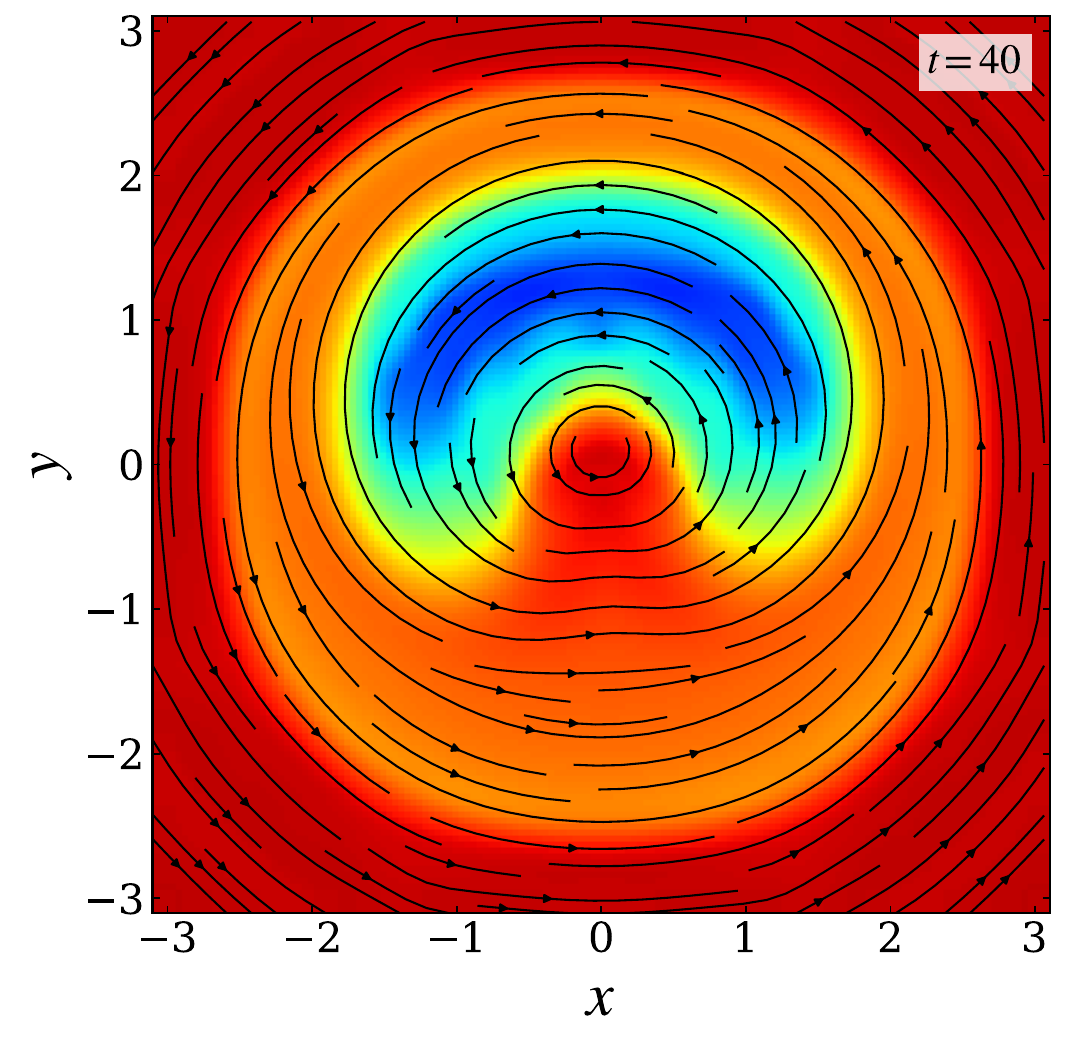}
    \includegraphics[width=0.195\linewidth]{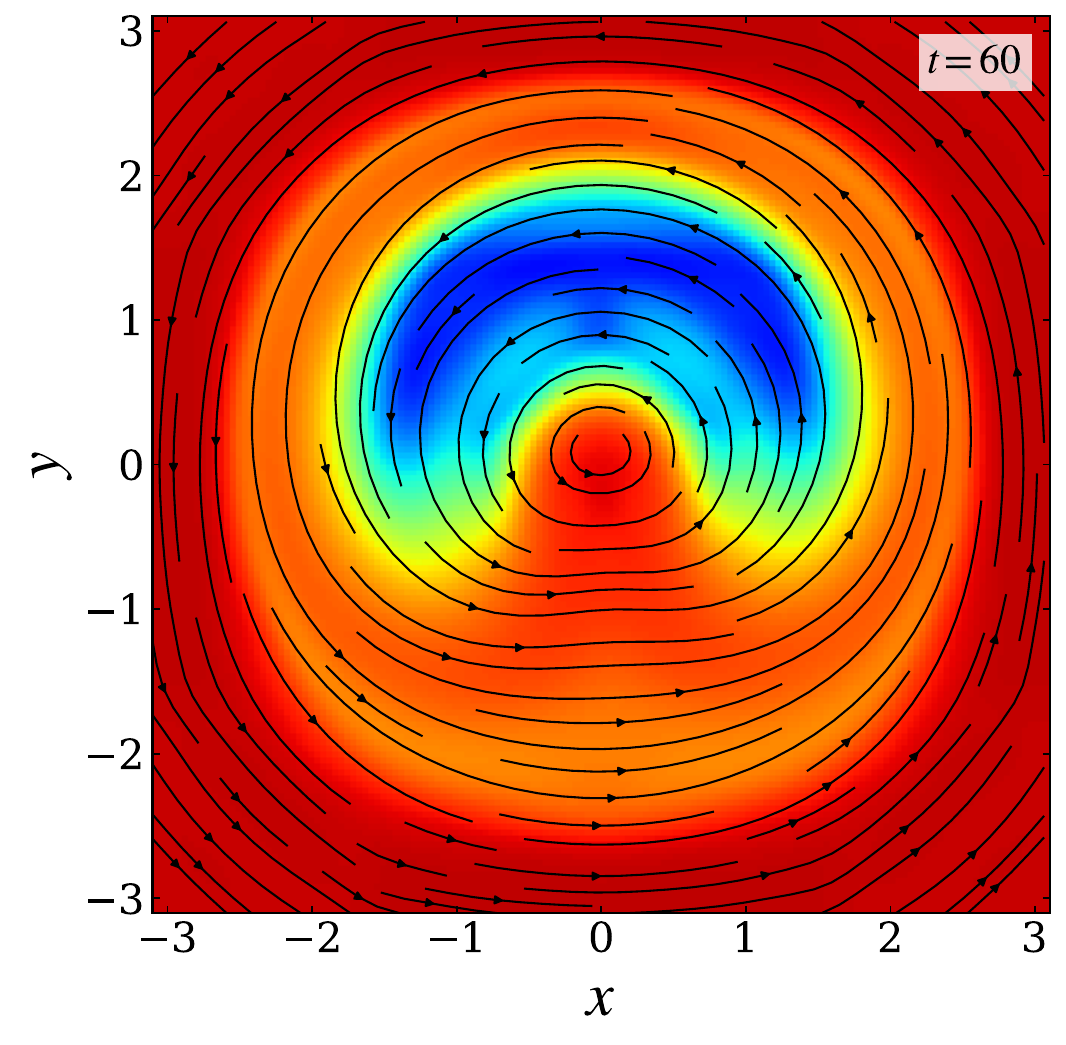}
    \includegraphics[width=0.195\linewidth]{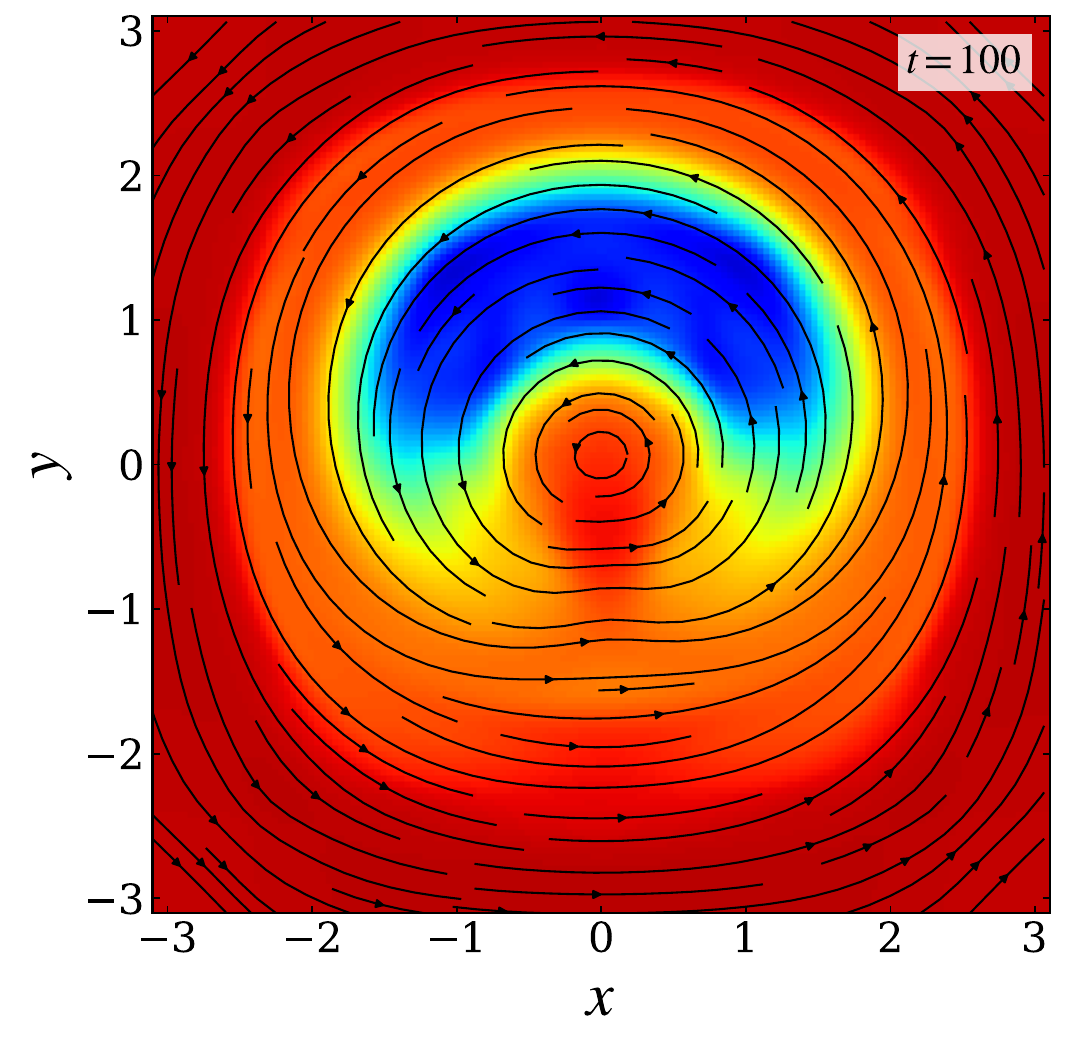}
    \includegraphics[width=0.195\linewidth]{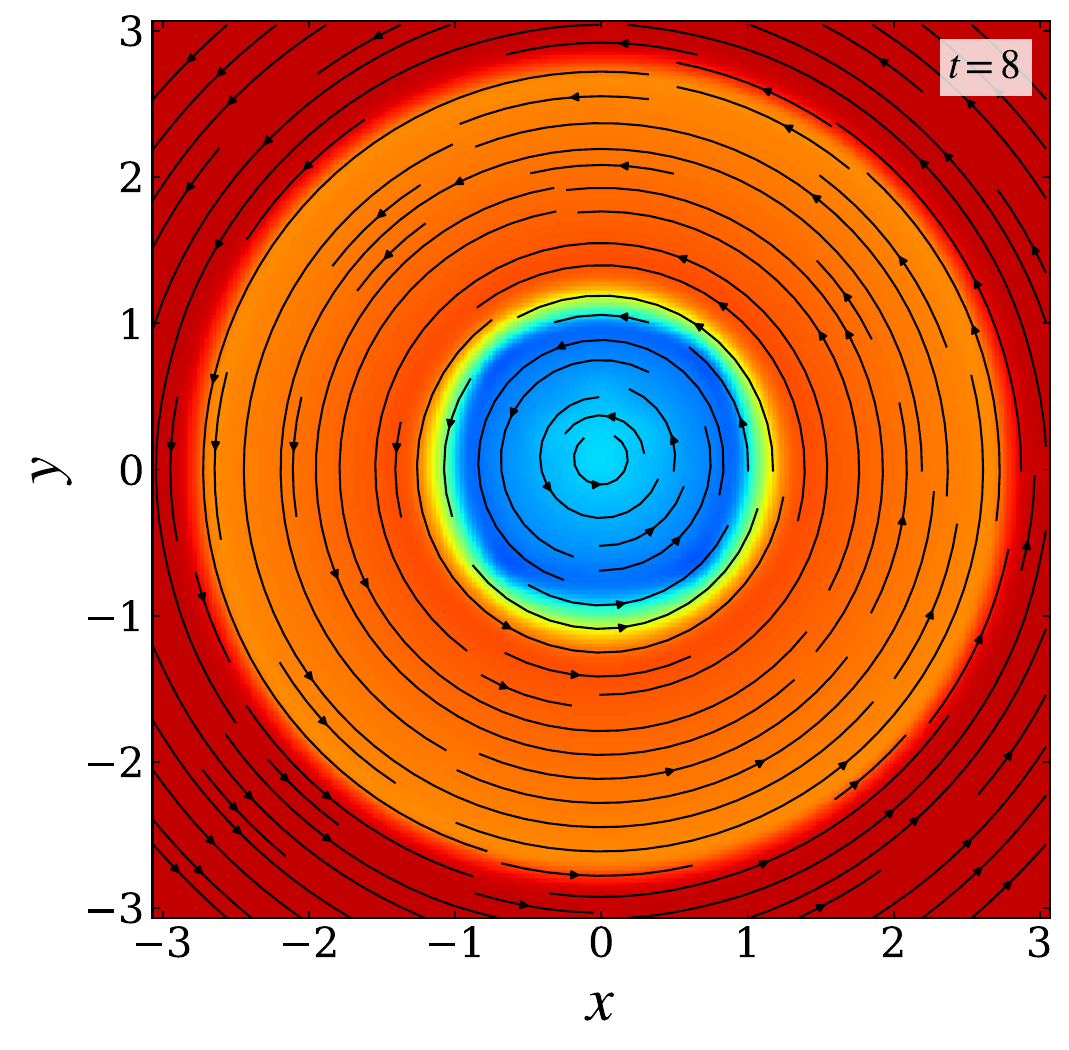}
    \includegraphics[width=0.195\linewidth]{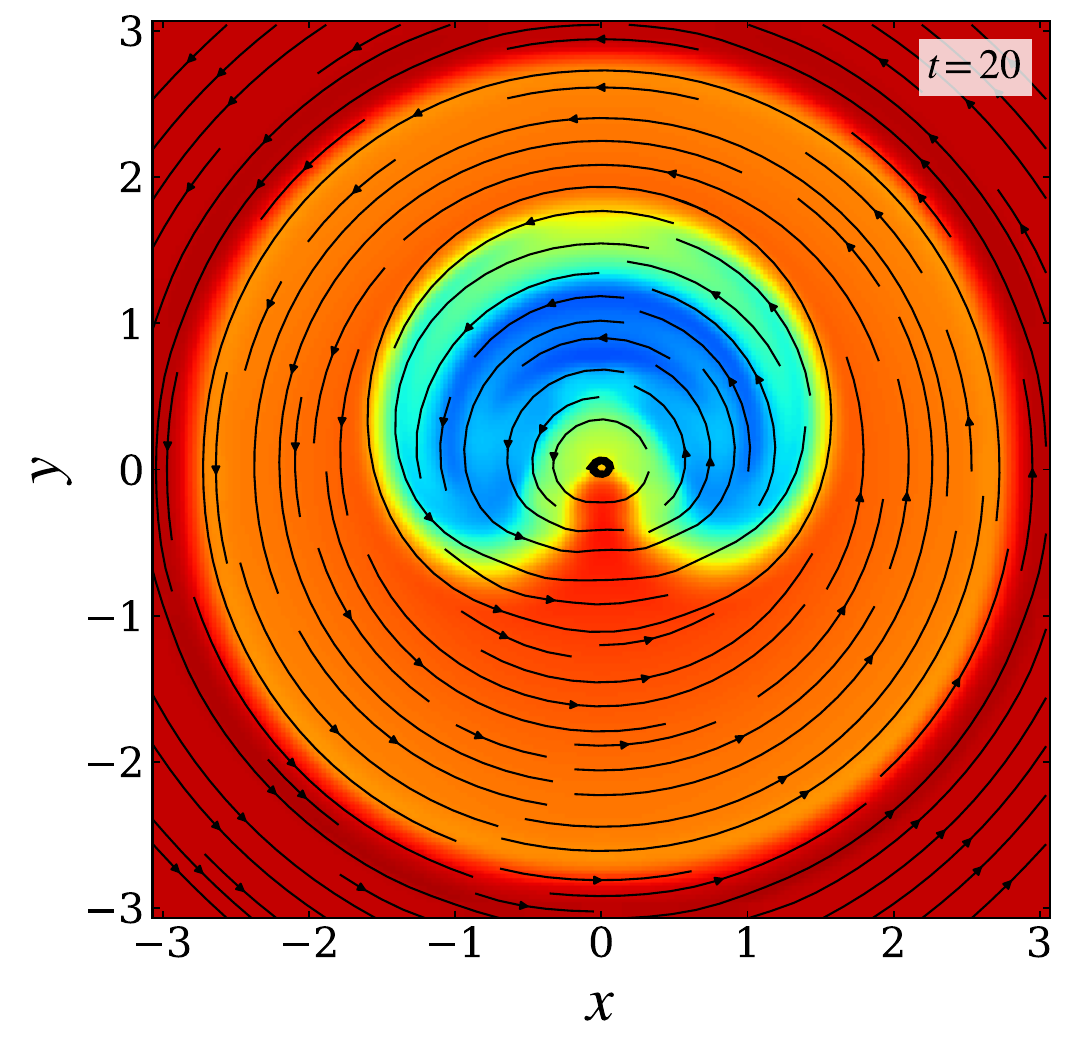}
    \includegraphics[width=0.195\linewidth]{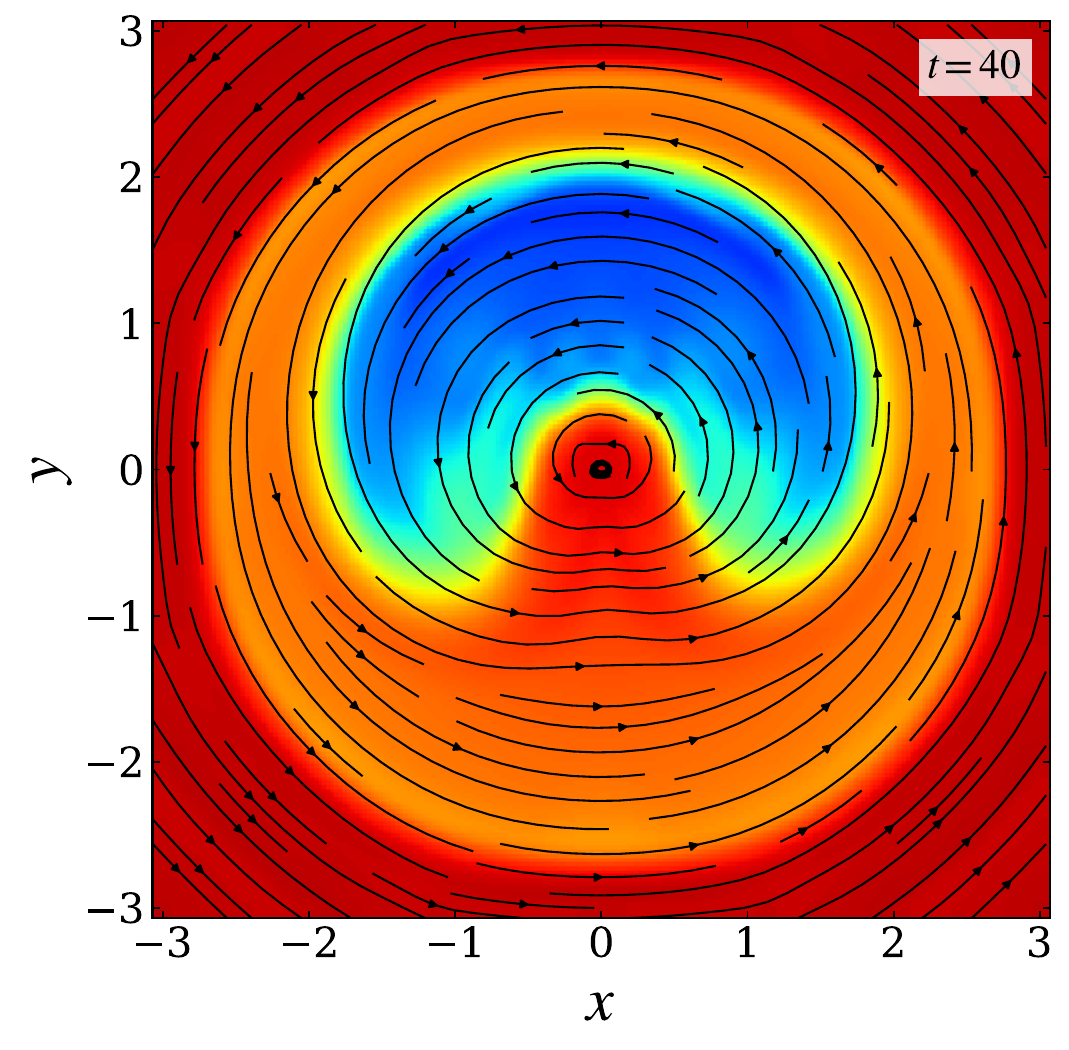}
    \includegraphics[width=0.195\linewidth]{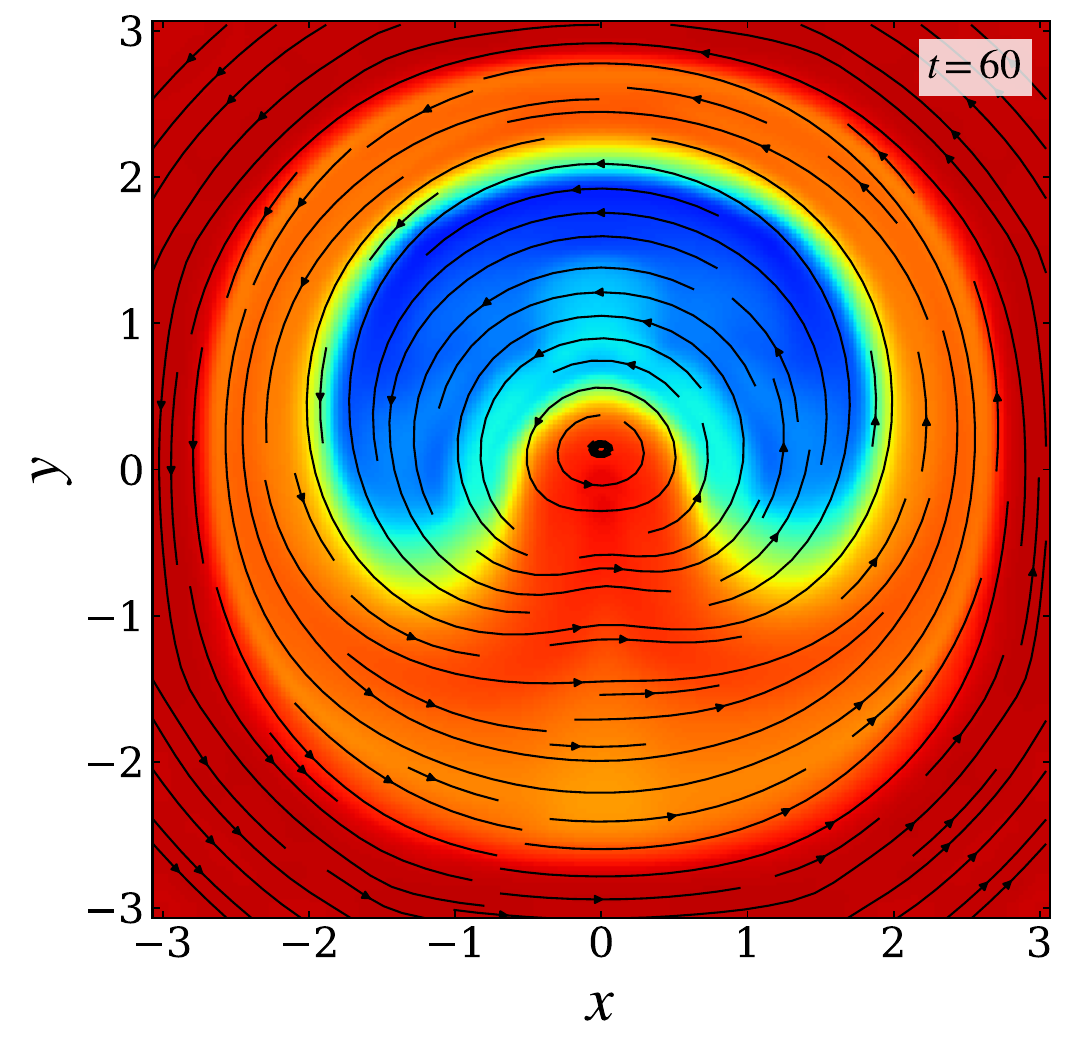}
    \includegraphics[width=0.195\linewidth]{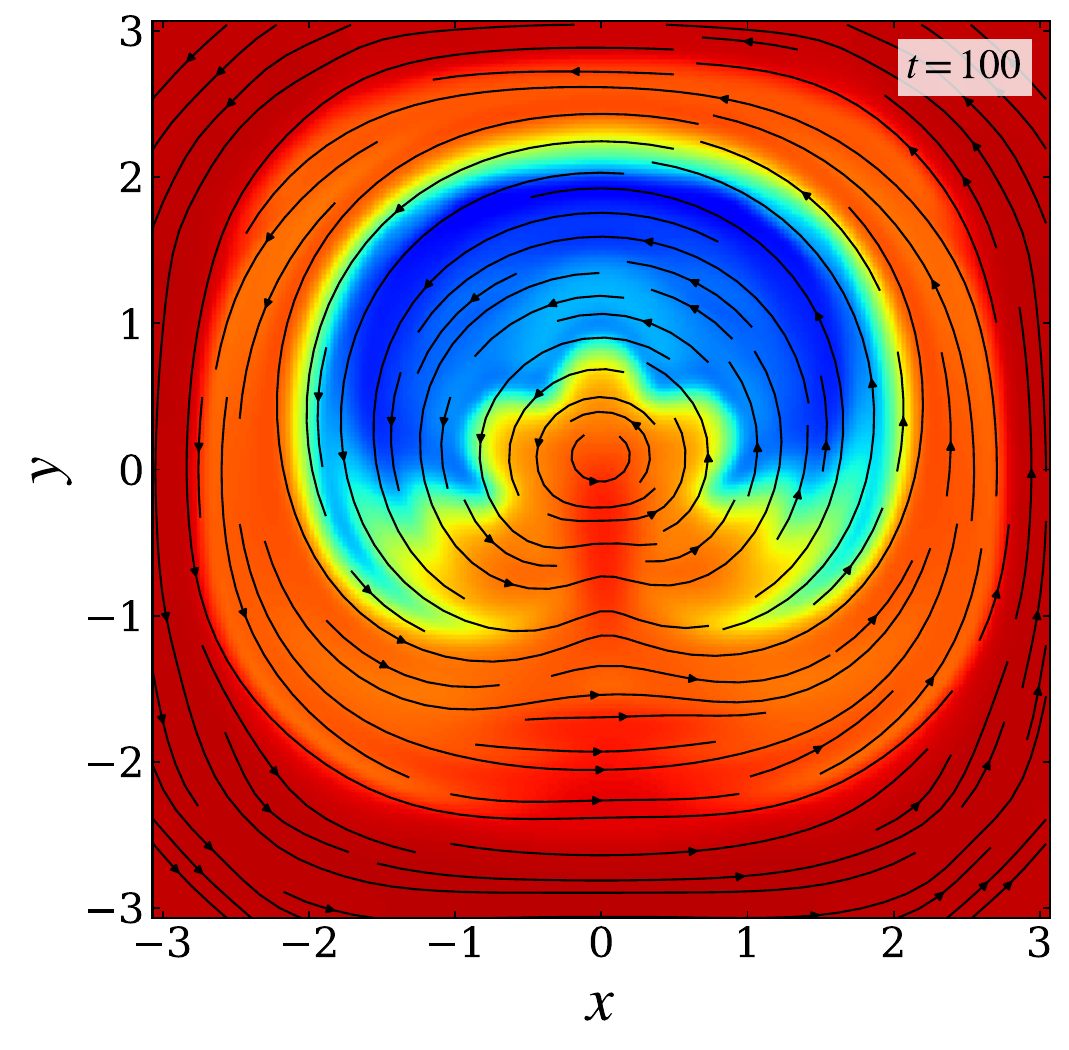}
    \includegraphics[width=0.195\linewidth]{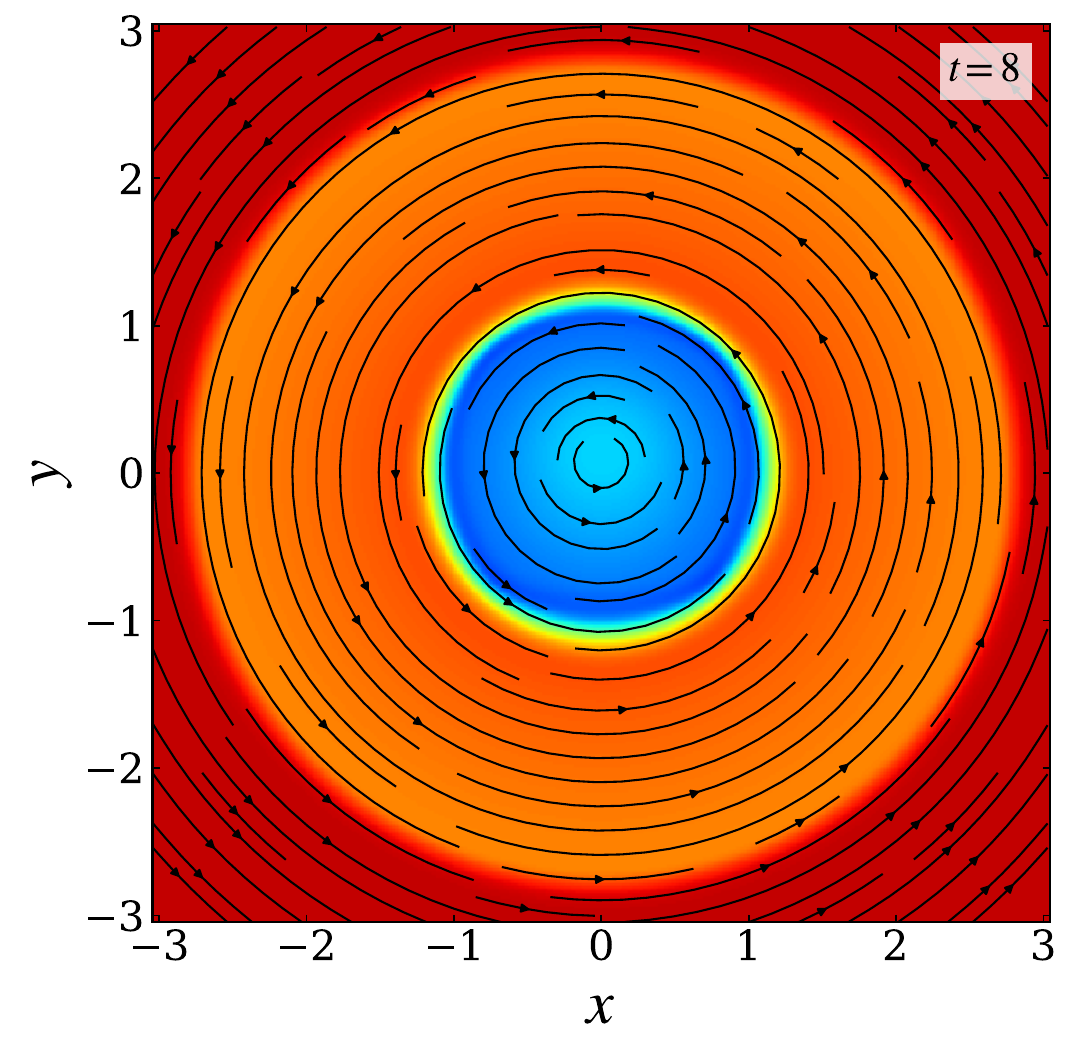}
    \includegraphics[width=0.195\linewidth]{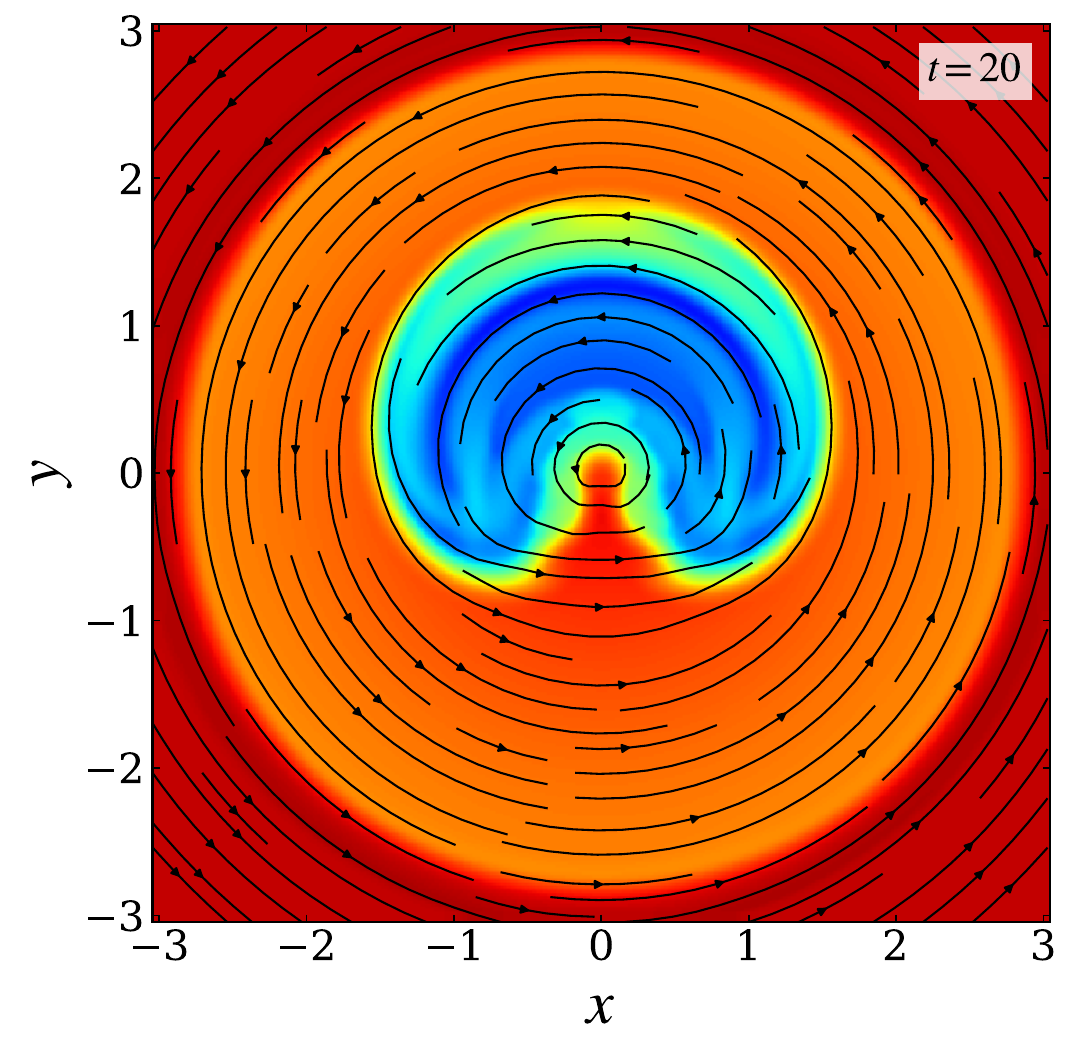}
     \includegraphics[width=0.195\linewidth]{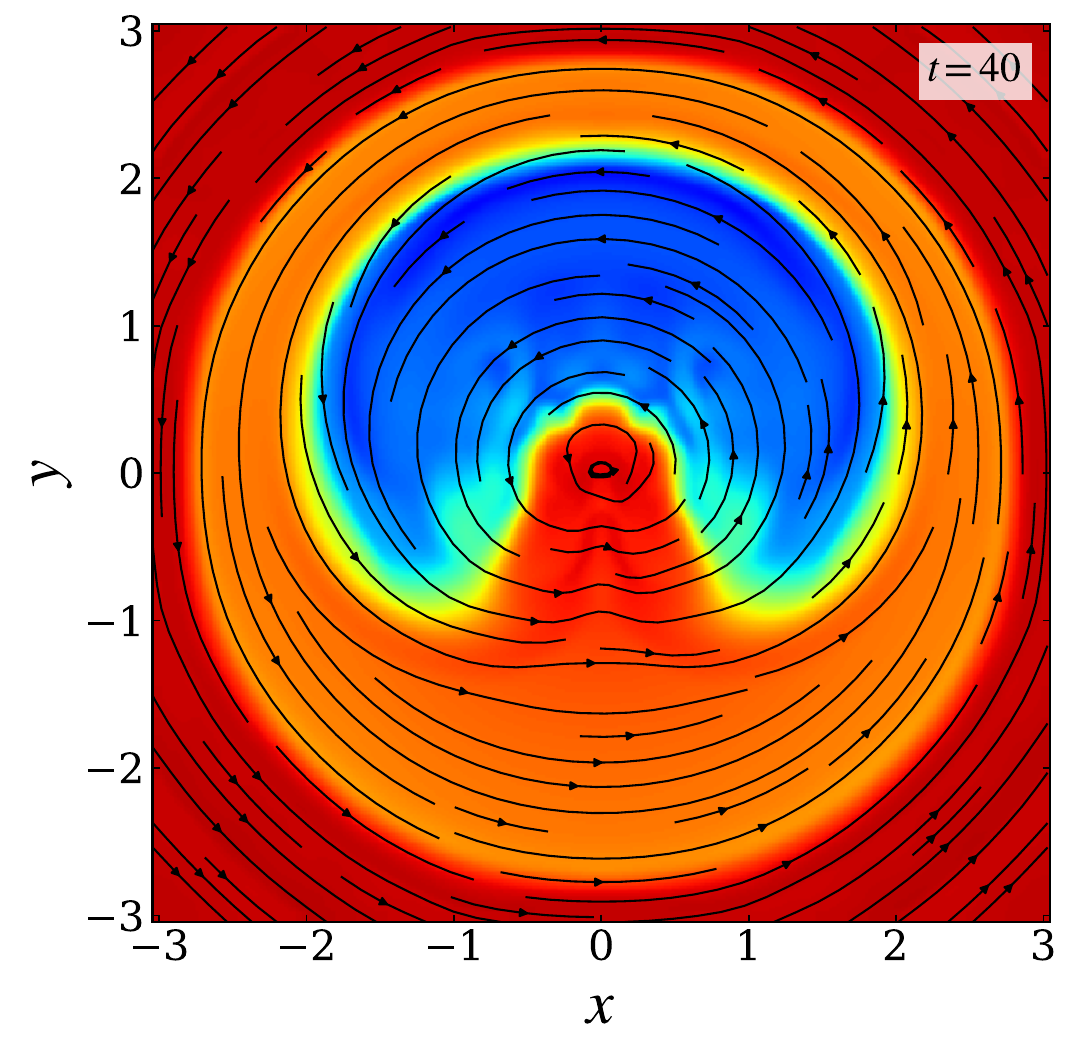}
    \includegraphics[width=0.195\linewidth]{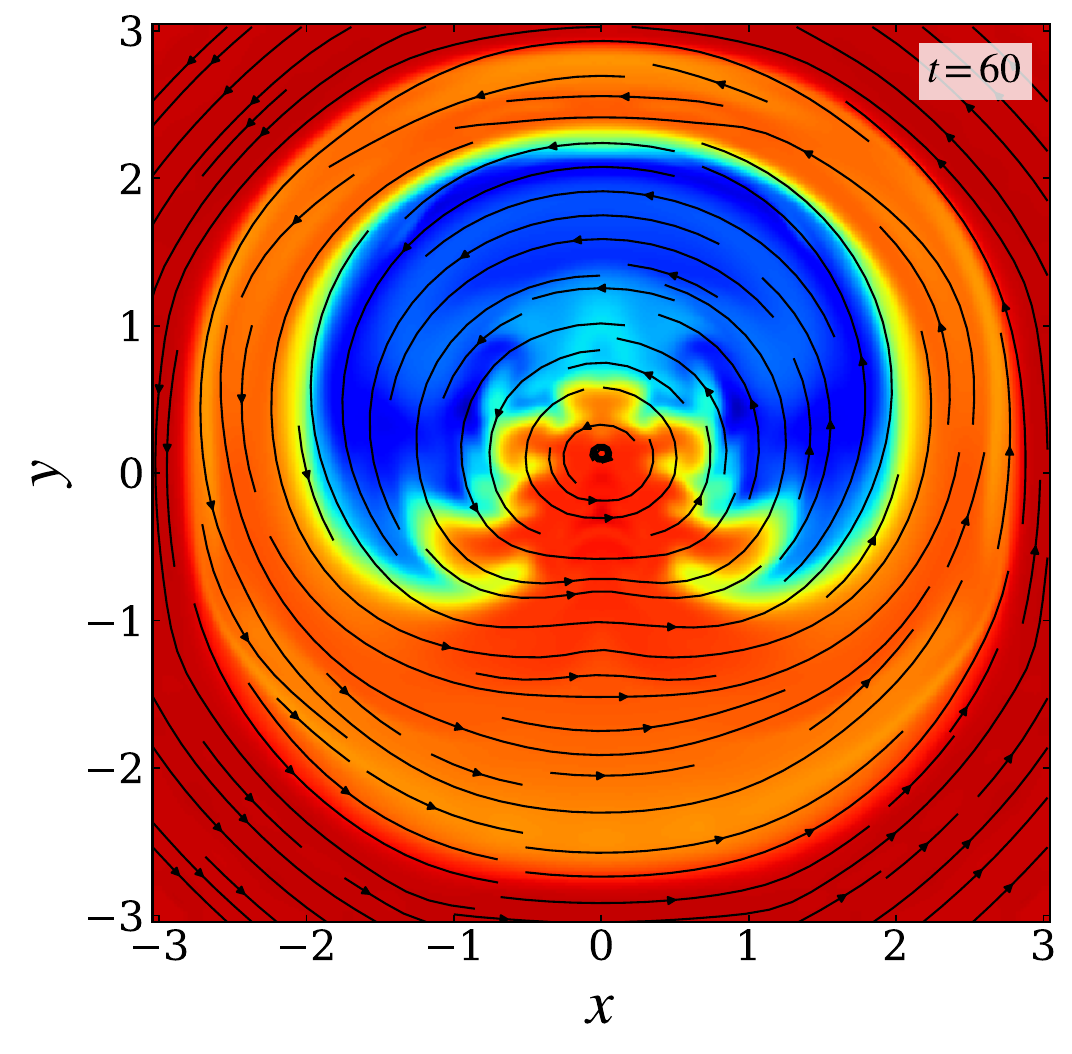}
    \includegraphics[width=0.195\linewidth]{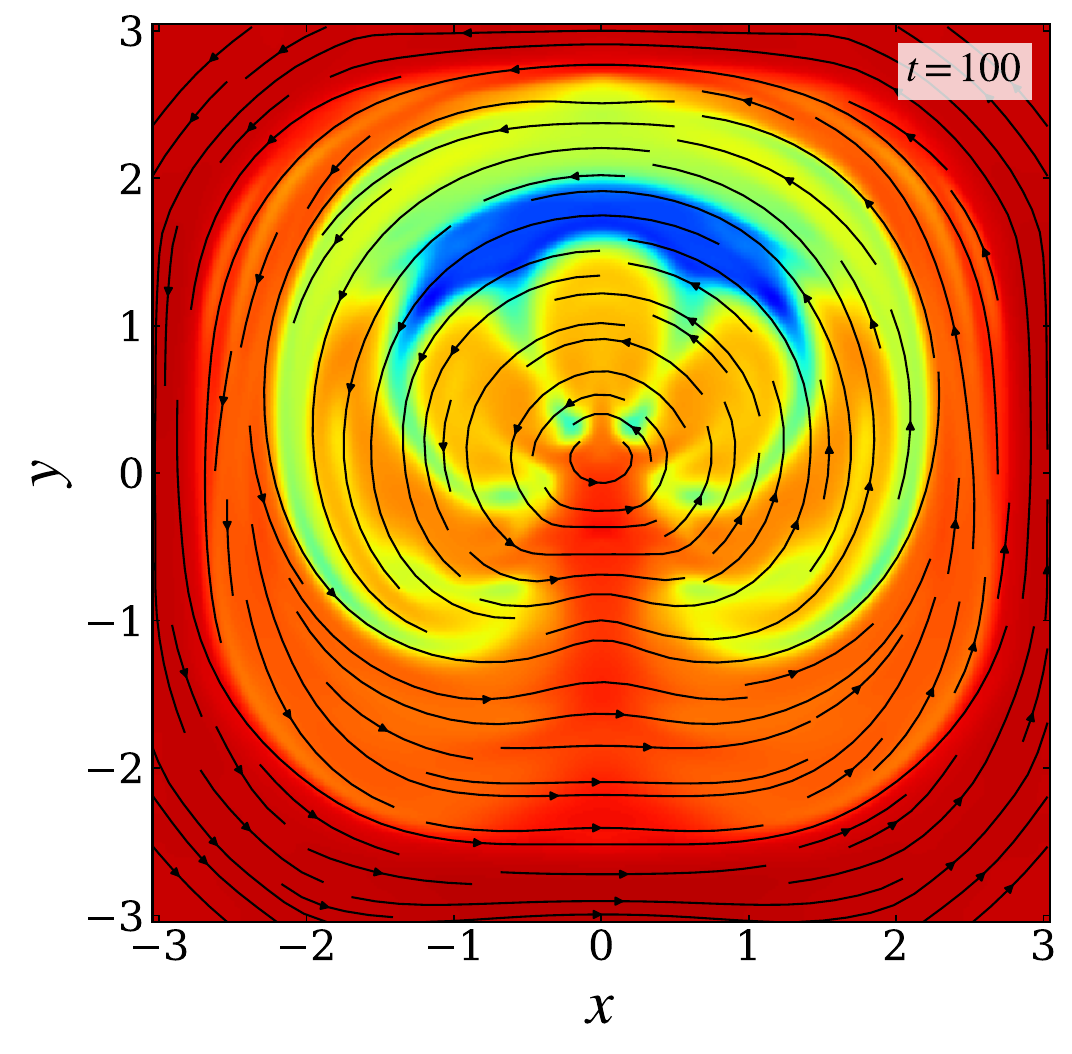}
    \includegraphics[width=0.925\linewidth]{Figures/CDKI/colorbar_rho_2D.pdf}

 \caption{Evolution of the rest-mass density ($\rho$) distribution in the $z=1.5$ mid-plane for the test models $T_1$-$T_5$ (top to bottom). Arrows show the direction of magnetic field lines in this plane.}
    \label{fig:CTests-rho}
\end{figure*}

 \begin{figure*}
     \centering
    \includegraphics[width=0.35\linewidth]{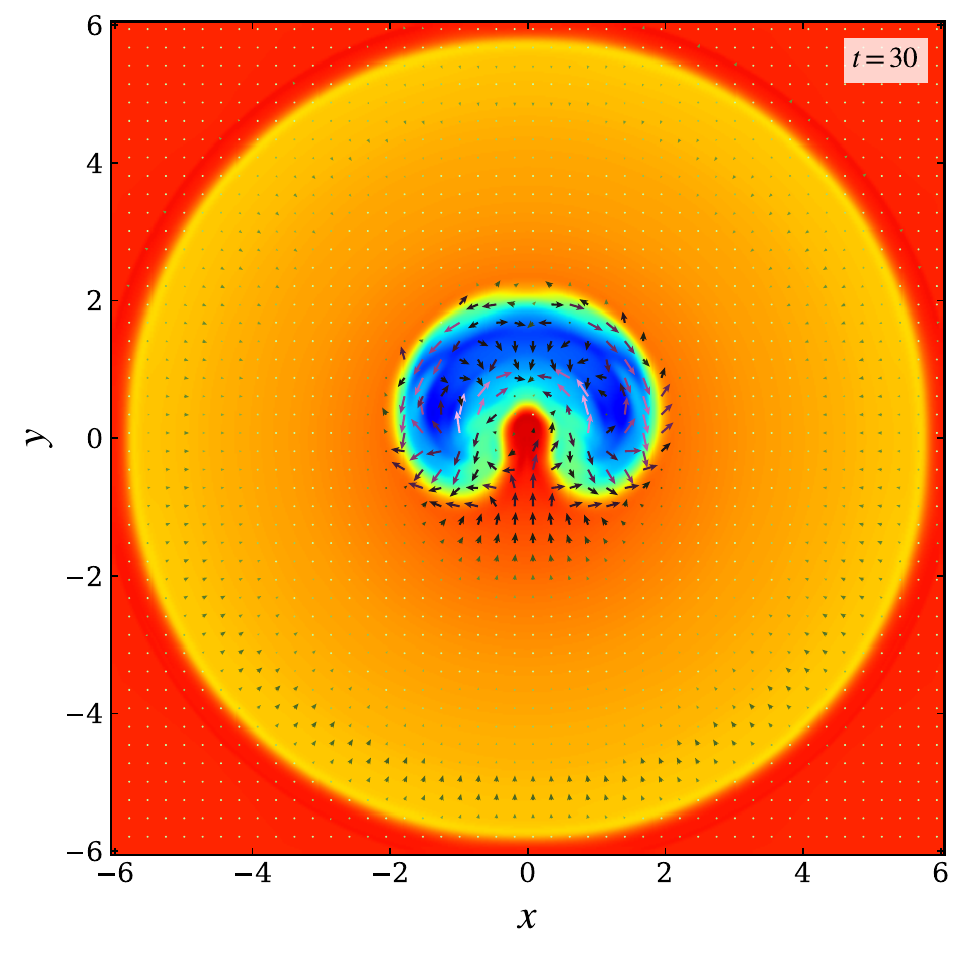}
    \includegraphics[width=0.35\linewidth]{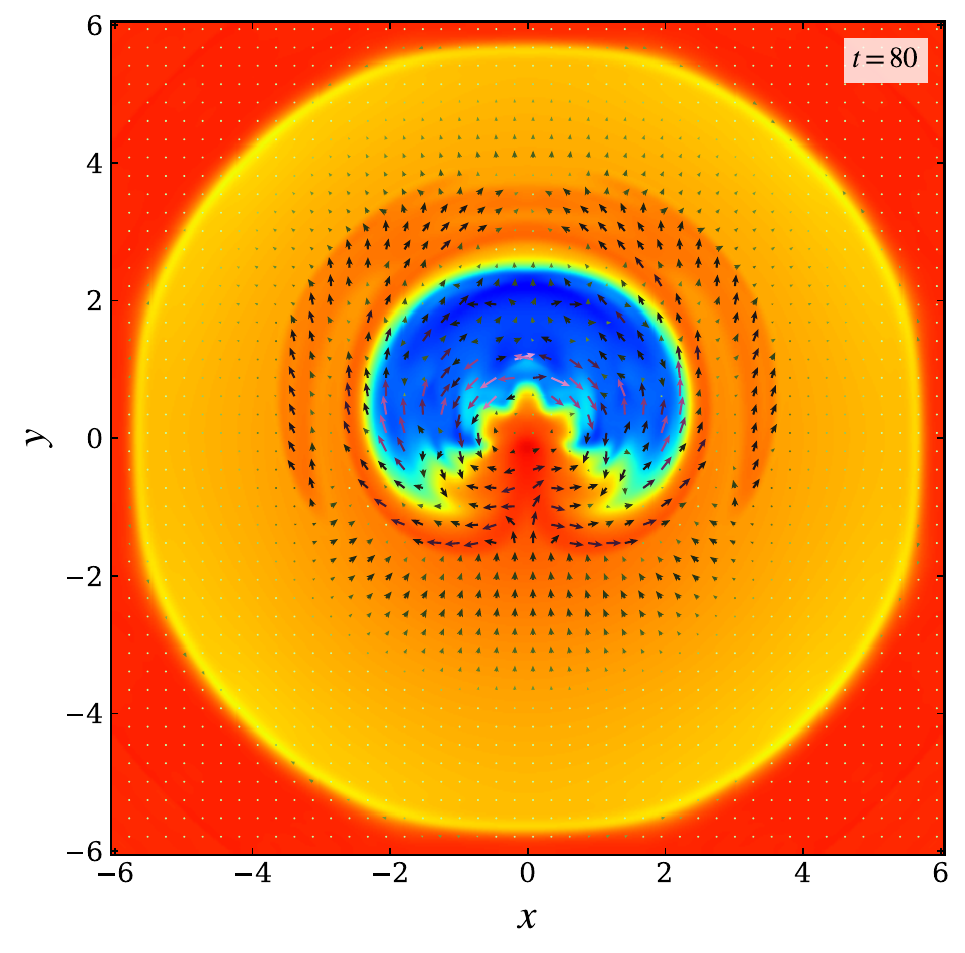}
    \includegraphics[width=0.7\linewidth]{Figures/CDKI/colorbar_v_2D.pdf}
     \includegraphics[width=0.69\linewidth]{Figures/CDKI/colorbar_rho_v_2D.pdf}\caption{The density distribution and velocity field in the $M_3$ model with extended computational domain. The image shows the solution in the mid-plane $z=1.5$.}
\label{fig:M3-ext-bound}
\end{figure*}

\subsection{Convergency test}
\label{sec:converge}

To study the effect of numerical resolution on the linear growth rate and nonlinear development of the CD kink instability, we selected the initial equilibrium solution $M_3$ and ran a series of simulations with progressively increasing numerical resolution.   The test runs are named as  $T_k$, $k=1\dots 5$,  with the number of grid cells per jet radius  $n_k=9,15,24,36,57$ respectively (For the $M_3$ run, it is 90 cells per jet radius). Since PERT is third-order accurate, the truncation error reduces as $n_k^{-3}$. Thus, it is $10^{-3}$ smaller in the $T_6$ ($M_3$) run compared to the $T_1$ run.

Figure \ref{fig:CTests-params} shows the temporal evolution of the perturbation-amplitude parameters $E_t$, $V_{r, max}$ and $\langle \hat{B}_r\rangle$ evaluated over the entire jet-wind region. In the lower resolution models, the initial equilibrium is not sustainable. This explains the rapid increase of $E_t$ at the very beginning of the simulations in the models $T_1$ and $T_2$. Starting from $T_3$, the linear growth rates of the instability are quite close to those obtained in the $M_3$ model.  
At the nonlinear phase, the amplitudes generally reduce with the resolution, although this is no longer true for the $E_t$ parameter starting from $T_3$.  The nonlinear phase is reached at more or less the same time in all models starting from $T_2$.  

Figure \ref{fig:CTests-rho} shows the evolution of the rest-mass density distribution in the midplane $z=1.5$ for the test models.  Qualitatively, the evolution is similar in all models, exhibiting the kidney-bean shape of the jet cross-section and its shift in the positive y direction.  However, the density contrast between the jet and wind is strongly reduced in the lowest resolution models $T_1$ and $T_2$, as well as the jet size. The secondary instability is suppressed in the $T_1$-$T_3$ models, although some hints of the instability can be seen in the t=100 image of $T_2$.   Interestingly, in terms of the density distribution at $t=100$, the $T_5$ (and the main $M_3$) model is closer to $T_1$ than $T_3$ and $T_4$. However, in $T_5$ the disappearance of the low density jet region is caused by the interpenetration of jet and wind plasma driven by the secondary instability, whereas in the $T_1$ models it is caused by the numerical diffusion.

\subsection{Effect of the domain size}
\label{sec:boundary}

As one can in the figures \ref{fig:isosurf} and \ref{fig:rho-slices}, the jet coil produced via the current-driven  kink instability comes rather close to the x and y-boundaries of the computational domain, rising questions about the potential influence of the boundary conditions on the nonlinear development of the instability.  In particular, the images suggest that the tightness of the domain may have been responsible for the lack of further expansion of the coil, starting from about $t=40$. 
To explore this issue, we ran test simulations for a model with the same parameters as the M3 model, but twice the size of the computational domain in x and y directions. To reduce the computational cost, the resolution was reduced to 45 cells per jet radius. This is half the resolution of the M3 run proper, and is between the resolutions of the T4 and T5 models.  


At the nonlinear phase, the solution exhibits the same features as the main $M_3$ model, and in most respects is not much different from it. Figure \ref{fig:M3-ext-bound} shows the density distribution and the velocity field at $t=30,80$. It can be compared against the $M_3$ solution at the same times shown in figure \ref{fig:M3-v}.  One can see that the overall structure is the same in both figures and that the displacement of the jet cross-section along the y-axis is almost the same. One new feature, not seen in the $M_3$ simulations, is the development of the envelope of compressed wind plasma around the jet.  The entropy jump at the outer edge of the envelope indicates that the edge is an outgoing shock wave.      

Based on this test, we conclude the boundaries of the computational domain in the main suite of the simulations for the mildly magnetised wind do have an effect on the solutions, but only a relatively small one.


\bsp	
\label{lastpage}
\end{document}